\documentclass[bibliography=totoc, a4paper, english, 12pt, twoside]{scrbook}

\usepackage[english]{babel}
\usepackage[a4paper]{geometry}
\usepackage[utf8]{inputenc}

\usepackage[pdftex,xdvi]{graphicx}
\usepackage{float}
\usepackage{epstopdf}
\graphicspath{{../figures/}{./}}

\usepackage[usenames,dvipsnames]{xcolor}

\usepackage[numbers,sort&compress]{natbib}

\usepackage{amsmath,amssymb,dsfont}
\usepackage{upgreek}
\usepackage{subdepth}
\usepackage{color}
\usepackage[plain]{fancyref}
\usepackage[makeroom]{cancel}
\usepackage{braket}
\usepackage{soul}
\usepackage{tabto}
\usepackage[colorinlistoftodos,prependcaption]{todonotes}
\usepackage{caption}
\usepackage{subcaption}
\usepackage{rotating}
\usepackage[boxsize=1em,aligntableaux=center]{ytableau}
\usepackage{booktabs}

\DeclareMathOperator{\diag}{diag}
\DeclareMathOperator{\adj}{ad}
\DeclareMathOperator{\rank}{rank}
\DeclareMathOperator{\ord}{ord}

\usepackage{colortbl}
\definecolor{fillcolor}{RGB}{196,196,196}

\usepackage{amssymb}
\usepackage{pifont}

\usepackage{geometry}
  \usepackage[
  			 hidelinks,
              pdftitle={Advancements in double \& exceptional field theory on group manifolds}, 
              pdfsubject={Promotion an der Fakultät für Physik der Ludwig-Maximilians-Universität München},
              pdfauthor={Pascal du Bosque}
              ]{hyperref}
\usepackage[toc,page]{appendix}

  \hypersetup{colorlinks=true}
  \hypersetup{allcolors=Black}

\allowdisplaybreaks
\addtokomafont{caption}{\footnotesize\sffamily}
\addtokomafont{captionlabel}{\footnotesize\sffamily\bfseries}

\usepackage{tikz}
\usetikzlibrary{arrows,chains,matrix,positioning,scopes,shapes}

\makeatletter
\tikzset{join/.code=\tikzset{after node path={\ifx\tikzchainprevious\pgfutil@empty\else(\tikzchainprevious)edge[every join]#1(\tikzchaincurrent)\fi}}}
\makeatother

\tikzset{>=stealth',every on chain/.append style={join}, every join/.style={->}}
\tikzstyle{labeled}=[execute at begin node=$\scriptstyle, execute at end node=$]
\tikzstyle{decision} = [diamond, draw, fill=lightgray, 
    text width=4.5em, text badly centered, node distance=3cm, inner sep=0pt]
\tikzstyle{block} = [rectangle, draw, fill=lightgray, 
    text width=5em, text centered, rounded corners, minimum height=4em]
\tikzstyle{line} = [draw, -latex']
\tikzstyle{cloud} = [draw, ellipse,fill=red!20, node distance=3cm,
    minimum height=2em]

\DeclareOldFontCommand{\bf}{\normalfont\bfseries}{\mathbf}
\DeclareOldFontCommand{\it}{\normalfont\itshape}{\mathit}
\DeclareOldFontCommand{\tt}{\normalfont\ttfamily}{\mathtt}

\newcommand{\bref}[1]{(\ref{#1})}
\newcommand{\DFTwzw}{DFT$_{\mathrm{WZW}}$}

\clearpage{}\newcommand{\LMUTitle}[9]{
  \thispagestyle{empty}
  \vspace*{\stretch{1}}
  {\parindent0cm
   \rule{\linewidth}{.25ex}
   }
  \begin{center}
    \vspace*{\stretch{1}}
    \sffamily\bfseries\huge
    #1 \\
    \vspace*{\stretch{1}}
    \sffamily\bfseries\large
    #2
    \vspace*{\stretch{1}}
  \end{center}
  \rule{\linewidth}{.25ex}
  \vspace*{\stretch{5}}
  \begin{center}
    \includegraphics[width=2in]{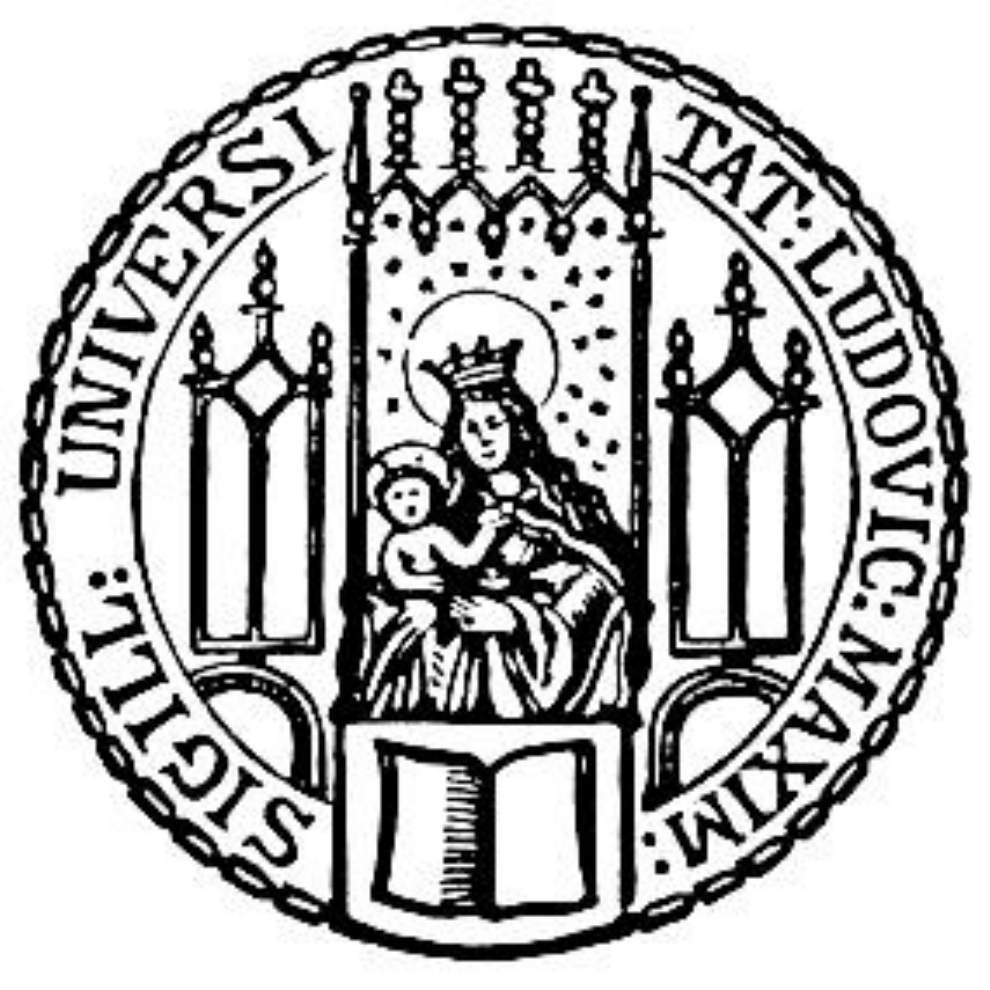}
  \end{center}
  \vspace*{\stretch{1}}
  \begin{center}\sffamily\large{#5}\end{center}
  \newpage
  \thispagestyle{empty}

  \cleardoublepage
  \thispagestyle{empty}

  \vspace*{\stretch{1}}
  {\parindent0cm
  \rule{\linewidth}{.25ex}
  }
  \begin{center}
    \vspace*{\stretch{1}}
    \sffamily\bfseries\huge
    #1 \\
    \vspace*{\stretch{1}}
    \sffamily\bfseries\large
    #2
    \vspace*{\stretch{1}}
  \end{center}
  \rule{\linewidth}{.25ex}

  \vspace*{\stretch{3}}
  \begin{center}
    \sffamily\large #4\\
    \sffamily\large an der Fakultät für Physik\\
    \sffamily\large der Ludwig-Maximilians-Universit\"at\\
    \sffamily\large M\"unchen\\
    \vspace*{\stretch{1}}
    \sffamily\large vorgelegt von\\
    \sffamily\large #2\\
    \sffamily\large aus #3\\
    \vspace*{\stretch{2}}
    \sffamily\large München, den #6
  \end{center}

  \newpage
  \thispagestyle{empty}

  \vspace*{\stretch{1}}

  \begin{flushleft}
    \sffamily\large Erstgutachter:\tab	  #7 \\[1mm]
    \sffamily\large Zweitgutachter:\tab	 #8 \\[1mm]
    \sffamily\large Tag der mündlichen Prüfung:\tab #9\\
  \end{flushleft}

  \cleardoublepage
}
\clearpage{}

\begin{document}
  \frontmatter
  \LMUTitle
      {\bf Advancements in  \\
      Double \& Exceptional Field Theory   \\
      on Group Manifolds }            
      {Pascal du Bosque}
      {Berlin}           
      {Dissertation}           
      {München, 2017}       
      {24. Juli 2017}          
      {Prof. Dr. Dieter Lüst}           
      {Priv.-Doz. Dr. Ralph Blumenhagen}           
      {18. September 2017}        

  \hypersetup{allcolors=NavyBlue}    
 
  \setlength{\parindent}{1.5em}   
  \clearpage{}\chapter*{Zusammenfassung}

Diese Dissertation beschäftigt sich mit neuen Hintergründen und Konzepten in ‘Double Field Theory’ (DFT) \cite{Hull:2009mi}, einer T-Dualität invarianten Reformulierung der Supergravitation (SUGRA). Es ist eine effektive Theorie, die die Dynamiken eines geschlossenen Strings auf einem Torus beschreibt. Für ein konsistentes Framework benötigt die Theorie das Hinzufügen von $D$ Windungskoordinaten zu den $D$ physischen Koordinaten und führt damit zu einem gedoppelten Raum. Eine wichtige Konsistenzbedingung für die Theorie ist die sogenannte ‘strong constraint’. Nach dem Fordern dieser ‘strong constraint’ reduziert sich die Abhängigkeit aller Felder auf die Hälfte der Koordinaten. Wir fangen damit an, die grundlegenden Konzepte und Ideen von DFT zu wiederholen. In diesem Zusammenhang betrachten wir die generalisierten Diffeomorphismen, welche die lokalen Diffeomorphismen und Eichtransformationen implementieren, sowie deren assoziierte Eichalgebra gegeben durch die C-Klammer. Hierbei untersuchen wir die Rolle der ‘strong constraint’ für das Schließen der Eichalgebra. Weiterhin analysieren wir die Wirkung, sowohl in der Generalisierten Metrik Formulierung als auch in der Flussformulierung, und die zugrundeliegenden Symmetrien.

Anschließend widmen wir uns der ‘Double Field Theory on group manifolds’ (DFT$_{\mathrm{WZW}}$) \cite{Blumenhagen:2014gva,Blumenhagen:2015zma,Bosque:2015jda,Blumenhagen:2017noc}, einer Verallgemeinerung von DFT, dessen ‘Worldsheet’-Darstellung durch ein Wess-Zumino-Witten Modell beschrieben wird. Um die Wirkung und die dazugehörigen Eichtransformationen zu erhalten, führt man Rechnungen mithilfe geschlossener String Feldtheorie (CSFT) auf ‘tree level’ bis zu kubischer Ordnung in den Feldern sowie führender Ordnung in $\alpha’$ durch. Hier setzen wir uns wieder mit den generalisierten Diffeomorphismen und deren Eichalgebra auseinander, welche nun mittels einer modifizierten ‘strong constraint’ schließen. In diesem Setup wird es offensichtlich, dass sich die originale DFT und DFT$_{\mathrm{WZW}}$ auf einem sehr fundamentalen Level unterscheiden. Allerdings sind sie miteinander verbunden. All diese Schritte erlauben es uns DFT$_{\mathrm{WZW}}$ durch gedoppelte, generalisierte Objekte mittels Extrapolation zu allen Ordnungen in den Feldern zu ersetzen. Dies führt zu einer Generalisierten Metrik Formulierung \cite{Blumenhagen:2015zma} und einer Flussformulierung \cite{Bosque:2015jda} der Theorie. Jedoch im Gegensatz zu originaler DFT spalten sich die Flüsse in einen Hintergrundanteil als auch einen Fluktuationsanteil auf, während das generalisierte Hintergrundvielbein die Rolle des Twist in dem generalisierten Scherk-Schwarz Ansatz übernimmt. In dieser Arbeit studieren wir die zugrundeliegenden Symmetrien und Feldgleichen beider Formulierungen. Ein entscheidender Unterschied zwischen DFT$_{\mathrm{WZW}}$ und originaler DFT liegt in dem Auftreten einer $2D$-Diffeomorphismen Invarianz unter der standard Lie-Ableitung. Außerdem tritt eine zusätzliche Nebenbedingung in Erscheinung, die ‘extended strong constraint’, welche falls gefordert DFT$_{\mathrm{WZW}}$ zu originaler DFT reduziert und beide Theorien werden äquivalent, wobei die $2D$-Diffeomorphismen Invarianz zusammenbricht. Folgt man weiteren Schritten, kann man mithilfe eines generalisierten Scherk-Schwarz Kompaktifizierungsansatz den bosonischen Subsektor von halb-maximaler, elektrisch geeichter Supergravitation reproduzieren. Ferner lösen wir das lang stehende Problem zur Konstruktion eines Twists bei vorgegebener Einbettungstensorlösung, indem wir eine Maurer-Cartan Form benutzen um das Hintergrundvielbein aufzubauen.

Zu guter Letzt verallgemeinern wir unsere Ideen und Konzepte von DFT$_{\mathrm{WZW}}$ zu geometrischen ‘Exceptional Field Theories’ (gEFT) \cite{Bosque:2016fpi,Bosque:2017dfc}. Im Anschluss präsentieren wir eine Prozedur, welche die Konstruktion von generalisierten, parallelisierbaren Räumen in $\dim M = 4$ SL($5$) ‘Exceptional Field Theory’ (EFT) erlaubt. Diese Räume lassen eine vereinheitlichte Behandlung von konsistenten, maximal supersymmetrischen Trunkierungen von zehn sowie elf dimensionaler Supergravitation zu, und ihre Konstruktion ist schon immer eine offene Frage gewesen. Hinzu gestatten sie ein generalisiertes ‘Frame’-Feld über einer Nebenklasse $M = G/H$, dass die Lie-Algebra $\mathfrak{g}$ von $G$ unter der generalisierten Lie-Ableitung reproduziert. Hierfür identifizieren wir die Gruppenmannigfaltigkeit $G$ mit dem erweiterten Raum der EFT. Im nächsten Schritt muss die ‘section condition’ (SC) gelöst werden, um unerwünschte, unphysische Richtungen von diesem erweiterten Raum zu entfernen. Schlussendlich konstruieren wir ein generalisiertes ‘Frame’-Feld mithilfe einer links-invarianten Maurer-Cartan Form auf $G$. All diese Schritte führen zu zusätzlichen Bedingungen auf die Gruppen $G$ und $H$.\clearpage{}     
  \clearpage{}\chapter*{Abstract} \label{Kap_0}

This thesis deals with new backgrounds and concepts in Double Field Theory (DFT)~\citep{Hull:2009mi}, a T-Duality invariant reformulation of supergravity (SUGRA). It is an effective theory capturing the dynamics of a closed string on a torus. For a consistent framework, the theory requires to add $D$ winding coordinates to the $D$ physical spacetime coordinates and gives rise to a doubled space. An important constraint for the consistency of the theory is the strong constraint. After imposing this constraint, all fields are only allowed to depend on half the coordinates.  We begin by reviewing the basic concepts and notions of DFT. With regard to this context, we consider generalized diffeomorphisms, implementing the local diffeomorphisms and gauge transformations from SUGRA, and their associated gauge algebra which is governed by the C-bracket. In this setting, we examine the importance of the strong constraint for the closure of the gauge algebra. Subsequently, we investigate the action, in both the generalized metric formulation and the flux formulation, and its underlying symmetries.

Afterwards, we turn to Double Field Theory on group manifolds (DFT$_{\mathrm{WZW}}$)~\citep{Blumenhagen:2014gva, Blumenhagen:2015zma,Bosque:2015jda,Blumenhagen:2017noc}, a generalization of DFT, whose worldsheet description is governed by a Wess-Zumino-Witten model on a group manifold. In order to obtain an action and the gauge transformations, Closed String Field Theory (CSFT) computations at tree level up to cubic order in fields and leading order in $\alpha'$ have to be performed. Again, we consider generalized diffeomorphisms and their gauge algebra, which closes under a modified strong constraint. From this setup, it is going to become clear that original DFT and DFT$_{\mathrm{WZW}}$ differ on a very fundamental level. However, they are related to each other.
All these steps allow us to recast DFT$_{\mathrm{WZW}}$ in terms of doubled generalized objects by extrapolating it to all orders in fields. It yields a generalized metric formulation~\citep{Blumenhagen:2015zma} and a flux formulation~\cite{Bosque:2015jda} of the theory. Although, in contrast to original DFT the fluxes split into a background and a fluctuation part, while the background generalized vielbein takes on the role of the twist in the usual generalized Scherk-Schwarz ansatz. In this thesis, we are going to study the underlying symmetries and field equations for both formulations. A striking difference between between DFT$_{\mathrm{WZW}}$ and original DFT lies in the appearance of an additional $2D$-diffeomorphism invariance under the standard Lie derivative. On top of this, we observe the emergence of an additional extended strong constraint, which when imposed, reduces DFT$_{\mathrm{WZW}}$ to original DFT and both theories become equivalent while the $2D$-diffeomorphism invariance breaks down. Following these steps, one can perform a generalized Scherk-Schwarz compactification ansatz to recover the bosonic subsector of half-maximal, electrically gauged supergravities. Moreover, we are going to solve the long standing problem of constructing a twist for each embedding tensor solution by using Maurer-Cartan forms to derive an appropriate background vielbein.

Last but not least, we generalize the ideas and notions from DFT$_\mathrm{WZW}$ to geometric Exceptional Field Theory (gEFT)~\cite{Bosque:2016fpi,Bosque:2017dfc}. Subsequently, we show a procedure which allows for the construction of generalized parallelizable spaces in dim $M = 4$ SL($5$) Exceptional Field Theory (EFT). These spaces permit a unified treatment of consistent maximally supersymmetric truncations of ten- and eleven-dimensional supergravity in Generalized Geometry (GG) and their construction has always been an open question. Furthermore, they admit a generalized frame field over the coset $M= G/H$ reproducing the Lie algebra $\mathfrak{g}$ of $G$ under the generalized Lie derivative. Therefore, we identify the group manifold $G$ with the extended space of the EFT. In the next step, the section condition (SC) needs to be solved to remove unwanted, unphysical directions from this extended space. Finally, we construct the generalized frame field using a left invariant Maurer Cartan form on $G$. All of these steps cast additional constraints on the groups $G$ and $H$.\clearpage{}    
  \clearpage{}\chapter*{Acknowledgements}

First of all, I want to thank my supervisor Prof. Dr. Dieter Lüst for the constant support and the many fruitful discussions during the course of my time as a PhD student. Furthermore, I am grateful for the opportunity he gave me by working in his group and by letting me participate in this very interesting project. I consistently enjoyed his vivid ideas, physical intuition and tremendous knowledge of the literature.

However, my foremost gratitude goes to Falk Hassler with whom I spent hours and hours of constructive discussions and who supported me throughout this entire time. Especially, I want to thank him for his dedication to share his knowledge and the opportunity to ask several questions at every point. In particular, his assistance in the constant fight against extremely technical computations.

Additionally, I would like to thank Ralph Blumenhagen for several useful discussions and his ability to spot hitches immediately. On top of that, I also thank (in alphabetical order) Enrico Brehm, Ilka Brunner, Daniel Jaud, Michael Fuchs, Sebastian Greiner, Michael Haack, Daniela Herschmann, Abhiram Kidambi, Cornelius Schmidt-Colinet, Marc Syvari, and Florian Wolf as well as all other members of the Ludwig Maximilian University and Max Planck Institute for Physics.

Moreover, I am deeply indebted to my close friend Arnau Pons Domenech for always answering my \LaTeX\;related questions and his continuing support. In this context, I show my gratitude to Thomas Hertle for proofreading parts of my thesis.

Furthermore, I particularly want to thank my parents for making this all possible and their ongoing support during all my life. \par
Last but not least, my special thanks goes to Kiki and Jürgen for printing this thesis.\clearpage{}     
  
  \hypersetup{allcolors=Black}  
  
  \tableofcontents
  \cleardoubleplainpage

  \hypersetup{allcolors=NavyBlue}
  \mainmatter
  \clearpage{}\chapter{Introduction} \label{Kap_0}

\section{Unification}

The hunt for a world formula has been on going since the antiquities. In a permanent effort to understand the universe, humanity attempts to dive further and further into the world of physics until there has been an explanation for everything. But it is not as easy as it may sound. The idea of unification seems to be more intricate than it appears. 

However, we first want to give a brief history of unification. So, let us go back in time and work ourselves back to the present. One of the first achieved unifications is classical electrodynamics~\cite{Maxwell:1865zz}. Maxwell combined electricity and magnetism into electromagnetism in 1865 following two remarkable observations by Faraday and Ørsted. His theory predicted amongst other things the existence of electromagnetic waves traveling at speed of light $c$. Their existence was subsequently been shown with experiments 20 years later by Hertz. Many years afterwards, based on the ideas of Lorentz and Poincaré, Einstein was able to unify space and time. A first offspring was the theory of special relativity~\cite{Einstein:1905ve} and his most famous formula $E= mc^2$. Next, Einstein was able to unite the idea of spacetime with gravity. It resulted in general relativity~\cite{Einstein:1916vd}. Although, he did not want to stop there and dedicated the rest of his life to the search of a world formula, unifying all four fundamental forces. Sadly, he failed in his attempt.

Thus far, we have made contact with two of the four fundamental forces. The remaining two forces have been experimentally observed during the last century after the discovery of quantum mechanics. They are called the weak and strong nuclear force. These two forces can be described through the means of Quantum Field Theory (QFT). Elementary particles like electrons and quarks as well as their interactions can be described by QFT. Again, electrodynamics pushed the way forward with its quantum formulation called Quantum Electrodynamics (QED). Nevertheless, observing these ideas made the development of particle colliders essential, as it requires extremely high energies for their detection. During a collision of subatomic particles massive amounts of energy are released and cause the creation of new particles. Their inherent properties such as charge and momentum are then analyzed by several detectors. This allows the reconstruction of the fundamental interaction between all the involved particles. In this context, one has to differentiate between two kinds of particles: fermions forming the matter content as we know it and bosons which mediate their interactions. Additionally to the photon, the mediator of the electromagnetic force, there have been found other bosons as well. The $W^+$, $W^-$, $Z^0$ bosons which mediate the weak force as well as eight gluons for the strong interaction. All of them emerge naturally in the concept of gauge theories and their corresponding QFT frameworks. Furthermore, they are connected to different generators of Lie groups which take on the role of the respective gauge group and therefore are symmetries of their theory.

Now, it is further possible to unify electrodynamics and the weak nuclear force to an electroweak interaction given by the gauge group SU($2$)$\times$U($1$)$_Y$. For low energies, it gets broken down to QED's gauge group U($1$) through the Higgs mechanism and the weak gauge bosons acquire a mass~\cite{Higgs:1964pj,Englert:1964et,Guralnik:1964eu}. The Higgs mechnism is based on an additional massive, spin-0 scalar field called the Higgs boson. Physicists have undertaken extreme efforts to detect this particle with the Large Hadron Collider (LHC). In 2012, it was finally made public by the ATLAS and CMS collaborations at CERN that they had observed a particle matching the properties of the Higgs particle~\cite{Aad:2012tfa,Chatrchyan:2012xdj}. It fixes the energy scale of $m_{EW}$, at which the unification occurs, to $m_{EW} = 246$ GeV. The full theory, containing the strong interaction as well, is called the standard model. However, it should be noted that it does not include gravity.

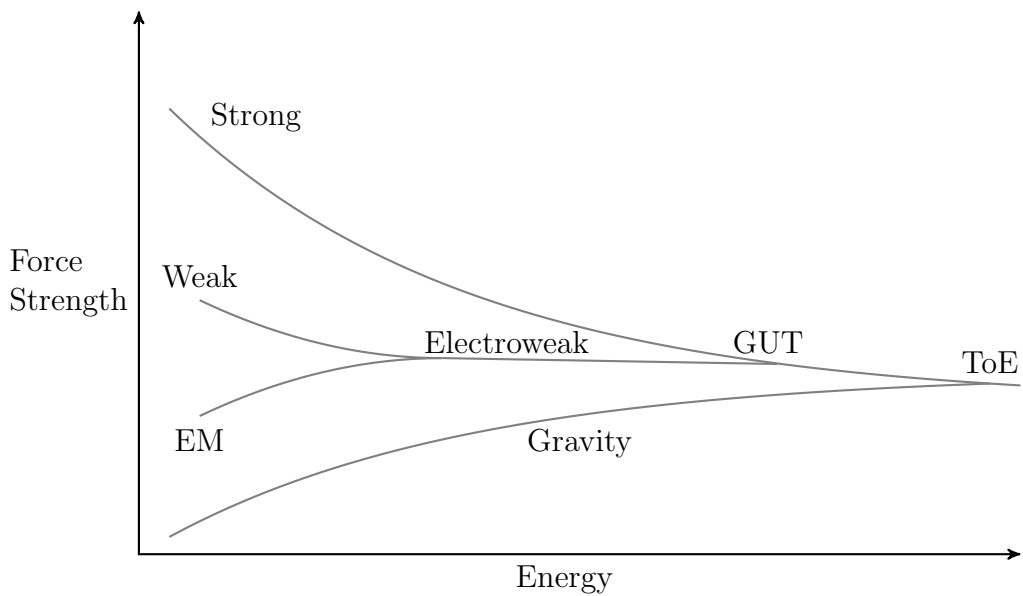
\begin{figure}[h]
\begin{center}
\begin{tikzpicture}[scale=4]
\draw[thick,gray] plot[domain=0.1:2.9,samples=200] (\x,{0.5+0.8*exp(-(\x-0.3))});
\draw[thick,gray] plot[domain=0.2:1,samples=200] (\x,{0.3*(\x-1)*(\x-1)+0.65});
\draw[thick,gray] plot[domain=0.2:1,samples=200] (\x,{-0.3*(\x-1)*(\x-1)+0.65});
\draw[thick,gray] plot[domain=1:2.1,samples=200] (\x,{0.65+(\x-1)*(-0.65+0.5+0.8*exp(-1.8)});
\draw[thick,gray] plot[domain=0.1:2.8,samples=200] (\x,{(-exp(-\x)+1)*(0.5+0.8*exp(-(2.8-0.3)))/(-exp(-2.8)+1)});
\node[above](EM) at (0.2,0.85){Weak};
\node[below](weak) at (0.2,0.45){EM};
\node[right](strong) at (0.2,1.45){Strong};
\node[below](Gravity) at (1.45,0.45){Gravity};
\node[above](EW) at (1.21,0.63){Electroweak};
\node[above](GUT) at (2.07,{0.65+1.07*(-0.65+0.5+0.8*exp(-1.8)}){GUT};
\node[above](ToE) at (2.8,{(-exp(-2.8)+1)*(0.5+0.8*exp(-(2.8-0.3)))/(-exp(-2.8)+1)}){ToE};
\draw [<->,thick,black] (2.9,0)--(0,0)--(0,1.8);
\node [below] at (1.4,0) {Energy};
\node [left, align=left] at (0,0.9) {Force\\ Strength};
\end{tikzpicture}
\caption{\label{int:unification}Energy scales for the unification of all four fundamental forces. Unifications above $m_{EW}$ are only conjectured.}
\end{center}
\end{figure}

All these results raise the hope that it might even be feasible to unify the four fundamental forces into a single one at a certain energy scale $m_{ToE}$, see fig. \ref{int:unification}. Under the assumption of a minimal supersymmetric standard model (MSSM), mathematical physicists conjecture the combination of the electroweak and strong forces at an energy scale of $m_{GUT} = 10 ^{16}$ GeV~\cite{Dimopoulos:1981zb,Dimopoulos:1981yj}. Clearly, this energy scale is far out of reach for present particle colliders such as the LHC which produce center of mass energies of around $10^4$ GeV. Nevertheless, it is extremely important for a theory describing physics shortly after the Big Bang. A viable energy scale for a theory of everything is the Planck scale $m_{Pl} = 1.22 \cdot 10^{19}$ GeV. At this scale, one assumes that all fundamental forces unify and yield the world formula as mentioned above. Fig. \ref{int:unification} visualizes this unification picture.

Due to a lack of empirical data it is hard to predict how such a theory, describing physics at the Planck scale, might look like. Although, there exist a number of possible candidates for it. string theory, quantum loop gravity\footnote{ It should be noted that loop quantum gravity is only a quantum theory of gravity and hence is not a true theory of everything. Although there are endeavors to introduce gauge interactions as well~\cite{BilsonThompson:2006yc}}~\cite{Nicolai:2006id,Rovelli:2011eq}, and non-commutative geometry are the most famous of them~\cite{connes1995noncommutative,Szabo:2001kg}. Yet, all of them try to address the common topics:
\begin{enumerate}
\item They want to reproduce standard model physics at low energies.
\item They attempt to conjecture new physics beyond the standard model.
\item They strive to make as little assumption as possible.
\end{enumerate}
At present, string theory appears to be the most probable candidate.

\section{String theory}

What is today known as string theory was initially an attempt to describe strong interactions in the late 1960s. As opposed to standard quantum field theories which consider point particles, string theory makes use of one-dimensional extended objects called strings. In general, one has to distinguish between open and closed strings, with closed strings satisfying additional boundary conditions. However, it was discarded very quickly as it required the existence of a critical dimension much larger than four. Furthermore, the existence of a spin two particle emerged in strong contradiction to the observations of quantum chromodynamics. During the year of 1974 two physicists named Scherk and Schwarz had the idea to use this unknown massless spin two particle, a massless string excitation, to their advantage by identifying it with the graviton. Moreover, they observed that this mysterious particle behaves at low energies in accordance to the covariance laws of general relativity. As a result, the theory became an immediate candidate for a possible description of quantum gravity and therefore might even be a suitable contender for a theory of everything~\cite{string2}. Nevertheless, it possesses many more string excitations as well. For instance, there exist even additional massless string excitations which can be interpreted as gauge bosons. Thus, it can be regarded as a theory unifying quantum gravity with the other gauge interactions and thereby highlights its significance as a possible true theory of everything. It comes in two descriptions, a worldsheet and a target space description which we are going to scrutinize in the next two subsections. 

During the course of this thesis, we are mainly interested in the bosonic sector of superstring theory but let us begin by giving some remarks about bosonic and superstring theory, including fermionic fields as well. Bosonic string theory is plagued by several major issues. One of them regards the existence of a tachyon, a negative mass squared excitation, appearing in the spectrum of the theory. It is a highly instable ground state. The second major disadvantage lies in the fact that it, thus far, only describes bosonic fields. Yet, in reality we detect fermions, too. Ergo, the fields describing the matter content are missing. The solution to these problems is given by superstring theory, a supersymmetric extension of bosonic string theory. Hence, it also considers fermionic degrees of freedom on the worldsheet and successively yields a supersymmetric theory with fermionic fields in target space~\cite{Gliozzi:1976jf,Gliozzi:1976qd,Green:1980zg,Green:1981xx,Green:1981yb}. Furthermore, the requirement of a vanishing Weyl anomaly reduces the critical spacetime dimension from bosonic string theory with $D=26$ down to $D=10$ for superstring theory. On top of that, a GSO projection removes the tachyonic degrees of freedom from the spectrum and leads to a modular invariant partition function.

\begin{figure}[t!]
\label{fig:dualities}
\begin{center}
\begin{tikzpicture}[scale=3]
\draw [thick]({sin(10)},{cos(10)}) to[out=-110, in=-170] ({sin{ 70}},{cos( 70})node[right]{\bf type IIA}
							to[out=-170, in=-230] ({sin{130}},{cos(130})node[right](tIIb){\bf type IIB}
							to[out=-230, in=-290] ({sin{190}},{cos(190})node[below]{\bf type I}
							to[out=-290, in=-350] ({sin{250}},{cos(250})node[left]{\bf het. SO($\mathbf{32}$)}
							to[out=-350, in=-410] ({sin{310}},{cos(310})node[left]{\bf het. E$_8$}
							to[out=-410, in=-110] ({sin{ 10}},{cos( 10})node[above]{\bf $\mathbf{11d}$ SUGRA};
\node at (0,0) {\bf M-theory};
\node [align=center] at ({0.9*sin( 40)},{0.9*cos( 40)}){compact.\\ on $S^1$};
\node at ({0.95*sin(100)},{0.95*cos(100)}){T-duality};
\node at ({0.95*sin(220)},{0.95*cos(220)}){S-duality};
\node at ({0.95*sin(280)},{0.95*cos(280)}){T-duality};
\node [align=center] at ({0.9*sin(340)},{0.9*cos(340)}){compact.\\ on I};
\draw [thick,<->]({1.2*sin(130)},{1.15*cos(130)}) arc (140:410:0.15);
\node [below] at ({1.2*sin(130)+0.15},{1.15*cos(130)-0.28}) {S-duality};
\end{tikzpicture}
\end{center}
\caption{\label{fig:dualities} Dualities connecting the five different superstring theories with M-theory. Here, $I$ denotes the compactification on a line interval.}
\end{figure}
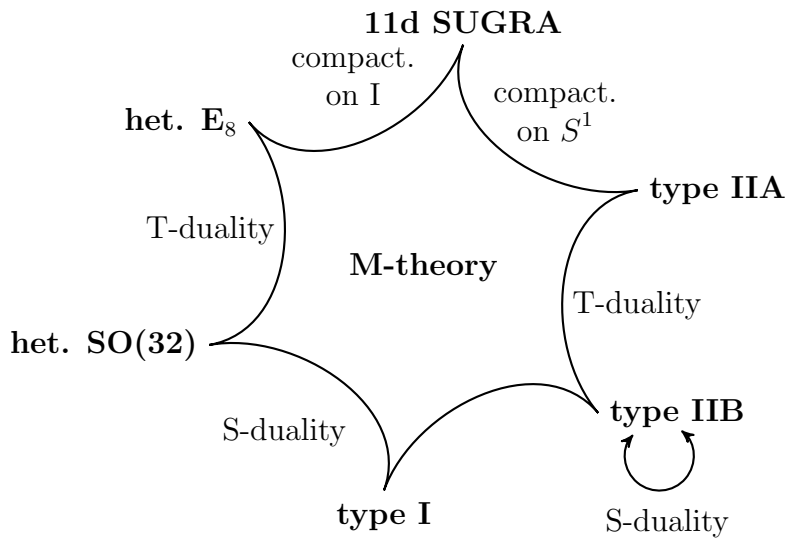

Although, there actually exist five different stable and consistent superstring theories. They are:
\begin{itemize}
\item {\bf Type I} It describes unoriented closed and open strings in ten dimensions. The low energy effective action is of $\mathcal{N} = 1$ super Yang-Mills with gauge group SO($32$) coupled to type I SUGRA.

\item {\bf Type IIA and IIB} These contain oriented closed strings in ten dimensions. The low energy effective descriptions are given by type IIA or IIB SUGRA.

\item {\bf Heterotic} They combine a bosonic string part in the left moving sector with a superstring in the right moving sector of a closed string. Again, its critical dimension is ten and its corresponding gauge groups are SO($32$) as well as E$_8$ $\times$ E$_8$ while their low energy effective descriptions are given through $\mathcal{N}=1$ super Yang-Mills coupled to type I SUGRA.
\end{itemize}
Additionally, we are left with two non-supersymmetric and unstable theories. These are type 0A and 0B. However, they are not viable to describe real world physics.

Last but not least, we have two theories related to string theory:
\begin{itemize}
\item {\bf M-theory} The strong coupling limit of type IIA superstring theory. Furthermore, it possesses eleven-dimensional Poincaré invariance.
\item {\bf F-theory} A geometric description of type IIB superstring theory formulated on $12$-dimensional space-time which is subsequently compactified on a elliptically fibered Calabi-Yau manifold.
\end{itemize}
All of the aforementioned superstring theories are conntected by dualities, see fig \ref{fig:dualities}.

\subsection{Worldsheet}

String theory can be cast as a two-dimensional conformal field theory on a Riemannian surface $\Sigma$ called the worldsheet. It describes open as well as closed strings, based on whether $\Sigma$ once all the punctures have been removed, is compact or not. Here, a puncture refers to a point which misses in the worldsheet. During the course of this thesis, we mainly consider the bosonic subsector of closed string theory. Its dynamics are derived from the Polyakov action
\begin{equation}
\label{int:Polyakov}
S_P = - \frac{1}{4\pi \alpha'} \int_{\Sigma} d^2 \sigma \sqrt{h} \, h^{\alpha \beta} \partial_\alpha x^i \partial_\beta x^j \left( g_{ij} + B_{ij} \right) + S_\chi
\end{equation}
with
\begin{equation}
S_\chi = \frac{1}{4\pi} \int_\Sigma d^2 \sigma \sqrt{h} \, \phi \, R = \phi \chi(\Sigma)
\end{equation}
being the Gauss-Bonnet term coupled to a dilaton field $\phi$. In the last equation, $R$ represents the curvature scalar $R$ of the metric $h^{\alpha \beta}$. It can also be expressed in terms of the topological invariant Euler number $\chi(\Sigma)$ which can be calculated by
\begin{equation}
\chi(\Sigma) = 2 - 2g - b
\end{equation}
with $g$ being the genus of the surface $\Sigma$ and b the number of boundaries. The Gauss-Bonnet term has locally the form of a total derivative and therefore does not contribute to the field equations. Yet, it plays a role in the string perturbation theory. Furthermore, the Polyakov action is the starting point for the path integral quantization procedure of string theory~\cite{string2}. However, this is beyond the scope of this thesis.

\subsection{Target space}

At this point, we can solve the field equations for the worldsheet metric $h^{\alpha \beta}$ of the Polyakov action \eqref{int:Polyakov}. When considering a two-punctured sphere i.e. a cylinder, $S_\chi$ vanishes and moreover dropping the $B$-field in the Polyakov action, we obtain the Nambu-Goto action
\begin{equation}
S_{NG} = - \frac{1}{2\pi \alpha'} \int_{\Sigma} d^2 \sigma \sqrt{\gamma}\,,
\end{equation}
where $\Sigma$ denotes the area of the worldsheet and $\gamma$ is the induced metric's determinant, given by
\begin{equation}
\gamma_{\alpha \beta} = \frac{\partial x^i}{\partial \sigma^\alpha} \frac{\partial x^j}{\partial \sigma^\beta} g_{ij}\,.
\end{equation}
It originates from the ambient $d$-dimensional Minkowski space the string moves through and subsequently $x^i(\tau, \sigma)$ with $i=0,\ldots,d-1$ maps the worldsheet into $d$-dimensional Minkowski space. Thus, the string propagates through a $d$-dimensional target space.

In this context, it becomes important to distinguish between open and closed strings. Furthermore, we identify the explicit worldsheet parametrization 
\begin{equation}
\sigma^0 = \tau \in \mathds{R}\, \quad \text{and} \quad \sigma^1 = \sigma \in [0,2\pi)\,.
\end{equation}
A closed string has to fulfill the additional boundary condition
\begin{equation}
x^i(\tau,\sigma) = x^i(\tau,\sigma+2\pi)\,.
\end{equation}
Therefore, its corresponding worldsheet has the form of a cylinder, whereas for open strings it has the shape of a strip~\cite{string2}.

\section{Low energy effective theory}

Ultimately, one should be able to derive the standard model at low energies from string theory if it truly is the theory of everything. The standard model is based on quantum field theory (QFT) with a finite particle content. However, all the particle masses, coupling constants etc. have to be introduced by hand. Now, returning back to string theory the particles correlate to different string excitations in target space. Clearly, this would produce an unlimited amount of particles. Although, for a low energy effective theory we are only focusing on the lightest of them. Therefore, we are interested in formulating a low energy description of string theory. One of the intrinsic choices for the required energy cut-off is given by the string mass
\begin{equation}
m_s = \frac{1}{\sqrt{\alpha'}}\,.
\end{equation}
Up to now, there exist only very few restrictions on its scale. Yet, it must be much higher than the currently available energy at the Large Hadron Collider (LHC) as no signatures have been detected. The string mass could even be as high as the Planck mass $m_{Pl}$

There exist two possible ways to obtain an effective action from string theory's worldsheet description~\cite{Hassler:2015pea}:
\begin{itemize}
\item We can derive the string amplitudes on worldsheets with different topologies and match them with a low energy effective field theory in target space whose Feynman diagrams reproduce the same amplitudes. Finding such a field theory which describes the string at classical level with weak coupling can be achieved by analyzing two and three punctured spheres. Then, the most general ansatz must consider terms quadratic, cubic, and quartic in fields with arbitrary coupling constant for the effective action. Subsequently, these constants can be fixed by comparing the amplitudes of the target space tree-level Feynman diagrams with the ones obtained from the worldsheet.

\item We compute the one-loop $\beta$-function for the coupling constants on the worldsheet. As a consequence, it is possible to perturb/fluctuate the coupling constants around a given background, e.g. a flat one, which coincides with massless string excitations. In the end, the $\beta$-function has to vanish, if the conformal symmetry of the worldsheet theory holds at quantum level. This allows to obtain the field equations of the effective field theory. Finally, one finds the accompanying action.
\end{itemize}
It should be mentioned that both procedures produce exactly the same results. Nevertheless, they only describe the string classically but for quantum effects, it becomes necessary to consider String Field Theory (SFT) calculations as well.

\subsection{Compactifications and T-Duality}
\label{sec:T-Duality}

For string theory to make contact with experimental observations we need more than just a low energy effective description. So far, we have only encountered four dimensions in nature. This raises the question to what happens with the remaining six dimensions required for a consistent superstring theory in $D=10$ dimensions? A possible explanation could be the existence of small extra dimensions which allow the strings to elude detection by particle accelerators currently at our disposal. The energies for detecting the string are simply too high for present colliders such as the Large Hadron Collider (LHC). The idea of small, compact, extra dimensions goes back to Kaluza and Klein who proposed a fifth dimension to unify electrodynamics with general relativity in 1921 \cite{Kaluza:1921tu,Klein:1926tv,Klein:1926fj}.

The procedure of going from a higher dimensional theory down to a lower dimensional theory by assuming small compact dimensions is called \textit{compactification}. Moreover, the shape of the compact space determines the properties of the effective theory in four dimensions. For instance, one tries to find a four dimensional theory with Minkowski vacuum and minimal supersymmetry to implement the Minimal Supersymmetric Standard Model (MSSM). Then, the internal space can be chosen to be a Calabi-Yau threefold~\cite{Hassler:2015pea,Blumenhagen:2008zz,Blumenhagen:2009yv}. There exists an infinite variety of these manifolds and they are distinguished by their \textit{moduli}. These are counted by their hodge numbers $h^{1,1}$ and $h^{2,1}$. Each moduli in the four-dimensional theory gives rise to a massless scalar field. However, this poses severe consequences to observations and predictions of cosmology, as even if they would decouple from three fundamental forces, they still couple to gravity and therefore affect the cosmology of our universe. As a consequence, in string phenomenology intense efforts are being made into giving mass to the moduli and stabilize them at certain vacuum expectation values. This technique is called \textit{moduli stabilization}.

One approach is to stabilize the moduli at tree-level by using flux compactifications~\cite{Giddings:2001yu,Grana:2005jc,Douglas:2006es,Denef:2008wq}. Now, giving non-vanishing vacuum expectation values to the fluxes, such as the $H$-flux, yields a scalar potential for the moduli. Ideally, this potential would have at least one minimum stabilizing all moduli. But generally, it is impossible to find tree-level fluxes which stabilize all moduli, and the scalar potential possesses at least one flat or runaway direction. However, it is possible to apply non-perturbative effects to the remaining moduli, but normally there exists no procedure to stabilize all moduli.

For a full grasp of Double/Exceptional Field Theory it is crucial to completely understand the notion of T/U-Duality, as Double/Exceptional Field Theory makes T/U-Duality a manifest symmetry of the theory. T-Duality is a symmetry that unfolds during certain compactifcations, mostly in context of circular and toroidal compactifications. It connects different background topologies with each other. We start with the demonstration of an illustrative example, a circular compactification on a circle i.e. $S^1$ in~\ref{S1compact}. In this context, we make contact with string winding, and the concept of T-Duality. Subsequently, we generalize this idea to toroidal compactifications in $D$-dimensions~\ref{toruscompact} and see the emergence of the Buscher rules~\ref{sec:buscherrules}. Moreover, we can combine T-duality with S-duality and obtain U-duality. We mainly follow~\cite{Pascal2015}.

\subsubsection{T-Duality: $S^1$ compactification}

\begin{figure}[h]
    		\centering
    		\begin{tikzpicture}[scale=1.5]
			\draw (-0.5,0) arc (270:90:0.3 and 1);
			\draw [dashed](-0.5,0) arc (-90:90:0.3 and 1);
			\draw (-0.5,0)--(4.5,0);
			\draw (-0.5,2)--(4.5,2);
			\draw (4.5,1) ellipse (0.3 and 1);
			\draw [->](3,0) arc (270:180:0.3 and 1);
			\draw (2.7,1) arc (180:90:0.3 and 1);
			\draw [dashed](3,0) arc (-90:90:0.3 and 1);
			\draw [dashed](2,0) arc (-90:90:0.3 and 1);
			\draw [dashed](1,0) arc (-90:90:0.3 and 1);
			\draw [->](2,0) arc (270:205:0.3 and 1)coordinate (c);
			\draw [->](1,0) arc (270:205:0.3 and 1) coordinate (a);
			\draw [-<](2,2) arc (90:155:0.3 and 1) coordinate (b);
			\draw [-<](1,2) arc (90:155:0.3 and 1) coordinate (d);
			\draw (a) to [in = 270,out = 90](b) ;
			\draw (c) to [in = 270,out = 90](d);
			\node [below] at (1.5,0) {(a)};
			\node [below] at (3,0) {(b)};
		\end{tikzpicture}
    		\caption{Closed string winding, when compactified on circle.}
    		\label{fig:StringPic}
\end{figure}
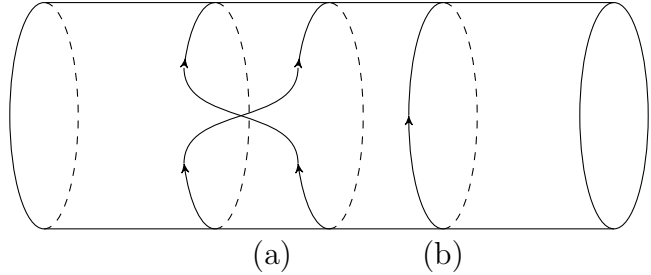

Understanding T-Duality thoroughly requires several important steps. In order for us to understand it properly, we begin by introducing the concept of circular compactifications, the most straight forward example~\citep{string2, string1}. We can identify the compact dimension of such a compactification with a circle
\begin{equation}
\label{S1compact}
S^1 = \mathbb{R} / (2\pi R \mathbb{Z})\,,
\end{equation}
while the total space takes the form
\begin{equation}
\mathbb{R}^{1,D-1} \to \mathbb{R}^{1,D-2} \times S^1\,.
\end{equation}
Here, $R$ denotes the radius of our compact direction. Note, if we have a closed string curled around the compact dimension, it is obviously not possible to contract said closed string to a point anymore. Let now $x$ be our compact coordinate, then the periodicity condition for this coordinate requires
\begin{equation}
\label{eqn:perdiodic}
x \sim x + 2 \pi R.
\end{equation}
(After going a full loop around the circle, we have to be in the initial point again.) 
On the other hand, this constrains our worldsheet coordinates to fulfill the equation
\begin{equation}
X(\tau, \sigma + 2\pi ) = X(\tau, \sigma) + 2\pi R\, \tilde{p}\,,
\end{equation}
where $\tilde{p} \in \mathbb{Z}$ represents the winding number and thus counts how many times the string wraps around the compact dimension, i.e. see figure \ref{fig:StringPic}. Here, string (a) wraps twice around the compact direction while string (b) winds only once. 
The winding numbers (conserved charges) generate so-called winding states. They possess no classical counterpart, and are topologically stable solitons~\citep{string2}. The reason for this lies in the existence of non-contractible loops i.e. that the closed strings can't be shrunk to a point anymore.
At this point, we can perform the usual mode expansion and complete some straightforward calculations~\citep{string1}. It allows us to obtain the mass formula
\begin{equation}
\label{massformula}
M^2 = \Big(\frac{p}{R}\Big)^2 + \Big(\frac{\tilde{p}R}{2}\Big)^2 + (N + \tilde{N} - 2)\,,
\end{equation}
where we choose $\alpha' = 2$. In this equation, $p$ and $\tilde{p}$ denote the quantized momentum and winding, while $N$ and $\tilde{N}$ count the number of oscillators. The first two terms emerge due to the compactification around the circle $R$, whereas the last term is a remnant of the uncompactified external directions. Additionally, the level matching $L_0 - \tilde{L}_0 = 0$ gives us the following condition
\begin{equation}
\label{levelmatching}
N - \tilde{N} = p\tilde{p}\,.
\end{equation}
If we apply the decompactification limit, i.e. $R \gg 2$, to equation~\eqref{massformula}, the winding modes become very heavy, as the energy to wrap around the compact dimension increases, and hence the mass spectrum becomes continuous.

In the opposite limit, where $R \ll 2$, the momentum modes become exceedingly heavy while the winding modes are very light (requires small energy to wrap around the compact dimension), and the spectrum becomes continuous~\citep{Aldazabal:2013sca} as well.

\begin{figure}
\centering
\begin{subfigure}{.3\textwidth}
  \centering
  \begin{sideways}
    		\centering
    		\begin{tikzpicture}[scale=0.5]
			\draw (-0.5,0) arc (270:90:0.3 and 2);
			\draw [dashed](-0.5,0) arc (-90:90:0.3 and 2);
			\draw (-0.5,0)--(4.5,0);
			\draw (-0.5,4)-- node[below=-10pt, rotate=-90]{\scriptsize $R$}(4.5,4);
			\draw (4.5,2) ellipse (0.3 and 2);
			\draw[Blue,semithick] [->](3,0) arc (270:180:0.3 and 2);
			\draw[Blue,semithick] (2.7,2) arc (180:90:0.3 and 2);
			\draw[Blue,semithick] [dashed](3,0) arc (-90:90:0.3 and 2);
			\draw[red,semithick] (0.8,2) ellipse (0.5cm and 0.25cm);
		\end{tikzpicture}
\end{sideways}	
\end{subfigure}
\begin{subfigure}{.3\textwidth}
 \centering
 \begin{equation*}
 \Longleftrightarrow
 \end{equation*}
\end{subfigure}
\begin{subfigure}{.3\textwidth}
  \centering
  \begin{sideways}
    		\centering
    		\begin{tikzpicture}[scale=0.5]
			\draw (-0.5,0) arc (270:90:0.3 and 1);
			\draw [dashed](-0.5,0) arc (-90:90:0.3 and 1);
			\draw (-0.5,0)-- node[below=25pt, rotate=-90]{\scriptsize $\tilde{R} = 2 / R$}(4.5,0);
			\draw (-0.5,2)--(4.5,2);
			\draw (4.5,1) ellipse (0.3 and 1);
			\draw[Red,semithick] [->](1,0) arc (270:180:0.3 and 1);
			\draw[Red,semithick] (0.7,1) arc (180:90:0.3 and 1);
			\draw[Red,semithick] [dashed](1,0) arc (-90:90:0.3 and 1);
			\draw[Blue,semithick] (3,1) ellipse (0.5cm and 0.25cm);
		\end{tikzpicture}
\end{sideways}	
\end{subfigure}
\caption{T-Duality between circle with radius $R$ and circle with dual radius $\tilde{R}$}
\label{fig:T-duality}
\end{figure}
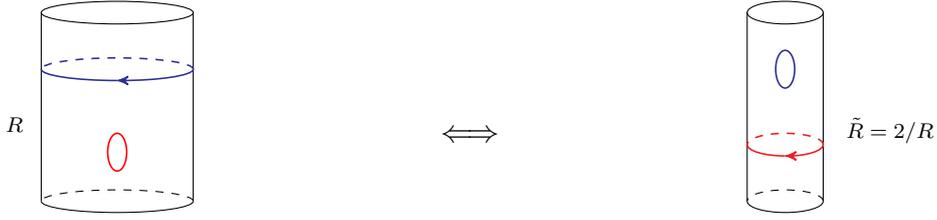

Already by examining equations \bref{massformula} and \bref{levelmatching} it should become quite obvious that there should exist a symmetry between momentum modes $p$ and winding modes $\tilde{p}$. This symmetry is called \textit{T-Duality}. It is given by the transformation, see figure~\ref{fig:T-duality},
\begin{equation}
\label{T-Duality}
R \leftrightarrow \tilde{R} = \frac{2}{R}\,,\quad p \leftrightarrow \tilde{p}\,.
\end{equation}

The fact that our equations are invariant under the interchange of momentum modes $p$ with radius $R$ and winding modes $\tilde{p}$ with dual radius $\tilde{R}$ is very astonishing. In fact, it implies that we are incapable to distinguish between small and large compact directions when compactifying on a circle. Much more, these two different compactifications are physically indistinguishable~\citep{string1} and as a result T-Duality relates different background topologies with each other. 

Thus, at the string scale ordinary geometric concepts and intuitions break down~\citep{string3}. However, at the self-dual radius $R^* = \sqrt{2}$ these two different compactifications coincide, and hence it marks a fixed point under the T-Duality transformation. It comes along with the occurrence of non-abelian gauge groups at this special point \cite{string3}, also called symmetry enhancement. The particular symmetry group depends on the excitation of the string.  As a result, the radius defines a continuous parameter of spacetimes which allow for a consistent string theory. In particular, it is a modulus which forms the one-dimensional \textit{moduli space} of this compactification.

Double Field Theory is currently restricted to massless states not in the decompactification limit. From \bref{massformula}, we obtain $N+\tilde{N}=2$, and the level matching condition \bref{levelmatching} cancels out the states $(N,\tilde{N})=(2,0)$, $(N,\tilde{N})=(0,2)$ which restricts us to $(N,\tilde{N})=(1,1)$. The state
\begin{equation}
\alpha_{-1}^{\mu} \tilde{\alpha}_{-1}^{\nu} |0\rangle\,,
\end{equation}
with $\alpha_{-1}^{\mu}$ and $\tilde{\alpha}_{-1}^{\nu}$ being oscillators of the mode expansion, allows us to acquire the following field content: a symmetric metric $g_{ij}$, a two-form field $B_{ij}$, and the dilaton $\phi$.

Now, we want to generalize this idea and turn to toroidal compactifications in $D$ dimensions.

\subsubsection{T-Duality: Toroidal compactification}

Generalizing the above discussion to a bosonic closed string compactified on a $D$-dimensional torus $T^D$~\citep{string2} yields the periodicity condition
\begin{equation}
\label{toruscompact}
X^{I} \sim X^{I} + 2\pi \tilde{P}^{I}\,,
\end{equation}
with
\begin{equation}
\tilde{P} = \sum\limits_{i=1}^D \tilde{p}^{i} e_i^{I}\,, \quad n_i \in \mathbb{Z}\,,
\end{equation}
where $\mathbf{e}_i$, $i \in \{1,...,D\}$ are the linear independent basis vectors spanning the lattice $\Lambda_D$, and $I \in \{1,...,D\}$ labels the internal directions. Hence, the winding $\mathbf{\tilde{P}} \in \Lambda_D$ becomes an integral lattice vector. Therefore, after compactification on the torus $T^D$, the lattice $\Lambda_D$ is given by
\begin{equation}
T^D = \mathbb{R}^D / (2\pi \Lambda_D)\,.
\end{equation}
In worldsheet coordinates the periodicity conditions takes on the form
\begin{equation}
X^{I}(\sigma +2\pi,\tau) = X^{I}(\sigma,\tau) + 2\pi \tilde{P}^I\,.
\end{equation}
Now, we can introduce the dual lattice $\Lambda^*_D$ by using the state condition
\begin{equation}
e^{i X^I P_I} = e^{i (X^I + 2\pi \tilde{P}^I) P_I} \implies \tilde{P}^I P_I \in \mathbb{Z}\,.
\end{equation}
This gives us
\begin{equation}
P_I = \sum\limits_{i=1}^D p_i\, {e^*_I}^i\,,
\end{equation}
 with $\mathbf{e^*}^{i} \in \Lambda^*_D$. Thus, the momentum $\mathbf{P}$ lies on the dual lattice $\Lambda_D^*$ and is integral.

The basis vectors $\mathbf{e}_i$ on the lattice $\Lambda_D$, and basis vectors $\mathbf{e^*}^i$ on the dual lattice $\Lambda_D^*$ satisfy the following properties
\begin{equation}
\mathbf{e}_i \cdot \mathbf{e^*}^j = \sum\limits_{I=1}^D e_i^I {e^*_I}^j = {\delta_i}^{j}\,,
\end{equation}
\begin{equation}
\sum\limits_{i=1}^D e_i^I {e^*_J}^i = \delta^I{}_{J}\,.
\end{equation}
Moreover, the metric on the lattice $\Lambda_D$ is given by
\begin{equation}
g_{ij} = \mathbf{e}_i \cdot \mathbf{e}_j = \sum\limits_{I,J=1}^D e_i^I e_j^J \delta_{IJ}\,,
\end{equation}
and for the dual lattice $\Lambda^*_D$ through
\begin{equation}
g^*_{ij} \equiv g^{ij} = \mathbf{e^*}^i \cdot \mathbf{e^*}^j = \sum\limits_{I,J=1}^D {e^*_I}^i {e^*_J}^j \delta^{IJ} = (g^{-1})_{ij}\,.
\end{equation}
Again, we execute some straightforward calculations, and make use of the well-known mode expansion for $D$ compact directions, we finally arrive at the mass formula
\begin{align}
\label{massformulatorus}
M^2 &= ( N + \tilde{N} - 2 ) + \sum\limits_{I=1}^D \Big( P^I P^J + \frac{1}{4} \tilde{P}^I \tilde{P}^J \Big) \delta_{IJ} \\
&= ( N + \tilde{N} - 2 ) + \sum\limits_{i,j=1}^D \Big( p_i\, g^{ij}\, p_j + \frac{1}{4} \tilde{p}^i\, g_{ij}\, \tilde{p}^j \Big) \nonumber \\
&= ( N + \tilde{N} - 2 ) + \mathbf{p}^{\mathrm{T}} \mathbf{g}^{-1}\, \mathbf{p} +  \frac{1}{4} \mathbf{\tilde{p}}^{\mathrm{T}} \mathbf{g}\, \mathbf{\tilde{p}}\,. \nonumber
\end{align}
Subsequently, the level matching condition \bref{levelmatching} takes on the form
\begin{equation}
\label{levelmatchingtorus}
N - \tilde{N} = \mathbf{P} \cdot \mathbf{\tilde{P}} = P_I \tilde{P}^I = \sum\limits_{i=1}^D p_i\, {\tilde{p}}^i \in \mathbb{Z}\,.
\end{equation}
We can now generalize equations \bref{massformulatorus} and \bref{levelmatchingtorus} further, by turning on an additional 2-form field $\mathbf{b}$. Equation \bref{massformulatorus} then becomes
\begin{equation}
M^2 = ( N + \tilde{N} - 2 ) + \mathbf{p}^{\mathrm{T}} \mathbf{g}^{-1}\, \mathbf{p} +  \frac{1}{4} \mathbf{\tilde{p}}^{\mathrm{T}}( \mathbf{g} - \mathbf{b} \mathbf{g}^{-1} \mathbf{b} )\, \mathbf{\tilde{p}} + \mathbf{\tilde{p}}^{\mathrm{T}} \mathbf{b} \mathbf{g}^{-1} \mathbf{p}\,,
\end{equation}
which can be rewritten in an equivalent form
\begin{equation}
\label{torodmass}
M^2 = ( N + \tilde{N} - 2 ) + \frac{1}{2} \mathcal{P}^{\mathrm{T}} \mathcal{\tilde{H}}^{-1}\, \mathcal{P}\,,
\end{equation}
by introducing a generalized vector $\mathcal{P}$
\begin{equation}
\mathcal{P} = \begin{pmatrix} \mathbf{\tilde{p}} \\ \mathbf{p}\end{pmatrix}\,,\quad \text{whose components are given by} \quad\mathcal{P}_M = \begin{pmatrix} \tilde{p}^i \\ p_i \end{pmatrix}\,,
\end{equation}
and the O($D,D$) valued generalized metric $\mathcal{\tilde{H}}^{-1} \in$ O$(D,D)$
\begin{equation}
\mathcal{\tilde{H}}^{-1} = \begin{pmatrix}
2( \mathbf{g} - \mathbf{b} \mathbf{g}^{-1}\mathbf{b}) &  \mathbf{b} \mathbf{g}^{-1}
\\ -\mathbf{g}^{-1} \mathbf{b} & \frac{1}{2} \mathbf{g}^{-1} \end{pmatrix}\,.
\end{equation}
Removing factors of 2 by $g \to g/2$ and $b \to b/2$, we obtain
\begin{equation}
\label{eqn:introgenmetr}
\mathcal{H}^{-1} = \begin{pmatrix}
\mathbf{g} - \mathbf{b} \mathbf{g}^{-1}\mathbf{b} &  \mathbf{b} \mathbf{g}^{-1}
\\ -\mathbf{g}^{-1} \mathbf{b} & \mathbf{g}^{-1} \end{pmatrix}\,, \quad
\mathcal{H}^{MN} = \begin{pmatrix}
g_{ij} - b_{ik} g^{kl}b_{lj} &  b_{ik} g^{kj}
\\ -g^{ik} b_{kj} & g^{ij} \end{pmatrix}\,.
\end{equation}
The indices $M$ and $N$ are raised and lowered with the O($D,D$) invariant metric
\begin{equation}
\eta = \begin{pmatrix}
0 & \;\;1_D \\ \;\;1_D & 0
\end{pmatrix}\,, \quad \eta_{MN} = \begin{pmatrix}
0 & \delta^i_{\:j} \\  \delta_i^{\:j} & 0
\end{pmatrix}\,.
\end{equation}
Subsequently, the metric and the generalized metric need to fulfill the following identities:
\begin{equation}
\mathcal{H}^T \eta \, \mathcal{H} = \eta\,,\quad \mathcal{H}^{MN} = \eta^{MP} \eta^{NQ} \mathcal{H}_{PQ}\,,\quad \mathcal{H}^{MN} \mathcal{H}_{NP} = {\delta^M}_{P}\,.
\end{equation}
Although, the level matching condition \bref{levelmatchingtorus} remains unaltered, and can also be written in terms of the generalized vector $\mathcal{P}$
\begin{equation}
N - \tilde{N} = \frac{1}{2} \mathcal{P}_M \mathcal{P}^M\,.
\end{equation}
Finally, the massless states are given by $N = \tilde{N} = 1$ in this case as well and reduce to the orthogonality condition $p_i\, \tilde{p}^i = 0$.

From equation \bref{torodmass} we can directly see the emergence of the T-Duality group O($D,D$)~\citep{string3}. In particular, equation \bref{torodmass} is invariant under exchange of
\begin{equation}
\tilde{p}^i \leftrightarrow p_i\,, \quad \mathcal{H}_{MN} \leftrightarrow \mathcal{H}^{MN}\,,
\end{equation}
and discrete shifts of an antisymmetric matrix $n_{ij}$
\begin{equation}
b_{ij} \mapsto b_{ij} + \frac{1}{2} n_{ij}\,,\quad \text{with} \quad \tilde{p}^i\mapsto\tilde{p}^i, p_i \mapsto	p_i + n_{ij} \tilde{p}^j\,.
\end{equation}
When combined, the inversion symmetry, and the shift symmetry generate the group O($D,D,\mathbb{Z}$) which acts geometrically on the torus \cite{Aldazabal:2013sca}.

\subsubsection{Buscher Rules}

It is possible, to reduce any element of the group O($D,D$) further as a product of the following transformations~\citep{Aldazabal:2013sca}:
\begin{align}
\text{Diffeomorphisms:}&\quad h_M^{\,\,\,N} = \begin{pmatrix}
E^i_{\,j} & 0 \\ 0 & E_i^{\,j}
\end{pmatrix}\,, \quad \quad \quad \;\, \quad E \in GL(D) \\
\text{Shifts:}&\quad h_M^{\,\,\,N} = \begin{pmatrix}
\delta^i_{\,j} & 0 \\ n_{ij} & \delta_i^{\,j}
\end{pmatrix}\,, \quad \quad \quad \;\;\; \quad n_{ij} = - n_{ji} \nonumber \\
\text{Factorized\,T-Dualities:}&\quad h_M^{\,\,\,N} = \begin{pmatrix}
\delta^i_{\,j} - t^i_{\,j} & t^{ij} \\ t_{ij} & \delta_i^{\,j} - t_i^{\,j}
\end{pmatrix}\,, \quad t_{ij} = \mathrm{diag}(0,...,1,...,0)\,. \nonumber
\end{align}
(T-Duality is applied in the $k$-th direction and thereby, $\eta_{MN}$ corresponds to the application of $D$ successive T-Duality transformations.) The diffeomorphisms correspond to a change in the basis of the lattice underlying the torus, whereas the factorized T-Dualities generalize the $R \leftrightarrow \tilde{R} = 2/R$ symmetry. Hence, carrying out T-Duality along the $k$-th direction, we obtain the well-known Buscher Rules~\citep{Buscher:1987sk, Buscher:1987qj}:
\begin{align}
\label{sec:buscherrules}
g_{kk} \mapsto &\frac{1}{g_{kk}}\,,\quad g_{ki} \mapsto \frac{b_{ki}}{g_{kk}}\,,\quad g_{ij} \mapsto g_{ij} - \frac{g_{ki} g_{kj} - b_{ki} b_{kj}}{g_{kk}}\,, \\
&b_{ki} \mapsto \frac{g_{ki}}{g_{kk}}\,,\quad b_{ij} \mapsto b_{ij} -  \frac{g_{ki} b_{kj} - b_{ki} g_{kj}}{g_{kk}}\,, \nonumber
\end{align}
where T-Duality is performed in an isometric direction. In the $k$-th direction $g$ is being exchanged with $g^{-1}$ and thus corresponds to a generalization of equation \bref{T-Duality}. These transformation rules were obtained by Buscher~\citep{Buscher:1987sk, Buscher:1987qj} through gauging an U($1$) isometry in the worldsheet action and successively obtaining a way in which T-Duality acts on the target space. They map solutions of the theory to other ones. However, they are not a manifest symmetry of the SUGRA action \eqref{SUSY} and motivated the development of Double Field Theory~\cite{Hull:2009mi,Hohm:2010jy,Hohm:2010pp}, which we are going to discuss in the next chapter.

Now, we want to obtain the moduli space generated by these toroidal compactifications~\citep{string3}. We find
\begin{equation}
\mathcal{M}^0_{n,n} = \text{O}(D,D;\mathbb{R})/[\text{O}(D;\mathbb{R}) \times \text{O}(D;\mathbb{R})]\,.
\end{equation}
Nevertheless, we still have physically identical states related by T-Duality $\text{O}(D,D,\mathbb{Z})$ which have to be divided out. This gives us the physical moduli space
\begin{equation}
\mathcal{M}_{n,n} = \mathcal{M}^0_{D,D} / \text{O}(D,D;\mathbb{Z})\,.
\end{equation}
As a result, we obtain fixed points under $\text{O}(D,D;\mathbb{Z})$ transformations which cause singularities. It implies that we must have special values $(g_{ij}, b_{ij})$ resulting in additional massless gauge bosons and therefore yielding a non-abelian gauge symmetry. In general, toroidal compactifications $T^D$ only allow for $U(1)^D$ isometries, and are non-chiral, i.e. they don't have a chiral matter content. Furthermore, they can't reproduce non-abelian gauge interactions. Consequently, it becomes impossible to explain extensions of the standard model~\citep{string2}. Yet, at the self dual radius we observe a symmetry enhancement, similar to circular compactifications, as well.

Another important duality in string theory is S-duality. It relates weakly and strongly coupled string theories with another. Later on, we are going to combine it with the T-duality group which gives rise to U-duality. U-duality plays a crucial role in the context of Exceptional Field Theories (EFTs).

\subsection{Supergravity}
\label{SUGRA}

First, we want to address the question, what exactly supergravity (SUGRA) is~\cite{Pascal2015}. SUGRA is an effective field theory that attempts to unify supersymmetry (SUSY) with general relativity, while the invariance under local SUSY transformations has been made manifest. (We are only interested in the bosonic part of the action. Hence, we do not consider SUSY transformations which exchange bosons and fermions.)
In general, supergravity can be seen as the low energy limit $(E \ll M_s)$ of superstring theory, i.e. in 11 dimension it is the low energy effective theory of M-Theory. One of the most natural questions in physics to ask is, whether there exists a generalization, i.e. a generalization of supergravity? However, let us first give a short review about supergravity. In this context, we want to consider $D$-dimensional supergravity in the type II bosonic sector where we only focus on massless fields, implying that we neglect fermionic fields such as gravitinos, and dilatinos as they would only impede the discussion. After integrating out the massive modes, all of the information and degrees of freedom of the theory are embedded in a symmetric metric $g_{ij}$, a two-form field $b_{ij}$, and the dilaton $\phi$. (The variables depend on the coordinates $x^i$ in $D$-dimensions.) As we have already seen in the previous section, these are the only allowed massless excitations. Thus, the supergravity action for the $\mathcal N=2$ NS/NS sector is given by
\begin{equation}\label{SUSY}
  S = \int d^Dx \sqrt{g} e^{-2\phi}\Big(R+4(\partial\phi)^2-\frac{1}{12}H^{ijk}H_{ijk}\Big)
\end{equation}
and involves the metric $g_{ij}$ making up the Ricci scalar $R$, the 3-form field $H_{ijk}$ consisting of the 2-form field $b_{ij}$, and the dilaton $\phi$.
Here, $g_{ij}$ and $b_{ij}$ are invariant under the usual diffeomorphisms, and gauge transformations  $b_{ij} \mapsto b_{ij} + \partial_i \tilde{\lambda} - \partial_j \tilde{\lambda}$. Additionally, the three-form field $H_{ijk}$ given by
\begin{equation}
 H_{ijk} = 3 \partial_{[i} b_{jk]}\,,
\end{equation} 
satisfies the Bianchi identity
\begin{equation}
\label{sugrabianchi}
 \partial_{[i} H_{jkl]} = 0\,.
\end{equation}

Varying the action \bref{SUSY} with respect to the three individual fields, one obtains the equations of motion
\begin{equation}
\label{sugraeom1}
 R_{ij} + 4 \partial_i \partial_j \phi -\frac{1}{4} {H_i}^{mn} H_{jmn} = 0\,,
\end{equation}
\begin{equation}
\label{sugraeom2}
 \frac{1}{2} \partial_k H^{ijk}-H^{ijk} \partial_k \phi = 0\,,
\end{equation}
\begin{equation}
\label{sugraeom3}
 R + 4\big(\partial^2\phi - (\partial\phi)^2 \big) - \frac{1}{12}H^2 = 0\,.
\end{equation}
In the first equation we used that the trace is zero.

At this point, let us revisit the notion of diffeomorphism invariance. This topic will become of great importance, when generalizing the concept to Double Field Theory later in this thesis and hence it is crucial to fully understand it~\citep{Aldazabal:2013sca}. We start by introducing well-known Lie derivative (Lie bracket)
 \begin{equation}
 \label{lie1}
 L_{\lambda} V^i = [\lambda, V]^i = \lambda^j \partial_j V^i - V^j \partial_j \lambda^i\,,
 \end{equation}
 which is antisymmetric under the exchange of fields and satisfies the Jacobi identity. It allows us to write the transformation properties under local coordinate changes for the three individual fields $g_{ij}$, $b_{ij}$, and $\phi$ as
 \begin{equation*}
 g_{ij} \mapsto g_{ij} + L_{\lambda}g_{ij}\,,\quad L_{\lambda}g_{ij} = \lambda^k \partial_k g_{ij} + g_{kj}\partial_i\lambda^k + g_{ik}\partial_j\lambda^k\,
 \end{equation*}
 \begin{equation*}
 b_{ij} \mapsto b_{ij} + L_{\lambda}b_{ij}\,,\quad L_{\lambda}b_{ij} = \lambda^k \partial_k b_{ij} + b_{kj}\partial_i\lambda^k + b_{ik}\partial_j\lambda^k\,
 \end{equation*}
 \begin{equation}
 \phi \mapsto \phi + L_{\lambda}\phi\,,\quad L_{\lambda} = \lambda^k \partial_k \phi \,.
 \end{equation}
 During the course of this thesis, we will see how these transformations are implemented in Double Field theory.
 
Unfortunately, T-Duality is not a manifest symmetry of the SUGRA action \eqref{SUSY}. In fact, we always have to be in the large volume limit to not violate the low energy limit of supergravity. This implies that we cannot consider small dimensions. As a result, we cannot encounter any winding modes. Thus, we are only left with momentum modes. We are going to see, how we can overcome these issues by introducing Double Field Theory. 

Furthermore, compactifying $11$-dimensional SUGRA and M-theory on an $d$-dimensional torus gives rise to the U-duality group $E_{d(d)}$~\cite{Hull:1994ys,Cremmer:1997ct,Obers:1998fb} and consequently Exceptional Field Theories~\cite{Hull:2007zu,Pacheco:2008ps,Berman:2010is,Coimbra:2012af,Coimbra:2011ky,Berman:2011jh,Berman:2011pe,Hohm:2013vpa,Hohm:2013uia,Hohm:2014fxa}.

\subsection{Non-geometric backgrounds}

An astonishing fact of gauged SUGRAs is that they provide more deformations than those of geometric compactifications, i.e. twisted tori with two-form flux~\cite{Aldazabal:2013sca}. In particular, it is generally not possible to turn on the following non-geometric components of the embedding tensor
\begin{equation}
Q_{a}{}^{bc} = f_{a}{}^{bc} \quad \text{and} \quad R^{abc} = f^{abc}
\end{equation}
through geometric Scherk-Schwarz compactifications of $10$-dimensional SUGRA. The other two components
\begin{equation}
H_{abc} = f_{abc} \quad \text{and} \quad \omega_{ab}{}^c = f_{ab}{}^c 
\end{equation}
are the geometric fluxes. At this point, it is natural to ask the question which backgrounds respectively compactifications would produce these gaugings. T-duality has a very explicit answer to this which we discuss later on.

However, it should be pointed out that this clearly does not answer the question of the necessity of non-geometric fluxes. E.g. in~\cite{Shelton:2005cf,Aldazabal:2006up}, the SUGRAs are compactified from $D=10,11$ dimensions down to four dimensions with a geometric approach. These higher dimensional theories, as we already discussed earlier, can be seen as the low energy effective limit of string theory. Everyone of these compactifications yields a flux action with only geometric fluxes.

Now, carrying out duality transformations at the level of the four dimensional effective theory, one observes that even though the original actions have been related by dualities, the effective theories are not connected anymore~\cite{Shelton:2005cf,Aldazabal:2006up}. As a result, we have to introduce new non-geometric fluxes, otherwise the two theories would not match. Thus, gaugings appearing geometric in one duality frame might be non-geometric in another.

For instance, it is possible to connect a toroidal background with a two-form flux $H_{123}$, by performing a $h^{(3)}$ T-duality in the $y^3$-direction using the Buscher rules~\ref{sec:buscherrules}, with a twisted torus with metric flux $\omega_{12}{}^3$~\cite{Aldazabal:2013sca}. Furthermore, one can perform an additional T-duality transformation $h^{(2)}$ in the $y^2$-direction as the latter case still possesses an isometry. 
Doing this, we find the globally non-geometric flux $Q_1{}^{23}$.
In general, this gives rise to the duality chain
\begin{equation}
H_{abc} \overset{h^{(c)}}{\longleftrightarrow} \omega_{ab}{}^c \overset{h^{(b)}}{\longleftrightarrow} Q_a{}^{bc}\,.
\end{equation}
Nevertheless, the background associated to the $Q_1{}^{23}$ flux depends only on the $y^1$-direction and subsequently when looking at the monodromy $y^1 \rightarrow y^1 + 1$ does not map onto itself. This non-trivial mixing of the metric and two-form is called a T-fold~\cite{Dabholkar:2005ve,Hull:2006va,Hull:2006qs}. These backgrounds are globally ill-defined from a SUGRA perspective, as the T-duality element required to glue the two different patches is not an element of the geometric subgroup of O($3,3$)~\cite{Aldazabal:2013sca}. However, from the doubled space perspective, which we are going to introduce in chapter~\ref{Kap_1}, this obstacle does not occur if one admits transition of the full O($3,3$) symmetry group. Then, the monodromy identifications of the coordinates also include the dual ones and the generalized vielbein is globally well defined.

If we were able to perform an additional T-duality in the $y^1$-direction, we would obtain the full duality chain
\begin{equation}
\label{int:dualitychain}
H_{abc} \overset{h^{(c)}}{\longleftrightarrow} \omega_{ab}{}^c \overset{h^{(b)}}{\longleftrightarrow} Q_a{}^{bc} \overset{h^{(a)}}{\longleftrightarrow} R^{abc}
\end{equation}
and would have found the locally non-geometric $R$-flux. In this case, the arising background would have to depend on a dual coordinate which results in a loss of locality in terms of the physical coordinates SUGRA is based on. Although, with a doubled space at hand, this is not an issue either.

All of these gaugings appearing in the duality chain \eqref{int:dualitychain} however belong to the same orbit and are therefore indistinguishable by the theory as they are all connected by T-duality. This implies that the backgrounds in this orbit can all be seen as geometric since we were always able to find a geometric uplift. The situation is different once geometric and non-geometric fluxes are turned on simultaneously. Then, T-duality would replace geometric by non-geometric fluxes and vice versa. As a consequence, we were never able to eliminate the non-geometric ones. These belong to the orbit of non-geometric fluxes~\cite{Dibitetto:2012rk} which cannot be reached by standard Scherk-Schwarz reductions. Such orbits are actually the most fascinating as they avoid all no-go theorems preventing moduli stabilization, de Sitter vacua, etc.~\cite{Aldazabal:2013sca,Blumenhagen:2015kja,Blumenhagen:2015xpa,Blumenhagen:2015jva,Blumenhagen:2016bfp,Blumenhagen:2017nmu,Danielsson:2012et,Danielsson:2012by,Blaback:2013ht,Damian:2013dwa,Damian:2013dq,deCarlos:2009fq,deCarlos:2009qm,Shelton:2006fd,Micu:2007rd,Palti:2007pm,Becker:2006ks}.

Moreover, as soon as we attempt to consider non-geometric backgrounds~\citep{Hassler:2014sba, Blumenhagen:2010hj, Lust:2010iy, Blumenhagen:2011ph, Condeescu:2012sp, Andriot:2012vb, Mylonas:2012pg, Bakas:2013jwa, Blumenhagen:2013zpa, Mylonas:2013jha}, the SUGRA action~\eqref{SUSY} becomes obscured due to the interplay between momentum and winding modes. However, if these non-geometric backgrounds are T-Dual to geometric ones, as we previously discussed, it is always possible to perform a field redefinition and lift these aberrations in order to acquire a well-behaved geometric description for them~\citep{Andriot:2011uh, Andriot:2012an, Andriot:2012wx, Blumenhagen:2012nk, Blumenhagen:2012nt, Blumenhagen:2013aia, Andriot:2013xca, Hassler:2014sba}. Unfortunately, traditional methods such as non-linear sigma models break down and cannot be applied to reproduce those backgrounds~\citep{Hull:2014mxa}. But this raises the question of, how we can approach these non-geometric representations.

This is one of the starting points of Double/Exceptional Field Theory. They attempt to overcome the issues with ill defined non-geometric backgrounds.

\section{Outline and Summary}

This thesis is based on the papers~\cite{Blumenhagen:2015zma,Bosque:2015jda,Bosque:2016fpi,Blumenhagen:2017noc,Bosque:2017dfc} and organized as follows
\begin{itemize}
\item In the chapter~\ref{Kap_1}, we review the basic ideas and principles underlying DFT. We start with the introduction of the doubled coordinates and its associated doubled space~\ref{doublecoordinates}. Afterwards, by defining the generalized diffeomorphisms, we implement the C- and D-bracket which govern the gauge algebra of DFT in~\ref{sec:C-bracket}-\ref{sec:D-bracket}. It will emphasize the importance and role of the strong constraint in this context, especially with regard to the closure of the algebra. All these steps lead to the DFT action in its generalized metric formulation~\ref{metricformulation} as well as in its flux formulation~\ref{sec:fluxformulation}. Subsequently, we analyze the corresponding symmetries and field equations.

\item The chapter~\ref{Kap_2} is designed to produce an overview of DFT$_{\mathrm{WZW}}$. Beginning with a WZW model on a group manifold, we examine the steps leading to DFT$_{\mathrm{WZW}}$. In order to fully grasp this framework it is essential to comprehend some basic concepts of Lie algebras~\ref{representationsDFT*} and Closed String Field Theory~\ref{sec:CSFT}. Then, one is able to evaluate the corresponding two-point and three-point functions which give rise to the cubic order action~\ref{sec:DFTWZWaction} and gauge transformations~\ref{sec:2gaugetrafo} of DFT$_{\mathrm{WZW}}$. Equivalently to original DFT, it is possible to introduce a gauge algebra, dictated by the C-bracket~\ref{sec:cbracket}, which closes under strong constraint for the fluctuations and Jacobi identity for the background.

\item We begin chapter~\ref{Kap_3} with a rescaling of the DFT$_{\mathrm{WZW}}$ action~\ref{sec:fieldredef}, as we want to get rid of an undesired $1/2$ factor. Consequently, we cast the theory into a more convenient form, the generalized metric formulation~\ref{dft*action*}, by introducing doubled generalized objects. This makes it easier to compare our result to those of original DFT. Thereafter, we derive the associated equation of motion and define a generalized Ricci scalar as well as a generalized curvature tensor~\ref{dft*eom}. At this point, we are able to show DFT$_{\mathrm{WZW}}$ action's invariance under generalized diffeomorphisms~\ref{GenDiffInv} and $2D$-diffeomorphisms~\ref{dft*2ddiffs}. Finally, we present an additional constraint, the extended strong constraint~\ref{subextendedstrongconstraint}, which relates DFT$_{\mathrm{WZW}}$ with the toroidal DFT formulation and analyze how they are connected~\ref{sec:reltodft}.

\item During chapter~\ref{Kap_4}, we introduce the covariant fluxes~\ref{sec:covfluxesWZW} and subsequently perform all steps required to recast the generalized metric formulation's action of DFT${}_\mathrm{WZW}$ through these fluxes~\ref{sec:FluxFormAction}. Conclusively, we argue why in DFT$_\mathrm{WZW}$ the strong violating term $1/6 F_{ABC} F^{ABC}$ known from original DFT is absent~\ref{sec:missingFABCFABCterm}. It was needed in the traditional flux formulation in order to reproduce the scalar potential of half-maximal, electrically gauged supergravities. Moreover, we show the invariance of the flux formulation under double Lorentz transformations~\ref{sec:doublelorentz}. Afterwards, we obtain the gauge transformations~\ref{sec:gaugetrafo} and field equations~\ref{sec:eom} in this formulation.

\item Chapter~\ref{sec:genSSWZW} is dedicated to generalized Scherk-Schwarz compactifications. We start with a short review of the embedding tensor formalism~\ref{sec:embeddingtensor}, in particular for n=3 dimensions, and thereafter we are going to discuss generalized Scherk-Schwarz compactifications in the context of original DFT~\ref{sec:traditionalSS}. Here, the problem of constructing the twist becomes evident. In section ~\ref{sec:genSSDFTWZW}, we introduce generalized Scherk-Schwarz compactifications for the flux formulation of DFT$_\mathrm{WZW}$. The generalized background vielbein takes on the role of the twist in our framework and can be chosen as the left invariant Maurer-Cartan form on the group manifold. We demonstrate the explicit construction procedure, beginning from an arbitrary embedding tensor solution, in section~\ref{sec:twistconstr}. We close this chapter by providing the background generalized vielbeins for all compact O($3,3$) embeddings in appendix~\ref{app:twists}.

\item In this chapter, we want to extend the DFT$_\mathrm{WZW}$ framework to gEFT. We begin by presenting an approach to implement generalized diffeomorphisms on group manifolds~\ref{sec:gendiffonG}. This question is tackled from a slightly different point of view 
than~\cite{Bosque:2016fpi}. Here, we try to keep the treatment as general as possible and only specify explicit U-duality groups when absolutely required. Subsequently, we highlight the important differences and similarities with \DFTwzw{}. Simultaneously, we introduce the relevant notation and provide a short review of the main results of DFT and EFT. In this context~\ref{sec:genLie}, we derive the corresponding two linear and the quadratic constraints from demanding closure of the gauge algebra once the SC is imposed. As we are interested in solving these constraints, we now have to fix a specific U-duality group for which we choose SL($5$)~\ref{sec:linconstsl5}. Thus, we observe how a detailed picture of the SL($5$) breaking into group manifolds with $\dim G<10$, as a result of the embedding tensor solutions in the $\overline{\mathbf{40}}$, emerges. The second part of this chapter is covered by~\ref{sec:solSC} where we want to solve the SC. To do so, we adapt the techniques known from \DFTwzw{}~\cite{Hassler:2016srl} to gEFT. The henceforth obtained SC solutions allow for a GG description which we discuss in~\ref{sec:gg}. As a consequence of these results, we know how to construct a generalized frame field $\mathcal{E}_A$~\ref{sec:genframe}. This however requires some additional linear constraint. Finally, we give some illustrative examples such as the four-torus with $G$-flux as well as the backgrounds contained in its duality chain, and the four-sphere with $G$-flux~\ref{sec:examples}
\end{itemize}\clearpage{}	
  \clearpage{}\chapter{Double Field Theory} \label{Kap_1}

In this chapter, we are going to review the most important aspects of Double Field Theory (DFT) and set the prerequisites for the upcoming chapters~\cite{Pascal2015,Blumenhagen:2017noc}. We start by introducing the doubled space with its gauge algebra \ref{sec:dftgaugealg}, governed by the C-bracket, and see how this setup neatly gives rise to the action in both formulations of DFT. Subsequently, we consider the action in its generalized metric formulation~\ref{metricformulation} with its associated symmetries and are going to discuss the equation of motions which arise after variation. Afterwards, we follow an analogous argumentation for the flux formulation~\ref{sec:fluxformulation} of DFT. The flux formulation allows for a relaxation of the strong constraint by replacing it with the weaker closure constraint.

\section{Double coordinates}
\label{doublecoordinates}

Let us start by extending all the notions and principles we introduced in the introduction~\ref{Kap_0} into a T-Duality invariant formulation of DFT. In order to make T-Duality a manifest symmetry of the theory, we have to introduce so-called doubled coordinates~\citep{Tseytlin:1990va, Siegel:1993th, Hull:2009mi} given on a toroidal background $\mathbb{R}^{2n-2,2}\times T^{2d}$. For closed string theory, this is by construction a Double Field Theory~\citep{Hohm:2010jy}. It means that in addition to our $D$ spacetime coordinates $x^i$, $D=n+d$, which are conjugate to the momentum modes, we incorporate $D$ new coordinates $\tilde{x}_i$ that are conjugate to the winding modes into the doubled space. For us to be able to write down a covariant Double Field Theory action, we combine these two coordinate types to $2D$-dimensional generalized coordinates by
\begin{equation}
X^M = \begin{pmatrix}
\tilde{x}_i \\ x^i
\end{pmatrix}\,,\quad X_M = \begin{pmatrix}
 x^i \\ \tilde{x}_i
\end{pmatrix}\,.
\end{equation}
We raise and lower the indices with the O($D,D$) invariant metric and its inverse
\begin{equation}
\label{eta}
\eta^{MN} = \begin{pmatrix}
0 & {\delta_i}^{j} \\ {\delta^i}_{j} & 0
\end{pmatrix}\,,\quad \eta_{MN} = \begin{pmatrix}
0 & {\delta^i}_{j} \\  {\delta_i}^{j} & 0
\end{pmatrix}\,.
\end{equation}
Furthermore, we have to define according generalized partial derivatives
\begin{equation}
\partial^M =
\begin{pmatrix}
 \partial_i \\ \tilde{\partial}^i 
\end{pmatrix}\,,\quad
\partial_M =
\begin{pmatrix}
 \tilde{\partial}^i \\  \partial_i
\end{pmatrix}
\end{equation}
as well. Naturally, we also have to consider a new generalized metric $\mathcal{H}_{MN}$~\eqref{eqn:introgenmetr}~\citep{Hohm:2010pp} made up of the metric $g_{ij}$ and the two-form $b_{ij}$ as well. It is given by
\begin{equation}
\label{metric}
\mathcal{H}_{MN} = \begin{pmatrix}
g^{ij} & -g^{ik}b_{kj} \\ b_{ik}g^{kj} & g_{ij} - b_{ik}g^{kl}b_{lj}
\end{pmatrix}\,.
\end{equation}
Clearly, the generalized metric lies in O($D,D$) ($\mathcal{H} \in $O$(D,D)$) and satisfies the following identities
\begin{equation}
\label{metric2}
\mathcal{H}^T \eta \, \mathcal{H} = \eta\,,\quad \mathcal{H}^{MN} = \eta^{MP} \eta^{NQ} \mathcal{H}_{PQ}\,,\quad \mathcal{H}^{MN} \mathcal{H}_{NP} = {\delta^M}_{P}\,.
\end{equation}
Moreover, the dilaton $\phi$ combined with the determinant of the metric $g$ transforms as an O($D,D$) scalar, particularly for the dilaton $d$ we have
\begin{equation}
e^{-2d} = \sqrt{g} \, e^{-2\phi}\,.
\end{equation}
As a consequence, our generalized fields are given by the field content $\mathcal{H}(X)$ and $d(X)$~\citep{Aldazabal:2013sca}. (With the additional restriction that they have to fulfill the strong constraint~\eqref{eqn:DFTstrongconstraint})

It is worth noting that the mass formula takes on the form
\begin{equation}
M^2 = \big( N + \tilde{N} - 2 \big) + \mathcal{P}^M \mathcal{H}_{MN} \mathcal{P}^N\,,
\end{equation}
while the level matching condition (LMC) becomes
\begin{equation}
N - \tilde{N} = \frac{1}{2} \mathcal{P}^M \mathcal{P}_M\,,
\end{equation}
where
\begin{equation}
\mathcal{P}^{M} = \begin{pmatrix}
\tilde{p}_i \\ p^i
\end{pmatrix}\,.
\end{equation}
\label{strongconstraint}This immediately raises the questions, whether we can formulate a consistent theory out of these constituents and whether there exists a procedure to recover supergravity? For a consistent formulation of DFT it is necessary to constrain the coordinate dependency of the doubled space. This constraint is called \textit{weak constraint} and is originating in the CSFT level matching condition $L_0 - \bar{L}_0 = 0$. It is a remainder of the toroidal background of the theory. (As we see later, we even have to impose a much more restrictive strong constraint.) 

A field at levels $N,\tilde{N}$ generally fulfills $\partial^M \partial_M A = N - \tilde{N}$~\citep{Hull:2014mxa}. Since we are only interested in massless states, the constraint reads
\begin{equation}
\partial^M \partial_M A = 0\,,\quad\forall\;\text{fields}\;A
\end{equation}
or for the components
\begin{equation}
\partial_i \tilde{\partial}^i (...) = 0\,,\quad\forall\;\text{fields}\;A\,.
\end{equation}
The weak constraint is invariant under T-Duality or any other O($D,D$) rotations as well, because $\eta_{MN}$ is an invariant under O($D,D$) transformations. It has always to be satisfied. One way to solve this constraint is $\tilde{\partial}^i (...) = 0$. 
In turn, it can be seen as if the fields are independent of the winding coordinates $\tilde{x}_i$, once the weak constraint is imposed. Thus, the fields live on a $D$-dimensional subspace of the $2D$-dimensional doubled space-time~\citep{Hohm:2010pp, Hassler:2014sba}. 

Another much more powerful constraint which has to be invoked is the \textit{strong constraint}. It takes on the following form for generic field products
\begin{equation}
\label{eqn:DFTstrongconstraint}
\partial^M \partial_M ( A \cdot B ) = 0\,,\quad\forall\;\text{fields}\;A\,,B\,.
\end{equation}
Clearly, this constraint is invariant under global O($D,D$) transformations, too. However, it highly truncates the theory and makes it possible to construct a Double Field Theory in all orders~\citep{Hull:2014mxa, Hohm:2010pp, Hull:2009zb, Hohm:2010jy}. It is a direct consequence of the level-matching condition during string scattering processes.  

We will use the strong constraint as a way to check consistency with SUGRA~\ref{SUGRA} in the remainder of this chapter. We are going to see that it reduces the DFT action to the well-known SUGRA action~\eqref{SUSY}. The importance of the strong constraint is going to become much more obvious during the course of this thesis. Nevertheless, it is worth noting that the strong constraint is generally not invariant under local O($D,D$) transformations. We will always comment whether the strong constraint is invoked or not.

\label{vielbeinsss}Furthermore, we can decompose the metric~\bref{metric} through generalized vielbeins ${E^A}_{M}$ in an O($D,D$) generalized frame~\citep{Hassler:2014sba}. We find
\begin{align}
\mathcal{H}_{MN} &= {E^A}_{M} \delta_{AB} {E^B}_{N}\,, \\
\eta_{MN} &= {E^A}_{M} \eta_{AB} {E^B}_{N}\,,
\end{align}
with $\eta$ defined as in~\bref{eta}, indicating ${E^A}_{M} \in O(D,D)$ as well. The delta $\delta_{AB}$ is given by
\begin{equation}
\delta_{AB} = \begin{pmatrix}
\eta^{ab} & 0 \\ 0 & \eta_{ab}
\end{pmatrix}\,.
\end{equation}
Without gauge fixing it is possible to express any vielbein in terms of a vielbein belonging to $g_{ij} = {e^a}_{i} \eta_{ab} {e^b}_{j}$, a two-form field $b_{ij}$, and an antisymmetric bi-vector $\beta^{ij}$
\begin{equation}
\label{vlbn}
{E^A}_{M} = \begin{pmatrix}
{e_a}^{i} & {e_a}^{l} b_{li} \\ {e^a}_{l} \beta^{li}  & {e^a}_{i} + {e^a}_{l} \beta^{lk}b_{ki}
\end{pmatrix}\,.
\end{equation}
The bi-vector can be gauged away by imposing local double Lorentz symmetry $H = O(1,D-1) \times O(1,D-1)$. Evaluating the coset $G/H$ reduces the number of generators for $G = O(D,D)$ elements from $d(2D-1)$ to $D^2$, and it casts equation~\bref{vlbn} into upper triangular form~\citep{Aldazabal:2013sca} by
\begin{equation}
\label{vielbeingauge}
{E^A}_{M} = \begin{pmatrix}
{e_a}^{i} & {e_a}^{l} b_{li} \\ 0  & {e^a}_{i}
\end{pmatrix}\,.
\end{equation}
Now, we turn to generalizing the SUGRA diffeomorphisms into the DFT framework. 

\section{Generalized diffeomorphisms}
\label{generalizeddiffs}

In the previous section~\ref{doublecoordinates} we introduced the concept of a generalized metric, consisting of a symmetric metric $g_{ij}$ and an antisymmetric $B$-field $B_{ij}$. Further, from the introduction~\ref{SUGRA} we are already familiar with how these two fields transform under diffeomorphisms, and gauge transformations of the $B$-Field. This raises the questions in what extend we can combine these two features into DFT. By introducing so-called \textit{generalized diffeomorphisms}~\citep{Aldazabal:2013sca, Hohm:2010pp} we can achieve this goal. Therefore, let us consider the following generalization of the gauge parameter~\bref{lie1}
\begin{equation}
\xi^M = \begin{pmatrix}
\tilde{\lambda}_i \\ \lambda^i
\end{pmatrix}\,,
\end{equation}
where through doubling of the underlying manifold the tangent bundles $TM$ and $T^*M$ are put on an equal standing~\citep{Hull:2009mi}. It is possible to cast the generalized diffeomorphisms of the generalized metric in an manifest O($D,D$) covariant way by
\begin{equation}
\label{metricdiff}
\delta_{\xi} \mathcal{H}^{MN} = \xi^P \partial_P \mathcal{H}^{MN} + \big(\partial^M \xi_P - \partial_P \xi^M \big) \mathcal{H}^{PN} + \big(\partial^N \xi_P - \partial_P \xi^N \big) \mathcal{H}^{MP}\,.
\end{equation}
We raised and lowered indices with the O($D,D$) metric $\eta^{MN}$~\bref{eta}.

From here it is quite obvious that these diffeomorphisms act in a fashion similar to a Lie derivative, suggesting the identification of a \textit{generalized Lie derivative} generating gauge transformations through
\begin{equation}
\delta_{\xi} \mathcal{H}^{MN} = \mathcal{\widehat{L}}_{\xi} \mathcal{H}^{MN}\,.
\end{equation}
This allows us to define a generalized Lie derivative acting on arbitrary \textit{generalized tensors} $A^{M...N}_{P...Q}$, i.e. for a tensor with one upper and one lower index through
\begin{equation}
\label{lie2}
 \mathcal{\widehat{L}}_{\xi} {\mathcal{A}_{M}}^{N} = \xi^P \partial_P {\mathcal{A}_{M}}^{N} + \big(\partial_M \xi^P - \partial^P \xi_M \big) {\mathcal{A}_{P}}^{N} + \big(\partial^N \xi_P - \partial_P \xi^N \big) {\mathcal{A}_{M}}^{P}\,.
\end{equation}
In fact, the change from the standard Lie derivative to the generalized Lie derivative is essential to preserve the O$(D,D)$ symmetry group. For example, just a term $- \partial^P \xi_M {\mathcal{A}_{P}}^{N}$ in~\bref{lie2} would be incompatible with the O($D,D$) symmetry. Hence, to protect the invariance, the term needs to be projected into the representation of O$(D,D)$.

Now, it is quite easy to verify that the generalized Lie derivative applied to the O($D,D$) metric $\eta^{MN}$ and the Kronecker delta ${\delta_M}^{N}$ vanish
\begin{equation}
\label{lieeta}
\mathcal{\widehat{L}}_{\xi} \eta^{MN} = 0\,,\quad \mathcal{\widehat{L}}_{\xi} {\delta_M}^{N} = 0\,,
\end{equation}
e.g.
\begin{align}
 \mathcal{\widehat{L}}_{\xi} {\delta_M}^{N} &= \xi^P \partial_P {\mathcal{\delta}_{M}}^{N} + \big(\partial_M \xi^P - \partial^P \xi_M \big) {\mathcal{\delta}_{P}}^{N} + \big(\partial^N \xi_P - \partial_P \xi^N \big) {\mathcal{\delta}_{M}}^{P} \\
&= \big(\partial_M \xi^N - \partial^N \xi_M \big) + \big(\partial^N \xi_M - \partial_M \xi^N \big) = 0\,. \nonumber
\end{align}
Furthermore, using the Leibniz rule, the product of two arbitrary tensors $A^{M...N}_{P...Q}$, and $B^{R...S}_{T...U}$  can be decomposed as
\begin{equation}
\mathcal{\widehat{L}}_{\xi} \big( A^{M...N}_{P...Q} B^{R...S}_{T...U} \big) = \mathcal{\widehat{L}}_{\xi} \big( A^{M...N}_{P...Q} \big) B^{R...S}_{T...U}  +  A^{M...N}_{P...Q} \mathcal{\widehat{L}}_{\xi} \big( B^{R...S}_{T...U} \big)\,.
\end{equation}
For vectors $A_M$, and $A^M$ this gives us
\begin{equation*}
\mathcal{\widehat{L}}_{\xi} A_M =  \xi^P \partial_P A_{M} + \big(\partial_M \xi^P - \partial^P \xi_M \big) A_{P}\,,
\end{equation*}
\begin{equation}
\mathcal{\widehat{L}}_{\xi} A^M = \xi^P \partial_P A^{M} + \big(\partial^M \xi_P - \partial_P \xi^M \big) A^{P}\,.
\end{equation}
Moreover, we also have to act with the generalized Lie derivative on the dilaton
\begin{equation}
\mathcal{\widehat{L}}_{\xi} e^{-2d} = \partial_M \big(\xi^M e^{-2d}\big)\,,
\end{equation}
which shows that $e^{-2d}$ transforms as a density. Subsequently, we can use this result to apply it onto the O($D,D$) condition
\begin{equation}
\mathcal{H}^T \eta \mathcal{H} = \eta^{-1}
\end{equation}
and obtain
\begin{equation}
\mathcal{\widehat{L}}_{\xi} \big(\mathcal{H}\big) \eta \mathcal{H} + \mathcal{H} \eta \mathcal{\widehat{L}}_{\xi}  \big(\mathcal{H}\big) = 0\,.
\end{equation}
This confirms that the O($D,D$) condition is preserved under generalized diffeomorphisms and that they are compatible with the gauge symmetries~\citep{Hohm:2010pp}.
It can be further shown, a fully equivalent way of writing equation~\bref{metricdiff} is
\begin{equation}
\label{eqn:genlieytensor}
\mathcal{\widehat{L}}_{\xi} \mathcal{H}_{MN} = L_{\xi} \mathcal{H}_{MN} + Y^R{}_M{}^P{}_Q \partial^Q \xi_P \mathcal{H}_{RN} + Y^R{}_N{}^P{}_Q \partial^Q \xi_P \mathcal{H}_{MR}\,,
\end{equation}
with the deviance of Riemann geometry $Y^M{}_P{}^N{}_Q$ given by
\begin{equation}
Y^M{}_P{}^N{}_Q= \eta^{MN} \eta_{PQ}\,,
\end{equation}
where $L_{\xi}$ labels the standard Lie derivative in $2D$ dimensions~\citep{Aldazabal:2013sca}. (This form is of great importance for the Exceptional Field Theory setup discussed in the last chapter, where the $Y$-tensor takes on a different form.)

In order not to spoil the O($D,D$) symmetry group it becomes necessary to introduce $Y^M{}_P{}^N{}_Q$, which projects onto the adjoint representation of O($D,D$). With the help of the strong constraint~\bref{strongconstraint} some quantities can be evaluated (Under the assumption that a vector field $A^M$, and the gauge parameters $\xi^P$ do not depend on the dual coordinates $\tilde{x}_i$). A short computation shows
\begin{equation}
\mathcal{\widehat{L}}_{\xi} A^M = \begin{pmatrix}
L_{\xi} \tilde{A}_i + (\partial_i \tilde{\xi}_j - \partial_j \tilde{\xi}_i) A^j \\ L_{\xi} A^i
\end{pmatrix}\,.
\end{equation}
We can also apply the generalized Lie derivative to the gauged vielbein~\bref{vielbeingauge}. This yields
\begin{align}
\label{lievielbein}
\mathcal{\widehat{L}}_{\xi} {E^{A}}_{M} &= \xi^P \partial_P {E^{A}}_{M} + \big(\partial_M \xi^P - \partial^P \xi_M \big) {E^{A}}_{P} \nonumber \\ 
&= \begin{pmatrix}
L_{\xi} {e_a}^{i} & \big(L_{\xi} {e_a}^{j}) b_{ji} + {e_a}^{j} \big[ L_{\xi} b_{ji} + \big( \partial_i \tilde{\xi}_j - \partial_j \tilde{\xi}_i \big) \big] \\ 0 & L_{\xi} {e^a}_{i}
\end{pmatrix}\,.
\end{align}
A rather involved calculation shows
\begin{equation}
\label{liemetric}
\mathcal{\widehat{L}}_{\xi} \mathcal{H}^{MN} = \begin{pmatrix}
L_{\xi} g_{ij} - \big(L_{\xi} b_{ik} + \partial_i \tilde{\xi}_k - \partial_k \tilde{\xi}_i\big) g^{kl} b_{lj} & \big(L_{\xi} b_{ik} + \partial_i \tilde{\xi}_k - \partial_k \tilde{\xi}_i \big) g^{kj} \\ -b_{ik} g^{kl} \big(L_{\xi} b_{lj} + \partial_l \tilde{\xi}_j - \partial_j \tilde{\xi}_l \big) & +b_{ik} \big( L_{\xi} g^{kj} \big)\\ & \\ -\big(L_{\xi} g^{ik}\big) b_{kj} & L_{\xi} g^{ij} \\ - g^{ik}\big(L_{\xi}b_{kj} + \partial_k \tilde{\xi}_j - \partial_j \tilde{\xi}_k \big) &
\end{pmatrix}\,.
\end{equation}
From which we can recover the local SUGRA diffeomorphisms~\bref{lie1}. For the metric
\begin{equation}
\mathcal{\widehat{L}}_{\xi} g_{ij} = L_{\xi} g_{ij}\,,
\end{equation}
and for the b-field
\begin{equation}
\mathcal{\widehat{L}}_{\xi} b_{ij} = L_{\xi} b_{ij} + \partial_i \tilde{\xi}_j-\partial_j \tilde{\xi}_i\,,
\end{equation}
by comparison with
\begin{equation}
\mathcal{\widehat{L}}_{\xi} \mathcal{H}^{MN} = \begin{pmatrix}
\mathcal{\widehat{L}}_{\xi} g_{ij} - \big( \mathcal{\widehat{L}}_{\xi} b_{ik} \big) g^{kl} b_{lj} - b_{ik} g^{kl} \big( \mathcal{\widehat{L}}_{\xi} b_{lj} \big) & \big( \mathcal{\widehat{L}}_{\xi} b_{ik} \big) g^{kj} + b_{ik} \big( \mathcal{\widehat{L}}_{\xi} g^{kj} \big) \\ & \\ - \big( \mathcal{\widehat{L}}_{\xi} g^{ik} \big)b_{kj} - g^{ik} \big( \mathcal{\widehat{L}}_{\xi} b_{kj} \big) & \mathcal{\widehat{L}}_{\xi} g^{ij}
\end{pmatrix}\,.
\end{equation}
Altogether, we retrieve the transformation properties known from the SUGRA frame, once we impose the strong constraint. 
Nevertheless, one should pay attention to the fact that certain quantities such as $\partial_M V^N$ transform non-covariantly~\citep{Aldazabal:2013sca} under the generalized Lie derivative. We can represent these transformations through
\begin{equation}
\delta_\xi \big( \partial_M V^N \big) = \partial_M \big( \mathcal{\widehat{L}}_{\xi} V^N  \big)\,,
\end{equation}
suggesting that generalized diffeomorphisms only act on tensorial quantities. As a consequence, it is possible to define the failure of an object to transform non-tensorically, specifically
\begin{equation}
\label{failure}
\Delta_\xi \equiv \delta_\xi - \mathcal{\widehat{L}}_{\xi}\,.
\end{equation}

\section{Gauge algebra}
\label{sec:dftgaugealg}

In this section, we want to focus on the gauge algebra generated through non-linear acting, generalized Lie derivatives. For the gauge algebra to close, we investigate the required consistency constraints, while the transformations itself should remain O($D,D$) covariant~\citep{Siegel:1993th, Hull:2009zb}. As can already be seen from equation~\bref{lie2} the ordinary Lie-bracket~\bref{lie1} has to be modified to $C$-and $D$-brackets~\ref{sec:C-bracket},~\ref{sec:D-bracket}, which are generalizations of the Courant-and Dorfman brackets~\citep{courant}.

\subsection{C-bracket}
\label{sec:C-bracket}

The C-bracket governs the gauge algebra on the doubled space generated by generalized Lie derivatives~\citep{Hull:2009zb, Siegel:1993th}. It can be seen as an O($D,D$) covariant extension of the Courant-bracket. Restricting the fields independent of the winding coordinates, the C-bracket reduces precisely to the Courant-bracket~\citep{Hohm:2010pp, courant}. Hence, the antisymmetric C-bracket is given by
\begin{equation}
\big[\xi_1, \xi_2 \big]_C^M \equiv \xi^N_{[1} \partial_N \xi^M_{2]} - \frac{1}{2} \xi^P_{[1} \partial^M \xi_{2]P}\,,
\end{equation}
with $[ij] = ij - ji$. Now, we evaluate the commutator algebra created by the generalized Lie derivative. Acting on an arbitrary test vector $A_M$ yields
\begin{equation}
\big[ \mathcal{\widehat{L}}_{\xi_1} , \mathcal{\widehat{L}}_{\xi_2} \big] A_M = \mathcal{\widehat{L}}_{[\xi_1, \xi_2]_C} A_M + F_M \big( \xi_1, \xi_2, A \big)\,,
\end{equation}
and
\begin{equation}
F_M \big( \xi_1, \xi_2, A \big) = \frac{1}{2} \xi_{[1 N} \partial^Q \xi_{2]}^N \partial_Q A_M - \partial^Q \xi_{[1M} \partial_{Q} \xi_{2]}^P A_P\,.
\end{equation}
Once the strong constraint~\bref{strongconstraint} is imposed, implying $F_M \big( \xi_1, \xi_2, A \big) = 0$, it closes
\begin{equation}
\label{c-bracket}
\big[ \mathcal{\widehat{L}}_{\xi_1} , \mathcal{\widehat{L}}_{\xi_2} \big] A_M = \mathcal{\widehat{L}}_{[\xi_1, \xi_2]_C} A_M\,.
\end{equation}
This relation holds as well for arbitrary tensor fields $A^{M...N}_{P...Q}$ by iterating this relation~\bref{c-bracket}. For their products we have
\begin{equation}
\big[ \mathcal{\widehat{L}}_{\xi_1} , \mathcal{\widehat{L}}_{\xi_2} \big] \big( A^{M...N}_{P...Q} B^{R...S}_{T...U} \big) = \mathcal{\widehat{L}}_{[\xi_1, \xi_2]_C} \big( A^{M...N}_{P...Q} B^{R...S}_{T...U} \big)\,,
\end{equation}
since $A^M = \eta^{MN} A_N$ as well as $\mathcal{\widehat{L}}_{\xi} \eta^{MN} = 0$ with
\begin{equation}
\big[ \mathcal{\widehat{L}}_{\xi_1} , \mathcal{\widehat{L}}_{\xi_2} \big] A^M = \big[ \mathcal{\widehat{L}}_{\xi_1} , \mathcal{\widehat{L}}_{\xi_2} \big] \big( \eta^{MN} A_N \big) = \eta^{MN} \mathcal{\widehat{L}}_{[\xi_1, \xi_2]_C} A_N = \mathcal{\widehat{L}}_{[\xi_1, \xi_2]_C} A^M\,.
\end{equation}
Subsequently, the gauge transformations close for $\xi_{12} = [\xi_1, \xi_2]_C$, and therefore under the strong constraint
\begin{equation}
\big[ \delta_{\xi_1} , \delta_{\xi_2} \big] A^{M...N}_{P...Q} = \delta_{\xi_{12}} A^{M...N}_{P...Q}\,, 
\end{equation}
for arbitrary tensor fields. However, the Jacobiator of the C-bracket
\begin{equation}
J(\xi_1, \xi_2, \xi_3) = \big[ {\xi_1} , \big[ {\xi_2}, {\xi_3} \big]_C \big]_C + \text{cyclic}
\end{equation}
unfortunately does not vanish and in turn this implies that the C-bracket cannot generate a Lie algebra. Nevertheless, the failure to fulfill the Jacobi identity is only a trivial gauge transformation that leaves all fields satisfying the strong constraint invariant.

\subsection{D-bracket}
\label{sec:D-bracket}
It is intuitive to introduce an analogon to the ordinary Lie bracket, which is given through
\begin{equation}
\big[X,Y \big] = L_{X} Y\,,
\end{equation}
by introducing an additional generalized Lie bracket that is invariant under O($D,D$) transformations, the so-called \textit{D-bracket}~\citep{Hull:2009zb, Hohm:2010pp}
\begin{equation}
\label{D-bracket}
\big[A,B \big]_D = \mathcal{\widehat{L}}_A B\,.
\end{equation}
As can be checked easily, we can recast~\bref{D-bracket}, using the C-bracket, in the following way
\begin{equation}
\big[A,B \big]_D^M = \big[A, B \big]_C^M + \frac{1}{2} \partial^M ( B_N A^N )\,.
\end{equation}
The structure of the last term is quite similar to a gauge parameter. When restricting the fields, as in the case for the C-bracket, the D-bracket reduces to the well-known Dorfman bracket.
Intriguingly, as opposed to the C-bracket which does not satisfy the Jacobi identity, the D-bracket fulfills the Jacobi like identity~\citep{Hohm:2010pp}
\begin{equation}
\big[A, \big[B, C\big]_D\big]_D = \big[\big[A, B\big]_D, C\big]_D + \big[B, \big[A, C\big]_D\big]_D\,,
\end{equation}
and much further, it satisfies the Jacobi identity~\citep{Aldazabal:2013sca}.

Now, that we are equipped with the gauge algebra we can turn to writing down and analyzing the Double Field Theory action.

\section{Action}

There exist two descriptions to write down an O($D,D$) invariant action incorporating generalized diffeomorphism invariance~\ref{generalizeddiffs}. Following~\citep{Hohm:2010pp}, it is possible to construct a gauge invariant action, subsection~\ref{metricformulation}, in terms of the generalized metric~\bref{metric}. Subsequently, this action must be manifestly O($D,D$) invariant while it also has to possess an additional $\mathbb{Z}_2$ symmetry. At this point, we see that a generalized scalar curvature is going to emerge. On the other hand, the Double Field Theory action can also be expressed in terms of so-called fluxes~\citep{Geissbuhler:2011mx,Hohm:2010xe}, as we observe in the latter part of this section~\ref{sec:fluxformulation}. In the end, both formulations turn out to be entirely equivalent up to a total derivative.

\subsection{Generalized metric formulation}
\label{metricformulation}

We are given the O($D,D$) transformation properties of $\mathcal{H}^{MN}$, the invariant metric $\eta^{MN}$, the partial derivatives $\partial_M$, and the dilaton $d$ out of which we have to build the action~\cite{Hohm:2010pp}. In order to write it down, we start by 
constructing the corresponding O($D,D$) scalars by contracting all indices appropriately. There are various scalars to write down by considering that the O($D,D$) transformations are acting globally.
As a result of the globality of the O($D,D$) transformations it is not very difficult to handle the transformation properties of the partial derivatives

Just to name two examples
\begin{equation}
\label{blablabla}
\mathcal{H}^{MN} \partial_M \partial_N d\,,\quad \partial^K \mathcal{H}^{MN} \partial_M \mathcal{H}_{KN}\,.
\end{equation}
Specifically, we need to find all the terms containing two partial derivatives and are gauge invariant. Furthermore, we can use our additional knowledge about the presence of an extra $\mathbb{Z}_2$ symmetry. This highly reduces the amount of viable terms. Originating in the antisymmetry of the 2-form $b_{ij} \mapsto -b_{ij}$, this symmetry results in $\tilde{x}_i \mapsto -\tilde{x}_i$ and $\tilde{\partial}^i \mapsto -\tilde{\partial}^i$~\citep{Hohm:2010pp}. In terms of double coordinates
\begin{equation}
\partial_M = \begin{pmatrix}
\tilde{\partial}^i \\ \partial_i
\end{pmatrix}\,,
\end{equation}
the $\mathbb{Z}_2$ symmetry can be expressed through the matrix representation
\begin{equation}
\partial_{\bullet} \mapsto Z \partial_{\bullet}\,,\quad Z = \begin{pmatrix}
-1 & 0 \\ 0 & 1
\end{pmatrix}\,.
\end{equation}
Here, $\partial_\bullet$ denotes the column vector corresponding to $\partial_M$. The matrix $Z$ satisfies the following additional conditions
\begin{equation}
Z = Z^T = Z^{-1}\,,\quad Z^2 = 1\,.
\end{equation}
Transforming $b_{ij} \mapsto -b_{ij}$ leads to a flip in the signature of the off-diagonal terms contained in the generalized metric $\mathcal{H}^{MN}$, as do the off-diagonal elements of $\mathcal{H}_{MN}$. We achieve this transformation behavior through the identifications
\begin{equation}
\mathcal{H}^{\bullet\bullet} \mapsto Z\mathcal{H}^{\bullet\bullet}Z\,,\quad \mathcal{H}_{\bullet\bullet} \mapsto Z\mathcal{H}_{\bullet\bullet}Z\,.
\end{equation}
For instance,
\begin{align}
Z\mathcal{H}^{MN}Z &= \begin{pmatrix}
-1 & 0 \\ 0 & 1
\end{pmatrix}
\begin{pmatrix}
g^{ij} & -g^{ik}b_{kj} \\ b_{ik}g^{kj} & g_{ij} - b_{ik}g^{kl}b_{lj}
\end{pmatrix}
\begin{pmatrix}
-1 & 0 \\ 0 & 1
\end{pmatrix} \nonumber \\
&= \begin{pmatrix}
g^{ij} & g^{ik}b_{kj} \\ -b_{ik}g^{kj} & g_{ij} - b_{ik}g^{kl}b_{lj}
\end{pmatrix}\,.
\end{align}
Clearly, $Z$ is not an element of O($D,D$) since
\begin{equation}
Z^T \eta^{\bullet \bullet} Z \neq \eta^{\bullet \bullet}\,,\quad Z^T \eta_{\bullet \bullet} Z \neq \eta_{\bullet \bullet}\,,
\end{equation}
i.e.
\begin{align}
Z^T \eta^{MN} Z &= \begin{pmatrix}
-1 & 0 \\ 0 & 1
\end{pmatrix}
\begin{pmatrix}
0 & 1 \\ 1 & 0
\end{pmatrix}
\begin{pmatrix}
-1 & 0 \\ 0 & 1
\end{pmatrix} \nonumber \\
&= \begin{pmatrix}
0 & -1 \\ -1 & 0
\end{pmatrix} = - \eta^{MN}\,.
\end{align}
Now, we want to write down all possible combinations of $\mathbb{Z}_2$ invariant terms, i.e. the last term in~\bref{blablabla} is not invariant under $\mathbb{Z}_2$. Moreover, terms with two terms of the generalized metric and more than two derivatives cannot exist without breaking the $\mathbb{Z}_2$ invariance.

The dilaton and generalized metric can be combined to give four possible terms, particularly
\begin{equation}
\partial_M d\, \partial_N \mathcal{H}^{MN}\,,\quad \mathcal{H}^{MN} \partial_M d\, \partial_N d\,,\quad \mathcal{H}^{MN} \partial_M \partial_N d\,,\quad \partial_M \partial_N \mathcal{H}^{MN}\,.
\end{equation}
From the multiplication of the last term with the dilaton factor $e^{-2d}$ it can be seen as an interaction term. The last two terms turn out to be related by partial integration to the first two.
Thus, it reduces the amount of possible combinations even more
\begin{equation}
\partial_M d\, \partial_N \mathcal{H}^{MN}\,,\quad \mathcal{H}^{MN} \partial_M d\, \partial_N d\,.
\end{equation}
As we already know that there do not exist terms containing two generalized metrics, we turn to search for terms containing three of them. Due to the $\mathbb{Z}_2$ constraint, they have to be built without $\eta$. Hence, the following two possible candidates present themselves
\begin{equation}
\mathcal{H}^{MN} \partial_M \mathcal{H}^{KL} \partial_{N} \mathcal{H}_{KL}\,,\quad \mathcal{H}^{MN} \partial_N \mathcal{H}^{KL} \partial_{L} \mathcal{H}_{MK}\,.
\end{equation}
Therefore, the DFT action must consist of a linear combination of the above mentioned terms multiplied with the dilaton prefactor $e^{-2d}$. We can express the gauge invariant action~\citep{Hohm:2010pp} by
\begin{equation}
S = \int d^{2D}X\,\mathcal{L}\,,
\end{equation}
with
\begin{align}
\label{dftactionbla}
\mathcal{L} = e^{-2d} \Big(& \frac{1}{8} \mathcal{H}^{MN} \partial_M \mathcal{H}^{KL} \partial_{N} \mathcal{H}_{KL} - \frac{1}{2} \mathcal{H}^{MN} \partial_N \mathcal{H}^{KL} \partial_{L} \mathcal{H}_{MK} \nonumber \\
&-2\partial_M d\, \partial_N \mathcal{H}^{MN} + 4\mathcal{H}^{MN} \partial_M d\, \partial_N d \Big)\,.
\end{align}
At this point, it is possible to define a \textit{generalized Ricci scalar}
\begin{align}
\mathcal{R} \equiv &\;\;\;\; 4\mathcal{H}^{MN} \partial_M d\, \partial_N d - \partial_M \partial_N \mathcal{H}^{MN} \nonumber \\
&-4\mathcal{H}^{MN} \partial_M d\, \partial_N d + 4 \partial_M \mathcal{H}^{MN} \partial_N d \\
&+\frac{1}{8} \mathcal{H}^{MN} \partial_M \mathcal{H}^{KL} \partial_{N} \mathcal{H}_{KL} - \frac{1}{2} \mathcal{H}^{MN} \partial_N \mathcal{H}^{KL} \partial_{L} \mathcal{H}_{MK} \nonumber\,,
\end{align}
which turns out to be a scalar under generalized diffeomorphisms. Up to total derivatives action~\bref{dftactionbla} can be rewritten in an equivalent way through
\begin{equation}
\label{dftaction}
S = \int d^{2D}X\, e^{-2d}\, \mathcal{R}\,.
\end{equation}
Using partial integration, we can show that~\citep{Hohm:2010pp}
\begin{equation}
\mathcal{L} = e^{-2d}\, \mathcal{R} + \partial_M \Big( e^{-2d}\,\big[ \partial_N \mathcal{H}^{MN} - 4 \mathcal{H}^{MN} \partial_N d \big] \Big)\,.
\end{equation}
The generalized Ricci scalar can then proven to be a gauge scalar. This is done by showing that the failure~\bref{failure} of the generalized Ricci scalar is zero~\citep{Hohm:2010pp}.
In fact,
\begin{equation}
\delta_\xi \mathcal{R} = \mathcal{\widehat{L}}_\xi \mathcal{R} = \xi^M \partial_M \mathcal{R}\,.
\end{equation}
Furthermore, we know that the dilaton exponential has to behave like a density
\begin{equation}
\delta_\xi e^{-2d} = \partial_M \big( \xi^M e^{-2d} \big)\,.
\end{equation}
Combining these two results, it follows that action~\bref{dftaction} must be gauge invariant, and specifically
\begin{equation}
\delta_\xi S = 0\,.
\end{equation}
In analogy to the generalized Ricci scalar, we can also find a \textit{generalized Ricci curvature} in Double Field Theory. Varying the action~\bref{dftaction} after the generalized metric $\mathcal{H}^{MN}$ it can be expressed through
\begin{equation}
\label{riccivariation}
\delta S = \int d^{2D}X \,e^{-2d} \delta \mathcal{H}^{MN} \mathcal{K}_{MN}\,,
\end{equation} 
with 
\begin{align}
\mathcal{K}_{MN} = &\;\;\;\,\, \frac{1}{8} \partial_M \mathcal{H}^{KL} \partial_N \mathcal{H}_{KL} - \frac{1}{4} \big[ \partial_L - 2(\partial_L d) \big] \mathcal{H}^{LK} \partial_K \mathcal{H}_{MN} + 2 \partial_M \partial_N d \\ &-\frac{1}{2} \partial_{(M} \mathcal{H}^{KL} \partial_L \mathcal{H}_{N)K} + \frac{1}{2}\big[ \partial_L - 2(\partial_L d) \big] \big( \mathcal{H}^{KL} \partial_{(M} \mathcal{H}_{N)K} + {\mathcal{H}^K}_{\;\;(M} \partial_K {\mathcal{H}^L}_{N)} \big)\,. \nonumber
\end{align}
However, the variation is still unconstrained and needs to be restricted. This comes from the fact that $\mathcal{H} \in O(D,D)$. Remembering the O($D,D$) constraint for the generalized metric $\mathcal{H} \eta \mathcal{H} = \eta^{-1}$, the equations of motions must preserve this constraint~\cite{Hohm:2010pp}. Thus, considering metric variations $\mathcal{H'} = \mathcal{H} + \delta \mathcal{H} $, which have to fulfill $\mathcal{H'} \eta \mathcal{H'} = \eta^{-1}$, yield the following condition
\begin{equation}
\label{variationbla}
\delta \mathcal{H} \eta \mathcal{H} + \mathcal{H} \eta \delta \mathcal{H} = 0\,.
\end{equation}
By defining
\begin{equation}
S = \mathcal{H} \eta,
\end{equation}
we can recast~\bref{variationbla}, with the use of the symmetry condition for the generalized metric, as
\begin{equation}
\delta \mathcal{H} S^T + S \delta \mathcal{H} = 0\,.
\end{equation}
Taking a closer look at~\bref{metric2}, we find
\begin{equation}
S^2 = \mathcal{H} \eta \mathcal{H} \eta = \mathcal{H} \mathcal{H}^{-1} = 1\,,
\end{equation}
and additionally derive
\begin{equation}
\label{solution111}
\delta \mathcal{H} = -S \delta \mathcal{H} S^T\,.
\end{equation}
At this point, we introduce projection operators $\frac{1}{2} ( 1 \pm S )$. These project by acting onto vectors $V = V^M$ with upper indices into the subspaces of $S$ with eigenvalues $\pm 1$. It allows us to regard any matrix $M^{MN}$ as a bivector which can be written as a projection into four different subspaces
\begin{align}
M = &\;\;\;\; \frac{1}{4} \big( 1 + S \big) M \big( 1 + S^T \big) + \frac{1}{4} \big( 1 + S \big) M \big( 1 - S^T \big) \nonumber \\
& +\frac{1}{4} \big( 1 - S \big) M \big( 1 + S^T \big) + \frac{1}{4} \big( 1 - S \big) M \big( 1 - S^T \big)\,.
\end{align}
The general solution of~\bref{solution111} is given by
\begin{equation}
\delta \mathcal{H} = \frac{1}{4} \big( 1 + S \big) \mathcal{M} \big( 1 - S^T \big) +\frac{1}{4} \big( 1 - S \big) \mathcal{M} \big( 1 + S^T \big)\,.
\end{equation}
As an immediate result of the symmetry of $\mathcal{H}$, the matrix $M$ has to be symmetric as well. Inserting the obtained result into equation~\bref{riccivariation} we find
\begin{align}
S &= \int dx d\tilde{x} e^{-2d}\, Tr\big( \delta \mathcal{H} \mathcal{K} \big) \\
&= \int dx d\tilde{x} e^{-2d}\, Tr\Big( \Big[ \frac{1}{4} \big( 1 + S \big) \mathcal{M} \big( 1 - S^T \big) +\frac{1}{4} \big( 1 - S \big) \mathcal{M} \big( 1 + S^T \big)\Big] \mathcal{K} \Big) \nonumber \\
&= \int dx d\tilde{x} e^{-2d}\, Tr\Big( \mathcal{M} \Big[ \frac{1}{4} \big( 1 - S^T \big) \mathcal{K} \big( 1 + S \big) +\frac{1}{4} \big( 1 + S^T \big) \mathcal{K} \big( 1 - S \big)\Big] \Big) \nonumber\,,
\end{align}
where we used the cyclicity of the trace in the last step. Finally, this yields the field equation 
\begin{equation}
\mathcal{R}_{MN} = 0.
\end{equation}
$\mathcal{R}_{MN}$ are the components of the matrix $\mathcal{R}$. It is defined through
\begin{equation}
\mathcal{R} \equiv \frac{1}{4} \big( 1 - S^T \big) \mathcal{K} \big( 1 + S \big) +\frac{1}{4} \big( 1 + S^T \big) \mathcal{K} \big( 1 - S \big)\,.
\end{equation}
Ultimately, we restore the individual indices 
\begin{equation}
\mathcal{R}_{MN} = \frac{1}{4} \big( \delta_M^{\;\;P} - S^P_{\;\;M} \big) \mathcal{K}_{PQ} \big( \delta^Q_{\,\,N} + S^Q_{\,\,N} \big) +\frac{1}{4} \big( \delta_M^{\;\;P} + S^P_{\;\;M} \big) \mathcal{K}_{PQ} \big( \delta^Q_{\,\,N} - S^Q_{\,\,N} \big)\,,
\end{equation}
or alternatively~\citep{Hassler:2014sba} we can express it as
\begin{equation}
\label{metricformulationeom}
\mathcal{R}_{MN} = {P_{(M}}^{P} {Q_{N)}}^{Q} \mathcal{K}_{PQ} = 0\,,
\end{equation}
with the operators
\begin{equation}
\label{projectionoperators}
P_{MN} = \frac{1}{2} \big( \eta_{MN} - \mathcal{H}_{MN} \big)\,,\quad Q_{MN} = \frac{1}{2} \big( \eta_{MN} + \mathcal{H}_{MN} \big)\,,
\end{equation}
projecting on the 'left-handed' and 'right-handed' subspaces~\citep{Hohm:2011si}. If we now impose the strong constraint $\tilde{\partial}^i = 0$ and recast the generalized metric~\bref{metric} in terms of the $D$-dimensional objects, in particular the metric $g_{ij}$, and the two-form $b_{ij}$, we can recover the SUGRA action~\eqref{SUSY} from the DFT action~\bref{dftaction}~\citep{Hull:2009mi}.

\subsection{Flux formulation}
\label{sec:fluxformulation}

We want to start by remembering the gauged vielbein ${E^A}_{M}$~\bref{vielbeingauge} along with the generalized Lie derivative~\bref{lievielbein}. This gives rise to the following decomposition
\begin{equation}
\mathcal{H}_{MN} = {E^A}_{M} \delta_{AB} {E^B}_{N}\,,\quad \eta_{MN} = {E^A}_{M} \eta_{AB} {E^B}_{N}\,,
\end{equation}
with the elements $\eta_{AB}$, and $\delta_{AB}$ given as in~\bref{vielbeinsss}. Specifically, the vielbein takes on the form
\begin{equation}
\label{vielbeingauge2}
{E^A}_{M} = \begin{pmatrix}
{e_a}^{i} & {e_a}^{l} b_{li} \\ 0  & {e^a}_{i}
\end{pmatrix}\,.
\end{equation}
In a very similar fahsion to~\citep{Siegel:1993th}, we can now construct \textit{covariant fluxes}~\citep{Aldazabal:2013sca, Hassler:2014sba, Geissbuhler:2011mx, Hohm:2010xe, Hohm:2014qga}, i.e.
\begin{equation}
\label{covariantflux}
\mathcal{F}_{ABC} = E_{CM} \mathcal{\widehat{L}}_{E_A} {E_{B}}^{M} = \big[ E_A, E_B \big]^M_C E_{CM} = 3 \Omega_{[ABC]}\,,
\end{equation}
\begin{equation}
\mathcal{F}_A = - e^{2d} \mathcal{\widehat{L}}_{E_A} e^{-2d} = \Omega^B_{\;\;BA} + 2 {E_A}^{M} \partial_M d\,.
\end{equation}
Here, we used the definition of the C-bracket~\bref{c-bracket} in equation~\bref{covariantflux}. Furthermore, we introduce the \textit{Weitzenböck connection}
\begin{equation}
\label{weitzenboeck}
\Omega_{ABC} = {E_A}^{M} \partial_M {E_B}^{N} E_{CN} = - \Omega_{ACB}\,,
\end{equation}
being antisymmetric in the last two components, and obtain
\begin{align}
\mathcal{F}_{ABC} = &\;\;\;\;\,{E_A}^{M} \partial_M {E_B}^{N} E_{CN} - \frac{1}{2} E_{AM} \partial^N {E_B}^{M} E_{CN} \nonumber \\
&-{E_B}^{M} \partial_M {E_A}^{N} E_{CN} + \frac{1}{2} E_{BM} \partial^N {E_A}^{M} E_{CN} \nonumber \\
&= \Omega_{ABC} - \frac{1}{2} \Omega_{CBA} - \Omega_{BAC} + \frac{1}{2}\Omega_{CAB} \nonumber \\
&= \Omega_{ABC} + \Omega_{BCA} + \Omega_{CAB} \nonumber \\
&= 3\Omega_{[ABC]}\,,
\end{align}
by using the antisymmetry in the last two components. Moreover, it is possible to work out the covariant fluxes in small indices in a somewhat lengthy but straightforward computation. However, we will omit these steps and instead refer to~\citep{Hassler:2014sba}.

Next, we have to figure out how we construct an O($D,D$) invariant action using these covariant fluxes~\citep{Geissbuhler:2011mx,Hohm:2014qga}. Flat indices $A,B,C, \ldots$ as they occur in the vielbein are manifestly O($D,D$) invariant and therefore any contraction of them will be as well. Currently, we consider the covariant fluxes as dynamical entities in our theory. However, when compactifying they reduce to the familiar constant fluxes, or gaugings.

Subsequently, we can express the gauge invariant DFT action using the generalized frame~\citep{Geissbuhler:2011mx,Hohm:2014qga} by
\begin{equation}
\label{fluxaction}
S = \int d^{2D}X \, e^{-2d}\, \mathcal{R}\,,
\end{equation}
with
\begin{align}
\label{fluxricci}
\mathcal{R} = &\;\;\;\; \mathcal{F}_{ABC} \mathcal{F}_{DEF} \Big[ \frac{1}{4} \delta^{AD} \eta^{BE} \eta^{CF} - \frac{1}{12} \delta^{AD} \delta^{BE} \delta^{CF} - \frac{1}{6} \eta^{AD} \eta^{BE} \eta^{CF} \Big] \nonumber \\
&+\mathcal{F}_{A}\mathcal{F}_{B} \Big[ \eta^{AB} - \delta^{AB} \Big]\,.
\end{align}
It has been shown~\citep{Geissbuhler:2011mx,Hohm:2014qga} that this action is invariant under generalized diffeomorphisms as well as double Lorentz transformations~\cite{Geissbuhler:2011mx,Hohm:2014qga, Geissbuhler:2013uka}. Additionally, it is equivalent to the frame formulation introduced in~\citep{Siegel:1993th}. Showing this requires to know the transformation behavior of the fluxes under generalized diffeomorphisms
\begin{equation}
\delta_\xi \mathcal{F}_{ABC} = \xi^D \partial_D \mathcal{F}_{ABC} + \Delta_{\xi} \mathcal{F}_{ABC}\,,
\end{equation}
\begin{equation}
\delta_\xi \mathcal{F}_{A} = \xi^D \partial_D \mathcal{F}_{A} + \Delta_{\xi} \mathcal{F}_{A}\,.
\end{equation}
Making use of the the closure constraint~\citep{Aldazabal:2013sca} and the definition of the failure to transform covariantly, we find~\bref{failure}
\begin{equation}
\label{consistency}
\Delta_{123}^{\;\;M} = - \Delta_{\xi_1} \big( \mathcal{\widehat{L}}_{\xi_2} \xi_3^M \big) = \big( \big[ \mathcal{\widehat{L}}_{\xi_1}, \mathcal{\widehat{L}}_{\xi_2} \big] - \mathcal{\widehat{L}}_{\xi_{12}} \big) \xi_3^M = 0\,,
\end{equation}
which always vanishes upon imposing the strong constraint. Here, we used
\begin{equation}
\xi_{12} = \mathcal{\widehat{L}}_{\xi_1} \xi_2\,.
\end{equation}
This allows us to derive
\begin{equation}
\label{fluxtransform1}
\Delta_{\xi} \mathcal{F}_{ABC} = E_{CM} \Delta_\xi \big( \mathcal{\widehat{L}}_{E_A} {E_B}^{M} \big) = 0\,,
\end{equation}
\begin{equation}
\label{fluxtransform2}
\Delta_{\xi} \mathcal{F}_{A} = - e^{2d} \Delta_\xi \big( \mathcal{\widehat{L}}_{E_A} e^{-2d} \big) = 0\,.
\end{equation}
Subsequently, the covariant fluxes transform as scalars under generalized diffeomorphisms. Moreover, the dilaton transforms again by
\begin{equation}
\delta_\xi e^{-2d} = \partial_M ( \xi^M e^{-2d} )\,.
\end{equation}
Altogether, from~\bref{fluxtransform1} and~\bref{fluxtransform2} we observe that the generalized curvature scalar~\bref{fluxricci} must transform as
\begin{equation}
\delta_\xi \mathcal{R} = \xi^M \partial_M \mathcal{R}\,,
\end{equation}
just as we observed in the previous section~\ref{metricformulation}.
Combining these two results leads to the invariance of the action under generalized diffeomorphisms. When expressing action~\bref{fluxaction} in terms of the generalized metric $\mathcal{H}^{MN}$ it takes up to total derivatives and a term modulo strong constraint the same form as action~\bref{dftaction}~\citep{Geissbuhler:2011mx,Hohm:2014qga}
\begin{equation}
S = \int d^{2D}X\, e^{-2d}\, \mathcal{R}\,,
\end{equation}
\begin{align}
\mathcal{R} \equiv &\;\;\;\; 4\mathcal{H}^{MN} \partial_M d\, \partial_N d - \partial_M \partial_N \mathcal{H}^{MN} \nonumber \\
&-4\mathcal{H}^{MN} \partial_M d\, \partial_N d + 4 \partial_M \mathcal{H}^{MN} \partial_N d \\
&+\frac{1}{8} \mathcal{H}^{MN} \partial_M \mathcal{H}^{KL} \partial_{N} \mathcal{H}_{KL} - \frac{1}{2} \mathcal{H}^{MN} \partial_N \mathcal{H}^{KL} \partial_{L} \mathcal{H}_{MK} \nonumber \\
&+\Delta_{(SC)} \mathcal{R}\,. \nonumber
\end{align}
Altogether, the flux formulation slightly extends the generalized metric formulation as it contains covariant terms which would vanish under the imposition of the strong constraint. Naturally, as is the case in the generalized metric formulation~\ref{metricformulation}, imposing the strong constraint leads back into the SUGRA frame again~\ref{SUGRA}. (From here, using~\ref{metricformulation}, it would have been trivial to see that this action must be unaffected by the gauge transformations.)

Varying the action~\bref{fluxaction} with respect to the vielbein ${E_A}^{M}$, and the dilaton $d$ yields
\begin{equation}
\delta_E S = \int d^{2D}X\, e^{-2d}\, \mathcal{G}^{AB} \delta E_{AB}\,,
\end{equation}
\begin{equation}
\delta_d S = \int d^{2D}X\, e^{-2d}\, \mathcal{G} \delta d\,.
\end{equation}
Incorporating the O($D,D$) conservation we get
\begin{equation}
\delta E_{AB} = \delta {E_{A}}^{M} E_{BM} = - \delta E_{BA}\,.
\end{equation}
The individual variations of the fluxes are now given by
\begin{equation}
\delta_E \mathcal{F}_{ABC} = 3 \Big( \partial_{[A} \delta E_{BC]} + \delta  {E_{[A}}^{D} \mathcal{F}_{BC]D} \Big)\,,
\end{equation}
\begin{equation}
\delta_d \mathcal{F}_{ABC} = 0\,,
\end{equation}
\begin{equation}
\delta_E \mathcal{F}_{A} = \partial^B \delta E_{BA} + \delta {E_A}^{B} \mathcal{F}_B\,,
\end{equation}
and
\begin{equation}
\delta_d \mathcal{F}_{A} = 2 \partial_A \delta d\,.
\end{equation}
It gives rise to the field equations
\begin{equation}
\mathcal{G}^{[AB]} = 2 \big( \delta^{D[A} - \eta^{D[A} \big) \partial^{B]} \mathcal{F}_D + \big( \mathcal{F}_D - \partial_D \big) \mathcal{\widetilde{F}}^{D[AB]} + \mathcal{\widetilde{F}}^{CD[A} {\mathcal{F}_{CD}}^{B]} = 0\,,
\end{equation}
\begin{equation}
\mathcal{G} = - 2 \mathcal{R} = 0\,,
\end{equation}
with
\begin{equation}
\mathcal{\widetilde{F}}^{ABC} = \frac{3}{2} {\mathcal{F}_D}^{BC} \delta^{AD} - \frac{1}{2} \mathcal{F}_{DEF} \delta^{AD} \delta^{BE} \delta^{CF} - \mathcal{F}^{ABC}\,.
\end{equation}
Decomposing the equations of motion with regard to the single fields $g_{ij}$,  $b_{ij}$, and imposing the strong constraint $\tilde{\partial}^i = 0$, we retrieve the SUGRA equations of motion~\bref{sugraeom1}-\bref{sugraeom3}.
\clearpage{}
  \clearpage{}\chapter{Double Field Theory on Group Manifolds} \label{Kap_2}

Starting from the results of the previous chapter~\ref{Kap_1} in which we introduced Double Field Theory over a torus, we will now examine Double Field Theory on a group manifold~\cite{Blumenhagen:2014gva, Pascal2015}. Starting from a left/right asymmetric Wess-Zumino-Witten (WZW) model one can employ Closed String Field Theory (CSFT) calculations to evaluate the two-point and three-point functions at tree level up to cubic order in fields as well as leading order in $\alpha'$ to derive a Double Field Theory on group manifolds (DFT$_{\mathrm{WZW}}$). The doubling of the coordinates refers to the left- and right-moving currents on the WZW model on a group manifold and its corresponding Ka$\check{c}$-Moody algebra. A related approach can be found in~\cite{Hohm:2015ugy}. Primary fields of the CFT are represented as scalar functions on the group manifold $G = G_L \times G_R$~\cite{Blumenhagen:2017noc}. This allows to derive the action~\ref{sec:DFTWZWaction} and the gauge transformations~\ref{sec:DFTWZWgauge} of DFT$_{\mathrm{WZW}}$. It opens up new intriguing features and possibilities. Furthermore, we are going to see how the weak and strong constraint emerge in this picture.

We mainly follow~\cite{Blumenhagen:2014gva} during this chapter.

\section [\texorpdfstring{$\text{DFT}_{\mathrm{WZW}}\, \text{origins}$}{}]      {$\textbf{DFT}_{\mathrm{wzw}}\, \text{origins}$}
In this section, we discuss the prerequisites and underlying concepts leading to the formulation of DFT$_{\mathrm{WZW}}$. We begin by procuring an explicit representation for semisimple\footnote{ For
simplicity we assume that $G$ is semisimple. However, the equations we discuss later also hold for a more
general case.} Lie algebras the theory is based upon~\ref{representationsDFT*}. Afterwards, we present the basic concepts and ideas behind Closed String Field Theory (CSFT)~\ref{sec:CSFT}. These CSFT computations allow to finally obtain an explicit form for the action and its associated gauge transformations. All of these notions are employed  on a Wess-Zumino-Witten model governed by an underlying group manifold.

\subsection{Lie algebra representation}
\label{representationsDFT*}

A Lie algebra $\mathfrak{g}$ is build up by its basis elements the generators $t_a$. For two arbitrary generators $t_a$ and $t_b$, the commutator algebra is given by
\begin{equation}
[t_a, t_b] = {F_{ab}}^c\,t_c\,.
\end{equation}
Moreover, it is useful to normalize these generators with regard to the Killing form~\cite{Blumenhagen:2014gva}
\begin{equation}
\label{liemetric}
\eta_{ab} = \mathcal{K}(t_a, t_b) = \frac{\text{Tr}(t_a t_b)}{2x_\lambda} = - \frac{1}{2 h^\vee} {F_{ad}}^c {F_{bc}}^d\,.
\end{equation}
Here, $x_\lambda$ marks the Dynkin index, whereas in the adjoint representation it is equal to the dual Coxeter number $h^\vee$.
For any semisimple Lie algebra $\mathfrak{g}$ it is now always possible to find a parametrization in which its Killing form $\eta_{ab}$ diagonalizes with entries $\pm 1$. E.g. the associated Lie algebra $\eta_{ab}$ for a compact Lie group $G$ has always a negative definite Killing form with the signature $\eta_{ab} = (-,\ldots,-)$. At this point, we are able to raise and lower indices using the Killing form $\eta_{ab}$ and its inverse $\eta^{ab}$.

There exists a convenient method to obtain the representation of a semisimple Lie algebra making use of the scalar functions on the group manifold $G = G_L \times G_R$. Hence, it is useful to switch from the very conceptual Maurer-Cartan forms to more favorable vielbeins~\citep{Blumenhagen:2014gva}. This relation is given through
\begin{equation}
\label{eqn:leftmaurercartan}
\omega_\gamma = t_a {e^a}_i dx^i\,, \quad \text{where} \quad {e^a}_i = \mathcal{K}(t^a, \gamma^{-1} \partial_i \gamma)\,.
\end{equation}
In this equation, we have to distinguish between two types of indices: the flat ones labeled with $a,b,c,\ldots$, and curved ones denoted by $i,j,k,\ldots$. As previously mentioned, flat indices can be raised and lowered with the flat metric $\eta_{ab}$ given through the Killing form of the Lie algebra. However, the curved indices have to be raised and lowered using the target space metric $g_{ij}$, which is related to the vielbein by
\begin{equation}
g_{ij} = {e^a}_i\, \eta_{ab}\, {e^b}_j\,.
\end{equation}
It allows us to write the $H$-flux in the following form
\begin{equation}
H_{ijk} = {e^a}_i {e^b}_j {e^c}_k F_{abc}\,.
\end{equation}
Returning to representations, we introduce the coordinates $x^i$ of the left-moving (chiral) part together with the flat derivative
\begin{equation}
D_a = {e^a}_i \partial_i\,,
\end{equation}
where we made use of the background vielbein $e_a{}^i$ on $G_L$.
Their commutation relations implement the Lie algebra, specifically
\begin{equation}
[D_a, D_b] = {F_{ab}}^c D_c\,,
\end{equation}
and when recast through vielbeins, they give rise to
\begin{equation}
{F_{ab}}^c =  {e_{[a}}^i \partial_i {e_{b]}}^j {e^{c}}_j = D_{[a} {e_{b]}}^i {e^{c}}_i\,.
\end{equation}
As a result, we obtained a representation for the flat derivatives $D_a$ which are spanned by the underlying generators $t_a$ of the Lie algebra. These flat derivatives act on patches of the group manifold in contrast to the generators $t_a$. They act on an abstract notion of the related vector space. Therefore, the functions on these patches, the flat derivatives are acting on, are a representation of the universal enveloping algebra of the associated Lie algebra~\cite{Blumenhagen:2014gva,Blumenhagen:2017noc}.

Furthermore, to implement integration by parts we require boundary terms such as
\begin{equation}
\int d^Dx \sqrt{g} D_a v = \int d^Dx \partial_i ( \sqrt{g} {e_a}^i v ) = 0
\end{equation}
to vanish. In this context, $v$ denotes an arbitrary scalar function on the target space. We assume that such boundary terms will always vanish at $\pm \infty$. Subsequently, we are allowed to make use partial integration, i.e.
\begin{equation}
\label{DFT*intbyparts}
\int d^Dx \sqrt{g} (D_a v) w = - \int d^Dx \sqrt{g} v (D_a w)\,. 
\end{equation}
This identity is equivalent to
\begin{equation}\label{eqn:nablaIe-2bard=0}
      \nabla_I e^{-2 \bar d} = \partial_I e^{-2\bar d} - \Gamma^J{}_{IJ} e^{-2\bar d} = 0
        \quad \text{or} \quad \Omega_{IJ}{}^J = 2 \partial_I \bar d\,.
\end{equation}
Here, we used that the dilaton factor $e^{-2 \bar d}$ transforms as a scalar density with weight $+1$.

More general, any Lie algebra fulfilling
\begin{equation}
\label{unimodular}
{F_{ab}}^b = 0\,,\quad \text{or equivalently} \quad \text{Tr}(\text{ad}_x) = 0
\end{equation}
called unimodular, solves this relation.

In the next step one would have to define an highest weight state. This is always possible for compact Lie algebras. For non-compact Lie algebras the discussion becomes much more elaborate. However, we will not go into any more detail and instead refer to~\citep{Blumenhagen:2014gva}.

\subsection{Effective theory}
\label{sec:CSFT}

Originating from a WZW model on a group manifold, the CSFT computations for the DFT$_{\mathrm{WZW}}$ action and corresponding gauge transformation require the evaluation of two-point and three-point functions. Therefore, it is necessary to derive the correlation functions of the Ka$\check{c}$-Moody primary fields. They can be found in~\citep{Blumenhagen:2014gva}.

It is worth noting that the chiral and anti-chiral currents possess the same underlying Ka$\check{c}$-Moody algebra. We can understand this through the relations connecting them: Inverting $\gamma$ and performing a complex conjugation. On the algebra level, an inversion is isomorph to multiplying the generators with $-1$. This modifies the structure coefficients to
\begin{equation}
[t_a, t_b] = {F_{ab}}^c t_c \quad \mapsto \quad [-t_a, -t_b] = {\bar{F}_{ab}}^{\;\;\;c} (-t_c) \quad \text{with} \quad {\bar{F}_{ab}}^{\;\;\;c} = - {{F}_{ab}}^c\,.
\end{equation}
As a consequence, this result makes it possible to use the operator product expansion (OPE) defining the chiral Ka$\check{c}$-Moody algebra, and we substitute $j_a(z)$ by $-\bar{j}_a(\bar{z})$, as well as replacing ${{F}_{ab}}^c$ through ${\bar{F}_{ab}}^{\;\;\;c}$.
Similarly, a flat derivative has to be introduced using the background vielbein on $G_R$ acting on the right-moving (anti-chiral) coordinates $\bar{x}^i$. It gives rise to
\begin{equation}
{e^{\bar{a}}}_{\bar{i}} = \mathcal{K}(t^a, \partial_{\bar{i}} \gamma \gamma^{-1})\,,\quad\text{and}\quad D_{\bar{a}} = {e_{\bar{a}}}^{\bar{i}} \partial_{\bar{i}}\,,
\end{equation}
where we used bared indices to differentiate between chiral and anti-chiral parts. By construction, these bared (anti-chiral) flat derivatives reproduce the according Lie algebra (For convenience ${\bar{F}_{ab}}^{\;\;\;c}$ replaced by ${F_{\bar{a}\bar{b}}}^{\bar{c}}$.)
\begin{equation}
[D_{\bar{a}}, D_{\bar{b}}] = {F_{\bar{a}\bar{b}}}^{\bar{c}} D_{\bar{c}}\,.
\end{equation}
At this point, it useful to note that the unbared flat derivative only acts on coordinates $x^i$, whereas the bared flat derivative only works on coordinates $x^{\bar{i}}$. Ergo, we treat the left-movers and right-movers independently of each other.

Now, we can combine the $D$ unbared coordinates with the newly introduced $D$ bared ones to $2D$ doubled coordinates $X^I = ( x^i, x^{\bar{i}} )$. Of course, it also allows to define an according doubled derivative by $\partial_I = (\partial_i, \partial_{\bar{i}})$, and the doubled vielbein
\begin{equation}
\label{DFT*vielbein}
{E_A}^I = \begin{pmatrix}
{e_a}^i & 0 \\ 0 & {e_{\bar{a}}}^{\bar{i}}
\end{pmatrix}\,.
\end{equation}
These are the so-called \textit{doubled generalized objects}. Furthermore, it also makes it possible to implement the commutation relations of the chiral and anti-chiral Lie algebras into doubled objects and obtain
\begin{equation}
D_A = {E_A}^I \partial_I\,, \quad \text{along with} \quad [D_A, D_B] = {F_{AB}}^C D_C\,,
\end{equation}
This form poses a striking resemblance to the flux formulation of DFT~\citep{Hohm:2010xe, Geissbuhler:2013uka, Blumenhagen:2014gva}. 
We will go into more detail about formulating DFT$_\mathrm{WZW}$ using doubled generalized objects in the next chapter.

All necessary tools to perform the CSFT computations can be found in~\citep{Blumenhagen:2014gva}.

Basis for the CSFT computations are two level-matched string fields $\ket{\Psi}$, and $\ket{\Lambda}$, which are put in Siegel gauge~\citep{Blumenhagen:2014gva, Siegel:1988yz}. As a result, they are annihilated by
\begin{equation}
L_0 - \bar{L}_0\,,\quad \text{and} \quad b_0^- = b_0 - \bar{b}_0\,,
\end{equation}
with ghost number two and one, respectively. Moreover, the combination $L_0 + \bar{L}_0$, being equivalent to the string field energy, should be small compared to the energy scale of the massive string excitations as we are focusing on low-energy excitations of the theory. Subsequently, we find for the Virasoro operator
\begin{equation}
L_m = -\frac{\alpha' \eta^{ab}}{4} \big( 1-h^\vee k^{-1} \big) \sum_n : j_{a\,n-m} j_{b-n} : +\, \mathcal{O}(k^{-3})\,,
\end{equation}
with the modes $j_a(z)$ fulfilling the Ka$\check{c}$-Moody algebra
\begin{equation}
[j_{a\,m}, j_{b\,n}] = {F_{ab}}^c j_{c\,m+n} - \frac{2}{\alpha'} m \eta_{ab} \delta_{m+n}\,.
\end{equation}
Due to the low energy condition $L_0 + \bar{L}_0 \ll 1$, $k$ has to be very large. In fact, this is equivalent to the large volume limit of the background geometry~\cite{Blumenhagen:2014gva}. We can now express the two string fields $\ket{\Psi}$ and $\ket{\Lambda}$ by
\begin{align}
\label{fieldsSFT}
\ket{\Psi} = &\sum_R \Big[\frac{\alpha'}{4} \epsilon^{a\bar{b}}(R) j_{a-1} \bar{j}_{\bar{b}-1} c_1 \bar{c}_1 + e(R) c_1 c_{-1} + \bar{e}(R) \bar{c}_1 \bar{c}_{-1} \nonumber \\ &\;\;\;\:+\frac{\alpha'}{2} \big( f^a(R) c_0^+ c_1 j_{a-1} + {f}^{\bar{b}}(R) {c}_0^+ \bar{c}_1 \bar{j}_{\bar{b}-1} \big) \Big] \ket{\phi_R}, \\
\ket{\Lambda} = &\sum_R \Big[ \frac{1}{2} \lambda^a(R) j_{a-1} c_1 - \frac{1}{2} \lambda^{\bar{b}}(R) \bar{j}_{\bar{b}-1} \bar{c}_1 + \mu(R) c_0^+ \Big] \ket{\phi_R}\,,
\end{align}
with
\begin{equation}
c_0^\pm = \frac{1}{2} ( c_0 \pm \bar{c}_0 )\,.
\end{equation}
These are very similar to the fields given in~\citep{Hull:2009mi}, and present the most general solution to the aforementioned compatibility conditions. Nevertheless, there is a striking difference. Equation~\eqref{fieldsSFT} sums over the different representations $R = ( \lambda q, \bar{\lambda} \bar{q} ) $~\ref{representationsDFT*} as opposed to~\citep{Hull:2009mi}, where they sum over the momentum and winding modes. Although, in the abelian limit the summation over the different representation reduces to the sum over the left- and right-moving momenta. These are a linear combination of the string's momentum and winding modes. Hence, they equal another. As a consequence, it results in a natural extension of toroidal DFT~\cite{Blumenhagen:2014gva}.

For a simply-connected group manifold $G$, we can express each $e(X) \in L^2(G)$ through
\begin{equation}
e(X) = \sum_R e(R) Y_R(X)\,.
\end{equation}
Specifically, the level matching condition~\bref{levelmatching} becomes
\begin{equation}
\label{levelmatchingDFT*}
\big( D_a D^a - D_{\bar{a}} D^{\bar{a}} \big) e = 0\,.
\end{equation}
We can recast this expression using doubled indexed objects and find
\begin{equation}
\label{DFT*strongconstraint}
\eta^{AB} D_A D_B \cdot = D_A D^A \cdot = 0\,,
\end{equation}
where used the constant tangent space metric
\begin{equation}
\label{DFT*diagonaleta}
\eta	^{AB} = \begin{pmatrix} \eta^{ab} & 0 \\ 0 & -\eta^{\bar{a}\bar{b}}
\end{pmatrix}\,,\quad \text{and it's inverse} \quad \eta_{AB}^{\,} = \begin{pmatrix}
{\eta_{ab}}^{\,} & 0 \\ 0 & -{\eta_{\bar{a}\bar{b}}}^{\,}
\end{pmatrix}\,,
\end{equation}
to raise and lower the doubled indices. Here, $\cdot$ is a placeholder for the physical fields $e, \bar{e}, \epsilon^{a\bar{b}}, f^a, f^{\bar{b}}$, and the gauge parameters $\lambda^a, \lambda^{\bar{b}}, \mu$. This notation might be a bit misleading and confuse somebody into mistakenly concluding it would be the weak constraint known from toroidal DFT. We are dealing in this context with flat indices and not with curved indices~\cite{Blumenhagen:2014gva}. For a proper comparison it would be necessary to switch into curved coordinates Therefore, let us make use of the following identities
\begin{equation}
{\Omega_b}^{ba} = -{\Omega_b}^{ab} + \partial_i g^{ij} {e^a}_j\,,
\end{equation}
with the anholonomy coefficients
\begin{equation}
{\Omega_{ab}}^{c} = {e_a}^i \partial_i {e_b}^j {e^c}_j\,.
\end{equation}
From the unimodularity of the Lie algebra $\mathfrak{g}$~\bref{unimodular} we obtain
\begin{equation}
{F_{ab}}^b = 0 = {\Omega_{[ab]}}^{b} \quad \to \quad {\Omega_{ab}}^{b} = {\Omega_{ba}}^{b}\,.
\end{equation}
On the other hand, a short calculation yields
\begin{equation}
2D^a \tilde{d} = {{\Omega^a}_b}^a\,, \quad \text{and} \quad \tilde{d} = \phi - \frac{1}{2} \text{log} \sqrt{G}\,,
\end{equation}
with $d$ being the generalized dilaton of DFT, while $\phi$ marks the string theory dilaton assumed to be constant in this situation. Combining these two results, we arrive at the relation
\begin{equation}
{\Omega_{ba}}^{b} = -2 D^a \tilde{d} + \partial_i g^{ij} {e^a}_j\,.
\end{equation}
Hence, we can recast~\bref{levelmatchingDFT*} through
\begin{equation}
D_a D^a \cdot = \big( {\Omega_b}^{ba} D_a + g^{ij} \partial_i \partial_j \big) \cdot = \big( -2\partial_i \tilde{d} \partial^i + g^{ij} \partial_i \partial_j \big) \cdot\,.
\end{equation}
The argumentation for bared indices follows analogously. Finally, with the curved metric
\begin{equation}
\label{DFT*metric}
\eta^{IJ} = {E_A}^I \eta^{AB} {E_B}^J = \begin{pmatrix}
g^{ij} & 0 \\ 0 & - g^{\bar{i}\bar{j}}
\end{pmatrix}\,,
\end{equation}
we derive
\begin{equation}
\label{levelmatchingDFT*2}
\big( \partial_I \partial^I - 2 \partial_I \tilde{d} \partial^I \big) \cdot = 0\,.
\end{equation}
Here, the curved doubled indices are raised and lowered with the non-constant metric $\eta^{IJ}$ and $\eta_{IJ}$, respectively. However, we need to be cautious as $\eta^{IJ}$ is coordinate dependent, and as a result cannot be pulled in or out of partial derivatives. In contrast to toroidal DFT we get an additional term $- 2 \partial_I \tilde{d} \, \partial^I$. This term comes from the background in DFT$_\mathrm{WZW}$. Specifically,
\begin{equation}
\nabla_I V^J = \partial_I V^J + {\Gamma_{IK}}^J V^K\,.
\end{equation}
Requiring compatibility with the dilaton (see also~\citep{Hohm:2011si, Blumenhagen:2014gva}) we obtain
\begin{equation}
\Gamma_I = {\Gamma_{JI}}^J = -2\partial_J \tilde{d}\,.
\end{equation}
Altogether, we get the result
\begin{equation}
\label{eqn:dftwzwweak}
\nabla_I \partial^I \cdot = \Delta \cdot = 0\,.
\end{equation}
It is consistent with the definition of the Laplace operator in Riemannian geometry. Subsequently, the newly derived weak constraint~\eqref{eqn:dftwzwweak} is invariant under local generalized transformations as well. This is in stark contrast to toroidal DFT where the weak constraint $\partial_I \partial^I \cdot = 0$ is only invariant under global O($D,D$) transformations.

Furthermore, this new constraint is also invariant under local generalized diffeomorphisms, as opposed to toroidal DFT where the constraint $\partial_I \partial^I \cdot = 0$ is only invariant under global O($D,D$) transformations. From metric compatibility $\nabla_I \eta^{JK} = 0$ we find $\nabla_I \partial^I = \nabla^I \partial_I$. 

Ultimately, one can evaluate the tree level action of Closed String Field Theory~\citep{Hull:2009mi, Blumenhagen:2014gva, Zwiebach:1992ie}
\begin{equation}
(2 \kappa^2) S = \frac{2}{\alpha'} \big( \{\Psi, Q\Psi\} + \frac{1}{3} \{\Psi, \Psi, \Psi \}_0 + \frac{1}{12} \{\Psi, \Psi, \Psi, \Psi \}_0 + \ldots \, \big)\,,
\end{equation}
with the already known string field $\Psi$. The whole calculation requires a successive expansion of the string functions $\{\cdot, \cdot, \cdot \}_0$ around the genus zero worldsheet $S^2$. Clearly, the computation becomes more challenging with an higher amount of slots for the string functions. In quadratic order we recover the free theory, whereas the cubic order gives rise to basic interaction terms.

Moreover, the gauge transformations can be obtained using CSFT as well, in particular
\begin{equation}
\delta_\Lambda \Psi = Q \Lambda + [\Lambda, \Psi]_0 + \frac{1}{2} [\Lambda, \Lambda, \Psi]_0 + \ldots\,.
\end{equation}
They are characterized by the ghost number one string field $\Lambda$. Further, the string product $[\cdot, \cdot]_0$ is related to the string functions by
\begin{equation}
[B_1, ..., B_n]_0 = \sum_s \ket{\phi_s} \{\phi_s^c, B_1, ..., B_n \}_0\,,
\end{equation}
where $\phi_s^c$ are the conjugate fields to $\phi_s$. When evaluating CSFT on the torus, the CFT on the sphere $S^2$ is free and its straightforward to derive the conjugate fields. However, in general this is not the case. For group manifolds the worldsheet theory typically interacts and therefore the concept of conjugate fields is more complicated. 

A more detailed discussion, including the entire computation of the action and the gauge transformations, can be found in~\citep{Blumenhagen:2014gva}.

\section{Action}
\label{sec:DFTWZWaction}

Following the elimination of all auxiliary fields, and performing field redefinitions the DFT$_{\mathrm{WZW}}$ action can be expressed to cubic order as~\citep{Blumenhagen:2014gva}
\begin{align}
\label{DFT*action}
(2\kappa^2) S = &\int d^{2D}X \sqrt{H} \Big( \frac{1}{4}\, \epsilon_{a\bar{b}} \Box \epsilon^{a\bar{b}} + \frac{1}{4} \big( D^a \epsilon_{a\bar{b}} \big)^2 \, + \frac{1}{4}\big( D^{\bar{b}} \epsilon_{a\bar{b}} \big)^2 -2\tilde{d} D_a D_{\bar{b}} \epsilon^{a\bar{b}} -4\tilde{d} \Box \tilde{d} \nonumber \\ &+\frac{1}{4} \epsilon_{a\bar{b}} \Big( D^a \epsilon_{c\bar{d}} D^{\bar{b}} \epsilon^{c\bar{d}} - D^a \epsilon_{c\bar{d}} D^{\bar{d}} \epsilon^{c\bar{b}} - D^c \epsilon^{a\bar{d}} D^{\bar{b}} \epsilon_{c\bar{d}} \Big) + 4\tilde{d}^2 \Box \tilde{d} + 4\tilde{d} \, \epsilon^{a\bar{b}} \big( D_a D_{\bar{b}} \tilde{d} \big) \nonumber \\ &-\frac{1}{4} \epsilon_{a\bar{b}} \Big( {F^{ac}}_d D^{\bar{e}} \epsilon^{d\bar{b}} \epsilon_{c\bar{e}} + {F^{\bar{b}\bar{c}}}_{\bar{d}} D^{e} \epsilon^{a\bar{d}} \epsilon_{e\bar{c}} \Big) - \frac{1}{12} F^{ace} F^{\bar{b}\bar{d}\bar{f}} \epsilon_{a\bar{b}}\epsilon_{c\bar{d}}\epsilon_{e\bar{f}}  \\ &+\frac{1}{2}\tilde{d} \Big( \big( D^a \epsilon_{a\bar{b}} \big)^2 \, + \big( D^{\bar{b}} \epsilon_{a\bar{b}} \big)^2 + \frac{1}{2} \big( D_c \epsilon_{a\bar{b}} \big)^2 + \frac{1}{2} \big( D_{\bar{c}} \epsilon_{a\bar{b}} \big)^2 + 2 \epsilon^{a\bar{b}} \big( D_a D^c \epsilon_{a\bar{b}} + D_{\bar{b}} D^{\bar{c}} \epsilon_{a\bar{c}} \big) \Big)\,. \nonumber
\end{align}
This action depends on the fluctuations $\epsilon_{a\bar b}$, the dilaton $\tilde{d}$, the background vielbeins $e_a{}^i$ and $e_{\bar a}{}^{\bar i}$, as well as the structure coefficients $F_{abc}$ and $F_{\bar a \bar b \bar c}$.  It looks already very similar to the action~\cite{Hull:2009mi} derived by Hull and Zwiebach for toroidal DFT. Imposing the abelian limit, which implies that all terms containing structure coefficients vanish, the DFT$_\mathrm{WZW}$ becomes identical to the DFT action. The emergence of an additional potential is one of the most remarkable features of DFT$_{\mathrm{WZW}}$. Specifically, it takes on the form
\begin{equation}
V = \frac{1}{12} F^{ace} F^{\bar{b}\bar{d}\bar{f}} \epsilon_{a\bar{b}}\epsilon_{c\bar{d}}\epsilon_{e\bar{f}}\,,
\end{equation}
in contrast to toroidal DFT which only admits kinetic terms.

At this point, several open questions need to be addressed. These concern the possibility of recasting action~\eqref{DFT*action} into  a generalized metric formulation and whether we need to implement covariant derivatives as well. Furthermore, this action is expected from CSFT to be invariant under its gauge transformations and we should check it explicitly. However, all of these open question are going to be answered in the next chapter.

\section{Gauge algebra}
\label{sec:DFTWZWgauge}

We start out by reviewing the DFT$_{\mathrm{WZW}}$ gauge transformations and are going to take a closer look at their corresponding gauge algebra. As we already noticed, the occurrence of structure coefficient terms in action~\eqref{DFT*action} makes it useful to introduce a covariant derivative~\ref{sec:covderv} to simplify the expressions. This allows for the study of the strong constraint simultaneously. Finally, we observe that the gauge algebra~\ref{sec:cbracket} governed by the C-bracket closes modulo strong and closure constraint.

\subsection{Gauge transformations}
\label{sec:2gaugetrafo}

After the execution of a field redefinition and the removal of all auxiliary fields the gauge transformations for the fluctuations and the dilaton are given by~\citep{Blumenhagen:2014gva}
\begin{align}
\label{DFT*gaugetransformations}
\delta_\lambda \epsilon_{a\bar{b}} &=\, D_{\bar{b}} {\lambda}_a + \frac{1}{2} \big[ \lambda_c D^c \epsilon_{a\bar{b}} + D_a \lambda^c \epsilon_{c\bar{b}} - D^c \lambda_a \epsilon_{c\bar{b}} + {F_{ac}}^d \lambda^c \epsilon_{d\bar{b}} \big] \nonumber \\ 
&\;+ D_{a} {\lambda}_{\bar{b}} + \frac{1}{2} \big[ \lambda_{\bar{c}} D^{\bar{c}} \epsilon_{a\bar{b}} + D_{\bar{b}} \lambda^{\bar{c}} \epsilon_{a\bar{c}} - D^{\bar{c}} \lambda_{\bar{b}} \epsilon_{a\bar{c}} + {F_{\bar{b}\bar{c}}}^{\bar{d}} \lambda^{\bar{c}} \epsilon_{a\bar{d}} \big]\,, \\
\delta_\lambda \tilde{d} &= - \frac{1}{4} D_a \lambda^a + \frac{1}{2} \lambda_a D^a \tilde{d} - \frac{1}{4} D_{\bar{a}} \lambda^{\bar{a}} + \frac{1}{2} \lambda_{\bar{a}} D^{\bar{a}} \tilde{d}\,.
\end{align}

On top of this, all fields altered by gauge transformations should still satisfy the level matching condition~\eqref{levelmatching}. For gauge transformations of cubic order this is generally not the case. As a result, we need to project all of those fields into the kernel of the level matching operator $\Delta$. However, for us to avoid the restraint of always having to perform this explicit projection, we impose the strong constraint. It ensures that the string product is always level matched~\cite{Blumenhagen:2017noc}.
Explicitly, the strong constraint becomes
\begin{equation}
D_A D^A \cdot = \nabla_I \partial^I \cdot = 0\,,
\end{equation}
where $\cdot$ marks fluctuations, gauge parameters, and arbitrary products of either. Originally, in toroidal DFT the strong constraint was only considered in the context of gauge algebra closure and generalized diffeomorphism invariance. But in DFT$_\mathrm{WZW}$ it is necessary to impose this constraint even for field redefinitions above the linear level.

In order to further simplify the handling with gauge transformations, we recast most entities into doubled generalized objects. Subsequently, we obtain for the gaugings and flat derivatives
\begin{equation}
\lambda^A = ( \lambda^a, \lambda^{\bar{a}} )\,,\quad \text{and} \quad D_A = ( D_a, D_{\bar{a}} )\,,
\end{equation}
and after raising and lowering with the flat Killing metric~\eqref{DFT*diagonaleta}
\begin{equation}
\lambda_A = ( \lambda_a, -\lambda_{\bar{a}} )\,,\quad \text{and} \quad D^A = ( D^a, -D^{\bar{a}} )\,.
\end{equation}
The structure coefficients can accordingly be expressed as
\begin{equation}
{F_{AB}}^C = \begin{cases}
{F_{ab}}^c \\ {F_{\bar{a}\bar{b}}}^{\bar{c}} \\ 0\quad \text{otherwise}
\end{cases}\quad \to \quad F_{ABC} = \begin{cases}
\;\;\;{F_{abc}} \\ -F_{\bar{a}\bar{b}\bar{c}} \\ \;\;\;0 \quad \quad \text{otherwise}
\end{cases}\,.
\end{equation}
Nevertheless, this recasting requires one non-trivial step. It is still not clear which doubled generalized object is corresponding to the fluctuations $\epsilon^{a\bar{b}}$.

Therefore, we consider the following symmetric O($D,D$) transformation $\mathcal{H}^{AB}$ 
\begin{equation}
\label{eqn:blablabla}
\mathcal{H}^{AC} \eta_{CD} \mathcal{H}^{DB} = \eta^{AB}\,.
\end{equation}
An example for a transformation of this form is the matrix $S^{AB}$
\begin{equation}
S^{AB} = \begin{pmatrix}
\eta^{ab} & 0 \\ 0 & \eta^{\bar{a}\bar{b}}\,.
\end{pmatrix}
\end{equation}
This now allows us to assess small perturbations $\epsilon^{AB}$ of equation~\eqref{eqn:blablabla}. Clearly, it still has to be consistent with the features of $\mathcal{H}^{AB}$. As a result, $\epsilon^{AB}$ is symmetric as well and needs to fulfill
\begin{equation}
\label{DFT*condition}
\epsilon^{AC} \eta_{CD} S^{DB} + S^{AC} \eta_{CD} \epsilon^{DB} + \mathcal{O}(\epsilon^2) = 0\,.
\end{equation}
The most general solution is given by
\begin{equation}
\epsilon^{AB} = \begin{pmatrix}
0 & - \epsilon^{a\bar{b}} \\ - \epsilon^{\bar{a}b} & 0
\end{pmatrix}\,, \quad \text{with} \quad \epsilon^{a\bar{b}} = (\epsilon^T)^{\bar{b}a}\,.
\end{equation}
Thus, we can express the generalized metric $\mathcal{H}^{AB}$ through
\begin{equation}
\label{DFT*generalizedmetric}
\mathcal{H}^{AB} = S^{AB} + \epsilon^{AB} + \frac{1}{2} \epsilon^{AC} S_{CD} \epsilon^{DB} + ... = \text{exp}(\epsilon^{AB})\,.
\end{equation}
It also allows for the introduction of a generalized metric in the form
\begin{align}
\label{DFT*generalizedlie}
\mathcal{L}_\lambda \epsilon^{AB} &\,= \lambda^C D_C \epsilon^{AB} + \big( D^A \lambda_C - D_C \lambda^A \big) \epsilon^{CB} + \big( D^B \lambda_C - D_C \lambda^B \big) \epsilon^{AC} \nonumber \\ &\;+{F^A}_{CD} \lambda^C \epsilon^{DB} +{F^B}_{CD} \lambda^C \epsilon^{AD}\,.
\end{align} 
As can be checked easily, this generalized Lie derivative leaves the target space metric invariant
\begin{equation}
\mathcal{L}_\lambda \eta^{AB} = 0\,,
\end{equation}
and as a result preserves the O($D,D$) structure. Furthermore, the imposition of the strong constraint yields a trivial transformation behavior
\begin{equation}
\mathcal{L}_{D^A \chi} V^{BC} = 0
\end{equation}
for a closed gauge parameter.
Moreover, we want to rewrite the gauge transformations~\bref{DFT*gaugetransformations} in terms of generalized Lie derivatives. As was worked out in~\citep{Blumenhagen:2014gva}
\begin{equation}
\delta_\lambda \epsilon^{AB} = \frac{1}{2} \big( \mathcal{L}_\lambda S^{AB} + \mathcal{L}_\lambda \epsilon^{AB} + \mathcal{L}_\lambda {S^{(A}}_C {S^{B)}}_D \epsilon^{CD} \big)\,,
\end{equation}
the gauge transformations only affect fluctuations $\epsilon^{a\bar{b}}$ around the background.
The background remains invariant under them, and as a result
\begin{equation}
\delta_\lambda S^{AB} = 0\,.
\end{equation}
At this point, it has become possible to apply the gauge transformations to the generalized metric. It results in
\begin{align}
\delta_\lambda \mathcal{H}^{AB} &= \delta_\lambda S^{AB} + \delta_\lambda \epsilon^{AB} + \frac{1}{2} \delta_\lambda \epsilon^{AC} S_{CD} \epsilon^{DB} + \frac{1}{2}  \epsilon^{AC} S_{CD} \delta_\lambda \epsilon^{DB} + \mathcal{O}(\epsilon^2) \nonumber \\ &= \frac{1}{2} \big( \mathcal{L}_\lambda S^{AB} + \mathcal{L}_\lambda \epsilon^{AB} + \mathcal{L}_\lambda {S^{(A}}_C {S^{B)}}_D \epsilon^{CD} + \epsilon^{C(A} S_{CD} \mathcal{L}_\lambda S^{B)D} \big) + \mathcal{O}(\epsilon^2) \nonumber \\ &= \frac{1}{2} \big( \mathcal{L}_\lambda S^{AB} + \mathcal{L}_\lambda \epsilon^{AB} \big) + \mathcal{O}(\epsilon^2) = \frac{1}{2} \mathcal{L}_\lambda \mathcal{H}^{AB} + \mathcal{O}(\epsilon^2)\,.
\end{align} 
Here, we make use of the relation
\begin{equation}
{S^A}_C \epsilon^{CB} = - {S^B}_C \epsilon^{CA}
\end{equation}
which originates in~\bref{DFT*condition}.

Similarly, we find for a density like object such as the dilaton
\begin{equation}
\label{DFT*generalizeddiffs}
\delta_\lambda \tilde{d} = \frac{1}{2} \mathcal{L}_\lambda \tilde{d}\,, \quad \text{while} \quad \mathcal{L}_\lambda \tilde{d} = \lambda^A D_A \tilde{d} - \frac{1}{2} D_A \lambda^A\,.
\end{equation}
The derivation in this subsection remind a lot of the approach used in the two papers by Hohm, Hull, and Zwiebach~\citep{Hohm:2010jy, Hohm:2010pp}. However, the striking difference between original DFT and DFT$_{\mathrm{WZW}}$ lies in the occurrence of terms containing structure coefficients~\citep{Blumenhagen:2014gva}. These arise from the background vielbein~\eqref{eqn:leftmaurercartan}.

\subsection{Covariant derivative}
\label{sec:covderv}

Following from the underlying group manifold, we observe the emergence of structure coefficients in the whole theory, i.e. the action, generalized Lie derivatives, and gauge transformations. Subsequently, this raises the question whether we can absorb these terms by introducing a covariant derivative. Indeed, it is possible by defining a covariant derivative~\citep{Blumenhagen:2014gva} through
\begin{equation}
\label{DFT*covariantder}
\nabla_A V^B = D_A V^B + \frac{1}{3} {F^B}_{AC} V^C, \quad \text{and} \quad \nabla_A V_B = D_A V_B + \frac{1}{3} {F_{BA}}^C V_C\,.
\end{equation} 
Using the antisymmetry of the structure coefficients $F_{ABC}$, we can rewrite the strong constraint as
\begin{equation}
\label{eqn:SC}
\nabla_A D^A \cdot = \big( D_A D^A + \frac{1}{3} {F^A}_{AB} D^B \big) \cdot = D_A D^A \cdot \,.
\end{equation}
Note that the equation implies $\cdot$ to be a scalar. Nevertheless, this arises naturally since the strong constraint acts on fluctuations which are scalars from the background's point of view. It allows us to apply the generalized Lie derivative~\bref{DFT*generalizedlie} on arbitrary vectors by
\begin{equation}
\label{eqn:generalizeddiffeomorphism}
\mathcal{L}_\lambda V^A = \lambda^C \nabla_C V^A + \big( \nabla^A \lambda_C - \nabla_C \lambda^A \big) V^C \,.
\end{equation}
The generalization to arbitrary tensors follows in the same fashion as in toroidal DFT~\ref{generalizeddiffs}. Thus, yielding the same structure as in the original DFT formulation. The only difference lies in the replacement of partial derivatives by covariant ones.

\subsection{C-bracket}
\label{sec:cbracket}

We now examine whether the gauge transformations close to form a gauge algebra~\citep{Blumenhagen:2014gva}. Therefore, it is necessary to introduce a C-bracket analogously to original DFT. It is given by
\begin{equation}
\label{DFT*cbracket2}
[\lambda_1, \lambda_2]_C^A \equiv \lambda^B_1 D_B \lambda_2^A - \frac{1}{2} \lambda_1^B D^A \lambda_{2B} + \frac{1}{2} {F^A}_{BC} \lambda_1^B \lambda_2^C - (1 \leftrightarrow 2)\,,
\end{equation}
and using the covariant derivative~\bref{DFT*covariantder} it becomes
\begin{equation}
\label{DFT*cbracket1}
[\lambda_1, \lambda_2]_C^A \equiv \lambda^B_1 \nabla_B \lambda_2^A - \frac{1}{2} \lambda_1^B \nabla^A \lambda_{2B} - (1 \leftrightarrow 2)\,.
\end{equation}
In the abelian limit this gives rise to the C-bracket~\bref{c-bracket} known from toroidal DFT again. The last term in~\bref{DFT*cbracket2} extends it from toroidal backgrounds to the group manifold level. Moreover, it coincides with the results for the C-bracket obtained in~\citep{Geissbuhler:2013uka} for the flux formulation of DFT. Nevertheless, it is worth to mention that in the flux formulation~\citep{Geissbuhler:2011mx, Hohm:2013nja} the O($D,D$) metric $\eta$ is constant in curved and flat indices. However, for the DFT$_\mathrm{WZW}$ framework this is not the case. In flat indices the Killing metric $\eta$ remains constant, whereas in curved coordinates it becomes coordinates dependent.

Subsequently, we want to check whether the algebra closes. Hence, we evaluate the Jacobiator
\begin{equation}
J(\lambda_1, \lambda_2, \lambda_3) = [\lambda_1, [\lambda_2, \lambda_3]_C ]_C^A + [\lambda_3, [\lambda_1, \lambda_2]_C ]_C^A + [\lambda_2, [\lambda_3, \lambda_1]_C ]_C^A\,,
\end{equation}
while we impose that it must vanish up to trivial gauge transformations. This in return implies
\begin{equation}
\mathcal{L}_{J(\lambda_1, \lambda_2, \lambda_3)} = 0\,,
\end{equation}
or equivalently
\begin{equation}
\mathcal{L}_{[\lambda_1, \lambda_2]_C} V^A = [\mathcal{L}_{\lambda_1}, \mathcal{L}_{\lambda_2}] V^A\,.
\end{equation}
But this expression needs to be verified explicitly. Finally, computing the generalized Lie derivative of the C-bracket yields
\begin{align}
\mathcal{L}_{[\lambda_1, \lambda_2]_C} V^A &\,=  \mathcal{L}_{\lambda_1} \mathcal{L}_{\lambda_2} V^A - \mathcal{L}_{\lambda_2} \mathcal{L}_{\lambda_1} V^A \nonumber \\ &\:-\frac{1}{3} \big( {F_{BC}}^F {F_{FD}}^A + {F_{DB}}^F {F_{FC}}^A + {F_{CD}}^F {F_{FB}}^A \big) \lambda_1^B \lambda_2^C V^D \nonumber \\ &\,= [\mathcal{L}_{\lambda_1}, \mathcal{L}_{\lambda_2}] V^A\,,
\end{align}
where the second line vanishes as a result of the Jacobi identity
\begin{equation}
\label{DFT*Jacobi}
{F_{AB}}^E {F_{EC}}^D + {F_{CA}}^E {F_{EB}}^D + {F_{BC}}^E {F_{EA}}^D = 0
\end{equation}
which is always fulfilled by the background vielbein.
Thus, the gauge algebra closes up to trivial gauge transformations.\clearpage{} 
  \clearpage{}\chapter{Generalized Metric Formulation of DFT on Group Manifolds}
\label{Kap_3}

In this chapter we are going to introduce a generalized metric formulation~\cite{Blumenhagen:2015zma, Pascal2015}, similar to the one presented in the previous subsection~\ref{metricformulation}. The purpose of this generalized metric formulation is to cast all entities into doubled generalized objects, such as the generalized metric $\mathcal{H}^{AB}$ etc. We use them to write the action and gauge transformations in an efficient form. As a result the underlying structure becomes manifest and we are able to extend the CSFT results from cubic order to all orders in the fields~\ref{dft*action*}. Afterwards, we are going to derive the equations of motion~\ref{dft*eom} in this formulation and study its symmetries~\ref{sec:localsymm}. Astonishingly, besides the expected generalized diffeomorphism invariance, DFT$_\mathrm{WZW}$ also possesses an additional invariance under $2D$-diffeomorphisms. This symmetry is missing in the original DFT framework. The reason for it is the so-called \textit{extended strong constraint} which we are going to discuss in the latter part of this chapter~\ref{subextendedstrongconstraint}. Under it DFT$_\mathrm{WZW}$ reduces to toroidal DFT. 

This chapter is based upon~\cite{Blumenhagen:2015zma}.

\section{Field redefinition and toy example}
\label{sec:fieldredef}

For us to later be able to compare DFT$_\mathrm{WZW}$ with original DFT, we start by performing the following field redefinition~\citep{Blumenhagen:2015zma}
\begin{equation}
\epsilon^{a\bar{b}} \mapsto - 2 \epsilon^{a\bar{b}}\,,\quad \lambda^a \mapsto 2 \lambda^a\,,\quad \lambda^{\bar{a}} \mapsto 2 \lambda^{\bar{a}}\,.
\end{equation}
We obtain
\begin{align}
\label{DFT*action999}
(2\kappa^2) S = &\int d^{2D}X \sqrt{H} \Big( \epsilon_{a\bar{b}} \Box \epsilon^{a\bar{b}} + \big( D^a \epsilon_{a\bar{b}} \big)^2 \, + \big( D^{\bar{b}} \epsilon_{a\bar{b}} \big)^2 + 4\tilde{d} D_a D_{\bar{b}} \epsilon^{a\bar{b}} -4\tilde{d} \Box \tilde{d} \nonumber \\ &-2 \epsilon_{a\bar{b}} \Big( D^a \epsilon_{c\bar{d}} D^{\bar{b}} \epsilon^{c\bar{d}} - D^a \epsilon_{c\bar{d}} D^{\bar{d}} \epsilon^{c\bar{b}} - D^c \epsilon^{a\bar{d}} D^{\bar{b}} \epsilon_{c\bar{d}} \Big) + 4\tilde{d}^2 \Box \tilde{d} - 8\tilde{d}\, \epsilon^{a\bar{b}} \big( D_a D_{\bar{b}} \tilde{d} \big) \nonumber \\ &+2 \epsilon_{a\bar{b}} \Big( {F^{ac}}_d D^{\bar{e}} \epsilon^{d\bar{b}} \epsilon_{c\bar{e}} + {F^{\bar{b}\bar{c}}}_{\bar{d}} D^{e} \epsilon^{a\bar{d}} \epsilon_{e\bar{c}} \Big) + \frac{2}{3} F^{ace} F^{\bar{b}\bar{d}\bar{f}} \epsilon_{a\bar{b}}\epsilon_{c\bar{d}}\epsilon_{e\bar{f}}  \\ &+\tilde{d} \Big( 2 \big( D^a \epsilon_{a\bar{b}} \big)^2 \, + 2 \big( D^{\bar{b}} \epsilon_{a\bar{b}} \big)^2 + \big( D_c \epsilon_{a\bar{b}} \big)^2 + \big( D_{\bar{c}} \epsilon_{a\bar{b}} \big)^2 + 4 \epsilon^{a\bar{b}} \big( D_a D^c \epsilon_{a\bar{b}} + D_{\bar{b}} D^{\bar{c}} \epsilon_{a\bar{c}} \big) \Big)\,. \nonumber
\end{align}
First, we want to illustrate the idea of recasting DFT$_\mathrm{WZW}$ through doubled generalized entities. It is going to be an essential part of this chapter and therefore understanding it entirely is crucial. The DFT$_\mathrm{WZW}$ potential serves as a perfect toy example to demonstrate this procedure

Let us start by reprising the DFT$_{\mathrm{WZW}}$ potential~\citep{Blumenhagen:2014gva, Blumenhagen:2015zma}
\begin{equation}
\label{dft*potential}
V = -\frac{2}{3} F^{ace} F^{\bar{b}\bar{d}\bar{f}} \epsilon_{a\bar{b}}\, \epsilon_{c\bar{d}}\, \epsilon_{e\bar{f}}\,.
\end{equation}
We are interested in rewriting it using terms that contain doubled indices e.g. $A, B, \ldots$ which combine left- and right-movers. Switching from bared and unbared indices to doubled coordinates requires the introduction of doubled generalized objects. This change can be achieved using a perturbative expansion of the fields up to cubic order. As a result, we start by reintroducing some doubled generalized objects known from the previous chapter~\ref{Kap_2}. The structure coefficients in doubled indices are given by
\begin{equation}
\label{dft*structurecoeff}
{F_{AB}}^C = \begin{cases}
{F_{ab}}^c \\ {F_{\bar{a}\bar{b}}}^{\bar{c}} \\ 0 \quad \quad \text{otherwise}
\end{cases}\,,
\end{equation}
defining the underlying Ka$\check{c}$-Moody algebra. For the upcoming computation it will be essential to work with structure coefficients having only lower indices. Hence, we have to use the associated doubled flat metric~\citep{Blumenhagen:2014gva, Blumenhagen:2015zma}
\begin{equation}
\eta^{AB} = \frac{1}{2} \begin{pmatrix}
\eta^{ab} & 0 \\ 0 & -\eta^{\bar{a}\bar{b}}
\end{pmatrix},\quad \eta_{AB} = 2 \begin{pmatrix}
\eta_{ab} & 0 \\ 0 & -\eta_{\bar{a}\bar{b}}
\end{pmatrix}\,,
\end{equation}
where the metrics $\eta^{ab}, \eta^{\bar{a}\bar{b}}$ are governed by the underlying Lie algebra~\bref{liemetric}.
Subsequently, the structure coefficients can be expressed through
\begin{equation}
F_{ABC} = 2 \begin{cases}
\phantom{-}{F_{abc}} \\ -F_{\bar{a}\bar{b}\bar{c}} \\ \phantom{-}0 \quad \quad \text{otherwise}
\end{cases}\,.
\end{equation}
At this point, we introduce the analogue of the DFT generalized metric~\citep{Hohm:2010pp}. Expanding to all order in fields $\epsilon$ yields
\begin{align}
\label{dft*generalizemetric}
\mathcal{H}^{AB} = \text{exp}(\epsilon^{AB}) = S^{AB} + \epsilon^{AB} + \frac{1}{2} \epsilon^{AC} S_{CD} \epsilon^{DB} + \frac{1}{6} \epsilon^{AC} S_{CD} \epsilon^{DE} S_{EF} \epsilon^{FB} + \mathcal{O}(\epsilon^4)\,,
\end{align}
with
\begin{equation}
\label{smatrix}
S_{AB} = 2 \begin{pmatrix}
\eta_{ab} & 0 \\ 0 & \eta_{\bar{a}\bar{b}}
\end{pmatrix}\,,\quad\text{as well as the inverse}\quad S^{AB} = \frac{1}{2} \begin{pmatrix}
\eta^{ab} & 0 \\ 0 & \eta^{\bar{a}\bar{b}}
\end{pmatrix}\,,
\end{equation}
and the doubled fields
\begin{equation}
\epsilon^{AB} = \begin{pmatrix}
0 & \epsilon^{a\bar{b}} \\ \epsilon^{\bar{a}b} & 0 
\end{pmatrix} , \quad \text{with their symmetry condition} \quad \epsilon^{a\bar{b}} = ({\epsilon^T})^{\bar{b}a}\,.
\end{equation}
The doubled notation is a striking tool to simplify the equation significantly.

Using the perturbative expansion of the generalized metric~\bref{dft*generalizemetric}, we can evaluate it up to cubic order
\begin{equation}
\label{eqn:genmetrexpansion}
\mathcal{H}^{AB} = \begin{pmatrix}
\frac{1}{2 }\eta^{ab} + \epsilon^{a\bar{c}} \eta_{\bar{c}\bar{d}} \epsilon^{\bar{d}b} & \epsilon^{a\bar{b}} + \frac{2}{3} \epsilon^{a\bar{c}} \eta_{\bar{c}\bar{d}} \epsilon^{\bar{d}e} \eta_{ef} \epsilon^{f\bar{b}} \\ \epsilon^{\bar{a}b} + \frac{2}{3} \epsilon^{\bar{a}c} \eta_{cd} \epsilon^{d\bar{e}} \eta_{\bar{e}\bar{f}} \epsilon^{\bar{f}b} & \frac{1}{2 }\eta^{\bar{a}\bar{b}} + \epsilon^{\bar{a}c} \eta_{cd} \epsilon^{d\bar{b}}
\end{pmatrix}\,.
\end{equation}
Now, rewriting~\bref{dft*potential} requires us to express the potential through terms containing the symmetric generalized metric 
$\mathcal{H}^{AB}$ and applying the perturbative expansion~\eqref{eqn:genmetrexpansion}. A first conjecture would be the potential
\begin{equation}
\label{doubledpotential1}
\tilde{V} = -\frac{1}{12} F_{ACE} F_{BDF} \mathcal{H}^{AB}\mathcal{H}^{CD}\mathcal{H}^{EF}\,.
\end{equation}
(We already fixed the constants for later convenience.) Under consideration of the symmetry~\bref{dft*generalizemetric}, we find
\begin{align}
\tilde{V} =& -\frac{1}{4} F_{acd} {F_b}^{cd} \epsilon^{a\bar{x}} \eta_{\bar{x}\bar{y}} \epsilon^{\bar{y}b} - \frac{1}{24} F_{ace} F_{bdf} \eta^{ab} \eta^{cd} \eta^{ef} \nonumber \\
&- \frac{1}{4} F_{\bar{a}\bar{c}\bar{d}} {F_{\bar{b}}}^{\bar{c}\bar{d}} \epsilon^{\bar{a}x} \eta_{xy} \epsilon^{y\bar{b}} - \frac{1}{24} F_{\bar{a}\bar{c}\bar{e}} F_{\bar{b}\bar{d}\bar{f}} \eta^{\bar{a}\bar{b}} \eta^{\bar{c}\bar{d}} \eta^{\bar{e}\bar{f}} \nonumber \\
&-\frac{2}{3} F_{ace} F_{\bar{b}\bar{d}\bar{f}} \epsilon^{a\bar{b}} \epsilon^{c\bar{d}} \epsilon^{e\bar{f}} + \mathcal{O}(\epsilon^4)\,.
\end{align}
Hence, we need more additional terms to exactly reproduce~\bref{dft*potential}. Bearing DFT's flux formulation in mind~\cite{Geissbuhler:2011mx,Hohm:2013nja}, let us analyze what the following term
\begin{equation}
\frac{1}{4} F_{ACE} F_{BDF} \mathcal{H}^{AB} S^{CD} S^{EF}
\end{equation}
would affect. A straight forward computation yields
\begin{align}
\frac{1}{4} F_{ACE} F_{BDF} \mathcal{H}^{AB} S^{CD} S^{EF} =& \, \frac{1}{4} F_{ace} F_{bdf} \Big( \frac{1}{2} \eta^{ab} + \epsilon^{a\bar{x}} \eta_{\bar{x}\bar{y}} \epsilon^{\bar{y}b} \Big) \eta^{cd} \eta^{ef} \nonumber \\
+& \, \frac{1}{4} F_{\bar{a}\bar{c}\bar{e}} F_{\bar{b}\bar{d}\bar{f}} \Big( \frac{1}{2} \eta^{\bar{a}\bar{b}} + \epsilon^{\bar{a}x} \eta_{xy} \epsilon^{y\bar{b}} \Big) \eta^{\bar{c}\bar{d}} \eta^{\bar{e}\bar{f}} + \mathcal{O}(\epsilon^4)\,.
\end{align}
This allows us to conclude
\begin{align}
\tilde{V} &= -\frac{1}{12} F_{ACE} F_{BDF} \mathcal{H}^{AB}\mathcal{H}^{CD}\mathcal{H}^{EF} + \frac{1}{4} F_{ACE} F_{BDF} \mathcal{H}^{AB} S^{CD} S^{EF} \\
&= -\frac{2}{3} F_{ace} F_{\bar{b}\bar{d}\bar{f}} \epsilon^{a\bar{b}} \epsilon^{c\bar{d}} \epsilon^{e\bar{f}} + \frac{1}{12} F_{ace} F_{bdf} \eta^{ab} \eta^{cd} \eta^{ef} + \frac{1}{12} F_{\bar{a}\bar{c}\bar{e}} F_{\bar{b}\bar{d}\bar{f}} \eta^{\bar{a}\bar{b}} \eta^{\bar{c}\bar{d}} \eta^{\bar{e}\bar{f}} + \mathcal{O}(\epsilon^4)\,.  \nonumber
\end{align}
We now only need to fix the last two constant terms occurring in this equation. It can be easily achieved by the term
\begin{equation}
\frac{1}{6} F_{ACE} F_{BDF} S^{AB} S^{CD} S^{EF} =  \frac{1}{12} F_{ace} F_{bdf} \eta^{ab} \eta^{cd} \eta^{ef} + \frac{1}{12} F_{\bar{a}\bar{c}\bar{e}} F_{\bar{b}\bar{d}\bar{f}} \eta^{\bar{a}\bar{b}} \eta^{\bar{c}\bar{d}} \eta^{\bar{e}\bar{f}} + \mathcal{O}(\epsilon^4)\,.
\end{equation}
Altogether, we obtain for~\bref{dft*potential}
\begin{equation}
\label{eqn:dftpotentialresult}
V = F_{ACE} F_{BDF} \Big( - \frac{1}{12} \mathcal{H}^{AB}\mathcal{H}^{CD}\mathcal{H}^{EF} + \frac{1}{4} \mathcal{H}^{AB} S^{CD} S^{EF} - \frac{1}{6} S^{AB} S^{CD} S^{EF} \Big)\,
\end{equation}
using doubled generalized objects.
As can be verified without effort, our result~\eqref{eqn:dftpotentialresult} is in perfect agreement with the flux formulation~\cite{Geissbuhler:2011mx, Hohm:2013nja}. We can view it as a natural extension to non-trivial backgrounds~\citep{Blumenhagen:2015zma}.

\section{Action}
\label{dft*action*}

In this subsection, we want to rewrite the DFT$_{\mathrm{WZW}}$ action using doubled generalized object~\citep{Blumenhagen:2015zma}. The best way to do this begins with analyzing the toroidal DFT action~\citep{Hohm:2010pp} and investigating whether we are able to cast our action into a related form. Our guiding simply is quite simple, we replace all partial derivatives in original DFT with flat derivatives and observe the outcome.

But first, we need to comment on the dilaton. It splits into two parts, a background and a fluctuation part
\begin{equation}
\label{dft*dilaton}
e^{-2d} = e^{-2(\bar{d} +\tilde{d} )} = \sqrt{H} e^{-2\tilde{d}}\,.
\end{equation}
Furthermore, it should be noted that we assume the dilaton to be covariantly constant $\nabla_A d = D_A \tilde{d}$, implying it transforms as a scalar density. As a result, the background dilaton $\bar{d}$ is undynamical.

Let us start by computing the expression
\begin{align}
\label{equationxyz2}
e^{-2d} \frac{1}{8} \mathcal{H}^{CD} D_C \mathcal{H}_{AB} D_D \mathcal{H}^{AB}\,.
\end{align}
The computation becomes more transparent and better traceable by first expanding the term
\begin{align}
e^{-2d} \frac{1}{8} S^{CD} D_C \mathcal{H}_{AB} D_D \mathcal{H}^{AB}\,,
\end{align}
with $S^{AB}$ and $\mathcal{H}^{AB}$ as given in~\bref{dft*generalizemetric},~\bref{smatrix}. For simplicity, we drop for the successive calculations the term $\sqrt{H}$ in front of the individual terms. Hence,
\begin{align}
\label{equationxyz}
\frac{e^{-2d}}{8} S^{CD} D_C \mathcal{H}_{AB} D_D \mathcal{H}^{AB} &= \frac{1}{8} \Big( \underbrace{D_c \mathcal{H}_{ab} D^c H^{ab} + D_{\bar{c}} \mathcal{H}_{ab} D^{\bar{c}} H^{ab} + D_c \mathcal{H}_{\bar{a}\bar{b}} D^c H^{\bar{a}\bar{b}} + D_{\bar{c}} \mathcal{H}_{\bar{a}\bar{b}} D^{\bar{c}} H^{\bar{a}\bar{b}}}_{= \mathcal{O}(\epsilon^4)} \Big) \nonumber \\
&-\frac{1}{4} \Big( D_c \epsilon_{\bar{a}b} D^c \epsilon^{\bar{a}b} + D_{\bar{c}} \epsilon_{\bar{a}b} D^{\bar{c}} \epsilon^{\bar{a}b} + D_c \epsilon_{a\bar{b}} D^c \epsilon^{a\bar{b}} + D_{\bar{c}} \epsilon_{a\bar{b}} D^{\bar{c}} \epsilon^{a\bar{b}} \Big) + \mathcal{O}(\epsilon^4) \nonumber \\ &= -\frac{1}{2} \Big( D_c \epsilon_{a\bar{b}} D^c \epsilon^{a\bar{b}} + D_{\bar{c}} \epsilon_{a\bar{b}} D^{\bar{c}} \epsilon^{a\bar{b}} \Big) + \mathcal{O}(\epsilon^4) \\ &= \frac{1}{2}\, \epsilon_{a\bar{b}}\, \big( D^2 + \bar{D}^2 \big) \epsilon^{a\bar{b}} - \epsilon_{a\bar{b}} D_c \tilde{d} D^c \epsilon^{a\bar{b}} -  2 \epsilon_{a\bar{b}} D_{\bar{c}} \tilde{d} D^{\bar{c}} \epsilon^{a\bar{b}} + \mathcal{O}(\epsilon^4) \nonumber \\ &= \epsilon_{a\bar{b}} \Box \epsilon^{a\bar{b}} + \tilde{d} \big( D_c \epsilon_{a\bar{b}} \big)^2 + \tilde{d} \big( D_{\bar{c}} \epsilon_{a\bar{b}} \big)^2 + 2 \tilde{d}\, \epsilon_{a\bar{b}} \Box \epsilon^{a\bar{b}}  + \mathcal{O}(\epsilon^4)\,, \nonumber
\end{align}
for which we used
\begin{equation}
\Box = \frac{1}{2} \big( D^2 + \bar{D}^2 \big)\,.
\end{equation}
In the step from line (i) to (ii) in~\bref{equationxyz} we applied the symmetry of $\epsilon_{a\bar{b}}$. From line (ii) to (iii) we used that all terms are standing under an integral and performed integration by parts. For convenience, we stop writing $\mathcal{O}(\ldots)$. Being now familiar with the calculation, we are able to read off the remaining terms in~\bref{equationxyz2}.
Subsequently,
\begin{align}
e^{-2d}\frac{1}{8} \mathcal{H}^{CD} D_C \mathcal{H}_{AB} D_D \mathcal{H}^{AB} &= \epsilon_{a\bar{b}} \Box \epsilon^{a\bar{b}} - 2 \epsilon^{c\bar{d}} D_c \epsilon_{a\bar{b}} D_{\bar{d}} \epsilon^{a\bar{b}} + \tilde{d} \big( D_c \epsilon_{a\bar{b}} \big)^2 + \tilde{d} \big( D_{\bar{c}} \epsilon_{a\bar{b}} \big)^2\,. \nonumber
\end{align}
In this equation we used again the symmetry of the fields $\epsilon_{a\bar{b}}$ and that the last term in~\bref{equationxyz} will cancel with the term originating in the action's expansion of $e^{-2\tilde{d}}$.
\\
\big(I.e. $e^{-2\tilde{d}} \epsilon_{a\bar{b}} \Box \epsilon^{a\bar{b}} = (1-2\tilde{d}) \epsilon_{a\bar{b}} \Box \epsilon^{a\bar{b}}$ \big)

The next term we want to evaluate is
\begin{equation}
-e^{-2d}\frac{1}{2} \mathcal{H}^{AB} D_{B} \mathcal{H}^{CD} D_D \mathcal{H}_{AC}\,.
\end{equation}
This case requires us to execute a straightforward but rather cumbersome calculation. Making use of the commutation relations for the flat derivatives
\begin{equation}
[ D_a, D_b ] = {F_{ab}}^c D_c,\quad [ D_{\bar{a}}, D_{\bar{b}} ] = {F_{\bar{a}\bar{b}}}^{\bar{c}} D_{\bar{c}}\,,
\end{equation}
and performing partial integration we find
\begin{align}
-e^{-2d}\frac{1}{2} \mathcal{H}^{AB} D_{B} \mathcal{H}^{CD} D_D \mathcal{H}_{AC} &= \big( D^a e_{a\bar{b}} \big)^2 \, + \big( D^{\bar{b}} e_{a\bar{b}} \big)^2 \nonumber \\ &\,- \Big( {F_d}^{ac} D_c \epsilon^{d\bar{b}}  \epsilon_{a\bar{b}}\,+ {F_{\bar{d}}}^{\bar{a}\bar{c}} D_{\bar{c}} \epsilon^{b\bar{d}} \epsilon_{b\bar{a}} \Big) \big( 1-2\tilde{d} \big) \nonumber \\ &\,+ 2\tilde{d}\, \epsilon^{a\bar{b}} \big( D_a D^c \epsilon_{c\bar{b}} \big) - 2\tilde{d} \big( D_a D^c \epsilon^{a\bar{b}} \big) \epsilon_{c\bar{b}} - 2\tilde{d} \big( D^c \epsilon^{a\bar{b}} \big) \big( D_a \epsilon_{c\bar{b}} \big) \nonumber \\ &\,+2d\, \epsilon^{a\bar{b}} \big( D_{\bar{b}} D^{\bar{c}} \epsilon_{a\bar{c}} \big) -2\tilde{d} \big( D_{\bar{b}} D^{\bar{c}} \epsilon^{a\bar{b}} \big) \epsilon_{a\bar{c}} - 2\tilde{d} \big( D^{\bar{c}} \epsilon^{a\bar{b}} \big) \big( D_{\bar{b}} \epsilon_{a\bar{c}} \big) \nonumber \\ &\,+2 \epsilon_{a\bar{b}} \Big( D^a \epsilon_{c\bar{d}} D^{\bar{d}} \epsilon^{c\bar{b}} + D^c \epsilon^{a\bar{d}} D^{\bar{b}} \epsilon_{c\bar{d}} \Big)\,,
\end{align}
where we used again that $e^{-2d}$ stands in front of it.
It allows us to combine our two results into one expression
\begin{align}
&\;\;\;\;\;e^{-2d}\Big(\frac{1}{8} \mathcal{H}^{CD} D_C \mathcal{H}_{AB} D_D \mathcal{H}^{AB} -\frac{1}{2} \mathcal{H}^{AB} D_{B} \mathcal{H}^{CD} D_D \mathcal{H}_{AC} \Big) \nonumber \\ &= \epsilon_{a\bar{b}} \Box \epsilon^{a\bar{b}} + \big( D^a e_{a\bar{b}} \big)^2 \, + \big( D^{\bar{b}} e_{a\bar{b}} \big)^2 - \Big( {F_d}^{ac} D_c \epsilon^{d\bar{b}}  \epsilon_{a\bar{b}}\,+ {F_{\bar{d}}}^{\bar{a}\bar{c}} D_{\bar{c}} \epsilon^{b\bar{d}} \epsilon_{b\bar{a}} \Big) \big( 1-2\tilde{d} \big) \nonumber \\ &\,- 2 \epsilon_{a\bar{b}} \Big( D_a \epsilon_{c\bar{d}} D_{\bar{b}} \epsilon^{c\bar{d}} - D^a \epsilon_{c\bar{d}} D^{\bar{d}} \epsilon^{c\bar{b}} - D^c \epsilon^{a\bar{d}} D^{\bar{b}} \epsilon_{c\bar{d}} \Big) + \tilde{d} \big( D_c \epsilon_{a\bar{b}} \big)^2 +  \tilde{d} \big( D_{\bar{c}} \epsilon_{a\bar{b}} \big)^2 \nonumber \\ &\,+ 2\tilde{d}\, \epsilon^{a\bar{b}} \big( D_a D^c \epsilon_{c\bar{b}} \big) - 2 \tilde{d} \big( D_a D^c \epsilon^{a\bar{b}} \big) \epsilon_{c\bar{b}} - 2 \tilde{d} \big( D^c \epsilon^{a\bar{b}} \big) \big( D_a \epsilon_{c\bar{b}} \big) \nonumber \\ &\,+2\tilde{d}\, \epsilon^{a\bar{b}} \big( D_{\bar{b}} D^{\bar{c}} \epsilon_{a\bar{c}} \big) -2 \tilde{d} \big( D_{\bar{b}} D^{\bar{c}} \epsilon^{a\bar{b}} \big) \epsilon_{a\bar{c}} - 2 \tilde{d} \big( D^{\bar{c}} \epsilon^{a\bar{b}} \big) \big( D_{\bar{b}} \epsilon_{a\bar{c}} \big)\,.
\end{align}
In the limit of vanishing structure coefficients and no dilaton $d$ this result is already remarkably similar to the DFT$_\mathrm{WZW}$ action~\eqref{DFT*action999}. However, we are interested in reproducing all terms. Therefore, we evaluate the term
\begin{align}
4 \mathcal{H}^{AB} D_A \tilde{d} D_B \tilde{d} =  8 \epsilon^{a\bar{b}} D_a \tilde{d} D_{\bar{b}} \tilde{d} + 2 D_a \tilde{d} D^a \tilde{d} + 2 D_{\bar{a}} \tilde{d} D^{\bar{a}} \tilde{d}\,.
\end{align}
Under consideration of the additional prefactor $e^{-2d}$ it gives rise to
\begin{equation}
e^{-2d}\,4 \mathcal{H}^{AB} D_A \tilde{d} D_B \tilde{d} = 8 \epsilon^{a\bar{b}} D_a \tilde{d} D_{\bar{b}} \tilde{d} - 4 \tilde{d} \Box \tilde{d} + 4 \tilde{d}^2 \Box \tilde{d}\,.
\end{equation}
In this step, we applied the following relation
\begin{align}
- 4 \tilde{d}^2 \Box d = - 2 \tilde{d}^2 \big( D^2 + \bar{D}^2 \big) \tilde{d} = 4\tilde{d} D_a \tilde{d} D^a \tilde{d} + 4\tilde{d} D_{\bar{a}} \tilde{d} D^{\bar{a}} \tilde{d}\,,
\end{align}
which is a result of integrating by parts.
The last term we need to expand is
\begin{align}
- e^{-2d} 2 D_A \tilde{d} D_B \mathcal{H}^{AB} &= 4\tilde{d} D_a D_{\bar{b}} \epsilon^{a\bar{b}} - 8 \epsilon^{a\bar{b}} D_a d D_{\bar{b}} \tilde{d} - 8\tilde{d} \epsilon^{a\bar{b}} \big( D_a D_{\bar{b}} \tilde{d} \big) \nonumber \\ &\, +2 \tilde{d}  D_a D_b \big( \epsilon^{a\bar{c}} \eta_{\bar{c} \bar{d}} \epsilon^{b\bar{d}} \big) +2 \tilde{d} D_{\bar{a}} D_{\bar{b}} \big( \epsilon^{c\bar{a}} \eta_{cd} \epsilon^{d\bar{b}} \big) \nonumber \\ &=4\tilde{d} D_a D_{\bar{b}} \epsilon^{a\bar{b}} - 8 \epsilon^{a\bar{b}} D_a \tilde{d} D_{\bar{b}} \tilde{d} - 8 \tilde{d} \epsilon^{a\bar{b}} \big( D_a D_{\bar{b}} \tilde{d} \big) \nonumber \\ &\,+ 2\tilde{d} \big( D^a \epsilon_{a\bar{b}} \big)^2 + 2\tilde{d} \big( D^{\bar{b}} \epsilon_{a\bar{b}} \big)^2 + 2\tilde{d} \, \epsilon^{a\bar{b}} \big( D_a D^c \epsilon_{a\bar{b}} \big) + 2 \tilde{d}\, \epsilon^{a\bar{b}} \big( D_{\bar{b}} D^{\bar{c}} \epsilon_{a\bar{c}} \big) \nonumber \\ &\,+ 2 \tilde{d} \big( D^c \epsilon^{a\bar{b}} \big) \big( D_a \epsilon_{c\bar{b}} \big) + 2 \tilde{d} \big( D^{\bar{c}} \epsilon^{a\bar{b}} \big) \big( D_{\bar{b}} \epsilon_{a\bar{c}} \big) + 2 \tilde{d} \big( D_a D^c \epsilon^{a\bar{b}} \big) \epsilon_{c\bar{b}} \nonumber \\ &\,+ 2 \tilde{d} \big( D_{\bar{b}} D^{\bar{c}} \epsilon^{a\bar{b}} \big) \epsilon_{a\bar{c}}\,.
\end{align}
Finally, we obtain the action
\begin{align}
&\int d^{2D} X e^{-2d} \Big( \frac{1}{8} \mathcal{H}^{CD} D_C \mathcal{H}_{AB} D_D \mathcal{H}^{AB} -\frac{1}{2} \mathcal{H}^{AB} D_{B} \mathcal{H}^{CD} D_D \mathcal{H}_{AC} + 4 \mathcal{H}^{AB} D_A \tilde{d} D_B \tilde{d} \nonumber \\ &\quad\quad\quad\quad\quad\quad - 2 D_A \tilde{d} D_B \mathcal{H}^{AB} \Big) \nonumber \\
= &\int d^{2D}X \sqrt{H} \Big( \epsilon_{a\bar{b}} \Box \epsilon^{a\bar{b}} +  \big( D^a e_{a\bar{b}} \big)^2 \, + \big( D^{\bar{b}} e_{a\bar{b}} \big)^2 +4\tilde{d} D_a D_{\bar{b}} \epsilon^{a\bar{b}} -4\tilde{d} \Box \tilde{d} \nonumber \\ &\quad\quad\;\;\;\,-2 \epsilon_{a\bar{b}} \Big( D_a \epsilon_{c\bar{d}} D_{\bar{b}} \epsilon^{c\bar{d}} - D^a \epsilon_{c\bar{d}} D^{\bar{d}} \epsilon^{c\bar{b}} - D^c \epsilon^{a\bar{d}} D^{\bar{b}} \epsilon_{c\bar{d}} \Big) + 4\tilde{d}^2 \Box \tilde{d} - 8\tilde{d} \epsilon^{a\bar{b}} \big( D_a D_{\bar{b}} \tilde{d} \big) \nonumber \\ &\quad\quad\;\;\;\,+\tilde{d} \Big( 2 \big( D^a e_{a\bar{b}} \big)^2 \, + 2 \big( D^{\bar{b}} e_{a\bar{b}} \big)^2 + \big( D_c \epsilon_{a\bar{b}} \big)^2 + \big( D_{\bar{c}} \epsilon_{a\bar{b}} \big)^2 + 4 \epsilon^{a\bar{b}} \big( D_a D^c \epsilon_{a\bar{b}} + D_{\bar{b}} D^{\bar{c}} \epsilon_{a\bar{c}} \big) \Big) \nonumber \\ &\quad\quad\;\;\;\,- \Big( {F_d}^{ac} D_c \epsilon^{d\bar{b}}  \epsilon_{a\bar{b}}\,+ {F_{\bar{d}}}^{\bar{a}\bar{c}} D_{\bar{c}} \epsilon^{b\bar{d}} \epsilon_{b\bar{a}} \Big) \big( 1-2\tilde{d} \big) + \mathcal{O}(\epsilon^4)\,.
\end{align}
In the abelian limit $F_{ABC} = 0$ this action already coincides with the results from toroidal DFT~\cite{Hull:2009mi}. However, we still have not recovered all terms appearing in~\eqref{DFT*action999}. On one hand, we already recovered some terms containing structure coefficients in this action, but not all of them. On the other hand, we used only flat derivatives so far. Let us see whether we can obtain the missing terms by replacing flat derivatives with covariant ones~\citep{Blumenhagen:2015zma}, as given in~\bref{DFT*covariantder}.

Thus, we find
\begin{equation}
4 \mathcal{H}^{AB} \nabla_A d \nabla_B d = 4 \mathcal{H}^{AB} D_A \tilde{d} D_B \tilde{d}\,,
\end{equation}
and for the term
\begin{align}
- 2 \nabla_A d \nabla_B \mathcal{H}^{AB} &= - 2 D_A \tilde{d} \Big( D_B \mathcal{H}^{AB} + \frac{1}{3} \big( {F^A}_{BC} \mathcal{H}^{CB} + \underbrace{{F^B}_{BC} \mathcal{H}^{AC}}_{=\,0}  \big) \Big) \nonumber \\ &= - 2 D_A \tilde{d} D_B \mathcal{H}^{AB} - \frac{2}{3} D_A \tilde{d}\, {F^A}_{BC} \mathcal{H}^{CB} \nonumber \\ &= - 2 D_A \tilde{d} D_B \mathcal{H}^{AB} + \frac{2}{3} D_A \tilde{d}\, {F^A}_{BC} \mathcal{H}^{CB} \nonumber \\ &= - 2 D_A \tilde{d} D_B \mathcal{H}^{AB}\,,
\end{align}
we exploited the unimodularity ${F^A}_{AB} = 0$ of the Lie group~\bref{unimodular}. Additionally, we applied the symmetry of $\mathcal{H}^{AB}$ and the antisymmetry of $F_{ABC}$ in the third line , whereas from the third to fourth line we relabeled the indices. As a result, the last term needs to vanish.

This takes us to the more tricky part. We now have to expand
\begin{equation}
\frac{1}{8} \mathcal{H}^{CD} \nabla_C \mathcal{H}_{AB} \nabla_D \mathcal{H}^{AB} -\frac{1}{2} \mathcal{H}^{AB} \nabla_{B} \mathcal{H}^{CD} \nabla_D \mathcal{H}_{AC}\,.
\end{equation}
For the following computation we can ignore all terms consisting of more than three fields and more than one derivative. (We already computed terms containing two derivatives above) The first term gives rise to 
\begin{align}
\frac{1}{8} \mathcal{H}^{CD} \nabla_C \mathcal{H}_{AB} \nabla_D \mathcal{H}^{AB} &\,= \frac{1}{12} \mathcal{H}^{CD} D_C \mathcal{H}_{AB} {F^{A}}_{DE} \mathcal{H}^{EB} + \frac{1}{12} \mathcal{H}^{CD} {F_{AC}}^{F} \mathcal{H}_{FB} D_D \mathcal{H}^{AB} \nonumber \\ &\;+\frac{1}{36} \mathcal{H}^{CD} {F_{AC}}^{F} \mathcal{H}_{FB} {F^{A}}_{DE} \mathcal{H}^{EB} + \frac{1}{36} \mathcal{H}^{CD} {F_{AC}}^{F} \mathcal{H}_{FB} {F^{B}}_{DE} \mathcal{H}^{AE}\,. \nonumber
\end{align}
The second term requires us to apply the symmetry of $\mathcal{H}^{MN}$, i.e.
\begin{align}
\mathcal{H}^{CD} {F_{AC}}^{F} \mathcal{H}_{FB} D_D \mathcal{H}^{AB} &= \mathcal{H}^{CD} {F^{A}}_{CF} \mathcal{H}^{FB} D_D \mathcal{H}_{AB} \nonumber \\
&= \mathcal{H}^{CD} {F^{A}}_{DE} \mathcal{H}^{EB} D_C \mathcal{H}_{AB}\,,
\end{align}
and for the fourth term, we exploit the antisymmetry and cyclicity of the structure coefficients
\begin{align}
\mathcal{H}^{CD} {F_{AC}}^{F} \mathcal{H}_{FB} {F^{B}}_{DE} \mathcal{H}^{AE} &= \mathcal{H}^{CD} \mathcal{H}^{FB} \mathcal{H}^{AE} F_{ACF} F_{BDE} \nonumber \\ 
&= - F_{ACE} F_{BDF} \mathcal{H}^{AB} \mathcal{H}^{CD} \mathcal{H}^{EF}\,.
\end{align}
Altogether, we find the result
\begin{align}
\frac{1}{8} \mathcal{H}^{CD} \nabla_C \mathcal{H}_{AB} \nabla_D \mathcal{H}^{AB} &\,= \frac{1}{6} \mathcal{H}^{CD} D_C \mathcal{H}_{AB} {F^{A}}_{DE} \mathcal{H}^{EB} \\ &\;+\frac{1}{36} \mathcal{H}^{CD} {F_{AC}}^{F} {F^{A}}_{DE} \mathcal{H}_{FB} \mathcal{H}^{EB} - \frac{1}{36} F_{ACE} F_{BDF} \mathcal{H}^{AB} \mathcal{H}^{CD} \mathcal{H}^{EF}\,. \nonumber
\end{align}
For the second term, we obtain in a similar way
\begin{align}
-\frac{1}{2} \mathcal{H}^{AB} \nabla_{B} \mathcal{H}^{CD} \nabla_D \mathcal{H}_{AC} 
&\,= \frac{1}{3} \mathcal{H}^{CD} D_C \mathcal{H}_{AB} {F^{A}}_{DE} \mathcal{H}^{EB} - \frac{1}{6} \mathcal{H}^{AB} D_B \mathcal{H}^{CD} {F_{CD}}^E \mathcal{H}_{AE} \nonumber \\ &\;-\frac{1}{6} \mathcal{H}^{AB} {F^D}_{BE} \mathcal{H}^{CE} D_D \mathcal{H}_{AC} - \frac{1}{18} F_{ACE} F_{BDF} \mathcal{H}^{AB} \mathcal{H}^{CD} \mathcal{H}^{EF}  \nonumber \\ &\;+\frac{1}{18} \mathcal{H}^{CD} {F_{AC}}^{F} {F^{A}}_{DE} \mathcal{H}_{FB} \mathcal{H}^{EB}\,.
\end{align}
Combining these two results yields
\begin{align}
&\;\;\;\; \frac{1}{8} \mathcal{H}^{CD} \nabla_C \mathcal{H}_{AB} \nabla_D \mathcal{H}^{AB} -\frac{1}{2} \mathcal{H}^{AB} \nabla_{B} \mathcal{H}^{CD} \nabla_D \mathcal{H}_{AC} \\ &= \frac{1}{2} \mathcal{H}^{CD} D_C \mathcal{H}_{AB} {F^{A}}_{DE} \mathcal{H}^{EB} - \frac{1}{6} \mathcal{H}^{AB} D_B \mathcal{H}^{CD} {F_{CD}}^E \mathcal{H}_{AE} \nonumber \\ &-\frac{1}{6} \mathcal{H}^{AB} {F^D}_{BE} \mathcal{H}^{CE} D_D \mathcal{H}_{AC} - \frac{1}{12} F_{ACE} F_{BDF} \mathcal{H}^{AB} \mathcal{H}^{CD} \mathcal{H}^{EF} \nonumber \\ &+\frac{1}{12} \mathcal{H}^{CD} {F_{AC}}^{F} {F^{A}}_{DE} \mathcal{H}_{FB} \mathcal{H}^{EB}\,. \nonumber
\end{align}
We now evaluate them up to third order and derive
\begin{align}
&\;\;\;\;\; e^{-2d} \Big(\frac{1}{8} \mathcal{H}^{CD} \nabla_C \mathcal{H}_{AB} \nabla_D \mathcal{H}^{AB} -\frac{1}{2} \mathcal{H}^{AB} \nabla_{B} \mathcal{H}^{CD} \nabla_D \mathcal{H}_{AC} \Big) \\ &\,= 2 \Big( {F_d}^{ac} D_c \epsilon^{d\bar{b}}  \epsilon_{a\bar{b}}\,+ {F_{\bar{d}}}^{\bar{a}\bar{c}} D_{\bar{c}} \epsilon^{b\bar{d}} \epsilon_{b\bar{a}} \Big) \big( 1-2\tilde{d} \big) \nonumber \\ &\;+2 \epsilon_{a\bar{b}} \big( {F^{ac}}_d D^{\bar{e}} \epsilon^{d\bar{b}} \epsilon_{c\bar{e}} + {F^{\bar{b}\bar{c}}}_{\bar{d}} D^{e} \epsilon^{a\bar{d}} \epsilon_{e\bar{c}} \big) + \frac{2}{3} F_{ace} F_{\bar{b}\bar{d}\bar{f}} \epsilon^{a\bar{b}} \epsilon^{c\bar{d}} \epsilon^{e\bar{f}} \nonumber \\ &\;-\frac{1}{6} \Big( F_{acd} {F_b}^{cd} \epsilon^{a\bar{x}} \eta_{\bar{x}\bar{y}} \epsilon^{\bar{y}b} + F_{\bar{a}\bar{c}\bar{d}} {F_{\bar{b}}}^{\bar{c}\bar{d}} \epsilon^{\bar{a}x} \eta_{xy} \epsilon^{y\bar{b}} \Big) \big( 1 - 2\tilde{d} \big) + \mathcal{O}(D^2, \epsilon^4)\,, \nonumber
\end{align}
where we used~\bref{dft*potential}. In this equation, the terms obtained in the first line cancel the structure coefficients we collected from partial integration and the commutation relations. From the second line we receive the missing terms of~\bref{DFT*action999}, we were interested in retrieving. However, successfully reproducing action~\eqref{DFT*action999} requires the terms emerging in the last line to vanish. This might appear difficult at first glance, but we already acquired the necessary knowledge in section~\ref{sec:fieldredef}.

We just need to add the terms
\begin{align}
\label{xxxxmen}
&\;\;\;\;\; \frac{1}{6} F_{ACE} F_{BDF} \mathcal{H}^{AB} S^{CD} S^{EF} - \frac{1}{6} F_{ACE} F_{BDF} S^{AB} S^{CD} S^{EF} \\ &\,= \frac{1}{6} \Big( F_{acd} {F_b}^{cd} \epsilon^{a\bar{x}} \eta_{\bar{x}\bar{y}} \epsilon^{\bar{y}b} + F_{\bar{a}\bar{c}\bar{d}} {F_{\bar{b}}}^{\bar{c}\bar{d}} \epsilon^{\bar{a}x} \eta_{xy} \epsilon^{y\bar{b}} \Big) \big( 1 - 2\tilde{d} \big) + \mathcal{O}(\epsilon^4)\,.
\end{align}
Finally, we obtain DFT$_{\mathrm{WZW}}$ action~\citep{Blumenhagen:2015zma} in the generalized metric formulation by
\begin{align}
S = &\int d^{2D} X e^{-2d} \Big( \frac{1}{8} \mathcal{H}^{CD} \nabla_C \mathcal{H}_{AB} \nabla_D \mathcal{H}^{AB} -\frac{1}{2} \mathcal{H}^{AB} \nabla_{B} \mathcal{H}^{CD} \nabla_D \mathcal{H}_{AC} + 4 \mathcal{H}^{AB} \nabla_A d \nabla_B d \nonumber \\ &\quad\quad\quad\quad\quad\quad - 2 \nabla_A d \nabla_B \mathcal{H}^{AB} + \frac{1}{6} F_{ACE} F_{BDF} \mathcal{H}^{AB} S^{CD} S^{EF} - \frac{1}{6} F_{ACE} F_{BDF} S^{AB} S^{CD} S^{EF} \Big) \nonumber \\
= &\int d^{2D}X \sqrt{H} \Big( \epsilon_{a\bar{b}} \Box \epsilon^{a\bar{b}} + \big( D^a e_{a\bar{b}} \big)^2 \, + \big( D^{\bar{b}} e_{a\bar{b}} \big)^2 +4\tilde{d} D_a D_{\bar{b}} \epsilon^{a\bar{b}} -4\tilde{d} \Box \tilde{d} \nonumber \\ &-2 \epsilon_{a\bar{b}} \Big( D^a \epsilon_{c\bar{d}} D^{\bar{d}} \epsilon^{c\bar{b}} + D^c \epsilon^{a\bar{d}} D^{\bar{b}} \epsilon_{c\bar{d}} \Big) + 4\tilde{d}^2 \Box \tilde{d} - 8\tilde{d} \epsilon^{a\bar{b}} \big( D_a D_{\bar{b}} \tilde{d} \big) \\ &+2 \epsilon_{a\bar{b}} \big( {F^{ac}}_d D^{\bar{e}} \epsilon^{d\bar{b}} \epsilon_{c\bar{e}} + {F^{\bar{b}\bar{c}}}_{\bar{d}} D^{e} \epsilon^{a\bar{d}} \epsilon_{e\bar{c}} \big) + \frac{2}{3} F^{ace} F^{\bar{b}\bar{d}\bar{f}} \epsilon_{a\bar{b}}\epsilon_{c\bar{d}}\epsilon_{e\bar{f}}  \nonumber  \\ &\tilde{d} \Big( 2 \big( D^a e_{a\bar{b}} \big)^2 \, + 2 \big( D^{\bar{b}} e_{a\bar{b}} \big)^2 + \big( D_c \epsilon_{a\bar{b}} \big)^2 + \big( D_{\bar{c}} \epsilon_{a\bar{b}} \big)^2 + 4 \epsilon^{a\bar{b}} \big( D_a D^c \epsilon_{a\bar{b}} + D_{\bar{b}} D^{\bar{c}} \epsilon_{a\bar{c}} \big) \Big) + \mathcal{O}(\epsilon^4)\,. \nonumber
\end{align}
If we ignore constant terms, we are able to recast the covariant DFT$_{\mathrm{WZW}}$ action as
\begin{align}
\label{dft*action}
S = &\int d^{2D} X e^{-2d} \Big( \frac{1}{8} \mathcal{H}^{CD} \nabla_C \mathcal{H}_{AB} \nabla_D \mathcal{H}^{AB} -\frac{1}{2} \mathcal{H}^{AB} \nabla_{B} \mathcal{H}^{CD} \nabla_D \mathcal{H}_{AC}  \\ &\quad\quad\quad\quad\quad\quad + 4 \mathcal{H}^{AB} \nabla_A d \nabla_B d - 2 \nabla_A d \nabla_B \mathcal{H}^{AB} + \frac{1}{6} F_{ACE} F_{BDF} \mathcal{H}^{AB} S^{CD} S^{EF} \Big) \nonumber
\end{align}
in the generalized metric formulation. Considering this action~\bref{dft*action} in curved indices is straightforward. We only have to replace flat indices by curved ones. It follows immediately from the identity $\nabla_I {E_A}^J = 0$~\citep{Blumenhagen:2015zma}.

\section{Equations of motion}
\label{dft*eom}

During the previous section we obtained the action in the generalized metric formulation of DFT$_{\mathrm{WZW}}$~\citep{Blumenhagen:2015zma}. In this section, we want to derive the generalized curvature scalar and a generalized Ricci tensor in said formulation. The procedure is totally analogous to the one in original DFT~\citep{Hohm:2010pp}. Subsequently, the section is divided into two parts. A part~\ref{gencurvscal} where we vary after the dilaton which gives rise to the generalized curvature scalar, and a part~\ref{genricten} where we consider the variation after the generalized metric $\mathcal{H}^{AB}$. However, the latter part requires us to perform an additional projection as a consequence of the O($D,D$) constraint. This allows us to find the generalized Ricci tensor $\mathcal{R}_{AB}$.

\subsection{Generalized curvature scalar}
\label{gencurvscal}

We use the same methods used in~\citep{Hohm:2010pp, Blumenhagen:2015zma} to construct the generalized curvature. Therefore, we vary action~\eqref{dft*action} after the dilaton $d$. It yields
\begin{equation}
\delta S = -2 \int d^{2D}X e^{-2d} \mathcal{R} \delta d\,, 
\end{equation}
with the generalized Ricci scalar defined by
\begin{align}
\label{dft*gencurvature}
  \mathcal{R} &= 4 \mathcal{H}^{AB} \nabla_A \nabla_B d - \nabla_A \nabla_B \mathcal{H}^{AB} - 4 \mathcal{H}^{AB} \nabla_A d \, \nabla_B d + 4 \nabla_A d \, \nabla_B \mathcal{H}^{AB} \nonumber \\ &\,+ \frac{1}{8} \mathcal{H}^{CD} \nabla_C \mathcal{H}_{AB} \nabla_D \mathcal{H}^{AB} - \frac{1}{2} \mathcal{H}^{AB} \nabla_B \mathcal{H}^{CD} \nabla_D \mathcal{H}_{AC} + \frac{1}{6} F_{ACD} F_B{}^{CD} \mathcal{H}^{AB}\,.
\end{align}
During the derivation, we applied the following identities
\begin{align}
\label{dft*identities}
&\nabla_A d \, \nabla_B \mathcal{H}^{AB} = \nabla_A d D_B \mathcal{H}^{AB}\,, \nonumber \\
&\nabla_A \nabla_B \mathcal{H}^{AB} = D_A \nabla_B \mathcal{H}^{AB}\,, \nonumber \\
&\mathcal{H}^{MN} \nabla_M \nabla_N d =  H^{MN} D_M D_N \tilde{d}\,.
\end{align}
In the next section, we are going to show that $\mathcal{R}$ is indeed a scalar under generalized diffeomorphisms~\citep{Blumenhagen:2014gva, Blumenhagen:2015zma}, as was already expected from the CSFT calculations.

For later convenience, let us perform an partial integration of action~\bref{dft*action}. As a result, we are able to prove that~\bref{dft*action} is equivalent to the scalar curvature~\bref{dft*gencurvature} up to a total derivative~\citep{Hohm:2010pp}. Thus, we can rewrite the following term
\begin{align}
&\;\;\;\;e^{-2d} \big[ - 2 \nabla_A d \, \nabla_B \mathcal{H}^{AB} + 4 \mathcal{H}^{AB} \nabla_A d \, \nabla_B d \big] \nonumber \\ &= \sqrt{H} D_A \Big[ e^{-2\tilde{d}} \Big( \nabla_B \mathcal{H}^{AB} - 4 \mathcal{H}^{AB} \nabla_B d \Big)\Big] \nonumber \\ &+ e^{-2d} \Big[ -\nabla_A \nabla_B \mathcal{H}^{AB} - 4 \mathcal{H}^{AB} \nabla_A d \, \nabla_B d + 4 \nabla_A \Big( \mathcal{H}^{AB} \nabla_B d \Big)\Big]\,.
\end{align}
Subsequently, we recast action~\bref{dft*action} as
\begin{align}
\label{dft*gencurvaction}
S &= \int d^{2D}X e^{-2d} \mathcal{R} + \int \underbrace{d^{2D}X \sqrt{H} D_A \Big[ e^{-2\tilde{d}} \Big( \nabla_B \mathcal{H}^{AB} - 4 \mathcal{H}^{AB} \nabla_B d \Big)}_{\text{boundary term}}\Big] \nonumber \\ &= \int d^{2D}X e^{-2d} \mathcal{R}\,.
\end{align}
The last term in equation~\eqref{dft*gencurvaction} is just a boundary and hence vanishes at infinity. Due to the vielbein compatibility all flat indices in the scalar curvature~\eqref{dft*gencurvature} can again be replaced by covariant ones. Finally, the corresponding field equation is given by
\begin{equation}
\mathcal{R} = 0\,.
\end{equation}

\subsection{Generalized Ricci tensor}

\label{genricten}

Previously, we have varied the action~\bref{dft*action} with regards to the dilaton and obtained the generalized scalar curvature. By obeying this idea, we consider the variation after the generalized metric $\mathcal{H}^{AB}$~\bref{dft*generalizemetric}.

Following the same method outlined in~\citep{Hohm:2010pp, Blumenhagen:2015zma}, we start by evaluating the variation after the change $\delta \mathcal{H}^{AB}$ and successively symmetrize the obtained result $\mathcal{K}_{AB}$.

Specifically,
\begin{equation}
\delta S = \int d^{2D}X e^{-2d} \delta \mathcal{H}^{AB} \mathcal{K}_{AB}\,, 
\end{equation}
with
\begin{align}
\label{Ktensor}
\mathcal{K}_{AB} &= \frac{1}{8} \nabla_A \mathcal{H}_{CD} \nabla_B \mathcal{H}^{CD} - \frac{1}{4} \big[ \nabla_C - 2 (\nabla_C d) \big] \mathcal{H}^{CD} \nabla_D \mathcal{H}_{AB} + 2 \nabla_{(A} \nabla_{B)} d \\ &\,- \nabla_{(A} \mathcal{H}^{CD} \nabla_D \mathcal{H}_{B)C} + \big[ \nabla_D - 2 (\nabla_D d) \big] \big[ \mathcal{H}^{CD} \nabla_{(A} \mathcal{H}_{B)C} + {\mathcal{H}^C}_{(A} \nabla_C {\mathcal{H}^D}_{B)} \big]  \nonumber \\ &\,+ \frac{h^\vee}{3\alpha' k} S_{AB} \nonumber\,.
\end{align}
We now have to impose that the O($D,D$) constraint
\begin{equation}
\mathcal{H}^{AC} \eta_{CD} \mathcal{H}^{BD} = \eta^{AB}
\end{equation}
is preserved under this variation~\citep{Hohm:2010pp}. It implies that only certain parts of this tensor contain the dynamical information given by the field equations. Therefore, let us introduce the following projectors
\begin{equation}
P_{AB} = \frac{1}{2} \big( \eta_{AB} - S_{AB} \big)\,,\quad\text{and}\quad\bar{P}_{AB} = \frac{1}{2} \big( \eta_{AB} + S_{AB} \big)\,,
\end{equation}
which are used to define the equation of motion and the generalized Ricci tensor
\begin{equation}
\mathcal{R}_{AB} = {P_{(A}}^C {\bar{P}_{B)}}^{\;\;D} \mathcal{K}_{CD} = 0\,.
\end{equation}
These projection operators are very similar to the ones of the first chapter~\eqref{projectionoperators}. As a consequence of this projection, the term containing $S_{AB}$ in~\bref{Ktensor} drops out and hence gives no contribution. Altogether, the generalized Ricci tensor looks very much like the one known from toroidal DFT. Only the partial derivatives have to be replaced by covariant ones~\citep{Blumenhagen:2015zma}.

\section{Local symmetries}
\label{sec:localsymm}

We already expect from the Closed String Field Theory (CSFT) framework that DFT$_\mathrm{WZW}$ should be invariant under the corresponding gauge transformations~\citep{Blumenhagen:2014gva}. Nevertheless, the derivation of DFT$_{\mathrm{WZW}}$ required some demanding and complex steps. A lot can go wrong by missing prefactors and terms. Thus, we regard this section as an additional consistency check for the results of the previous chapter~\ref{Kap_2}. The previous two subsections~\ref{dft*action*},~\ref{dft*eom} have already provided an indication that everything went nicely. All of the bared and unbared entities integrated pleasingly into doubled generalized objects. Nevertheless, an explicit verification of the action's~\eqref{dft*action} invariance under the gauge transformations~\eqref{DFT*gaugetransformations} provides an even more compelling argument. As already mentioned, CSFT predicts this invariance up to cubic order. Better, we are able to show in section~\ref{GenDiffInv} that the generalized metric formulation of DFT$_{\mathrm{WZW}}$ stays invariant under its gauge transformations up to all orders in fields. This shows the entire power of our generalized metric formulation. Despite being invariant under generalized diffeomorphisms, action~\bref{dft*action} is manifestly invariant under $2D$-diffeomorphisms as well. We are going to demonstrate this in more detail during subsection~\ref{dft*2ddiffs}.

\subsection{Generalized diffeomorphism invariance}
\label{GenDiffInv}

At this point, we are interested in showing the invariance of action~\eqref{dft*action} under generalized diffeomorphisms~\citep{Blumenhagen:2015zma}. One can either use action~\bref{dft*action} to prove this invariance or work with the generalized curvature $\mathcal{R}$~\bref{dft*gencurvaction}. We choose the latter, as it significantly simplifies the computations. The subsequent proof consists of two steps: First, we start by showing that the generalized curvature behaves indeed like a scalar under generalized diffeomorphisms~\eqref{eqn:generalizeddiffeomorphism}. In this context, we apply the gauge transformations to each individual term of~\eqref{dft*gencurvature} and analyze their corresponding transformation behavior. Subsequently, we combine these results to obtain the failure~\eqref{failure} of the generalized curvature to transform covariantly~\citep{Hohm:2010pp, Blumenhagen:2015zma}. Secondly, we show that the dilaton prefactor $e^{-2d}$ transforms as a weight $+1$ scalar density under these generalized diffeomorphisms.

We start by recalling DFT$_{\mathrm{WZW}}$'s gauge transformations~\citep{Blumenhagen:2014gva, Blumenhagen:2015zma} from chapter~\ref{Kap_2}
\begin{equation}
\label{eqn:gendiffHAB&tilded}
\delta_\xi V^A = \mathcal{L}_\xi V^A,\quad \delta_\xi \tilde{d} =  \mathcal{L}_\xi \tilde{d}\,,
\end{equation}
with 
\begin{equation}
\mathcal{L}_\xi V^A = \xi^B \nabla_B V^A + \big( \nabla^A \xi_B - \nabla_B \xi^A \big) V^B\,,
\end{equation}
where the usual generalization to higher order tensors applies. For the dilaton, we have
\begin{equation}
\mathcal{L}_\xi \tilde{d} = \xi^A D_A \tilde{d} - \frac{1}{2} D_A \xi^A\,. 
\end{equation}
Showing the invariance under generalized diffeomorphisms for the generalized curvature~\bref{dft*gencurvature} is achieved best by comparing the results after application of the gauge transformations with the results we obtain from applying the generalized diffeomorphisms. This comparison lets us read off the failure to transform covariantly, specifically
\begin{equation}
\label{dft*failure}
\Delta_\xi V = \delta_\xi V - \mathcal{L}_\xi V\,,
\end{equation}
and it satisfies the product rule
\begin{equation}
\Delta_\xi \big( V W \big) = \big( \Delta_\xi V \big) W + V \big( \Delta_\xi W \big)\,.
\end{equation}
It should be noted that generalized diffeomorphisms $\delta_\xi$ only act on fields as opposed to the generalized Lie derivative $\mathcal{L}_\xi$ which acts on the full tensorial structure, e.g. 
\begin{align}
\Delta_\xi \big( D_A \mathcal{H}^{BC} \big) &= \delta_\xi \big( D_A \mathcal{H}^{BC} \big) - \mathcal{L}_\xi \big( D_A \mathcal{H}^{BC} \big) \nonumber \\ &= D_A \big( \mathcal{L}_\xi \mathcal{H}^{BC} \big) - \mathcal{L}_\xi \big( D_A \mathcal{H}^{BC} \big)\,.
\end{align} 
Trivially, from the definition~\bref{dft*failure} it directly follows
\begin{equation}
\Delta_\xi \mathcal{H}^{AB} = 0\,, \quad \text{and} \quad \Delta_\xi \tilde{d} = 0 \,.
\end{equation}
Now, we pursue a similar route as in~\citep{Hohm:2010pp, Blumenhagen:2015zma}. After imposing the strong constraint~\bref{strongconstraint}, we compute the failure $\Delta_\xi$ for each individual term occurring in the generalized curvature~\eqref{dft*gencurvature}. This allows us to combine all the individual results and use them to evaluate $\Delta_\xi \mathcal{R}$, by exploiting the product rule and linearity of the failure.

We start out by computing the term
\begin{equation}
\Delta_\xi \big( \nabla_A d \big) = \Delta_\xi \big( D_A \tilde{d} \big) = - \frac{1}{2} D_A \big( D_D \xi^D \big)\,,
\end{equation}
and since equation~\bref{dft*identities} holds, we only have to consider
\begin{equation}
\Delta_\xi \big( D_A D_B \tilde{d} \big) = \big( D_A D_B \xi^D \big) D_D \tilde{d} - \frac{1}{2} D_A D_B \big( D_D \xi^D \big) + {F_{BD}}^{C} \big( D_A \xi^D \big) \big( D_C \tilde{d} \big)\,.
\end{equation}
Additionally, we find
\begin{align}
\Delta_\xi \big( \nabla_A \mathcal{H}^{BC} \big) &= 2 D_A D^{(B} \xi_D \mathcal{H}^{C)D} - 2 D_A D_D \xi^{(B} \mathcal{H}^{C)D} + \frac{2}{3} {F^{(B}}_{AE} H^{C)D} \Big( D^E \xi_D - D_D \xi^E \Big) \nonumber \\ &+\frac{4}{3} {F^{(B}}_{DE} \mathcal{H}^{C)E} D_A \xi^D + \frac{2}{3} {F^{(B}}_{DE} \mathcal{H}^{C)E} D^D \xi_A \nonumber \\ &+\frac{2}{3} {F^D}_{AE} \big( D_D \xi^{(B} \big) \mathcal{H}^{C)E} - \frac{2}{3} {F^D}_{AE} \big( D^{(B} \xi_D \big) H^{C)E},
\end{align}
\clearpage
while
\begin{align}
\Delta_\xi \big( \nabla_A \nabla_B \mathcal{H}^{AB} \big) &= \frac{2}{3} F_{ACE} {F^{E}}_{BD} \xi^C D^B \mathcal{H}^{AD} + \frac{4}{3} F_{ACE} {F^{E}}_{BD} \mathcal{H}^{AB} D^C \xi^D - \frac{1}{3} F_{ACE} \mathcal{H}^{AB} D_B D^C \xi^E \nonumber \\ &\,+\frac{2}{3} F_{ACE} \xi^A D^C D_F \mathcal{H}^{EF} + \frac{10}{3} F_{ACE} \mathcal{H}^{AB} D^C D_B \xi^E + 2 F_{ACE} D^A \xi^C D_D \mathcal{H}^{DE} \nonumber \\ &\,+F_{ACE} D^A \mathcal{H}^{DE} D_D \xi^C - \frac{2}{3} F_{ACE} \xi^A D_D D^C \mathcal{H}^{DE} - D_A D_B \xi^C D_C \mathcal{H}^{AB} \nonumber \\ &\,- 2D_A \mathcal{H}^{AB} D_C D_B \xi^C - 2 \mathcal{H}^{AB} D_C D_A D_B \xi^C +\frac{2}{27} F_{ACE} F_{BDF} F^{EDF} \mathcal{H}^{BC} \xi^A \,.
\end{align}
We canceled all terms of the form
\begin{equation}
\label{DFT*advancedstrongconstraint}
F_{ABC} \big(D^B \cdot \big) \big(D^C \cdot \big) = \big( D^B \cdot \big) \big( [D_A,D_B] \cdot \big) = 0
\end{equation}
in the last equation, as they vanish as a result of the strong constraint~\bref{DFT*strongconstraint}.
We can now combine these results to calculate the failure of the curvature~\bref{dft*gencurvature} to transform covariantly without the term $F_{ACE} F_{BDF} \mathcal{H}^{AB} S^{CD} S^{EF}$. Let us denote it by $\mathcal{\tilde{R}}$ and the failure yields
\begin{align}
\Delta_\xi \mathcal{\tilde{R}} &= \frac{1}{6} \Big( - F_{ABH} F_{CDF} {F_{EG}}^H + \frac{1}{3} F_{AFH} F_{CGI} {F_{E}}^{HI} \eta_{BD} - \frac{1}{3} F_{ACH} F_{EFI} {F_{G}}^{HI} \eta_{BD}  \Big) \mathcal{H}^{BC} \mathcal{H}^{DE} \mathcal{H}^{FG} \xi^A \nonumber \\ &\,+\frac{1}{3}F_{ACD} {F^{CD}}_E \mathcal{H}^{AB} D_B \xi^E + \frac{1}{6} F_{ACD} {F^{CD}}_E \mathcal{H}^{AB} D^E \xi_B \\ &\,+\frac{1}{6} \Big( F_{IAG} {F^G}_{CD} + F_{CIG} {F^G}_{AD} + F_{ACG} {F^G}_{ID} \Big) \mathcal{H}^{BC} \mathcal{H}^{DE} \xi^A D_E \mathcal{H}^{FI} \nonumber \\ &\,+F_{ACD} D^A \xi_B D^C \mathcal{H}^{BD} - \frac{1}{2} F_{ACD} D^A \mathcal{H}^{BD} D_B \xi^C + F_{ACD} \mathcal{H}^{AB} D^D D^C \xi_B \nonumber \\ &\,-\frac{1}{2} F_{ACD} \mathcal{H}^{AB} \mathcal{H}^{EF} D_F \xi^D D^C \mathcal{H}_{BE} +\frac{1}{2} F_{ACD} \mathcal{H}^{AB} \mathcal{H}^{EF} D^C \xi_E D^D \mathcal{H}_{BF}\,. \nonumber
\end{align}
We ordered all terms according to the number derivatives, and used that terms with three flat derivatives vanish in the same way as in  toroidal DFT~\citep{Hohm:2010pp, Blumenhagen:2015zma}. In the third line all terms disappear as a consequence of the Jacobi identity~\bref{DFT*Jacobi}.

For the zeroth order terms, we apply the following recasting
\begin{align}
&= \frac{1}{6} \Big( - F_{ABH} F_{CDF} {F_{EG}}^H + \frac{1}{3} F_{AFH} F_{CGI} {F_{E}}^{HI} \eta_{BD} - \frac{1}{3} F_{ACH} F_{EFI} {F_{G}}^{HI} \eta_{BD}  \Big) \mathcal{H}^{BC} \mathcal{H}^{DE} \mathcal{H}^{FG} \xi^A \nonumber \\ &= -\frac{1}{36} \mathcal{H}^{AB} \xi^G \Big[ {F_{EA}}^P \mathcal{H}_{PF} + {F_{FA}}^P \mathcal{H}_{PE} \Big] \times \nonumber \\ &\quad \Big[\Big( {F_B}^{EJ} F_{GHJ} + {F_G}^{EJ} F_{HBJ} + {F_H}^{EJ} F_{BGJ} \Big) \mathcal{H}^{FH} \nonumber \\ &\,+ \Big( {F_B}^{FJ} F_{GHJ} + {F_G}^{FJ} F_{HBJ} + {F_H}^{FJ} F_{BGJ} \Big) \mathcal{H}^{EH} \Big]\,.
\end{align}
Thus, following the the Jacobi identity they vanish. Hence,
\begin{align}
\Delta_\xi \mathcal{\tilde{R}} &= \frac{1}{3}F_{ACD} {F^{CD}}_E \mathcal{H}^{AB} D_B \xi^E + \frac{1}{6} F_{ACD} {F^{CD}}_E \mathcal{H}^{AB} D^E \xi_B \\ &\,+F_{ACD} D^A \xi_B D^C \mathcal{H}^{BD} - \frac{1}{2} F_{ACD} D^A \mathcal{H}^{BD} D_B \xi^C + F_{ACD} \mathcal{H}^{AB} D^D D^C \xi_B \nonumber \\ &\,-\frac{1}{2} F_{ACD} \mathcal{H}^{AB} \mathcal{H}^{EF} D_F \xi^D D^C \mathcal{H}_{BE} +\frac{1}{2} F_{ACD} \mathcal{H}^{AB} \mathcal{H}^{EF} D^C \xi_E D^D \mathcal{H}_{BF} \nonumber\,.
\end{align}
Furthermore, we exploit the O($D,D$) property of $\mathcal{H}^{AB}$
\begin{equation}
\mathcal{H}_{AB}\mathcal{H}^{BC} = {\delta_A}^C\, \quad \text{and subsequently} \quad D_D \mathcal{H}_{AB} \mathcal{H}^{BC} = - \mathcal{H}_{AB} D_D \mathcal{H}^{BC}\,,
\end{equation}
which allows us to simplify the remaining terms to
\begin{align}
\Delta_\xi \mathcal{\tilde{R}} &= \frac{1}{3}F_{ACD} {F^{CD}}_E \mathcal{H}^{AB} D_B \xi^E + \frac{1}{6} F_{ACD} {F^{CD}}_E \mathcal{H}^{AB} D^E \xi_B \\ &\,+\underbrace{F_{ACD} D^A \xi_B D^C \mathcal{H}^{BD}}_{(1)} - \underbrace{\frac{1}{2} F_{ACD} D^A \mathcal{H}^{BD} D_B \xi^C}_{(2)} + F_{ACD} \mathcal{H}^{AB} D^D D^C \xi_B \nonumber \\ &\,+\underbrace{\frac{1}{2} F_{ACD} D^C \mathcal{H}^{AB} D_B \xi^D}_{(2)} -\underbrace{\frac{1}{2} F_{ACD} D^D \mathcal{H}^{AB} D^C \xi_B}_{(1)} \nonumber\,.
\end{align}
The terms $(1)$ can be combined, and the terms $(2)$ cancel each other, leaving us only with
\begin{align}
\Delta_\xi \mathcal{\tilde{R}} &= \frac{1}{3}F_{ACD} {F^{CD}}_E \mathcal{H}^{AB} D_B \xi^E + \frac{1}{6} F_{ACD} {F^{CD}}_E \mathcal{H}^{AB} D^E \xi_B \\ &\,+\frac{1}{2} F_{ACD} D^A \xi_B D^C \mathcal{H}^{BD} + F_{ACD} \mathcal{H}^{AB} D^D D^C \xi_B \nonumber\,.
\end{align}
Moreover, through using the antisymmetry of the structure coefficients we identify
\begin{align}
F_{ACD} \mathcal{H}^{AB} D^D D^C \xi_B &= \frac{1}{2} F_{ACD} \mathcal{H}^{AB} \big[D^D,D^C\big] \xi_B = \frac{1}{2} F_{ACD} \mathcal{H}^{AB} {F^{DC}}_E D^E \xi_B \nonumber \\ &= -\frac{1}{2} F_{ACD} \mathcal{H}^{AB} {F^{CD}}_E D^E \xi_B\,,
\end{align}
and subsequently
\begin{align}
\Delta_\xi \mathcal{\tilde{R}} &= \frac{1}{3}F_{ACD} {F^{CD}}_E \mathcal{H}^{AB} D_B \xi^E - \frac{1}{3} F_{ACD} {F^{CD}}_E \mathcal{H}^{AB} D^E \xi_B \\ &\,+\frac{1}{2} F_{ACD} D^A \xi_B D^C \mathcal{H}^{BD} \nonumber.
\end{align}
Here, the last term vanishes as a result of~\bref{DFT*advancedstrongconstraint}. For the terms in the first line we apply the normalization of the structure coefficients regarding the Killing metric~\bref{liemetric}.
\begin{equation}
F_{ACD} {F_B}^{CD} = \frac{2h^\vee}{\alpha'k} S_{AB}\,.
\end{equation}
Under consideration of the generalized metric's $\mathcal{H}$ symmetry, this finally yields
\begin{align}
\Delta_\xi \mathcal{\tilde{R}} &= \frac{1}{3}F_{ACD} {F^{CD}}_E \mathcal{H}^{AB} D_B \xi^E - \frac{1}{3} F_{ACD} {F^{CD}}_E \mathcal{H}^{AB} D^E \xi_B \\ &= \frac{2h^\vee}{3\alpha' k} \Big[ S_{AE} \mathcal{H}^{AB} D_B \xi^E - S_{AE} \mathcal{H}^{AB} D^E \xi_B \Big].
\end{align}
The up to now non-vanishing $\Delta_\xi \mathcal{\tilde{R}}$ should cancel with the remaining term
\begin{align}
\frac{1}{6} F_{ACE} F_{BDF} \mathcal{H}^{AB} S^{CD} S^{EF} &= \frac{1}{6} F_{ACE} F_{BDF} \mathcal{H}^{AB} \eta^{CD} \eta^{EF} \\ &= \frac{1}{6} F_{ACD} {F^{CD}}_B \mathcal{H}^{AB} \nonumber \\ &= \frac{h^\vee}{3\alpha'k} S^{AB} \mathcal{H}_{AB}\,.
\end{align}
In fact, we find
\begin{align}
\frac{h^\vee}{3\alpha'k} \Delta_\xi \big( S^{AB} \mathcal{H}_{AB} \big) &= \phantom{-} \frac{h^\vee}{3\alpha'k} \Delta_\xi \big( S^{AB} \big) \mathcal{H}_{AB}  \\ &= -\frac{h^\vee}{3\alpha'k} \mathcal{L}_\xi \big( S^{AB} \big) \mathcal{H}_{AB} \nonumber \\ &= - \frac{2h^\vee}{3\alpha' k} \Big[ S_{AE} \mathcal{H}^{AB} D_B \xi^E - S_{AE} \mathcal{H}^{AB} D^E \xi_B \Big].
\end{align}
Here, we used $\delta_\xi S^{AB} = 0$ (Gauge transformations act on fluctuations, they do not act on background fields.~\citep{Blumenhagen:2014gva})
Ultimately, we arrive at the desired result
\begin{equation}
\label{dft*dilatonequation}
\Delta_\xi \mathcal{R} = \Delta_\xi \mathcal{\tilde{R}} + \frac{1}{6} \Delta_\xi \big( F_{ACD} {F_B}^{CD} \big) = 0\,.
\end{equation}
Consequently, we have proven the generalized curvature behaves as a  scalar~\bref{dft*gencurvature} under generalized diffeomorphisms~\bref{DFT*generalizeddiffs}~\citep{Blumenhagen:2015zma}.

Last but not least, we need to consider the dilaton factor $e^{-2d}$ and analyze its behavior under generalized diffeomorphisms.
Therefore, we recast the generalized Lie derivative applied to dilaton fluctuations $\tilde{d}$ through covariant derivatives
\begin{equation}
\mathcal{L}_\xi \tilde{d} = \xi^A \nabla_A \tilde{d} - \frac{1}{2} \nabla_A \xi^A = \xi^A D_A \tilde{d} - \frac{1}{2} D_A \xi^A - \frac{1}{6} {F^A}_{AB} \xi^B\,.
\end{equation}
As a result of the unimodularity of the underlying Lie algebra, the last term vanishes in this equation.
Moreover, we obtain
\begin{equation}
\delta_\xi e^{-2d} = -2 e^{-2d} \delta_\xi d = -2 e^{-2d} \mathcal{L}_\xi \tilde{d}.
\end{equation}
It implies that gauge transformations do not affect the background dilaton $\bar{d}$. By writing $\mathcal{L}_\xi \tilde{d}$ with covariant derivatives we can simply replace flat by curved indices. Plugging this into~\bref{dft*dilatonequation}, we arrive at the equation
\begin{align}
\delta_\xi e^{-2d} &= \xi^I \partial_I e^{-2d} + e^{-2d} \big( \nabla_I \xi^I + \xi^I 2\partial_I \bar{d} \big) = \xi^I \partial_I e^{-2d} + e^{-2d} \big( \nabla_I \xi^I - {\Gamma_{JI}}^J \xi^I \big) \nonumber \\ &= \xi^I \partial_I e^{-2d} + e^{-2d} \partial_I \xi^I = \partial_I \big( \xi^I e^{-2d} \big)\,,
\end{align}
and we applied
\begin{equation}
{\Gamma_{JI}}^J = -2 \partial_I \bar{d}\,,
\end{equation}
as has been given in~\citep{Blumenhagen:2014gva}. As a result, $e^{-2d}$ transforms as a scalar density with weight $+1$. Therefore, action $e^{-2d} \, \mathcal{R}$~\eqref{dft*gencurvaction} remains invariant under generalized diffeomorphisms.

Moreover, we can show that the generalized Lie derivative~\bref{DFT*generalizedlie} is transforming covariantly under generalized diffeomorphisms as well. Implicitly, it has already been proven by the closure of the gauge algebra~\citep{Blumenhagen:2014gva}
\begin{equation}
\label{dft*xyz222}
\big[ \mathcal{L}_{\xi_1}, \mathcal{L}_{\xi_2} \big] V^A = \mathcal{L}_{[\xi_1, \xi_2]_C} V^A\,.
\end{equation}
Nevertheless, we consider
\begin{equation}
\Delta_\xi \mathcal{L}_\lambda V^A = \mathcal{L}_\xi \big( \mathcal{L}_\lambda	V^A \big) - \mathcal{L}_{\mathcal{L}_\xi \lambda} V^A - \mathcal{L}_\lambda \big( \mathcal{L}_\xi	V^A \big)\,.
\end{equation}
Subsequently, in conjunction with~\bref{dft*xyz222}, we find
\begin{equation}
\Delta_\xi \mathcal{L}_\lambda V^A = \mathcal{L}_{[\xi, \lambda]_C} V^A - \mathcal{L}_{\mathcal{L}_\xi \lambda} V^A = 0\,,
\end{equation}
since we are able to express the C-bracket by
\begin{equation}
\big[\xi, \lambda \big]^A_C = \mathcal{L}_\xi \lambda^A - \frac{1}{2} \nabla^A \big( \xi_B \lambda^B \big)\,,
\end{equation}
through a generalized Lie derivative (first term) and a trivial gauge transformation (last term), which does not affect the generalized Lie derivative~\citep{Blumenhagen:2015zma}.

\subsection{Local 2D-diffeomorphism invariance}
\label{dft*2ddiffs}

As opposed to the previous subsection~\ref{GenDiffInv}, we can also use standard Lie derivatives to parametrize a change of fields in DFT$_\mathrm{WZW}$. It gives rise to the well-known diffeomorphism invariance. However, instead of $D$-dimensions for $2D$-dimensions  as a consequence of the doubled space~\citep{Blumenhagen:2015zma}. We prove this by investigating the individual fields appearing in the DFT$_\mathrm{WZW}$ action~\eqref{dft*action}. The strategy of this proof works similar as in the previous subsection~\ref{GenDiffInv}. But in contrast to the previous subsection, we do not impose the strong constraint in any of the presented steps.

Again, we start by defining the failure to transform covariantly. Although, we have to replace the generalized diffeomorphism with the standard diffeomorphism. The same goes for the Lie derivatives as well. This yields
\begin{equation}
\Delta_\xi V = \delta_\xi V - L_\xi V\,.
\end{equation}
In this context, the standard Lie derivative~\bref{lie1} is given by
\begin{equation}
\delta_\xi V^I = L_\xi V^I = \xi^P \partial_P V^I - V^P \partial_P \xi^I\,.
\end{equation}
For the generalized dilaton fluctuations $\tilde{d}$ and the generalized vielbein ${E_A}^I$ we obtain the following results
\begin{equation}
\delta_\xi \tilde{d} = L_\xi \tilde{d} = \xi^P \delta_P \tilde{d}\,,
\end{equation}
\begin{equation}
\delta_\xi {E_A}^I = L_\xi {E_A}^I = \xi^J \partial_J {E_A}^I - {E_A}^J \partial_J \xi^I \,.
\end{equation}
As a result, the generalized dilaton transforms as a scalar while the generalized vielbein transforms like a vector.

In order to prove the invariance under standard diffeomorphisms we only need to consider three different terms.
Let us start by introducing the covariant derivative~\citep{Blumenhagen:2014gva}
\begin{equation}
\nabla_I V^J = \partial_I V^J + {\Gamma^J}_{IL} V^L\,.
\end{equation}
Remembering the relation $\Delta_\xi V^I = 0$, we can reduce the failure of this term to
\begin{equation}
\label{dft*xyz333}
\Delta_\xi \big( \nabla_I V^J \big) = \Delta_\xi \big( \partial_I V^J \big) + \Delta_\xi \big( {\Gamma^J}_{IL} \big) V^L\,.
\end{equation}
Clearly, this term should vanish as a consequence of the covariant derivative's structure. Straightforward, the first term in equation~\bref{dft*xyz333} yields
\begin{equation}
\Delta_\xi \big( \partial_I V^J \big) = - V^L \partial_L \partial_I \xi^J\,.
\end{equation}
Computing the second term is slightly more involved. Using the definition of the Christoffel symbols~\citep{Blumenhagen:2014gva, Blumenhagen:2015zma}
\begin{equation}
{\Gamma_{IJ}}^K = - \frac{1}{3} \big( 2 {\Omega_{IJ}}^K + {\Omega_{JI}}^K \big)\,
\end{equation}
with the coefficients of anholonomy in curved indices
\begin{equation}
\label{dft*anholonomy}
\Omega_{IJK} = {E^A}_I {E^B}_J {E^A}_K \Omega_{ABC} = - \partial_I {E^A}_J E_{AK}
\end{equation}
we find for the failure
\begin{equation}
\label{dft*equation999}
\Delta_\xi \big( {\Omega_{IJ}}^K \big) = - \partial_I \partial_J \xi^K\,, \quad \text{and} \quad \Delta_\xi \big( {\Gamma_{IJ}}^K \big) = \partial_I \partial_J \xi^K\,.
\end{equation}
Finally, it allows us to obtain the result
\begin{equation}
\Delta_\xi \big( \nabla_I V^J \big) = 0\,,
\end{equation}
and hence $\nabla_I V^J$ is indeed the covariant derivative under $2D$-diffeomorphisms.
Subsequently, we can generalize this computation to arbitrary tensorial structures, especially $\mathcal{H}^{IJ}$. It yields
\begin{equation}
\Delta_\xi \big( \nabla_I \mathcal{H}^{JK} \big) = 0\,.
\end{equation}
Furthermore, we need to evaluate the following term
\begin{equation}
\Delta_\xi \big( \nabla_I d \big) = \Delta_\xi \big( \partial_I \tilde{d} \big) = 0\,.
\end{equation}
Last but not least, the failure of the structure coefficients $F_{IJK}$ has to be derived. Thankfully, it also vanishes
\begin{equation}
\Delta_\xi {F_{IJ}}^K = {\Omega_{[IJ]}}^K = \partial_{[I} \partial_{J]} \xi^K = 0\,. 
\end{equation}
Combining all of these results and applying them to the full action, we arrive at the result
\begin{equation}
\Delta_\xi \mathcal{L} = 0\,.
\end{equation}
It proves that the action~\eqref{dft*action} is truly invariant under $2D$ diffeomorphisms. However, this is only the case if the dilaton prefactor $e^{-2\bar{d}}$ transforms as a weight $+1$ scalar density. In fact,
\begin{equation}
e^{-2\bar{d}} = \sqrt{H}\,.
\end{equation}
Therefore, action~\bref{dft*action} possesses a manifest $2D$-diffeomorphism invariance under standard diffeomorphisms.
Because the generalized Lie derivative only contains covariant derivatives, it has to transform covariantly under standard diffeomorphisms as well. We find
\begin{equation}
\Delta_\xi \mathcal{L}_\lambda V^I = 0\,.
\end{equation}
Recasting this equation, we have
\begin{equation}
\label{eqn:stdliegenlie}
\Delta_\xi \mathcal{L}_\lambda V^I = L_\xi \big( \mathcal{L}_\lambda V^I \big) - \mathcal{L}_{L_\xi \lambda} V^I - L_\lambda \big( \mathcal{L}_\xi V^I \big)\,,
\end{equation}
with the associated gauge algebra given by
\begin{equation}
\big[L_\xi, \mathcal{L}_\lambda] V^I = \mathcal{L}_{L_\xi \lambda} V^I\,.
\end{equation}
Consequently, equation~\eqref{eqn:stdliegenlie} connects $2D$-diffeomorphisms with generalized diffeomorphisms. This implies that DFT$_\mathrm{WZW}$ possesses an algebra extending the DFT gauge algebra proposed by Cederwall in~\citep{Cederwall:2014kxa, Cederwall:2014opa}. Although, as opposed to Cederwall's idea to consider a torsionless covariant derivative on arbitrary pseudo Riemannian manifolds to define a generalized Lie derivative, in this case formally resembling the one of DFT$_{\mathrm{WZW}}$, we consider a torsionful covariant derivative on a group manifold. Hence, Cederwall has to apply the Bianchi identity without torsion
\begin{equation}
{R_{[IJK]}}^L = 0\,,
\end{equation}
to show that the gauge algebra closes. For the case of a torsionful  covariant derivative, a special type of a pseudo Riemannian manifold, the Bianchi identity becomes
\begin{equation}
{R_{[IJK]}}^L + \nabla_{[I} {T^L}_{JK]} - {T^M}_{[IJ} {T^L}_{K]M} = \frac{2}{9} \big( {F_{IJ}}^M {F_{MK}}^L + {F_{KI}}^M {F_{MJ}}^L + {F_{JK}}^M {F_{MI}}^L \big) = 0 \,.
\end{equation}
Thus, the Bianchi identity exactly reproduces the Jacobi identity~\bref{DFT*Jacobi}. Furthermore, there exists the possibility that the here presented formalism could be extended to pseudo Riemannian manifolds as well and is not only limited to group manifolds~\citep{Blumenhagen:2015zma}. 

\section{Relation to original DFT}
\label{sec:reltodft}

In this section, we are going to analyze the connection between DFT$_{\mathrm{WZW}}$, which is based on an underlying group manifold, and its toroidal counterpart DFT~\citep{Blumenhagen:2015zma}. A link between both theories has already been conjectured in~\citep{Blumenhagen:2014gva}, however no direct evidence for such a relation was provided. As a consequence of the generalized metric formulation of DFT$_{\mathrm{WZW}}$ we are finally able to provide the missing connection between these two theories. They are associated with each other under the imposition of an additional constraint the so-called \textit{extended strong constraint}. For this purpose, we start by introducing a special generalized vielbein~\ref{dft*subgenvielbein}. Afterwards, in~\ref{subextendedstrongconstraint} we discuss the extended strong constraint, linking background and fluctuation fields with another. Through this additional constraint, the covariant fluxes $\mathcal{F}_{ABC}$ known from toroidal DFT's flux formulation ~\citep{Hassler:2014sba, Geissbuhler:2013uka, Andriot:2011uh, Hohm:2010xe, Geissbuhler:2011mx, Hohm:2013nja} and the structure coefficients $F_{ABC}$ of the group manifold become equivalent and can be exchanged. At last, we show in subsection~\ref{relatingtheactions} how the application of the extended strong constraint reduces DFT$_\mathrm{WZW}$ to toroidal DFT and utters the action as well as the gauge transformations identical. Subsequently, we briefly discuss the background independence of DFT.

\subsection{The generalized vielbein}
\label{dft*subgenvielbein}

Let us start by considering a background vielbein ${E_A}^I$ fulfilling the strong constraint~\bref{strongconstraint} of the original DFT formulation~\citep{Blumenhagen:2015zma}. As a result of the $2D$-diffeomorphism invariance of DFT$_{\mathrm{WZW}}$ we can choose whether we want to use left/right moving coordinates $x^i$ and $x^{\bar{i}}$ or not. In this context, we rather want to work with momentum and winding coordinates instead, $x^i$ and $\tilde{x}_i$, which are employed in the generalized metric formulation of DFT~\citep{Hohm:2010pp}. Thus, we use
\begin{equation}
X^I = \big( \tilde{x}_i, x^i \big)\,,\quad \partial_I = \big( \tilde{\partial}^i, \partial_i \big)\,,\quad \text{and}\quad \eta_{IJ} = \begin{pmatrix}
0 & \delta^i_j \\ \delta_i^j & 0
\end{pmatrix}\,.
\end{equation}
In the flux formulation of DFT, a viable choice for the generalized vielbein is~\citep{Hassler:2014sba, Aldazabal:2013sca, Geissbuhler:2013uka, Andriot:2011uh, Hohm:2010xe, Geissbuhler:2011mx, Hohm:2013nja}
\begin{equation}
\label{dft*newgeneralizedvielbein}
{E_{\hat{A}}}^I = \begin{pmatrix}
{e^a}_i & 0 \\ -{e_a}^j B_{ji} & {e_a}^i
\end{pmatrix}\,.
\end{equation}
After applying the strong constraint from original DFT the coordinate dependency reduces to half the number of coordinates. Therefore, without any loss of generality one can parametrize the generalized vielbein ${E_{\hat{A}}}^I$ using only momentum coordinates $x^i$. This is the 'gauge' we are going to choose. We work with the following metric to raise and lower hatted indices
\begin{equation}
\eta_{\hat{A}\hat{B}} = \begin{pmatrix}
0 & \delta^a_b \\ \delta_a^b & 0
\end{pmatrix}\,,\quad \text{and the according inverse} \quad \eta^{\hat{A}\hat{B}} = \begin{pmatrix}
0 & \delta_a^b \\ \delta^a_b & 0
\end{pmatrix}\,.
\end{equation}
Next, we want to identify this $\eta$-representation with the diagonal one known from DFT$_{\mathrm{WZW}}$~\bref{DFT*diagonaleta}. As a consequence, we make use of the following coordinate independent O($2D$) rotation
\begin{equation}
\label{dft*hatrotation}
{M_A}^{\hat{B}} = \begin{pmatrix}
\phantom{-}\eta_{ab} & \delta_a^b \\ -\eta_{\bar{a}\bar{b}} & \delta_{\bar{a}}^b
\end{pmatrix}\,,\quad \text{with}\quad {M_A}^{\hat{C}}  {M_B}^{\hat{D}} \eta_{\hat{C}\hat{D}} = \eta_{AB}\,.
\end{equation}
Under this transformation, the background metric $S_{AB}$ remains left invariant
\begin{equation}
{M_A}^{\hat{C}}  {M_B}^{\hat{D}} S_{\hat{C}\hat{D}} = S_{AB}\,,\quad\text{where}\quad S_{\hat{A}\hat{B}} = \begin{pmatrix}
\eta_{ab} & 0 \\ 0 & \eta_{ab}
\end{pmatrix}\,.
\end{equation}
Moreover, by employing the generalized vielbein we can switch to curved indices through
\begin{equation}
H^{IJ} = {E_{\hat{A}}}^I S^{\hat{A}\hat{B}} {E_{\hat{B}}}^J = \begin{pmatrix}
g_{ij} - B_{ik} g^{kl}B_{lj} &  B_{ik} g^{kj}
\\ -g^{ik} B_{kj} & g^{ij} \end{pmatrix}\,.
\end{equation}
However, the DFT$_{\mathrm{WZW}}$ generalized vielbein~\bref{DFT*vielbein} is usually not an element of O($D,D$), as it induces different metric representations in flat and curved indices, specifically
\begin{equation}
{E_{A}}^I \eta^{AB} {E_{B}}^J = \eta^{IJ} = 2 \begin{pmatrix}
g^{ij} & 0 \\ 0 & - g^{\bar{i}\bar{j}}
\end{pmatrix}\,.
\end{equation}
This is an evident problem, when trying to compare both theories. Nevertheless, the newly introduced generalized vielbein~\bref{dft*newgeneralizedvielbein} cures this problem by fulfilling the identity
\begin{equation}
\label{dft*ODDproperty}
 {E_{\hat{A}}}^I \eta^{\hat{A}\hat{B}} {E_{\hat{B}}}^J = \eta^{IJ} = \begin{pmatrix}
 0 & \delta_i^j \\ \delta^i_j & 0
 \end{pmatrix}\,,
\end{equation}
and hence transforms as an element of O($D,D$).

One would expect that a generalized vielbein of this form gives rise to the constant structure coefficients
\begin{equation}
\label{eqn:FABC&OmegaABC}
  F_{ABC} = 2 \Omega_{[AB]C} \quad \text{with the coefficients of anholonomy} \quad
  \Omega_{ABC} = D_A E_B{}^I E_{C I}
\end{equation}
from which the derivation in~\citep{Blumenhagen:2014gva} begins. Unfortunately, the resulting structure coefficients turn out to be non constant. An approach to avoid this issue is given by the covariant fluxes
\begin{equation}
\mathcal{F}_{\hat{A}\hat{B}\hat{C}} = 3 \Omega_{[\hat{A}\hat{B}\hat{C}]}\,.
\end{equation}
The backgrounds of DFT$_{\mathrm{WZW}}$ are by definition given through constant covariant fluxes. These correspond to a	generalized Scherk-Schwarz ansatz. From the results in~\citep{Hassler:2014sba}, we already know that if the vielbein ${e_a}^i$ and the B-field $B_{ij}$ are independent of the momentum coordinates, we are able to derive the following relations for the fluxes
\begin{equation}
\mathcal{F}_{abc} = -3 {e_a}^i{e_b}^j{e_c}^k \partial_{[i}B_{jk]} = - H_{abc} = - F_{abc}\,,
\end{equation}
\begin{equation}
{\mathcal{F}^a}_{bc} = 2 {e_{[b}}^i	\partial_i {e_{c]}}^j {e^a}_j = 2\Omega_{[bc]}^a = {F^a}_{bc}\,.
\end{equation}
All other remaining components {${\mathcal{F}^{ab}}_c$, and $\mathcal{F}_{abc}$ vanish. Now, let us switch to the covariant fluxes with hatted indices. It is achieved by applying the rotation ${M_A}^{\hat{B}}$~\bref{dft*hatrotation} onto the fluxes. We obtain
\begin{equation}
\label{dft*covariantfluxes}
  \mathcal{F}_{ABC} = \begin{cases}
    \mathcal{F}_{abc} + \eta_{ad} \mathcal{F}^d{}_{bc} +  \eta_{bd} \mathcal{F}_a{}^d{}_c +  \eta_{cd} \mathcal{F}_{ab}{}^d = \phantom{-}2 F_{abc} \\
    \mathcal{F}_{\bar a b c} - \eta_{\bar a\bar d} \mathcal{F}^{\bar d}{}_{b c} +  \eta_{b d} \mathcal{F}_{\bar a}{}^{d}{}_c +  \eta_{cd} \mathcal{F}_{\bar a b}{}^d = \phantom{-}0 \\
    \mathcal{F}_{\bar a\bar b c} - \eta_{\bar a\bar d} \mathcal{F}^{\bar d}{}_{\bar b c} -  \eta_{\bar b\bar d} \mathcal{F}_{\bar a}{}^{\bar d}{}_c +  \eta_{cd} \mathcal{F}_{\bar a\bar b}{}^d =  -2 F_{\bar a\bar b c} \\
    \mathcal{F}_{\bar a\bar b\bar c} - \eta_{\bar a\bar d} \mathcal{F}^{\bar d}{}_{\bar b\bar c} -  \eta_{\bar b\bar d} \mathcal{F}_{\bar a}{}^{\bar d}{}_{\bar c} -  \eta_{\bar c\bar d} \mathcal{F}_{\bar a\bar b}{}^{\bar d} =  -4 F_{\bar a\bar b\bar c}\,.
  \end{cases}
\end{equation}
In their present form, they are already constant. Nevertheless, they violate the strict left/right seperation of the structure coefficients required for the formulation of DFT$_{\mathrm{WZW}}$.
Fortunately, we still have a few tricks left up our sleeves. We execute the subsequent procedure to recover the O($D,D$) property~\bref{dft*ODDproperty}. Therefore, let us introduce the coordinate dependent O$(D)$ $\times$ O$(D)$ transformation acting on
\begin{equation}
{E_A}^I = {M_A}^{\hat{B}} {E_{\hat{B}}}^I = \begin{pmatrix}
\phantom{-}e_{ai} + {e_a}^j B_{ji} & {e_a}^i \\ -e_{ai} + {e_a}^j B_{ji} & {e_a}^i 
\end{pmatrix}\,,\quad\text{as}\quad {\tilde{E}_A}{}^I = {T_A}^B(x^i) {E_B}^I\,.
\end{equation}
Here,  we dropped the bar over $e_{ai}$ and ${e_a}^i$ in the second row of ${E_A}^I$ to stress that we only work with the left-moving vielbein. It is connected to the right-moving vielbein through the transformation
\begin{equation}
{e_{\bar{a}}}^i = {t_{\bar{a}}}^b {e_b}^i\,,\quad\text{with}\quad {t_{\bar{a}}}^b = \mathcal{K}(t_{\bar{a}}, g\, t^b g^{-1})\,,
\end{equation}
where $ {t_{\bar{a}}}^b$ designates an O$(D)$ transformation, and $\mathcal{K}$ is the Killing form
\begin{equation}
\mathcal{K}(x,y) = - \frac{\alpha'k}{\,4h^\vee} \text{Tr}(\text{ad}_x \text{ad}_y)\,,\quad\text{with}\quad x,y \in \mathfrak{g}\,,
\end{equation}
which has been introduced in~\eqref{liemetric}~\citep{Blumenhagen:2014gva}. On top of that, the group element $g$ depends only on the coordinates $x^i$. We can embed this transformation into
\begin{equation}
\label{dft*embedding}
{T_A}^B = \begin{pmatrix}
\delta_a^b & 0 \\ 0 & {t_{\bar{a}}}^b
\end{pmatrix}\quad\text{and generate}\quad {\tilde{E}_A}{}^I = \begin{pmatrix}
\phantom{-}e_{ai} + {e_a}^j B_{ji} & {e_a}^i \\ -e_{\bar{a}i} + {e_{\bar{a}}}^j B_{ji} & {e_{\bar{a}}}^i 
\end{pmatrix}\,.
\end{equation}
Hence, it allows us to obtain the desired matrix structure. As a consequence of the strict coordinate dependence of this transformation, the coefficients of anholonomy need to be modified through
\begin{equation}
\tilde{\Omega}_{ABC} = {T_A}^D {T_B}^E {T_C}^F \big( \Omega_{DEF} - {E_D}^I \partial_I T_{HE} {T^H}_F \big)\,.
\end{equation}
Completing some algebra and making use of the definition $t_a = - t_{\bar{a}}$, we find
\begin{equation}
\partial_i t_{\bar{d}b} {t^{\bar{d}}}_c = \mathcal{K}([t_b, t_c], t_a) {e^a}_i = {e^a}_i F_{abc}\,,
\end{equation}
which finally gives rise to the result
\begin{equation}
{E_A}^I \partial_I T_{DB} {T^D}_C = 2 {E_A}^I \begin{pmatrix}
0 & 0 \\ 0 & -\partial_I t_{\bar{d}b} {t^{\bar{d}}}_c
\end{pmatrix} = -2 \begin{cases} F_{a\bar{b}\bar{c}} \\ F_{\bar{a}\bar{b}\bar{c}} \\ 0 \quad \quad \text{otherwise}
\end{cases}\,.
\end{equation}
Thus, the O$(D)$ $\times$ O$(D)$ rotated generalized vielbein ${\tilde{E}_A}{}^I$ produces additional contributions to the structure coefficients and therefore eliminates the problems we encountered with the covariant fluxes~\bref{dft*covariantfluxes} before.  We arrive at the transformed covariant fluxes after antisymmetrizing the coefficients of anholonomy $\tilde{\Omega}_{ABC}$
\begin{equation}
\label{dft*finalfluxes}
\tilde{F}_{ABC} = 2 \begin{cases}
\phantom{-}F_{abc} \\ -F_{\bar{a}\bar{b}\bar{c}} \\ \phantom{-}0\quad \text{otherwise}
\end{cases}\quad\text{or in the standard form}\quad\begin{cases} {F_{ab}}^c \\ {F_{\bar{a}\bar{b}}}^{\bar{c}} \\ 0\quad \text{otherwise} 
\end{cases}\,.
\end{equation}
Through this procedure, the covariant fluxes of original DFT have become consistent with the left/right segregation we know from DFT$_{\mathrm{WZW}}$~\citep{Blumenhagen:2014gva}, and therfore the WZW background has been successfully embedded~\bref{dft*embedding} into the DFT flux formulation~\citep{Andriot:2011uh, Hohm:2010xe, Blumenhagen:2015zma}.

\subsection{Extended strong constraint}
\label{subextendedstrongconstraint}

The two definitions for the structure coefficients $F_{ABC}$ and the covariant fluxes $\mathcal{F}_{ABC}$ show a minor but distinctive difference. Specifically, they are defined through
\begin{align}
\label{eqn:originaldftstruct}
\text{Structure coefficients:}&\quad\quad F_{ABC} = 2 \Omega_{[AB]C}\,, \\
\text{Covarant fluxes:}&\quad\quad \mathcal{F}_{ABC} = 3 \Omega_{[ABC]}\,.
\end{align}
At this point, let us recall the antisymmetry of $\Omega_{ABC}$ in its last two indices, a direct consequence of the O($D,D$) property~\bref{dft*ODDproperty}. It allows us to relate the covariant fluxes and the structure coefficients with one another through the identification~\citep{Blumenhagen:2015zma}
\begin{equation}
\mathcal{F}_{ABC} = \Omega_{ABC} + \Omega_{CAB} + \Omega_{BCA} = F_{ABC} + \Omega_{CAB}\,.
\end{equation}
Moreover, the Lie algebra and its structure coefficients dictate the DFT$_{\mathrm{WZW}}$ commutation relations and the flat derivatives by
\begin{equation}
\big[D_A, D_B \big] = {F_{AB}}^C D_C\,.
\end{equation}
Therefore, we rewrite equation~\eqref{eqn:originaldftstruct}
\begin{equation}
\label{dft*fluxconnection}
{\mathcal{F}_{AB}}^C D_C\, \cdot = {F_{AB}}^C D_C \, \cdot + ( D^C {E_A}^I ) E_{BI} D_C \, \cdot\,,
\end{equation}
where the $\cdot$ indicates arbitrary products of fluctuations $\epsilon^{AB}$, $\tilde{d}$, and the gauge parameter $\xi^A$, also deemed a fluctuation in the context of the theory.
However, it should be noted that the strong constraint of DFT$_{\mathrm{WZW}}$ only acts on these fluctuations, the background or the relations between background and fluctuations are not affected. Although, we are able to implement an additional condition, linking background fields $b$ and fluctuations $f$ with each other, the so-called \textit{extended strong constraint}
\begin{equation}
\label{eqn:extendedSC}
D_A b \, D^A f = 0\,.
\end{equation}
This constraint restricts all viable field configurations to a certain subset, causing the last term in~\bref{dft*fluxconnection} to vanish and thus $F_{ABC} = \mathcal{F}_{ABC}$. Furthermore, the extended strong constraint cancels the background term in the strong constraint~\bref{levelmatchingDFT*2}, leaving us with
\begin{equation}
\label{dft*extendedstrongconstraint}
\big( \partial_I \partial^I - 2 \partial_I d \partial^I \big) \, \cdot = \partial_I \partial^I \, \cdot = 0\,.
\end{equation}
As a result, the strong constraint of DFT$_{\mathrm{WZW}}$ becomes equivalent to the strong constraint in the original DFT formulation~\citep{Blumenhagen:2015zma}. In the next subsection, we are going to investigate how the extended strong constraint affects the entire theory.

\subsection{Relating the theories}
\label{relatingtheactions}

Replacing the structure coefficients $F_{ABC}$ by covariant fluxes $\mathcal{F}_{AB}$ requires a computation of the Christoffel symbols, again. We follow~\citep{Blumenhagen:2015zma}. It results in having to solve the frame field compatibility condition
\begin{equation}
\label{dft*newconnection}
\nabla_A {E_B}^I = D_A {E_B}^I + \frac{1}{3} {\mathcal{F}_{BA}}^C {E_C}^I + {E_A}^K {\Gamma_{KJ}}^I {E_B}^J = 0\,, 
\end{equation}
and subsequently
\begin{equation}
\label{dft*connection2}
{\Gamma_{IJ}}^K = - {\Omega_{IJ}}^K + \Omega_{[IJL]} \eta^{LK} = \frac{1}{3} \big( - 2 {\Omega_{IJ}}^K + {\Omega^K}_{IJ} + {{\Omega_J}^K}_I \big)\,.
\end{equation}
The torsion of this generalized connection vanishes
\begin{equation}
{\mathcal{T}^I}_{JK} = 2 {\Gamma_{[JK]}}^I + {\Gamma^I}_{[JK]} = 0\,.
\end{equation}
Furthermore, the C-brackets~\bref{c-bracket},\bref{DFT*cbracket1} of both theories are connected through the torsion by
\begin{equation}
\big[ \xi_1, \xi_2 \big]_C^I = \big[ \xi_1, \xi_2 \big]_{\text{DFT},C}^I + {\mathcal{T}^I}_{JK} \xi^J_1 \xi^K_2\,.
\end{equation}
Both theories are governed by the same gauge algebra, however the strong constraints~\bref{dft*extendedstrongconstraint} need to be exchanged. The same argumentation holds for the generalized Lie derivative as well
\begin{equation}
\mathcal{L}_\xi V^I = \big[ \xi, V \big]_C^I + \frac{1}{2} \nabla^I \big( \xi_J V^J \big) = \big[ \xi, V \big]_{\text{DFT},C}^I + \frac{1}{2} \partial^I \big( \xi_J V^J \big) =	\mathcal{L}_{\text{DFT},\xi} V^I\,,
\end{equation}
as it can be expressed through the C-bracket. Therefore, any modification of the Christoffel symbols does not change their transformation behavior~\bref{dft*equation999} under $2D$-diffeomor\-phisms. Hence, the action and gauge transformations retain their $2D$-diffeomorphism invariance. Unfortunately, the O($D,D$) preserving constraint~\bref{dft*ODDproperty}
\begin{equation}
L_\xi \eta^{IJ} = 0 = \partial^I \xi^J + \partial^J \xi^I\,,
\end{equation}
partially violates the $2D$-diffeomorphism symmetry. On top of that, the extended strong constraint of DFT$_\mathrm{WZW}$ and strong constraint of toroidal DFT for the generalized vielbein  ${E_A}^I$ should transform covariantly. This leads to the following auxiliary constraints
\begin{equation}
\Delta_\xi \big( \partial_I {E_A}^J \partial^I f \big) = - {E_A}^K \partial_K \partial_I \xi^J \partial^I f = 0\,,
\end{equation}
\begin{equation}
\Delta_\xi \big( \partial_I {E_A}^J \partial^I {E_B}^K \big) = - {E_A}^L \partial_L \partial_I \xi^J \partial^I {E_B}^K - \partial_I {E_A}^J {E_B}^L \partial_L \partial^I \xi^K = 0\,,
\end{equation}
and consequently require
\begin{equation}
\partial_I \xi^J \partial^I f = 0\,,\quad\text{and}\quad \partial_I \xi^J \partial^I {E_A}^K = 0\quad\text{or}\quad \partial_I \xi^K = \text{const}\,.
\end{equation}
In this equation, the latter term allows for global O($D,D$) rotations~\citep{Blumenhagen:2015zma}.	Except them, only the following transformations are allowed for the generalized vielbein
\begin{equation}
L_\xi {E_A}^I = \xi^J \partial_J {E_A}^I + {E_A}^J \partial_J \xi^I = {E_A}^J \begin{pmatrix}
0 & 0 \\ \partial_{[j} \tilde{\xi}_{i]} & 0
\end{pmatrix}\,,
\end{equation}
corresponding to $B$-Field gauge transformations
\begin{equation}
B_{ij} \to B_{ij} + \partial_{[i} \xi_{j]}\,,
\end{equation}
which we can express through generalized diffeomorphisms, as in the case of global O($D,D$) rotations. This implies that the extended strong constraint~\bref{dft*extendedstrongconstraint} and the O($D,D$) generalized background vielbein break the DFT$_{\mathrm{WZW}}$ $2D$-diffeomorphism completely.

Moreover, the newly introduced generalized connection~\bref{dft*newconnection} also has an affect on the background dilaton $\bar{d}$~\bref{dft*dilaton} in a non-trivial fashion. Correspondingly, the background dilaton is required to satisfy the compatibility condition for partial integration~\bref{eqn:nablaIe-2bard=0}
\begin{equation}
{\Gamma_{IJ}}^I = {\Omega^I}_{IJ} = - 2 \partial_J \bar{d}\,.
\end{equation}
In flat indices it takes on the form
\begin{equation}
\label{dft*consistencyconstraint}
\mathcal{F}_A = {\Omega^A}_{AB} + 2 D_A \bar{d} = 0\,.
\end{equation}
The DFT$_{\mathrm{WZW}}$ backgrounds fulfill this relation by default, as they originate from a generalized Scherk-Schwarz ansatz which inherits~\bref{dft*consistencyconstraint} as a consistency requirement.

At this point, we want to recast the DFT$_\mathrm{WZW}$ action in the following way
\begin{equation} 
\label{dft*dftrelation}
S = S_{DFT} + S_{\Delta}\,,
\end{equation}
with $S_{DFT}$ representing the original DFT action
\begin{align}
S_{DFT} = \int d^{2D}X e^{-2d} \Big( &\frac{1}{8} \mathcal{H}^{KL} \partial_K \mathcal{H}_{IJ} \partial_L \mathcal{H}^{IJ} - \frac{1}{2} \mathcal{H}^{IJ} \partial_J \mathcal{H}^{KL} \partial_L \mathcal{H}_{IK} \nonumber \\
&-2\partial_I d \, \partial_J \mathcal{H}^{IJ} + 4 \mathcal{H}^{IJ} \partial_I d \, \partial_J d \Big)\,,
\end{align}
and the auxiliary term
\begin{equation}
S_\Delta = - \frac{2}{3} \int d^{2D} X e^{-2d} \, \mathcal{H}^{IJ} \mathcal{K}_{IJ}\,.
\end{equation}
However, this separation of the action is only valid under application of the extended strong constraint~\bref{dft*extendedstrongconstraint}, with $\mathcal{K}_{IJ}$ arising from the DFT field equations. (See~\citep{Hohm:2010pp} for more details). After performing the necessary projections~\bref{projectionoperators}, which are required due to the undetermined components of this tensor, it allows us to obtain the generalized Ricci tensor $\mathcal{R}_{IJ}$ and the corresponding the equations of motion. For the choice of our Scherk-Schwarz background, we find
\begin{equation}
\mathcal{K}_{IJ} = \frac{1}{4} F_{IKL} F_{JMN} \big( \eta^{KM} \eta^{LN} - H^{KM} H^{LN} \big)\,.
\end{equation}
The difference between all the metrics $\mathcal{H}^{IJ}$, combining background and fluctuations, and $H^{IJ}$, describing the background only, should be kept in mind. If equation~\bref{dft*dftrelation} holds, the remaining term
\begin{equation}
S - S_{DFT} - S_\Delta = \int d^{2D} X e^{-2d} \Delta
\end{equation}
must vanish. We now start to replace all covariant derivatives by partial derivatives and the according generalized connection~\bref{dft*connection2}. It yields
\begin{align}
\Delta = &\mathcal{H}^{IJ} \Big( \Omega_{IKL} {\Omega^{KL}}_J - {\Omega^K}_{KI} {\Omega^L}_{LJ} + \frac{1}{2} \Omega_{KLI} {\Omega^{KL}}_J \Big) \\
&\quad\,-{\Omega_{IJ}}^K \partial_K \mathcal{H}^{IJ} + 2 {\Omega^K}_{KI} \mathcal{H}^{IJ} \partial_J \tilde{d} - {\Omega^K}_{KI} \partial_J \mathcal{H}^{IJ} + 2\mathcal{H}^{IJ} {\Omega_{IJ}}^K \partial_K \tilde{d}\,. \nonumber
\end{align}
As a result of the strong constraint for background fields, the last term in the first line becomes zero. Furthermore, we perform a partial integration~\bref{DFT*intbyparts} and split the dilaton according to~\bref{dft*dilaton}. We obtain
\begin{equation}
\label{dft*ident1}
-{\Omega_{IJ}}^K \partial_K \mathcal{H}^{IJ} = -2 \mathcal{H}^{IJ} {\Omega_{IJ}}^K \partial_K \tilde{d} + \mathcal{H}^{IJ} {\Omega_{IJ}}^K {\Omega^L}_{LK} + \partial_K {\Omega_{IJ}}^K \mathcal{H}^{IJ} \,,
\end{equation}
\begin{equation}
\label{dft*ident2}
-{\Omega^K}_{KI} \partial_J \mathcal{H}^{IJ} = - 2 {\Omega^K}_{KI} \mathcal{H}^{IJ} \partial_J \tilde{d} + \mathcal{H}^{IJ} {\Omega^K}_{KI} {\Omega^L}_{LJ} + \mathcal{H}^{IJ} \partial_I {\Omega^K}_{KJ}\,.
\end{equation}
In order to get rid of the derivatives acting on $\bar{d}$, we applied equation~\bref{dft*consistencyconstraint}\,. Next, we utilize the newly introduced identities~\bref{dft*ident1},~\bref{dft*ident2}. As an immediate consequence, $\Delta$ reduces to
\begin{equation}
\Delta = \mathcal{H}^{IJ} \Big( \Omega_{IKL} {\Omega^{KL}}_J + {\Omega_{IJ}}^K {\Omega^L}_{LK} + \partial_K {\Omega_{IJ}}^K + \partial_I {\Omega^K}_{KJ} \Big)\,.
\end{equation}
During the last step, we use the definition of the coefficients of anholonomy~\bref{dft*anholonomy}. A straight forward computation gives rise to
\begin{equation}
\partial_K {\Omega_{IJ}}^K + \partial_I {\Omega^K}_{KJ} = - {\Omega_{IJ}}^K {\Omega^L}_{LK} - \Omega_{IKL} {\Omega^{KL}}_J\,.
\end{equation}
Finally, we arrive at the desired result
\begin{equation}
\Delta = 0\,.
\end{equation}
Moreover, we are interested in evaluating $\mathcal{K}_{IJ}$. Therefore, we switch back to flat indices and exploit the strict left/right segregation of the structure coefficients~\bref{dft*structurecoeff}. We obtain
\begin{equation}
\label{dft*auxillary}
F_{ACE} F_{BDF} \eta^{CD} \eta_{EF} = F_{ACE} F_{BDF} S^{CD} S^{EF}\,.
\end{equation}
Thus, we immediately find
\begin{equation}
\mathcal{K}_{AB} = \frac{1}{4} F_{ACE} F_{BDF} \Big( \eta^{CD} \eta^{EF} - S^{CD} S^{EF} \Big ) = 0\,.
\end{equation}
Right away, it follows $\mathcal{K}_{IJ} = 0$, in curved indices, and the main result of this section
\begin{equation}
S = S_{DFT}\,,
\end{equation}
that the action of DFT$_{\mathrm{WZW}}$ reduces to the one of original DFT under the imposition of the extended strong constraint~\citep{Blumenhagen:2015zma}.

At last, we show that its possible to choose an arbitrary realization of the generalized vielbein $E_A{}^I$ as long as the strong constraint is fulfilled. In this context, we argue why the O$(D)$ $\times$ O$(D)$ gauge fixing of section~\ref{dft*subgenvielbein} is very convenient. Therefore, we show that $S = S_{DFT}$ is invariant under local O$(D)$ $\times$ O$(D)$ transformations. Clearly, these transformations leave
\begin{equation}
\delta_\Lambda \eta^{AB} = {\Lambda^A}_C \eta^{CB} + {\Lambda^B}_C \eta^{AC} = 0\,,\quad\text{and}\quad \delta_\Lambda S^{AB} = {\Lambda^A}_C S^{CB} + {\Lambda^B}_C S^{AC} = 0
\end{equation}
invariant. A suitable way to show this requires us to compute the failure of $\mathcal{K}_{AB}$ to transform covariantly
\begin{equation}
\Delta_\Lambda \mathcal{K}_{AB} = 0\,.
\end{equation}
With regards to this, it is useful to take the failure of the covariant fluxes~\citep{Geissbuhler:2013uka} into account
\begin{equation}
\Delta_\Lambda \mathcal{F}_{ABC} = 3 D_{[A} \Lambda_{BC]}\,.
\end{equation}
From which we obtain
\begin{equation}
\Delta_\Lambda \mathcal{K}_{AB} = \frac{3}{4} \big( D_{[A} \Lambda_{CE]} F_{BDF} + D_{[B} \Lambda_{DF]} F_{ACE} \big)\big( \eta^{CD} \eta^{EF} - S^{CD} S^{EF} \big)\,.
\end{equation}
Making use of the O($D,D$) condition
\begin{equation}
S^{AC} \Lambda_{CD} S^{DB} = \Lambda^{AB}\,,
\end{equation}
analogous to equation~\bref{dft*auxillary}, we acquire the wanted result
\begin{equation}
\Delta_\Lambda \mathcal{K}_{AB} = 0\,.
\end{equation}
Thus, we can freely choose an arbitrary realization of the generalized background $E_A{}^I$, as long as it satisfies the strong constraint. For instance, we could choose the bivector $\beta^{ij}$ instead of the $B$-field $B_{ij}$, or a vielbein which neither lies in the left- nor right-moving Maurer-Cartan form.

Summarizing, we can view the entire computation executed in this subsection as kind of a generalization of the steps performed to find a background independent action~\citep{Hohm:2010jy} of the cubic DFT action~\citep{Hull:2009mi}. The proof of the background independence has been accomplished by absorbing the constant part of the fluctuations  $\epsilon_{ij}$ into a change of the background field $E_{ij}$, where the dilaton does not contribute in any way~\citep{Hohm:2010jy}. In our case, it is a very similar situation. The generalized metric~\bref{dft*generalizemetric} is split into a background field $H^{IJ}$ and fluctuations $h^{IJ}$ through
\begin{equation}
\mathcal{H}^{IJ} = H^{IJ} + h^{IJ}\,,\quad\text{with}\quad h^{IJ} = \epsilon^{IJ} + \frac{1}{2} \epsilon^{IK} H_{KL} \epsilon^{LJ} + \dots\,.
\end{equation}
Although, in contrast to~\citep{Hohm:2010jy} we have to consider generalized dilaton contributions as well. Moreover, the background field $H^{IJ}$ is generally not constant for arbitrary group manifolds. This indicates that we are not solely restricted to constant background fields. Only for an underlying group manifold such as the torus we obtain a constant background. Furthermore, the field equations must always hold for a consistent background. However, we are still able to reproduce the background independence known from original DFT~\citep{Hohm:2010jy, Blumenhagen:2015zma}. 

Therefore, this subsection shows that in order for DFT$_{\mathrm{WZW}}$ to be background independent, we must impose the extended strong constraint~\bref{dft*extendedstrongconstraint}. It rules out all solutions beyond the SUGRA regime. Additionally, DFT$_{\mathrm{WZW}}$ might give insights into new non-geometric background and physics which are going beyond the SUGRA/DFT regime despite inheriting the same background independence as in original DFT. On top of that, DFT breaks DFT$_{\mathrm{WZW}}$'s $2D$-diffeomorphism invariance as the derivation in this subsection clearly shows.\clearpage{}
  \clearpage{}\chapter{Flux Formulation of DFT on Group Manifolds}
\label{Kap_4}
\label{sec:FluxFormWZW}

During the past two chapters, we derived DFT$_{\mathrm{WZW}}$ for a closed string propagating on a group manifold~\cite{Blumenhagen:2014gva,Blumenhagen:2015zma}. Subsequently, we obtained a generalized metric formulation for DFT$_{\mathrm{WZW}}$ in a very similar fashion to the procedure known from original DFT~\cite{Hohm:2010pp}. It gives rise to an action which is simultaneously invariant under generalized and $2D$-diffeomorphisms. In chapter~\ref{Kap_1} we already observed the presence of a flux formulation for toroidal DFT with the covariant fluxes $\mathcal{F}_A$ and $\mathcal{F}_{ABC}$. We now want to generalize this approach to DFT$_{\mathrm{WZW}}$.

Starting from the generalized metric formulation~\ref{Kap_3}, we are interested in deriving the corresponding flux formulation~\cite{Bosque:2015jda}. Therefore, we begin by identifying the covariant fluxes of the DFT$_{\mathrm{WZW}}$ framework~\ref{sec:covfluxesWZW}. This makes it possible to recast the generalized metric action~\eqref{dft*action} through these covariant fluxes. It results in the desired flux formulation of DFT$_{\mathrm{WZW}}$~\ref{sec:FluxFormAction}. In this context, we discuss the action's symmetries and equations of motion

This chapter follows~\cite{Bosque:2015jda}.

\section{Covariant fluxes}\label{sec:covfluxesWZW}

Before we are able to cast the DFT${}_\mathrm{WZW}$ action~\eqref{dft*action} into its flux formulation, we need to determine its basic constituents, the covariant fluxes~\cite{Bosque:2015jda}. Therefore, let us consider the following composite generalized vielbein
\begin{equation}\label{eqn:fluxepsilon}
  \mathcal{E}_{\hat A}{}^I = \tilde{E}_{\hat A}{}^B {E_B}^I\,.
\end{equation}
It combines the background vielbein $E_A{}^I$ with a new fluctuation vielbein $\tilde{E}_{\hat A}{}^B$, seizing the field dynamics around a given background. While the former vielbein is only restricted to lie in GL($2D$) and thus not an element of O($D,D$), the former vielbein capturing the fluctuations is. As a consequence, it satisfies the O($D,D$) condition
\begin{equation}
  \eta_{AB} = \tilde{E}^{\hat C}{}_A \, \eta_{\hat C\hat D} \, \tilde{E}^{\hat D}{}_B\,,
\end{equation}
where $\eta_{AB}$ and $\eta_{\hat A\hat B}$ are defined exactly in the same way. On top of that, it allows to split the generalized metric according to
\begin{equation}\label{eqn:fluxgenmetric}
  \mathcal{H}_{AB} = \tilde{E}^{\hat C}{}_A \, S_{\hat C\hat D} \, \tilde{E}^{\hat D}{}_B\,.
\end{equation}
At this point, it is quite important to differentiate between all the different indices appearing in the vielbeins. We already encountered curved indices $I, J, K, L, \ldots$ and their flat versions $A, B, C, \ldots$ before. Now, we also introduce hatted indices such as $\hat A, \hat B, \hat C, \ldots$. These indices are related to the double Lorentz symmetry, discussed in subsection ~\ref{sec:doublelorentz}. At first, the introduction of additional indices and the presence of two generalized vielbeins might seem confusing, as the original formulation~\cite{Geissbuhler:2011mx,Hohm:2013nja} only required one. Let us demonstrate the extra structure induced by the background generalized vielbein through the following diagram:
\begin{equation}\label{eqn:groupindices}
  \tikz[baseline]{\matrix (m) [ampersand replacement=\&, matrix of math nodes, column sep=4em] {\text{O}(1,D-1)\times\text{O}(D-1,1) \& \text{O}(D,D) \& \text{GL}(2 D) \\};
    \draw[->] (m-1-3) -- (m-1-2) node[midway, above] {$\eta_{IJ}$} node[midway, below] {$E_B{}^I$};
    \draw[->] (m-1-2) -- (m-1-1) node[midway, above] {$\mathcal{H}_{AB}$} node[midway, below] {${\tilde E}_{\hat A}{}^B$};}\,.
\end{equation}
We start from a $2D$-dimensional smooth manifold $M$ with a pseudo Riemannian metric $\eta$. This metric possesses a split signature. Furthermore, it reduces the manifold's structure group from GL($2D$) to O($D,D$). The generalized background vielbein $E_A{}^I$ then induces the corresponding frame bundle on $M$. Moreover, we are equipped with the generalized metric $\mathcal{H}_{AB}$. It causes a further reduction of the structure group to the double Lorentz group O($1,D-1$)$\times$O($D-1,1$) and is represented by the fluctuation frame $\tilde E_{\hat A}{}^B$. The original DFT formulation lacks the information encoded in the background $\eta_{IJ}$.

We now want to become a bit more familiar with the new composite generalized vielbein $\mathcal{E}_{\hat A}{}^I$. It allows us to compute the C-bracket
\begin{align}\label{eqn:fluxcbracketepsilon1}
  \big[ \mathcal{E}{}_{\hat A}, \mathcal{E}{}_{\hat B} \big]_C^J \, \mathcal{E}_{\hat CJ} &=  2 \mathcal{E}_{[\hat A}{}^I \partial_I \mathcal{E}_{\hat B]}{}^J \mathcal{E}_{\hat CJ} - \mathcal{E}_{[\hat A}{}^I \partial^J \mathcal{E}_{\hat B]I} \mathcal{E}_{\hat CJ} + \mathcal{T}^J{}_{IK} \mathcal{E}_{\hat A}{}^I \mathcal{E}_{\hat B}{}^K \mathcal{E}_{\hat CJ} \nonumber\\ 
  &= F_{\hat A\hat B\hat C} +2 D_{[\hat A} \, \tilde{E}_{\hat B]}{}^D \tilde{E}_{\hat CD} - D_{\hat C} \, \tilde{E}_{[\hat B}{}^D \tilde{E}_{\hat A]D}\,.
\end{align}
We made use of the generalized torsion~\cite{Blumenhagen:2014gva}
\begin{equation}\label{eqn:gentorsion}
  \mathcal{T}^I{}_{JK} = - \Omega^I{}_{[JK]}
\end{equation}
in the first line. Through it, we can rewrite the covariant derivative $\nabla_I$ in terms of partial derivatives instead of covariant derivatives. On a similar note to the application of the background vielbein $E_A{}^I$ to switch between flat and curved indices, we use $\tilde{E}_{\hat A}{}^B$ to obtain the structure coefficients
\begin{equation}
  F_{\hat A\hat B\hat C} = \tilde{E}_{\hat A}{}^D \tilde{E}_{\hat B}{}^E \tilde{E}_{\hat C}{}^F F_{DEF}
\end{equation}
in hatted indices. Subsequently, we are able to define the coefficients of anholonomy
\begin{equation}\label{eqn:fluxanholonomy}
  \tilde{\Omega}_{\hat A\hat B\hat C} = \tilde{E}_{\hat A}{}^D D_D \tilde{E}_{\hat B}{}^E \tilde{E}_{\hat CE} = D_{\hat A} \tilde{E}_{\hat B}{}^E \tilde{E}_{\hat CE}
\end{equation}
with
\begin{equation}
  D_{\hat A} = \tilde{E}_{\hat A}{}^B D_B
\end{equation}
for the fluctuation vielbein equivalently to~\eqref{eqn:FABC&OmegaABC}. As a consequence of the constant flat metric $\eta_{AB}$, the coefficients of anholonomy are antisymmetric in the last two indices
\begin{equation}
  \tilde{\Omega}_{\hat A\hat B\hat C} = -\tilde{\Omega}_{\hat A\hat C\hat B}\,.
\end{equation}
Finally, we introduce the fluxes
\begin{equation}\label{eqn:fluxfluxes}
  \tilde{F}_{\hat A\hat B\hat C} = 3 \tilde{\Omega}_{[\hat A\hat B\hat C]} =
    \tilde{\Omega}_{\hat A\hat B\hat C} +\tilde{\Omega}_{\hat B\hat C\hat A} + \tilde{\Omega}_{\hat C\hat A\hat B}
\end{equation}
in the same fashion as they are defined in the flux formulation of original DFT~\cite{Geissbuhler:2011mx,Hohm:2013nja}. This simplifies equation~\eqref{eqn:fluxcbracketepsilon1} to
\begin{equation}
  \label{eqn:fluxcbracketepsilon2}
  \big[ \mathcal{E}{}_{\hat A}, \mathcal{E}{}_{\hat B} \big]_C^M \,\mathcal{E}_{\hat CM} =
    F_{\hat A\hat B\hat C} + 2\tilde{\Omega}_{[\hat A\hat B]\hat C} - \tilde{\Omega}_{\hat C[\hat B\hat A]} = F_{\hat A\hat B\hat C} + \tilde{F}_{\hat A\hat B\hat C} := \mathcal{F}_{\hat A\hat B\hat C}
\end{equation}
while we introduced the covariant fluxes $\mathcal{F}_{\hat A\hat B\hat C}$. The fluxes decompose into a background part $F_{\hat A\hat B\hat C}$ and a fluctuation part $\tilde{F}_{\hat A\hat B\hat C}$. Alternatively, the covariant fluxes can be constructed through the generalized Lie derivative by
\begin{equation}
  \mathcal{E}_{\hat CM} \, \mathcal{L}_{\mathcal{E}_{\hat A}} \mathcal{E}_{\hat B}{}^{M} =
    \big[ \mathcal{E}_{\hat A}, \mathcal{E}_{\hat B} \big]_C^M \mathcal{E}_{\hat CM} + \frac{1}{2} \nabla^M \big( \mathcal{E}_{\hat AN} \mathcal{E}_{\hat B}{}^N \big) = 
    \big[ \mathcal{E}_{\hat A}, \mathcal{E}_{\hat B} \big]_C^M \mathcal{E}_{\hat CM} =
    \mathcal{F}_{\hat A\hat B\hat C}\,.
\end{equation}
These fluxes are already covariant under generalized diffeomorphisms and $2D$-diffeomor\-phisms by construction. Under both transformations they behave as scalars.

Moreover, the original flux formulation~\cite{Geissbuhler:2011mx,Aldazabal:2011nj,Hohm:2013nja, Geissbuhler:2013uka} contains the covariant fluxes $\mathcal{F}_A$ on top of $\mathcal{F}_{ABC}$. We can embed them into the DFT${}_\mathrm{WZW}$ framework according to their definition
\begin{align}
  \mathcal{F}_{\hat A} &= - e^{2d} \mathcal{L}_{\mathcal{E}_{\hat A}} e^{-2d} = 
    - e^{2d} \nabla_B \big( \mathcal{E}_{\hat A}{}^B e^{-2d} \big) =
    \tilde{\Omega}^{\hat B}{}_{\hat B\hat A} + 2 D_{\hat A} \, \tilde{d} - \mathcal{E}_{\hat A}{}^B e^{2 \bar d} \nabla_B e^{-2\bar d} \nonumber\\
  &= 2 D_{\hat A} \, \tilde{d} + \tilde{\Omega}^{\hat B}{}_{\hat B\hat A} = \tilde F_{\hat A}\,.\label{eqn:defFA}
\end{align}
Here, we decomposed the generalized dilaton $d$ according to
\begin{equation}
  d = \bar d + \tilde d\,.
\end{equation}
It splits in a fluctuation and background part, $\tilde d$ and $\bar d$.  From the first to the second line, we exploited~\eqref{eqn:nablaIe-2bard=0}, which follows from the covariant derivative's compatibility with partial integration. Just like the covariant fluxes $\mathcal{F}_{\hat A \hat B \hat C}$ derived in the last paragraph, $\mathcal{F}_{\hat A}$ transforms as a scalar under generalized and $2D$-diffeomorphisms as well.

\section{Action}\label{sec:FluxFormAction}

At this point, we are ready to obtain the action of DFT${}_\mathrm{WZW}$'s flux formulation. Mainly following the steps in~\cite{Geissbuhler:2011mx,Hohm:2013nja, Bosque:2015jda}, we start by recasting the generalized curvature scalar,~\eqref{dft*gencurvature} using the generalized metric~\eqref{eqn:fluxgenmetric}, in terms of the generalized vielbein $\mathcal{E}_{\hat A}{}^I$

We begin by computing the term
\begin{align}\label{eqn:fluxmetrictrafo}
  \nabla_{\hat A} \mathcal{H}^{\hat B\hat C} &= \tilde{E}_{\hat A}{}^A \tilde{E}^{\hat B}{}_B \tilde{E}^{\hat C}{}_C \nabla_{A} \mathcal{H}^{BC} \nonumber \\
  &= \tilde{\Omega}_{\hat A\hat D}{}^{\hat B} S^{\hat D\hat C} + \tilde{\Omega}_{\hat A\hat D}{}^{\hat C} S^{\hat B\hat D} + \frac{1}{3} F^{\hat B}{}_{\hat A\hat D} S^{\hat D\hat C} + \frac{1}{3} F^{\hat C}{}_{\hat A\hat D} S^{\hat B\hat D}
\end{align}
which we are going to use several times in the rest of this section. Subsequently, we can derive the following two terms~\eqref{dft*gencurvature}
\begin{align}
  \frac{1}{8} \mathcal{H}^{CD} \nabla_C \mathcal{H}_{AB} \nabla_D \mathcal{H}^{AB}
  &= \frac{1}{36} F_{\hat A\hat C\hat D}\, F_{\hat B}{}^{\hat C\hat D} S^{\hat A\hat B} - \frac{1}{36} F_{\hat A\hat C\hat E}\, F_{\hat B\hat D\hat F}\, S^{\hat A\hat B} S^{\hat C\hat D} S^{\hat E\hat F} \nonumber\\
  & \quad + \frac{1}{4} \tilde{\Omega}_{\hat A\hat C\hat D}\, \tilde{\Omega}_{\hat B}{}^{\hat C\hat D} S^{\hat A\hat B} - \frac{1}{4} \tilde{\Omega}_{\hat A\hat C\hat E}\, \tilde{\Omega}_{\hat B\hat D\hat F}\, S^{\hat A\hat B} S^{\hat C\hat D} S^{\hat E\hat F} \nonumber \\
  & \quad + \frac{1}{6} F_{\hat A\hat C\hat D}\, \tilde{\Omega}_{\hat B}{}^{\hat C\hat D} S^{\hat A\hat B} - \frac{1}{6} F_{\hat A\hat C\hat E}  \,\tilde{\Omega}_{\hat B\hat D\hat F}\, S^{\hat A\hat B} S^{\hat C\hat D} S^{\hat E\hat F}
\end{align}
and
\begin{align}
  - \frac{1}{2} \mathcal{H}^{AB} &\nabla_B \mathcal{H}^{CD} \nabla_D \mathcal{H}_{AC} = \frac{1}{18} F_{\hat A\hat C\hat D}\, F_{\hat B}{}^{\hat C\hat D} S^{\hat A\hat B}  - \frac{1}{18} F_{\hat A\hat C\hat E}\, F_{\hat B\hat D\hat F}\, S^{\hat A\hat B} S^{\hat C\hat D} S^{\hat E\hat F} \nonumber \\
  &+ \frac{1}{2} \tilde{\Omega}_{\hat A\hat C\hat E}\, \tilde{\Omega}_{\hat D\hat B\hat F}\, S^{\hat A\hat B} S^{\hat C\hat D} S^{\hat E\hat F} - \frac{1}{2} \tilde{\Omega}_{\hat C\hat A\hat D}\, \tilde{\Omega}_{\hat B}{}^{\hat C\hat D} S^{\hat A\hat B} - \frac{1}{2} \tilde{\Omega}_{\hat A\hat C\hat D}\, \tilde{\Omega}^{\hat C}{}_{\hat B}{}^{\hat D} S^{\hat A\hat B} \nonumber \\ 
  &- \frac{1}{2} \tilde{\Omega}_{\hat C\hat D\hat A}\, \tilde{\Omega}^{\hat D}{}_{\hat B}{}^{\hat C} S^{\hat A\hat B} + \frac{1}{3} F_{\hat A\hat C\hat D}\, \tilde{\Omega}_{\hat B}{}^{\hat C\hat D} S^{\hat A\hat B} - \frac{1}{3} F_{\hat A\hat C\hat E}  \,\tilde{\Omega}_{\hat B\hat D\hat F}\, S^{\hat A\hat B} S^{\hat C\hat D} S^{\hat E\hat F}
\end{align}
using this result. For the remaining third term in this line, we find
\begin{equation}\label{eqn:fluxterm3}
  \frac{1}{6} F_{ACD} F_B{}^{CD} \mathcal{H}^{AB} =
    \frac{1}{6} F_{\hat A\hat C\hat D}\, F_{\hat B}{}^{\hat C\hat D} S^{\hat A\hat B}\,.
\end{equation}
Summing up these three individual terms and combining the corresponding terms into covariant fluxes $\mathcal{F}_{\hat A\hat B\hat C}$ yields
\begin{gather}
  \frac{1}{8} \mathcal{H}^{CD} \nabla_C \mathcal{H}_{AB} \nabla_D \mathcal{H}^{AB} - \frac{1}{2} \mathcal{H}^{AB} \nabla_B \mathcal{H}^{CD} \nabla_D \mathcal{H}_{AC} +\frac{1}{6} F_{ACE} F_{BDF} \mathcal{H}^{AB} \eta^{CD} \eta^{EF} = \nonumber \\ 
  \frac{1}{4} \mathcal{F}_{\hat A\hat C\hat E} \mathcal{F}_{\hat B\hat D\hat F} S^{\hat A\hat B} \eta^{\hat C\hat D} \eta^{\hat E\hat F} - \frac{1}{12} \mathcal{F}_{\hat A\hat C\hat E} \mathcal{F}_{\hat B\hat D\hat F} S^{\hat A\hat B} S^{\hat C\hat D} S^{\hat E\hat F} \nonumber \\ - \frac{1}{2} \tilde{\Omega}_{\hat C\hat D\hat A}\, \tilde{\Omega}^{\hat C\hat D}{}_{\hat B}\, S^{\hat A\hat B} - \tilde{\Omega}_{\hat C\hat D\hat A}\, \tilde{\Omega}^{\hat D}{}_{\hat B}{}^{\hat C} S^{\hat A\hat B} - F_{\hat A\hat C\hat D}\, \tilde{\Omega}^{\hat C\hat D}{}_{\hat B}\, S^{\hat A\hat B}\,. \label{eqn:fluxresults1}
\end{gather}
This result looks already very promising, except for the last line. As a result, we evaluate the terms in the first line of the generalized curvature~\eqref{dft*gencurvature}. We obtain for them the following
\begin{align}\label{eqn:fluxterm4}
  4\mathcal{H}^{AB} \nabla_A \nabla_B d &= 4 S^{\hat A\hat B} D_{\hat A} D_{\hat B} \, \tilde{d} - 4 S^{\hat A\hat B} \, \tilde{\Omega}_{\hat A\hat B}{}^{\hat C} D_{\hat C} \, \tilde{d}\,, \\
  \label{eqn:fluxterm5}
  -4 \mathcal{H}^{AB} \nabla_A d \, \nabla_B d &= -4 S^{\hat A\hat B} D_{\hat A} \tilde{d} \, D_{\hat B} \tilde{d}\,, \\
  \label{eqn:fluxterm6}
  4 \nabla_A d \, \nabla_B \mathcal{H}^{AB} &= - 4 D_{\hat A} \tilde{d} \, \tilde{\Omega}^{\hat C}{}_{\hat C\hat B} \, S^{\hat A\hat B} + 4 S^{\hat A\hat B} \, \tilde{\Omega}_{\hat A\hat B}{}^{\hat C} D_{\hat C} \, \tilde{d} \\
\intertext{and}
  - \nabla_A \nabla_B \mathcal{H}^{AB} &= - S^{\hat A\hat B}\, \tilde{\Omega}^{\hat C}{}_{\hat C\hat A} \, \tilde{\Omega}^{\hat D}{}_{\hat D\hat B} + S^{\hat A\hat B} D_{\hat A} \, \tilde{\Omega}^{\hat C}{}_{\hat C\hat B} \nonumber\\ 
  &\quad + \tilde{\Omega}_{\hat B\hat C}{}^{\hat A} S^{\hat B\hat C} \, \tilde{\Omega}^{\hat D}{}_{\hat D\hat A} - D_{\hat A}\, \tilde{\Omega}_{\hat B\hat C}{}^{\hat A} S^{\hat B\hat C}\,. \label{eqn:fluxterm7}
\end{align}
Now, we express the last two terms in~\eqref{eqn:fluxterm7} as
\begin{equation}
  - \tilde{E}_{\hat A}{}^A \tilde{E}_{\hat B}{}^B \big( D_A D_B \tilde{E}_{\hat C}{}^M \big) \tilde{E}^{\hat A}{}_M \, S^{\hat B\hat C} + \tilde{\Omega}_{\hat C\hat D\hat A}\, \tilde{\Omega}^{\hat D}{}_{\hat B}{}^{\hat C} S^{\hat A\hat B}\,.
\end{equation}
For the last term in the first line of this equation, we have
\begin{equation}
  - \tilde{E}_{\hat A}{}^A \tilde{E}_{\hat B}{}^B \big( D_A D_B \tilde{E}_{\hat C}{}^M \big) \tilde{E}^{\hat A}{}_M \, S^{\hat B\hat C} - F_{\hat A\hat C\hat D}\, \tilde{\Omega}^{\hat C\hat D}{}_{\hat B}\, S^{\hat A\hat B}\,.
\end{equation}
Putting these two results together, we arrive at
\begin{align}
  - \nabla_A \nabla_B \mathcal{H}^{AB} &= - S^{\hat A\hat B}\, \tilde{\Omega}^{\hat C}{}_{\hat C\hat A} \, \tilde{\Omega}^{\hat D}{}_{\hat D\hat B} + 2 S^{\hat A\hat B} D_{\hat A} \, \tilde{\Omega}^{\hat C}{}_{\hat C\hat B} \\
  &+ \tilde{\Omega}_{\hat C\hat D\hat A}\, \tilde{\Omega}^{\hat D}{}_{\hat B}{}^{\hat C} S^{\hat A\hat B} + F_{\hat A\hat C\hat D}\, \tilde{\Omega}^{\hat C\hat D}{}_{\hat B}\, S^{\hat A\hat B}\,. \label{eqn:fluxterm7*}
\end{align}
In total, we derive for the terms in the first line of~\eqref{dft*gencurvature}
\begin{gather}
  4 \mathcal{H}^{AB} \nabla_A \nabla_B d - \nabla_A \nabla_B \mathcal{H}^{AB} - 4 \mathcal{H}^{AB} \nabla_A d\, \nabla_B d + 4 \nabla_A d \,\nabla_B \mathcal{H}^{AB} = \nonumber \\
  \qquad 2 S^{\hat A\hat B} D_{\hat A}\,\mathcal{F}_{\hat B} - S^{\hat A\hat B} \mathcal{F}_{\hat A} \mathcal{F}_{\hat B} + \tilde{\Omega}_{\hat C\hat D\hat A}\, \tilde{\Omega}^{\hat D}{}_{\hat B}{}^{\hat C} S^{\hat A\hat B} + F_{\hat A\hat C\hat D}\, \tilde{\Omega}^{\hat C\hat D}{}_{\hat B}\, S^{\hat A\hat B}\,. \label{eqn:fluxresults2}
\end{gather}
Finally, we obtain
\begin{align}\label{eqn:fluxresults3}
  \mathcal{R} =& \frac{1}{4} \mathcal{F}_{\hat A\hat C\hat D} \mathcal{F}_{\hat B}{}^{\hat C\hat D} S^{\hat A\hat B} - \frac{1}{12} \mathcal{F}_{\hat A\hat C\hat E} \mathcal{F}_{\hat B\hat D\hat F}  S^{\hat A\hat B} S^{\hat C\hat D} S^{\hat E\hat F} \nonumber \\ 
  & - \frac{1}{2} \tilde{\Omega}_{\hat C\hat D\hat A}\, \tilde{\Omega}^{\hat C\hat D}{}_{\hat B}\, S^{\hat A\hat B} + 2 S^{\hat A\hat B} D_{\hat A}\,\mathcal{F}_{\hat B} - S^{\hat A\hat B} \mathcal{F}_{\hat A} \mathcal{F}_{\hat B}
\end{align}
by taking~\eqref{eqn:fluxresults1} and~\eqref{eqn:fluxresults2} into account. After applying the strong constraint
\begin{equation}
  D_{\hat C} \tilde{E}_{\hat D}{}^A  D^{\hat C} \tilde{E}^{\hat D}{}_B = 0
\end{equation}
for fluctuations, the first term in the second line of the last equation vanishes.  As was the case for the generalized metric of DFT${}_\mathrm{WZW}$, discussed in section~\ref{Kap_3}, the strong constraint only needs to hold for fluctuations. The background which is governed by the structure coefficients $F_{\hat A\hat B\hat C}$ only needs to fulfill the Jacobi identity~\eqref{DFT*Jacobi}. Again, executing a partial integration~\eqref{DFT*intbyparts}
\begin{equation}\label{eqn:ibpDhatA}
  \int d^{2 D}X \, e^{-2d} D_{\hat A} v \, w = \int d^{2 D}X\, ( \mathcal{F}_{\hat A} v \, w - v\, D_{\hat A} w )\,,
\end{equation}
ultimately gives rise to the action
\begin{align}\label{eqn:actionfluxform}
  S = \int d^{2 D}X \, e^{-2d} \big( S^{\hat A\hat B} \mathcal{F}_{\hat A} \mathcal{F}_{\hat B} + \frac{1}{4} \mathcal{F}_{\hat A\hat C\hat D} \, \mathcal{F}_{\hat B}{}^{\hat C\hat D} \, S^{\hat A\hat B} - \frac{1}{12} \mathcal{F}_{\hat A\hat C\hat E} \, \mathcal{F}_{\hat B\hat D\hat F} \, S^{\hat A\hat B} S^{\hat C\hat D} S^{\hat E\hat F} \big)\,. 
\end{align}
This action possesses a manifest invariance under generalized diffeomorphisms and $2D$-diffeomorphisms due to the form of the fluxes and since it does not contain any additional flat derivatives. It appears equivalent to the form of the original flux formulation~\cite{Geissbuhler:2011mx, Geissbuhler:2013uka, Hohm:2013nja, Bosque:2015jda}. However, there do not occur any strong constraint violating terms. We explain their absence in more detail during the next subsection. Nevertheless, the here obtained covariant fluxes 
$\mathcal{F}_{\hat A\hat B\hat C}$ differ significantly from the results found in original DFT. They now exhibit an explicit segregation into a fluctuation and a background part.

Let us further go into a bit more detail about the transition to the toroidal formulation of DFT. After imposing the extended strong constraint~\eqref{eqn:extendedSC} and restricting the background vielbein to lie in O($D,D$), the splitting must vanish. Noting that these two conditions allow us to replace~\cite{Blumenhagen:2015zma}
\begin{equation}
  F_{ABC} = 2\Omega_{[AB]C} \quad \text{with} \quad F_{ABC} = 3\Omega_{[ABC]}\,,
\end{equation}
yielding
\begin{equation}
  \mathcal{F}_{\hat A\hat B\hat C} = 3 ( \tilde \Omega_{[\hat A\hat B\hat C]} + \Omega_{[\hat A\hat B\hat C]} ) =
    3 D_{[\hat A} \mathcal{E}_{\hat B}{}^I \mathcal{E}_{\hat C] I}
\end{equation}
we can observe what happens. It breaks the strict distinction between background and fluctuations occurring in DFT$_\mathrm{WZW}$ and only the O($D,D$) valued composite vielbein remains. Yet, its dynamics are still given by action~\eqref{eqn:actionfluxform}.

\subsection{Strong constraint violating terms}\label{sec:missingFABCFABCterm}

As already mentioned before, the action~\eqref{eqn:actionfluxform} reproduces all terms known from DFT's flux formulation~\cite{Geissbuhler:2011mx, Geissbuhler:2013uka, Hohm:2013nja, Bosque:2015jda}
\begin{align}
  S_\mathrm{DFT} = \int d^{2D} X\, e^{-2d} &\big( \mathcal{F}_A \mathcal{F}_B S^{AB} + \frac{1}{4} \mathcal{F}_{ACD} \mathcal{F}_B{}^{CD} S^{AB} - \frac{1}{12} \mathcal{F}_{ABC} \mathcal{F}_{DEF} S^{AD} S^{BE} S^{CF} \nonumber \\
    & \quad - \frac{1}{6} \mathcal{F}_{ABC} \mathcal{F}^{ABC} - \mathcal{F}_A \mathcal{F}^A \big)\,,
\end{align}
except for the strong constraint violating ones seen in the second line. All of the fluctuations are required to satisfy the strong constraint. In turn, they cannot contribute to these missing terms. Nevertheless, one would expect to observe at least background contributions of the form
\begin{equation}\label{eqn:candiateterms}
  F_A F^A \quad \text{or} \quad \frac{1}{6} F_{ABC} F^{ABC}\,.
\end{equation}
For the reason why these terms do not appear in the action either, we have to go back to the origins of the CSFT computation needed to derive DFT${}_\mathrm{WZW}$. During these calculations only CFTs with constant dilaton were considered~\cite{Blumenhagen:2014gva}. As a consequence, $F_A = 0$ has to hold and therefore the first term in~\eqref{eqn:candiateterms} becomes zero. On top of that, we have the following relation for the central charge~\cite{Blumenhagen:2014gva}
\begin{equation}\label{eqn:centralcharge}
  c  = \frac{k D}{k + h^\vee}
\end{equation}
of the closed string's left moving part. It is given by the level $k$ and the dual Coxeter number $h^\vee$. Subsequently, total central charge yields
\begin{equation}\label{eqn:ctotleft}
  c_\mathrm{tot} = c + c_\mathrm{gh} = D - \frac{D h^\vee}{k} + c_\mathrm{gh} + \mathcal{O}(k^{-1})\,,
\end{equation}
after adding the ghost contribution $c_\mathrm{gh}$. Terms of order $k^{-2}$ and higher were neglected in the derivation of DFT${}_\mathrm{WZW}$. Thus, we exclude them when computing the central charge. We now use the Killing form~\eqref{liemetric}
\begin{equation}
  \eta_{ab} =  -\frac{\alpha' k}{4 h^\vee} F_{ad}{}^c F_{bc}{}^d \,,
\end{equation}
as defined in~\cite{Blumenhagen:2014gva}. It allows us to write the second term in~\eqref{eqn:ctotleft} as
\begin{equation}
  -\frac{D h^\vee}{k} = \frac{\alpha'}{4} F_{ad}{}^c F_{bc}{}^d \eta^{ab}\,,
\end{equation}
i.e. through the unbared structure coefficients\footnote{ Note that this identification only works for semisimple Lie algebras whose Killing form is non-degenerate. It was one of the assumptions to derive DFT$_\mathrm{WZW}$.}. Furthermore, the analogous relations have to hold for the central charge of the anti-chiral, right-moving part as well. Hence, we arrive at
\begin{equation}
  c_\mathrm{tot} - \bar c_\mathrm{tot} = \frac{\alpha'}{4} \big( F_{ad}{}^c F_{bc}{}^d \eta^{ab} - F_{\bar{a}\bar{d}}{}^{\bar{c}} F_{\bar{b}\bar{c}}{}^{\bar{d}} \eta^{\bar{a}\bar{b}} \big) = - \frac{\alpha'}{2} F_{ABC} F^{ABC}
\end{equation}
with the according identifications
\begin{equation}
  \eta^{AB} = \frac{1}{2} \begin{pmatrix}
    \eta^{ab} & 0 \\
    0 & -\eta^{\bar a\bar b}
  \end{pmatrix}
  \quad \text{and} \quad
  F_{AB}{}^C = \begin{cases}
    F_{ab}{}^c & \\
    F_{\bar a\bar b}{}^{\bar c} & \\
    0 & \text{otherwise\,.}
  \end{cases}
\end{equation}
This result is proportional to the second term occurring in~\eqref{eqn:candiateterms}. Moreover, the CSFT derivations require both central charges $c_\mathrm{tot}$ and $\bar c_\mathrm{tot}$ to vanish independently and hence we do not observe any strong constraint violating terms. Another interesting by-product of this computation is that the scalar curvature
\begin{equation}
  R = \frac{2}{9} F_{ABC} F^{ABC} = R_{ABC}{}^B \eta^{AC} = 0\,,
\end{equation}
which arises by contracting the Riemann curvature tensor
\begin{equation}
  R_{ABC}{}^D = \frac{2}{9} F_{AB}{}^E F_{EC}{}^D
\end{equation}
induced by the covariant derivative $\nabla_A$, is vanishing as well.

\subsection{Double Lorentz invariance}\label{sec:doublelorentz}

Despite the generalized and 2D-diffeomorphism invariance of the action~\eqref{eqn:actionfluxform}, it also possesses a local double Lorentz symmetry. This symmetry acts on hatted indices, as the ones of the generalized fluctuation vielbein, through
\begin{equation}\label{eqn:localodxodsym}
  \tilde E_{\hat A}{}^B \rightarrow T_{\hat A}{}^{\hat C} \tilde E_{\hat C}{}^B
\end{equation}
where the tensor $T_{\hat A}{}^{\hat B}$ satisfies the relation
\begin{equation}
  T_{\hat A}{}^{\hat C} \eta_{\hat C\hat D} T_{\hat B}{}^{\hat D} = \eta_{\hat A\hat B}
    \quad \text{and} \quad
  T_{\hat A}{}^{\hat C} S_{\hat C\hat D} T_{\hat B}{}^{\hat D} = S_{\hat A\hat B}\,.
\end{equation}
Whereas the local double Lorentz symmetry is manifest in the generalized metric formulation, due to the non-existence of hatted indices, in the flux formulation it is not that obvious and needs to be explicitly checked. In this context, we consider their infinitesimal counterpart~\eqref{eqn:localodxodsym} which are denoted through
\begin{equation}
  \delta_\Lambda \mathcal{E}_{\hat A}{}^I = \Lambda_{\hat A}{}^{\hat B} \mathcal{E}_{\hat B}{}^I\,.
\end{equation}
By generating these doubled Lorentz transformations $\Lambda_{\hat A\hat B}$ has to fulfill the following identities
\begin{equation}\label{eqn:odxodlambda}
  \Lambda_{\hat A\hat B} = - \Lambda_{\hat B\hat A} \quad \text{and} \quad
  \Lambda_{\hat A\hat B} = S_{\hat A\hat C} \Lambda^{\hat C\hat D} S_{\hat D\hat B}\,.
\end{equation}
After a short computation, we obtain the transformation behaviors
\begin{align}
  \delta_\Lambda \mathcal{F}_{\hat A\hat B\hat C} &= 3 \big( D_{\hat{[A}} \Lambda_{\hat B\hat C]} + \Lambda_{[\hat A}{}^{\hat D} \mathcal{F}_{\hat B\hat C]\hat D} \big)\,, \\
  \delta_\Lambda \mathcal{F}_{\hat A} &= D^{\hat B} \Lambda_{\hat B\hat A} + \Lambda_{\hat A}{}^{\hat B} \mathcal{F}_{\hat B}
\end{align}
of the covariant fluxes. Note that the last terms in both equations spoil covariance under double Lorentz transformations. With the use of these equations, it is straightforward to calculate the change of the action
\begin{equation}\label{eqn:odxodaction2}
  \delta_\Lambda S = - \int d^{2n} X \, e^{-2d} \Lambda_{\hat A}{}^{\hat C} \delta^{\hat A\hat B} \mathcal{Z}_{\hat B\hat C}
\end{equation}
with
\begin{equation}\label{eqn:odxodz1}
  \mathcal{Z}_{\hat A\hat B} = D^{\hat C} \mathcal{F}_{\hat C\hat A\hat B} + 2 D_{[\hat A} \mathcal{F}_{\hat B]} - \mathcal{F}^{\hat C} \mathcal{F}_{\hat C\hat A\hat B}\,.
\end{equation}
We do not demonstrate all the detailed steps of this computation. It is completely analogous to the derivation presented in the original flux formulation of DFT~\cite{Geissbuhler:2013uka}.
The evaluation of $\mathcal{Z}_{\hat A\hat B}$ requires a segregation of the covariant fluxes $\mathcal{F}_{\hat A\hat B\hat C}$ into their individual fluctuation and background parts according to equation~\eqref{eqn:fluxcbracketepsilon2}.
As a result, we have to compute the following terms
\begin{align}
  D^{\hat C} \tilde{F}_{\hat C\hat A\hat B} &= D^C \big( D_C \tilde{E}_{[\hat A}{}^D \tilde{E}_{\hat B]D} \big) + \tilde{\Omega}^{\hat C}{}_{\hat C\hat D} \, \tilde{\Omega}^{\hat D}{}_{\hat A\hat B} + \underline{2 D^{\hat C} \tilde{\Omega}_{[\hat A\hat B]\hat C}} \nonumber \\
  D^{\hat C} F_{\hat C\hat A\hat B} &= \tilde{E}_{\hat A}{}^A \tilde{E}_{\hat B}{}^B D^C F_{CAB} + \tilde{\Omega}^{\hat D}{}_{\hat D}{}^{\hat C} F_{\hat C\hat A\hat B} + \underline{2 F_{[\hat A\hat C\hat D} \tilde{\Omega}^{\hat C\hat D}{}_{\hat B]}} \nonumber \\
  2 D_{[\hat A} \tilde F_{\hat B]} &= 2 F_{\hat A\hat B}{}^{\hat C} D_{\hat C} \, \tilde{d} + 4 \tilde{\Omega}_{[\hat A\hat B]}{}^{\hat C} D_{\hat C} \, \tilde{d} + \underline{2 D_{[\hat A} \tilde{\Omega}^{\hat C}{}_{\hat C\hat B]}} \nonumber \\
  - \tilde F^{\hat C} F_{\hat C\hat A\hat B} &= - 2 F_{\hat A\hat B}{}^{\hat C} D_{\hat C} \, \tilde{d} - \tilde{\Omega}^{\hat D}{}_{\hat D}{}^{\hat C} F_{\hat C\hat A\hat B} \nonumber \\
  - \tilde F^{\hat C} \tilde{F}_{\hat C\hat A\hat B} &= - 2 \tilde{\Omega}^{\hat C}{}_{\hat A\hat B} D_{\hat C} \, \tilde{d} - 4 \tilde{\Omega}_{[\hat A\hat B]}{}^{\hat C} D_{\hat C} \, \tilde{d} - \tilde{\Omega}^{\hat D}{}_{\hat D}{}^{\hat C} \tilde{\Omega}_{\hat C\hat A\hat B} - \underline{2 \tilde{\Omega}^{\hat D}{}_{\hat D}{}^{\hat C} \tilde{\Omega}_{[\hat A\hat B]\hat C}} \nonumber\,.
\end{align}
The underlined terms cancel due to the relation
\begin{equation}
  2 D^{\hat C} \tilde{\Omega}_{[\hat A\hat B]\hat C} - 2 \tilde{\Omega}^{\hat D}{}_{\hat D}{}^{\hat C} \, \tilde{\Omega}_{[\hat A\hat B]\hat C} =
    - 2 F_{[\hat A\hat C\hat D} \tilde{\Omega}^{\hat C\hat D}{}_{\hat B]} - 2 D_{[\hat A} \tilde{\Omega}^{\hat C}{}_{\hat C\hat B]}
\end{equation}
which arises after exchanging two flat derivatives. Hence, equation~\eqref{eqn:odxodz1} takes on the form
\begin{equation}\label{eqn:odxodz2}
  \mathcal{Z}_{\hat A\hat B} = D^C \big( D_C \tilde{E}_{[\hat A}{}^D \tilde{E}_{\hat B]D} \big) -2 \tilde{\Omega}^{\hat C}{}_{\hat A\hat B} D_{\hat C} \, \tilde{d} + \tilde{E}_{\hat A}{}^A \tilde{E}_{\hat B}{}^B D^C F_{CAB}\,,
\end{equation}
where the first two terms vanish under the strong constraint. We obtain the result
\begin{equation}\label{eqn:odxodxz3}
  \mathcal{Z}_{\hat A\hat B} = \tilde{E}_{\hat A}{}^A \tilde{E}_{\hat B}{}^B D^C F_{CAB}\,.
\end{equation}
The structure coefficients $F_{ABC}$ in the context of DFT$_\mathrm{WZW}$ are constant since we only consider group manifolds. Finally, we are left with $\mathcal{Z}_{\hat A\hat B}=0$. Thus, we have shown the invariance of DFT$_\mathrm{WZW}$'s action~\eqref{eqn:actionfluxform} under double Lorentz transformations in the flux formulation.

\section{Gauge symmetries}\label{sec:gaugetrafo}

The flux formulation requires all indices to carry hatted labels. Thus, we need to check how the gauge transformations~\eqref{eqn:gendiffHAB&tilded} of DFT${}_\mathrm{WZW}$ affect these index structures. Therefore, we introduce an arbitrary vector in the canonical form
\begin{equation}\label{eqn:hatsplitting}
  V^{\hat A} = \tilde{E}^{\hat A}{}_B V^B\,.
\end{equation}
The parameters appearing in the gauge transformations are of course vectors as well. These are given totally analogous. Applying this splitting, we can compute the generalized Lie derivative to be
\begin{align}\label{eqn:hatliederivative}
  \mathcal{L}_{\xi} V^{\hat A} &= \xi^{\hat B} D_{\hat B} V^{\hat A} +  \big( D^{\hat A} \xi_{\hat B} - D_{\hat B} \xi^{\hat A} \big) V^{\hat B} + \mathcal{F}^{\hat A}{}_{\hat B\hat C} \, \xi^{\hat B} V^{\hat C} \quad \text{and} \\
  \mathcal{L}_{\xi} d &= \frac{1}{2} \xi^{\hat A} \mathcal{F}_{\hat A} - \frac{1}{2} D_{\hat A} \xi^{\hat A}\,,
\end{align}
where $\mathcal{F}_{\hat A\hat B\hat C}$ and $\mathcal{F}_{\hat A}$ represent the covariant fluxes defined in~\eqref{eqn:fluxcbracketepsilon2} and~\eqref{eqn:defFA}. This result agrees with the original flux formulation~\cite{Geissbuhler:2011mx, Hohm:2013nja}, but the covariant fluxes themselves are defined differently again, as they split into a background and fluctuation part in DFT$_\mathrm{WZW}$.

Subsequently, we can evaluate the C-bracket~\eqref{DFT*cbracket1} under the splitting~\eqref{eqn:hatsplitting}. It yields
\begin{equation}\label{eqn:hatcbracket}
  \big[ \xi_1,\xi_2 \big]_C^{\hat A} = \xi_1^{\hat B} D_{\hat B} \xi_2^{\hat A} - \frac{1}{2} \xi_1^{\hat B} D^{\hat A} \xi_{2\,\hat B} + \frac{1}{2}\mathcal{F}^{\hat A}{}_{\hat B\hat C} \xi_1^{\hat B} \xi_2^{\hat C} - (1 \leftrightarrow 2)\,.
\end{equation}
The same remarks as for the action and the generalized Lie derivative hold in this context as well.

\section{Field equations}\label{sec:eom}

In the following, we are going to derive the equations of motion. The procedure mainly follows in a similar fashion to~\cite{Aldazabal:2013sca, Geissbuhler:2013uka}. We can now vary the action~\eqref{eqn:actionfluxform} after the dilaton fluctuations $\tilde d$ and the fluctuation vielbein $\tilde{E}_{\hat A}{}^B$. One obtains
\begin{equation}\label{eqn:vardtilde}
  \delta_{\tilde d} \, S = \int d^{2n} X  e^{-2d} \, \mathcal{G} \, \delta\tilde{d}
\end{equation}
and
\begin{equation}\label{eqn:varEtilde}
  \delta_{E} \, S = \int d^{2n} X  e^{-2d} \, \mathcal{G}^{\hat A\hat B} \, \delta\tilde{E}_{\hat A\hat B} 
    \quad \text{with} \quad
  \delta\tilde{E}_{\hat A\hat B} = \delta\tilde{E}_{\hat A}{}^C \tilde{E}_{\hat BC}\,.
\end{equation}
Due to the antisymmetry of $\delta\tilde{E}_{\hat A\hat B}$, a result of
\begin{equation}
  \delta ( \tilde E_{\hat A}{}^C \tilde E_{\hat B C} ) = \delta \eta_{\hat A\hat B} = 0\,,
\end{equation}
only the antisymmetric part of $\mathcal{G}^{\hat A\hat B}$ gives contributions to the equation of motion. A straight forward calculation for the variations~\eqref{eqn:vardtilde} and~\eqref{eqn:varEtilde} gives rise to
\begin{equation}
  \mathcal{G} = - 2 \mathcal{R} 
    \quad \text{and} \quad
  \mathcal{G}^{[\hat A\hat B]} = 2 S^{\hat D[\hat A} D^{\hat B]} \mathcal{F}_{\hat D} + \big( \mathcal{F}_{\hat D} - D_{\hat D} \big) \check{\mathcal{F}} ^{\hat D[\hat A\hat B]} + \check{\mathcal{F}} ^{\hat C\hat D[\hat A} \mathcal{F}_{\hat C\hat D}{}^{\hat B]}
\end{equation}
with
\begin{equation}
  \check{\mathcal{F}}^{\hat A\hat C\hat E} = \Big( - \frac{1}{2} S^{\hat A\hat B}S^{\hat C\hat D}S^{\hat E\hat F} + \frac{1}{2} S^{\hat A\hat B}\eta^{\hat C\hat D}\eta^{\hat E\hat F} + \frac{1}{2} \eta^{\hat A\hat B}S^{\hat C\hat D}\eta^{\hat E\hat F} + \frac{1}{2} \eta^{\hat A\hat B}\eta^{\hat C\hat D}S^{\hat E\hat F} \Big) \mathcal{F}_{\hat B\hat D\hat F}\,.
\end{equation}
As a consequence, the equation of motion take on the form
\begin{equation}
  \mathcal{G} = 0 \quad \text{and} \quad 
  \mathcal{G}^{[\hat A\hat B]} = 0\,.
\end{equation}
This result is in one to one correspondence with the original flux formulation of DFT~\cite{Geissbuhler:2013uka} as well. But again, it is important to keep in mind that the covariant fluxes differ significantly between both theories. However, the term $\check{\mathcal{F}}^{\hat A\hat B\hat C}$ seems at first sight quite artificial. Its role becomes more obvious, if we replace it by a projector $\mathcal{P}_{\hat A\hat C\hat E}{}^{\hat B\hat D\hat F}$, i.e. 
\begin{equation}
  \check{\mathcal{F}}_{\hat A\hat C\hat E} = \mathcal{P}_{\hat A\hat C\hat E}{}^{\hat B\hat D\hat F} \mathcal{F}_{\hat B\hat D\hat F}\,,
\end{equation}
which is composed out of eight different projection operators
\begin{align}
\mathcal{P}_{\hat A\hat C\hat E}{}^{\hat B\hat D\hat F} =&\;\;\;\:\, P_{\hat A}{}^{\hat B} P_{\hat C}{}^{\hat D} P_{\hat E}{}^{\hat F} - \bar{P}_{\hat A}{}^{\hat B}\bar{P}_{\hat C}{}^{\hat D}\bar{P}_{\hat E}{}^{\hat F}
+ \bar{P}_{\hat A}{}^{\hat B} P_{\hat C}{}^{\hat D} P_{\hat E}{}^{\hat F} + P_{\hat A}{}^{\hat B} \bar{P}_{\hat C}{}^{\hat D} P_{\hat E}{}^{\hat F} \\ &+ P_{\hat A}{}^{\hat B} P_{\hat C}{}^{\hat D} \bar{P}_{\hat E}{}^{\hat F} - \bar{P}_{\hat A}{}^{\hat B} \bar{P}_{\hat C}{}^{\hat D} P_{\hat E}{}^{\hat F} - \bar{P}_{\hat A}{}^{\hat B} P_{\hat C}{}^{\hat D} \bar{P}_{\hat E}{}^{\hat F} - P_{\hat A}{}^{\hat B} \bar{P}_{\hat C}{}^{\hat D} \bar{P}_{\hat E}{}^{\hat F} \nonumber
\end{align}
with the individual projectors
\begin{equation}
  P_{\hat A}{}^{\hat B} = \frac{1}{2} \big( S_{\hat A}{}^{\hat B} + \delta_{\hat A}{}^{\hat B} \big) \quad\text{and}\quad \bar{P}_{\hat A}{}^{\hat B} = -\frac{1}{2} \big( S_{\hat A}{}^{\hat B} - \delta_{\hat A}{}^{\hat B} \big)\,.
\end{equation}
These are the two well known projectors~\eqref{projectionoperators} appearing in the equation of motion for original DFT's generalized metric formulation~\cite{Hohm:2010pp,Blumenhagen:2015zma}.

\chapter{Generalized Scherk-Schwarz Compactifications in DFT on Group Manifolds}\label{sec:genSSWZW}

During the previous chapter~\ref{Kap_4}, we obtained the flux formulation of DFT$_\mathrm{WZW}$. It allows us to connect DFT${}_\mathrm{WZW}$ with generalized Scherk-Schwarz compactifications~\ref{sec:T-Duality}. Conjectures for such a link have already been made in~\cite{Blumenhagen:2014gva,Blumenhagen:2015zma}. In this chapter, we want to make this relationship manifest by introducing a slightly adapted generalized Scherk-Schwarz ansatz. Consequently, we can derive the low-energy, effective theory in this chapter~\ref{sec:genSSWZW}. As was expected, the theory we obtain describes a bosonic subsector of a half-maximal, electrically gauged supergravity. Further, it is possible to classify all emerging gauged supergravities using an embedding tensor, section~\ref{sec:embeddingtensor}. Afterwards, we discuss all explicit solutions of the embedding tensor for $n=3$ internal dimensions by following~\cite{Dibitetto:2012rk,Dibitetto:2015bia}. However, before presenting any new results, we start by reviewing generalized Scherk-Schwarz compactifications in the context of original DFT and we observe the issues with constructing an appropriate twist, capturing all properties of the compactification, inherent in this framework. Generally, the original description of DFT lacks a procedure to obtain a twist from a given embedding tensor solution. As a result, one is left with guessing them. Subsequently, this raises the question whether there exist twists for all solutions of the embedding tensor at all. Due to the formalism presented in this chapter, these issues have been avoided entirely. Thus, we present a detailed procedure to derive the background generalized vielbeins, taking on the role of the twist, for arbitrary embedding tensor solutions in section~\ref{sec:twistconstr}.

The basis for the following chapter is~\cite{Bosque:2015jda}.

\section{Embedding tensor}\label{sec:embeddingtensor}

Before going into more detail about generalized Scherk-Schwarz compactifications, we discuss a crucial instrument to classify all of the maximal/half-maximal gauged supergravities arising from these compactification schemes~\cite{Bosque:2015jda}. It is called the embedding tensor $\Theta_I{}^{\alpha}$. For a detailed review, see~\cite{Samtleben:2008pe}. This tool illustrates the embedding of the supergravity's gauge group into the global symmetry group of the ungauged theory. In the case of DFT we are interested in the embedding lying in O($D,D$), the T-Duality group~\ref{sec:T-Duality} of a $D$-dimensional torus. The structure coefficients of the Lie algebra are then connected to the embedding tensor by
\begin{equation}
\label{eqn:embeddingtensor}
  F_{AB}{}^C = \Theta_A{}^{\alpha} \big( t_{\alpha} \big)_B{}^C = \big(X_A\big)_B{}^C\,.
\end{equation}
Here, $t_{\alpha}$ denotes the different O($D,D$) generators.
Their corresponding vector representation, by acting on arbitrary doubled vectors $V^A$ through
\begin{equation}
  t_\alpha V^A = V^B (t_\alpha)_B{}^A\,,
\end{equation}
is given by $\big(t_{\alpha}\big)_B{}^C$. In general, the embedding tensor is required to satisfy two conditions: a linear and a quadratic constraint. A consistent supergravity solution is specified by solving both of them.

Solving these constraints in higher dimensions can become quite involved. That is why we restrict ourselves to the discussion of $n=3$ internal dimensions. Following~\cite{Dibitetto:2012rk,Dibitetto:2015bia}, we obtain twelve different solutions in total, each of them is governed by a continuous parameter $\alpha$. Specifically, $t_\alpha$ in~\eqref{eqn:embeddingtensor} portrays six different generators of $\mathfrak{o}(3,3)$. In the vector representation, their indices are labeled by $A, B, \ldots$ running from $1, \ldots, 6$. From a group theoretic point of view, the embedding tensor lives in the tensor product
\begin{equation}\label{eqn:embtensordecomp}
  \mathbf{6} \otimes \mathbf{15} = \mathbf{6} \oplus \mathbf{\overline{10}} \oplus \mathbf{10} \oplus \mathbf{64}\,,
\end{equation}
where the first factor classifies the vector representation, while the second one represents the adjoint representation denoted by the subscript $\alpha$ in $t_\alpha$. Following the application of the linear constraint certain irreps are projected out. In our discussion, we are only interested in the irreps $\mathbf{\overline{10}} \oplus \mathbf{10}$ of the decomposition~\eqref{eqn:embtensordecomp}. We set all other components of the embedding tensor to zero. At this point, $F_{ABC}$ is in direct correspondence with the vacuum expectation value (or background part) of the covariant fluxes $\mathcal{F}_{ABC}$. These have indeed the right amount ($6\cdot 5 \cdot 4 / 3! = 20$) of independent components.

According to~\cite{Dibitetto:2012rk}, we are able to express $\big(X_A\big)_B{}^C$ through irreps of $\mathfrak{sl}(4)$ as opposed to using $\mathfrak{so}(3,3)$. This is possible since both algebras are isomorphic to each other and hence the decomposition~\eqref{eqn:embtensordecomp} does not change. We introduce the fundamental $\mathfrak{sl}(4)$ indices $p,q,r =1,\dots,4$ as we are interested in distinguishing between these two different algebras.
Subsequently, the relevant part of the embedding tensor  $\mathbf{\overline{10}} \oplus \mathbf{10}$ takes on the form~\cite{Dibitetto:2012rk}
\begin{equation}\label{eqn:embsl41}
  \big(X_{mn}\big)_p{}^q = \frac{1}{2} \delta^q{}_{[m} M_{n]p} - \frac{1}{4}  \varepsilon_{mnpr} \tilde{M}^{rq}.
\end{equation}
Here, $M_{np}$ and $\tilde{M}^{rq}$ are symmetric matrices and $\varepsilon$ represents the Levi-Civita tensor in $4$-dimensions. These symmetric matrices possess in total $4 \cdot 5 / 2 = 10$ independent components, each. As a consequence, we can identify $M_{pq}$ with the irrep $\mathbf{10}$, whereas $\tilde M^{rp}$ lies in the dual irrep $\mathbf{\overline{10}}$. On top of that, we have the indices $m$ and $n$ in $\big(X_{mn}\big)_p{}^q$. These are antisymmetric and denote the $4 \cdot 3 / 2 = 6$ independent components of the $\mathfrak{sl}(4)$ irrep $\mathbf{6}$. Its dual representation with two upper antisymmetric indices is defined through
\begin{equation}
  X_{mn} = \frac{1}{2} \varepsilon_{mnpq} X^{pq}\,.
\end{equation}

Thus, the irreps $\mathbf{\overline{10}} \oplus \mathbf{10}$ are embedded into the product $\mathbf{6} \otimes \mathbf{15}$ by equation~\eqref{eqn:embsl41}. But the structure coefficients are a rank 3 tensor living in $\mathbf{6} \otimes \mathbf{6} \otimes \mathbf{6}$. Hence, we need to embed $\big(X_{mn}\big)_p{}^q$ into this product using the following correlation
\begin{equation}
  \label{eqn:embsl42}
  \big(X_{mn}\big)_{pq}{}^{rs} = 2 \big( X_{mn} \big)_{[p}{}^{[r} \delta_{q]}{}^{s]}\,.
\end{equation}
Finally, we have to take a step back from $\mathfrak{sl}(4)$ to $\mathfrak{so}(3,3)$ again. Therefore, we introduce the 't Hooft symbols $\big( G_A \big)^{mn}$ to switch between both representations.  For $n=3$, they are given by
\begin{align}
  \big(G_1\big)^{mn} &= \frac{1}{\sqrt{2}} \begin{pmatrix}
    0 && -1 && 0 && \phantom{-}0 \\
    1 && \phantom{-}0 && 0 && \phantom{-}0 \\
    0 && \phantom{-}0 && 0 && -1 \\
    0 && \phantom{-}0 && 1 && \phantom{-}0
  \end{pmatrix}\,, &
  \big(G_2\big)^{mn} &= \frac{1}{\sqrt{2}} \begin{pmatrix}
    0 && \phantom{-}0 && -1 && 0 \\
    0 && \phantom{-}0 && \phantom{-}0 && 1 \\
    1 && \phantom{-}0 && \phantom{-}0 && 0 \\
    0 && -1 && \phantom{-}0 && 0
  \end{pmatrix}\,, \nonumber \\
  \big(G_3\big)^{mn} &= \frac{1}{\sqrt{2}} \begin{pmatrix}
    0 && 0 && \phantom{-}0 && -1 \\
    0 && 0 && -1 && \phantom{-}0 \\
    0 && 1 && \phantom{-}0 && \phantom{-}0 \\
    1 && 0 && \phantom{-}0 && \phantom{-}0
  \end{pmatrix}\,, &
  \big(G_{\bar 1}\big)^{mn} &= \frac{1}{\sqrt{2}} \begin{pmatrix}
    \phantom{-}0 && 1 && 0 && \phantom{-}0 \\
    -1 && 0 && 0 && \phantom{-}0 \\
    \phantom{-}0 && 0 && 0 && -1 \\
    \phantom{-}0 && 0 && 1 && \phantom{-}0
  \end{pmatrix}\,, \nonumber \\
  \big(G_{\bar 2}\big)^{mn} &= \frac{1}{\sqrt{2}} \begin{pmatrix}
    \phantom{-}0 && \phantom{-}0 && 1 && 0 \\
    \phantom{-}0 && \phantom{-}0 && 0 && 1 \\
    -1 && \phantom{-}0 && 0 && 0 \\
    \phantom{-}0 && -1 && 0 && 0
  \end{pmatrix}\,, &
  \big(G_{\bar 3}\big)^{mn} &= \frac{1}{\sqrt{2}} \begin{pmatrix}
    \phantom{-}0 && 0 && \phantom{-}0 && 1 \\
    \phantom{-}0 && 0 && -1 && 0 \\
    \phantom{-}0 && 1 && \phantom{-}0 && 0 \\
    -1 && 0 && \phantom{-}0 && 0
  \end{pmatrix}
\end{align}
and fulfill the following relations
\begin{align}
\big( G_{A} \big)_{mn} \big( G_{B} \big)^{mn} &= 2 \eta_{AB}\,, \\
\big( G_{A} \big)_{mp} \big( G_{B} \big)^{pn} + \big( G_{B} \big)_{mp} \big( G_{A} \big)^{pn} &= - \delta_m{}^n \, \eta_{AB}
\end{align}
with the flat O($D,D$) invariant metric
\begin{equation}
  \eta_{AB} = \begin{pmatrix}
    \delta_{ab} & 0 \\ 0 & -\delta_{\bar{a}\bar{b}}
  \end{pmatrix}
\end{equation}
of DFT${}_\mathrm{WZW}$. Ultimately, they allow us to derive the covariant fluxes
\begin{equation}\label{eqn:SL(4)toSO(3,3)}
  F_{ABC} = \left( X_{mn} \right)_{pq}{}^{rs}
    \left( G_A \right)^{mn} \left( G_B \right)^{pq} \left( G_C \right)_{rs}
\end{equation}
in their familiar form.

In the framework we work in, the quadratic constraint for the embedding tensor is the analogue of the Jacobi identity~\eqref{DFT*Jacobi} for the structure coefficients $F_{ABC}$ of the background vielbein. For the $\mathfrak{sl}(4)$ representation~\eqref{eqn:embsl42} discussed above, the Jacobi identity yields~\cite{Dibitetto:2012rk}
\begin{equation}\label{eqn:embsl4jacobi}
  M_{mp} \tilde{M}^{pn} = \frac{1}{4} \delta_m{}^n M_{pq} \tilde{M}^{pq}\,.
\end{equation}
Now, it is always possible to find an SO($4$) rotation diagonalizing the matrix $M_{np}$ since it is symmetric. Furthermore, SO($4$) is the maximal subgroup of SL($4$) and is isomorphic to SO($3$)$\times$SO($3$), the maximal compact subgroup of SO($3,3$), up to a $\mathds{Z}_2$ transformation. Thus, there always exists a double Lorentz transformation we can apply to the structure coefficients in order to diagonalize $M_{np}$. Once $M_{np}$ is diagonal, the dual $\tilde{M}^{rq}$ is diagonal as well. If it were not, equation~\eqref{eqn:embsl4jacobi} would be violated. As a result, we are able to solve the quadratic constraint.
\begin{table}[t!]
\centering
\begin{tabular}{| c | r r r r | r r r r | c | c |}
\hline
\textrm{ID} & \multicolumn{4}{|c|}{$\diag M_{mn}/\,\cos\alpha\,$} & \multicolumn{4}{|c|}{$\diag \tilde{M}^{mn}/\,\sin\alpha\,$} & range of $\alpha$ & gauging \\[1mm]
\hline \hline \rowcolor{fillcolor}
$1$ & $1$ & $1$ & $1$ & $1$ & 
      $1$ & $1$ & $1$ & $1$ & $-\frac{\pi}{4}\,<\,\alpha\,\le\,\frac{\pi}{4}$ & $\left\{
\setlength{\arraycolsep}{2pt}\begin{array}{cc}\textrm{SO}($4$)\ , & \alpha\,\ne\,\frac{\pi}{4}\ ,\\ \textrm{SO}(3)\ , & \alpha\,=\,\frac{\pi}{4}\ .\end{array}\right.$\\[4mm]
\hline
$2$ & $1$ & $1$ & $1$ & $-1$ & 
      $1$ & $1$ & $1$ & $-1$ & $-\frac{\pi}{4}\,<\,\alpha\,\le\,\frac{\pi}{4}$ & SO($3,1$)\\[1mm]
\hline
$3$ & $1$ & $1$ & $-1$ & $-1$ &
      $1$ & $1$ & $-1$ & $-1$ & $-\frac{\pi}{4}\,<\,\alpha\,\le\,\frac{\pi}{4}$ & $\left\{
\setlength{\arraycolsep}{2pt}\begin{array}{cc}\textrm{SO}($2,2$)\ , & \alpha\,\ne\,\frac{\pi}{4}\ ,\\ \textrm{SO}(2,1)\ , & \alpha\,=\,\frac{\pi}{4}\ .\end{array}\right.$\\[2mm]
\hline \hline \rowcolor{fillcolor}
$4$ & $1$ & $1$ & $1$ & $0$ & 
      $0$ & $0$ & $0$ & $1$ & $-\frac{\pi}{2}\,<\,\alpha\,<\,\frac{\pi}{2}$ & ISO($3$)\\[1mm]
\hline
$5$ & $1$ & $1$ & $-1$ & $0$ & 
      $0$ & $0$ & $0$  & $1$ & $-\frac{\pi}{2}\,<\,\alpha\,<\,\frac{\pi}{2}$ & ISO($2,1$)\\[1mm]
\hline \hline \rowcolor{fillcolor}
$6$ & $1$ & $1$ & $0$ & $0$ & 
      $0$ & $0$ & $1$ & $1$ & $-\frac{\pi}{4}\,<\,\alpha\,\le\,\frac{\pi}{4}$ & $\left\{
\setlength{\arraycolsep}{2pt}\begin{array}{cc}\textrm{CSO}(2,0,2)\ , & \alpha\,\ne\,\frac{\pi}{4}\ ,\\ \mathfrak{f}_{1}\quad(\textrm{Solv}_{6}) \ , & \alpha\,=\,\frac{\pi}{4}\ .\end{array}\right.$\\[4mm]
\hline
$7$ & $1$ & $1$ & $0$ & $0$ & 
      $0$ & $0$ & $1$,& $-1$ & $-\frac{\pi}{2}\,<\,\alpha\,<\,\frac{\pi}{2}$ & $\left\{
\setlength{\arraycolsep}{2pt}\begin{array}{cc}\textrm{CSO}(2,0,2)\ , & |\alpha|\,<\,\frac{\pi}{4}\ ,\\ \textrm{CSO}(1,1,2)\ , & |\alpha|\,>\,\frac{\pi}{4}\ ,\\ \mathfrak{g}_{0}\quad(\textrm{Solv}_{6}) \ , & |\alpha|\,=\,\frac{\pi}{4}\ .\end{array}\right.$\\[4mm]
\hline \rowcolor{fillcolor}
$8$ & $1$ & $1$ & $0$ & $0$ & 
      $0$ & $0$ & $0$ & $1$ & $-\frac{\pi}{2}\,<\,\alpha\,<\,\frac{\pi}{2}$ & $\mathfrak{h}_{1}\quad(\textrm{Solv}_{6})$\\[1mm]
\hline
$9$ & $1$ & $-1$ & $0$ & $0$ & 
      $0$ & $0$  & $1$ & $-1$ & $-\frac{\pi}{4}\,<\,\alpha\,\le\,\frac{\pi}{4}$ & $\left\{
\setlength{\arraycolsep}{2pt}\begin{array}{cc}\textrm{CSO}(1,1,2)\ , & \alpha\,\ne\,\frac{\pi}{4}\ ,\\ \mathfrak{f}_{2}\quad(\textrm{Solv}_{6}) \ , & \alpha\,=\,\frac{\pi}{4}\ .\end{array}\right.$\\[4mm]
\hline
$10$ & $1$ & $-1$ & $0$ & $0$ & 
       $0$ & $0$  & $0$ & $1$ & $-\frac{\pi}{2}\,<\,\alpha\,<\,\frac{\pi}{2}$ & $\mathfrak{h}_{2}\quad(\textrm{Solv}_{6})$\\[1mm]
\hline \hline \rowcolor{fillcolor}
$11$ & $1$ & $0$ & $0$ & $0$ & 
       $0$ & $0$ & $0$ & $1$ &
$-\frac{\pi}{4}\,<\,\alpha\,\le\,\frac{\pi}{4}$ &
$\left\{
\setlength{\arraycolsep}{2pt}\begin{array}{cc}\mathfrak{l}\quad(\textrm{Nil}_{6}(3)\,)\ , & \alpha\,\ne\,0\ ,\\
\textrm{CSO}(1,0,3)\ , &
\alpha\,=\,0\ .\end{array}\right.$\\[4mm]
\hline \hline \rowcolor{fillcolor}
$12$ & $0$ & $0$ & $0$ & $0$ & 
       $0$ & $0$ & $0$ & $0$ &
$\alpha = 0$ & 
$\textrm{U}(1)^6$ \\
\hline
\end{tabular}
\caption{Solutions of the embedding tensor for half-maximal, electrically gauged supergravity in $n=3$ dimensions. All shaded entries give rise to compact groups. Details about $\mathfrak{f}_{1}$, $\mathfrak{f}_{2}$, $\mathfrak{g}_{0}$, $\mathfrak{h}_{1}$ and $\mathfrak{h}_{2}$ can be found in~\cite{Dibitetto:2012rk}. All compact solution are also discussed in appendix~\ref{app:twists} in detail~\cite{Bosque:2015jda}.}\label{tab:solembedding}
\end{table}
We find eleven different non-trivial embedding tensor solutions~\cite{Dibitetto:2012rk} in total. These are illustrated in table~\ref{tab:solembedding}. They all depend on a real parameter $\alpha$. The shaded ones are compact \footnote{ Note that in general groups such as ISO($3$) or CSO($2,0,2$) are clearly not compact. However, one is able to make them compact by identifying various points. In the same fashion a compact $D$-tours arises from the non-compact plane $\mathds{R}^D$. As discussed e.g. in~\cite{Hassler:2014sba}, this procedure puts restrictions on the background fluxes and quantizes them.} and therefore represent a suitable starting point for a compactification. For completeness, we added the trivial solution $12$ for a compactification on a $\mathrm{T}^3$ with vanishing structure coefficients as well. However, only the solutions $1, 2$ and $3$ result in semisimple Lie groups. All others correspond to solvable or nilpotent Lie groups. In appendix ~\ref{app:twists}, we present an explicit construction scheme of the generalized background vielbein $E_A{}^I$ for all shaded, compact solutions~\cite{Bosque:2015jda}.

\section{Original DFT}\label{sec:traditionalSS}

During this subsection, we want to review generalized Scherk-Schwarz compactifications in the original flux formulation as a preliminary for the next subsection. Performing a compactification requires to distinguish between internal, compact, and external, extended directions~\cite{Bosque:2015jda}. Therefore, we assume that there are $n$ internal and $D-n$ external directions. In order to make this situation manifest, we split the flat and curved doubled indices, used in original DFT, componentwise according to
\begin{equation}
  V^{\bar A} = \begin{pmatrix} V_a & V^a & V^A \end{pmatrix}
    \quad \text{and} \quad
  W^{\bar I} = \begin{pmatrix} W_\mu & W^\mu & W^I \end{pmatrix}\,.
\end{equation}
Here, lowercase indices such as $a$ and $\mu$ describe external directions running from $0$ to $D-1$, whereas $A$ and $I$ represent the internal $2n$-dimensional doubled space. As a result, the O($D,D$) invariant metric takes on the form
\begin{equation}
  \eta_{\bar M\bar N}=\begin{pmatrix}
    0 & \delta^\mu_\nu & 0 \\
    \delta_\mu^\nu & 0 & 0 \\
    0 & 0 & \eta_{MN} 
  \end{pmatrix}\,, \qquad
  \eta^{\bar M \bar N}=\begin{pmatrix}
    0 & \delta_\mu^\nu & 0\\
    \delta^\mu_\nu & 0 & 0\\
    0 & 0 & \eta^{MN} 
  \end{pmatrix}
\end{equation}
and the flat generalized metric is defined through
\begin{equation}
  S_{\bar A\bar B}=\begin{pmatrix}
    \eta^{ab} & 0 & 0 \\
    0 & \eta_{ab} & 0 \\
    0 & 0 & S_{AB}
  \end{pmatrix}\,, \qquad
  S^{\bar A\bar B}=\begin{pmatrix}
    \eta_{ab} & 0 & 0 \\
    0 & \eta^{ab} & 0 \\
    0 & 0 & S^{AB}
  \end{pmatrix}\,.
\end{equation} 
The curved version of the generalized metric can be obtained after applying the twisted generalized vielbein~\cite{Geissbuhler:2011mx,Aldazabal:2013sca,Geissbuhler:2013uka}
\begin{equation} \label{eqn:twistscherkschw}
  E^{\bar A}{}_{\bar M}(X) = 
    \widehat E^{\bar A}{}_{\bar N}(\mathds{X}) 
    U^{\hat N}{}_{\hat M}(\mathds{Y}) 
  \quad \text{with} \quad
  U^{\hat N}{}_{\hat M} = 
  \begin{pmatrix}
    \delta^\mu_\nu & 0 & 0 \\
      0 & \delta_\mu^\nu & 0 \\
      0 & 0 & U^N{}_M
  \end{pmatrix}
\end{equation}
to the flat version $S_{\bar A\bar B}$. It results in
\begin{equation}
  \mathcal{H}_{\bar M\bar N} = E^{\bar A}{}_{\bar M} S_{\bar A\bar B} E^{\bar B}{}_{\bar N}\,.
\end{equation}
This twisted vielbein implements a generalized Scherk-Schwarz ansatz and is a special case of a generalized Kaluza-Klein ansatz~\cite{Hohm:2013nja}. It consists of two parts: The generalized vielbein
\begin{equation}\label{eqn:vielbeinKKansatz}
  \widehat{E}^{\bar A}{}_{\bar M} = \begin{pmatrix}
    e_\alpha{}^\mu & - e_\alpha{}^\rho C_{\mu\rho}
      & - e_\alpha{}^\rho \widehat{A}_{M\rho} \\
    0 & e^\alpha{}_\mu & 0 \\
    0 & \widehat{E}^A{}_L \widehat{A}^L{}_\mu & 
    \widehat{E}^A{}_M
  \end{pmatrix} \quad \text{with} \quad C_{\mu\nu} = B_{\mu\nu} + 
  \frac{1}{2} \widehat{A}^L{}_\mu \widehat{A}_{L\nu} \,,
\end{equation}
combines all field dynamics of the effective theory, while it only depends on the external coordinates $\mathds{X}$. The twist $U^N{}_M$ on the other hand just depends on the internal coordinates $\mathds{Y}$. All quantities it acts in a non-trivial fashion on are induced by a hat. For simplicity, the generalized dilaton $d$ is assumed to be constant in the internal space~\cite{Bosque:2015jda}. 

Additionally, the twist is required to satisfy the following conditions
\cite{Geissbuhler:2011mx,Dibitetto:2012rk,Aldazabal:2011nj,Aldazabal:2013sca}:
\begin{itemize}
  \item We only allow O($n,n$)-valued twists with the defining property
    \begin{equation}\label{eqn:twistprop1}
      U_I{}^K \eta_{KL} U_J{}^L = \eta_{IJ}\,.
    \end{equation}
  \item The structure coefficients of the effective theory's gauge algebra need to be constant
    \begin{equation}
      F_{IJK} = 3 U_{[I}{}^L \partial_L U_J{}^M U_{K]M} = \text{const.}
    \end{equation}
    
  \item The structure coefficients are required to fulfill the Jacobi identity
    \begin{equation}\label{eqn:twistprop3}
      F_{M[IJ} F^M{}_{K]L} = 0\,.
    \end{equation}
\end{itemize}
It should be noted that these properties imply that the structure coefficients $F_{IJK}$ are embedding tensor solutions. We discussed this topic in the last subsection.

Hence, we can compute all the components of the covariant fluxes
\begin{align}
  \mathcal{F}_{\bar A\bar B\bar C} &= 3 E_{[\bar A}{}^I \partial_I E_{\bar B}{}^J E_{\bar C]J} \quad \text{and} \\
  \mathcal{F}_{\bar A} &= E^{\bar BI}\partial_I E_{\bar B}{}^J E_{\bar A J} + 2 E_{\bar A}{}^I \partial_I d \quad \text{with} \quad d = \phi - \frac{1}{2} \log \det e^a{}_\mu\,.
\end{align}
It should be further kept in mind that these two definitions differ notably from the fluxes used in DFT${}_\mathrm{WZW}$. Following some algebra, we derive the non-vanishing flux components~\cite{Geissbuhler:2011mx,Hohm:2013nja}
\begin{align}
  \mathcal{F}_{abc} &= e_a{}^\mu e_b{}^\nu e_c{}^\rho \widehat{G}_{\mu\nu\rho} &
  \mathcal{F}_{ab}{}^c &= 2 e_{[a}{}^\mu \partial_\mu e_{b]}{}^\nu e^c{}_\nu = f_{ab}^c \nonumber \\
  \mathcal{F}_{abC} &= -e_a{}^\mu e_b{}^\nu \widehat{E}_{CM} \widehat{F}^M{}_{\mu\nu} &
  \mathcal{F}_{aBC} &= e_a{}^\mu \widehat{D}_\mu \widehat{E}_B{}^M \widehat{E}_{CM} \nonumber \\
  \mathcal{F}_{ABC} &= 3 \Omega_{[ABC]} &
  \mathcal{F}_{a}   &= f^{b}_{ab} + 2 e_a{}^\mu \partial_\mu \phi\,. \label{eqn:fluxcomponents}
\end{align}
Here, we used the gauge covariant derivative
\begin{equation}
  \widehat{D}_\mu \widehat{E}_A{}^M = \partial_\mu \widehat{E}_A{}^M  - \mathcal{F}^M{}_{JI} \widehat{A}^J{}_\mu \widehat{E}_A{}^I
\end{equation}
to write the equation in a manifest gauge covariant way.
The associated field strength is given by
\begin{equation}
  \widehat{F}^M{}_{\mu\nu} = 2 \partial_{[\mu}
  \widehat{A}^M{}_{\nu]} -
    \mathcal{F}^M{}_{NL} \widehat{A}^N{}_\mu
    \widehat{A}^L{}_\nu\,,
\end{equation}
the usual definition in Yang-Mills theories. It further satisfies the Bianchi identity
\begin{equation}
  D_{[\mu} F^M{}_{\nu\rho]} = 0\,.
\end{equation}
Moreover, we need to extend the canonical field strength for the $B$-field, $B_{ij}$, by a Chern-Simons term in order to be invariant under gauge transformations. It results in the 3-form
\begin{equation}
  \widehat{G}_{\mu\nu\rho} = 
  3\partial_{[\mu} B_{\nu\rho]} + 3\partial_{[\mu} 
    \widehat{A}^M{}_\nu
    \widehat{A}_{M \rho ]} - \mathcal{F}_{MNL} 
    \widehat{A}^M{}_\mu
    \widehat{A}^N{}_\nu
    \widehat{A}^L{}_\rho
\end{equation}
which fulfills a Bianchi identity as well, specifically
\begin{equation}
  \partial_{[\mu} G_{\nu\rho\lambda]} = 0\,.
\end{equation}
Inserting the covariant fluxes~\eqref{eqn:fluxcomponents} now into the action of the original flux formulation
\begin{align}
  S = \int d^{2D} X\, e^{-2d} &\big( \mathcal{F}_A \mathcal{F}_B S^{AB} + \frac{1}{4} \mathcal{F}_{ACD} \mathcal{F}_B{}^{CD} S^{AB} - \frac{1}{12} \mathcal{F}_{ABC} \mathcal{F}_{DEF} S^{AD} S^{BE} S^{CF} \nonumber \\ \label{eqn:Sdftfluxform}
    & \quad - \frac{1}{6} \mathcal{F}_{ABC} \mathcal{F}^{ABC} - \mathcal{F}_A \mathcal{F}^A \big)
\end{align}
and shifting to curved indices, we finally obtain the effective action~\cite{Geissbuhler:2011mx,Hohm:2013nja} 
\begin{align}\label{eqn:lowdimeffaction}
  S_\mathrm{eff} = \int d^{D-n}x \sqrt{-g} \, e^{-2\phi} \Big( &R + 4 \partial_\mu \phi \, \partial^\mu \phi - \frac{1}{12} \widehat{G}_{\mu\nu\rho} \widehat{G}^{\mu\nu\rho} \nonumber\\
  &-\frac{1}{4} \widehat{\mathcal H}_{MN} \widehat{\mathcal F}^{M \mu\nu} \widehat{\mathcal F}^N{}_{\mu\nu} + \frac{1}{8} \widehat{\mathcal D}_\mu \widehat{\mathcal H}_{MN} \widehat{\mathcal D}^\mu \widehat{\mathcal H}^{MN} - V \Big)\,,
\end{align}
with the scalar potential
\begin{equation}
  V = - \frac{1}{4} {F_I}^{KL} F_{JKL} \widehat{\mathcal H}^{IJ} + \frac{1}{12} F_{IKM} F_{JLN} \widehat{\mathcal H}^{IJ} \widehat{\mathcal H}^{KL} \widehat{\mathcal H}^{MN} + \frac{1}{6} F_{IJK} F^{IJK}\,.
\end{equation}
In this context, $R$ is the standard Riemannian scalar curvature in the external space. Furthermore, due to the generalized Scherk-Schwarz ansatz, the Lagrangian density of DFT becomes constant in the internal directions. Therefore, solving the action's integral is trivial in these directions. We neglected the global factor here. As expected before, the action~\eqref{eqn:lowdimeffaction} describes a bosonic subsector of a half-maximal, electrically gauged supergravity~\cite{Bosque:2015jda}. It is identical to the approach demonstrated in~\cite{Aldazabal:2011nj}.

All of the previous derivations in this subsection only took the properties~\eqref{eqn:twistprop1}-\eqref{eqn:twistprop3}  of the twist $U_I{}^J$ into account. Nevertheless, in general it is unclear whether there exists a twist with these exact properties for every embedding tensor solution. In original DFT, there does not exist a systematic way to construct these twists and as a result one is left with guessing solutions of partial differential equations~\eqref{eqn:twistprop1} which are constrained to be elements of O($n,n$) as well. This task proves to be highly difficult. Some of these solutions have been discussed in~\cite{Dibitetto:2012rk,Hassler:2014sba} and more recently in the context of Exceptional Field Theory (EFT)~\cite{Hohm:2014qga}. The problem regarding the twist is one of the major differences between geometric Scherk-Schwarz compactifications~\cite{Scherk:1978ta,Scherk:1979zr}, which have been known for several years in the context of supergravity compactifications, and their generalization in DFT. For the former compactifications, there exists a straightforward, systematic approach to construct their twists. One either uses the left or right invariant Maurer-Cartan form on the group manifold the compactification is considered on. However, for original DFT this systemic procedure is not applicable anymore, as it requires a geometry governed by ordinary diffeomorphisms as opposed to the O($n,n$) preserving generalized diffeomorphisms appearing in DFT. In the remainder of this chapter, we are going to show that DFT$_\mathrm{WZW}$ is able to cure this problem. It can be understood from the fact that all background fields transform covariantly under $2D$-diffeomorphisms and we therefore recover the common notion of geometry. As a consequence, subsection~\ref{sec:twistconstr} shows that in DFT$_\mathrm{WZW}$ one can construct a twist/background vielbein by either using a left or right invariant Maurer-Cartan form~\cite{Bosque:2015jda}.

\section{DFT on group manifolds}\label{sec:genSSDFTWZW}

Starting from the flux formulation of DFT${}_\mathrm{WZW}$, we are now in the position to perform a generalized Scherk-Schwarz compactification. Subsequently, we are going to derive the resulting low energy effective action~\cite{Bosque:2015jda}. Throughout the following steps and computations, we have to differentiate between $n$ compact, internal directions and $D-n$ extended, external directions. They correspond to the internal coordinates $\mathds{Y}$ and the external coordinates $\mathds{X}$. In order to make this separation manifest, we split the coordinates into three different types:
\begin{align}\label{eqn:genSSindexconv}
  V^{\hat{\tilde A}} &= \begin{pmatrix} V_{\hat a} & V^{\hat a} & V_{\hat A} \end{pmatrix} &
  W^{\tilde B} &= \begin{pmatrix} W_b & W^b & W^B \end{pmatrix} &
  X^{\tilde M} &= \begin{pmatrix} X_\mu & X^\mu & X^M \end{pmatrix}
\end{align}  
which are relevant for the flux formulation derived in the last chapter~\ref{Kap_4}. This step is analogous to the procedure in DFT. The only difference is that we have to treat three different kind of indices (hatted, flat, and curved) in this splitting, as opposed to DFT where only two different indices structures appear. The reason for this lies in the fact that DFT$_{\mathrm{WZW}}$ possesses an additional background vielbein. In our context, the external indices $\hat a$, $a$ and $\mu$ run from $0$ to $D-n-1$ and their internal counterparts $\hat A$, $A$ and $M$ parameterize a $2n$-dimensional, doubled space. As a consequence, we obtain the following three different versions of the $\eta$-metric
\begin{align}
  \eta_{\hat{\tilde A}\hat{\tilde B}} &= \begin{pmatrix} 0 & \delta^{\hat a}{}_{\hat b} & 0 \\  
    \delta_{\hat a}{}^{\hat b} & 0 & 0 \\ 0 & 0 & \eta_{\hat A\hat B} \end{pmatrix} &
  \eta_{\tilde A\tilde B} &= \begin{pmatrix} 0 & \delta^a{}_b & 0 \\ 
    \delta_a{}^b & 0 & 0 \\ 0 & 0 & \eta_{AB} \end{pmatrix} &
  \eta_{\tilde M\tilde N} &= \begin{pmatrix} 0 & \delta^\mu{}_\nu & 0 \\
    \delta_\mu{}^\nu & 0 & 0 \\ 0 & 0 & \eta_{MN} \end{pmatrix},
\end{align}
we are employing to lower the indices defined in~\eqref{eqn:genSSindexconv}. Furthermore, we work with the flat, background generalized metric
\begin{equation}
  S_{\hat{\tilde A}\hat{\tilde B}}=\begin{pmatrix}
    \eta^{\hat a\hat b} & 0 & 0 \\
    0 & \eta_{\hat a\hat b} & 0 \\
    0 & 0 & S_{\hat A\hat B}
  \end{pmatrix} \quad \text{and its inverse} \quad
  S^{\hat{\tilde A}\hat{\tilde B}}=\begin{pmatrix}
    \eta_{\hat a\hat b} & 0 & 0 \\
    0 & \eta^{\hat a\hat b} & 0 \\
    0 & 0 & S^{\hat A\hat B}
  \end{pmatrix}\,.
\end{equation}
In the next step, we need to specify the Scherk-Schwarz ansatz of the composite generalized vielbein
\begin{equation}\label{eqn:WZWSSansatz}
  \mathcal{E}_{\hat{\tilde A}}{}^{\tilde M} = \tilde E_{\hat{\tilde A}}{}^{\tilde B}(\mathbb{X}) \, E_{\tilde B}{}^{\tilde M}(\mathbb{Y})\,. 
\end{equation}
Here, we use the ansatz that the fluctuation part only depends on the external coordinates $\mathds{X}$, whereas the background part only depends upon the internal ones $\mathds{Y}$. Comparing our ansatz with the one in~\cite{Geissbuhler:2011mx,Aldazabal:2011nj,Aldazabal:2013sca,Hohm:2013nja}, we observe that the background generalized vielbein $E_{\tilde B}{}^{\tilde M}$ plays the role of the twist $U^{\hat N}{}_{\hat M}$. In contrast to the twist appearing in original DFT, our background vielbein is not restricted to lie in O($D,D$). Therefore, our framework allows to solve the problem of constructing an appropriate twist: We always possess a straight forward procedure to construct $E_{\tilde B}{}^{\tilde M}$ by using the left invariant Maurer Cartan form on a group manifold~\cite{Bosque:2015jda}. One possible example to apply this technique would be the $S^3$ with $H$-flux presented in~\cite{Hassler:2014sba}.

Subsequently, we adapt the generalized Kaluza-Klein ansatz~\cite{Geissbuhler:2011mx,Aldazabal:2013sca,Hohm:2013nja} for the fluctuation vielbein $\tilde{E}_{\hat{\tilde A}}{}^{\tilde B}$ and its associated index structure. It yields
\begin{equation}\label{eqn:KKansatzWZW}
  \tilde{E}_{\hat{\tilde A}}{}^{\tilde B}(\mathbb{X}) = \begin{pmatrix}
    {e_b}{}^{\hat a} & 0 & 0 \\ - {e_{\hat a}}^c C_{bc} & {e_{\hat a}}^b & -{e_{\hat a}}^c \widehat{A}^B{}_c \\ \widehat{E}_{\hat A}{}^C \widehat{A}_{C b} & 0 & \widehat{E}_{\hat A}{}^B
\end{pmatrix}
    \quad \text{with} \quad
    C_{ab} = B_{ab} + \frac{1}{2} \widehat{A}^D{}_a \widehat{A}_{D b}\,.
\end{equation}
During this ansatz, $B_{ab}$ represents the two-form field arising in the effective theory while
\begin{equation}
  \widehat{\mathcal H}^{CD} = \widehat{E}_{\hat A}{}^C S^{\hat A\hat B} \widehat{E}_{\hat B}{}^D
\end{equation}
denotes the $n^2$ independent scalar fields forming the moduli of the internal space. In the same fashion as for the twist, the background vielbein has only non-trivial components in the internal space. This gives rise to
\begin{equation}
\label{eqn:scherkbackgroundvielbein}
  E_{\tilde B}{}^{\tilde M}(\mathbb{Y}) = \begin{pmatrix}
    \delta^b{}_\mu & 0 & 0 \\ 0 & \delta_b{}^\mu & 0 \\ 0 & 0 & E_B{}^M
  \end{pmatrix}\,.
\end{equation}
Employing the Kaluza-Klein ansatz~\eqref{eqn:KKansatzWZW} and using the partial derivative
\begin{equation}
  \partial^{\tilde M} = \begin{pmatrix} \partial_\mu & \partial^\mu & \partial^M \end{pmatrix}\,
\end{equation}
it is now straightforward to compute the fluxes $\tilde{F}_{\hat{\tilde A}\hat{\tilde B}\hat{\tilde C}}$ and $\tilde{F}_{\hat{\tilde A}}$ defined in~\eqref{eqn:fluxfluxes} and~\eqref{eqn:defFA}. After some algebra, we find the following non-vanishing components
\begin{align}
  \tilde{F}_{\hat a\hat b\hat c} &= e_{\hat a}{}^{d} e_{\hat b}{}^{e} e_{\hat c}{}^{f} \, 3 \big( D_{[d} B_{e f]} + \widehat{A}^D{}_{[d} D_e \widehat{A}_{D f]} \big) & 
  \tilde{F}_{\hat a\hat b}{}^{\hat c} &= 2 e_{[\hat a}{}^d D_{d} e_{\hat b]}{}^e e_{e}{}^{\hat c} = \tilde{f}_{\hat a\hat b}^{\hat c} \nonumber \\ 
  \tilde{F}_{\hat a\hat b\hat C} &= - e_{\hat a}{}^d e_{\hat b}{}^e \widehat{E}_{\hat CD} \, 2 D_{[d} \widehat{A}^D{}_{e]} &
  \tilde{F}_{\hat a\hat B\hat C} &=  e_{\hat a}{}^d D_d \widehat{E}_{\hat B}{}^D \widehat{E}_{\hat CD} \nonumber \\
  \tilde{F}_{\hat a} &= \tilde f_{\hat a\hat c}^{\hat c} + 2 e_{\hat a}{}^b D_b \phi\,. \label{eqn:genSSfluxfluxes}
\end{align}
As we want to determine the full covariant fluxes $\mathcal{F}_{\hat{\tilde A}\hat{\tilde B}\hat{\tilde C}}$, we also have to take the background contribution $F_{\hat{\tilde A}\hat{\tilde B}\hat{\tilde C}}$ into account. Since the background vielbein~\eqref{eqn:scherkbackgroundvielbein} only depends on internal coordinates, the only non-vanishing components of $F_{\tilde A\tilde B\tilde C}$ are given by
\begin{equation}
  F_{ABC} = 2 \Omega_{[AB]C}\,.
\end{equation}
Thus, they yield the non-vanishing components
\begin{align}
  F_{\hat a\hat b\hat c} &= -e_{\hat a}{}^{d} e_{\hat b}{}^{e} e_{\hat c}{}^{f} \widehat{A}_d{}^D \widehat{A}_e{}^E \widehat{A}_f{}^F F_{DEF} &
  F_{\hat a\hat b\hat C}  &= e_{\hat a}{}^{d} e_{\hat b}{}^{e} \widehat{A}_c{}^D \widehat{A}_d{}^E \widehat{E}_{\hat C}{}^F F_{DEF} \nonumber \\ 
  F_{\hat a\hat B\hat C}  &= -e_{\hat a}{}^{b} \widehat{A}_b{}^D \widehat{E}_{\hat B}{}^E \widehat{E}_{\hat C}{}^F F_{DEF} &
  F_{\hat A\hat B\hat C}  &= E_{\hat A}{}^D E_{\hat B}{}^E E_{\hat C}{}^F F_{DEF} \,.
\end{align}
Combining these results with~\eqref{eqn:genSSfluxfluxes} and keeping the gauge covariant quantities in mind
\begin{align}
  \widehat{D}_\mu \widehat{E}_{\hat A}{}^B &= \partial_\mu \widehat{E}_{\hat A}{}^B - F^{B}{}_{CD} \widehat{A}_\mu{}^C \widehat{E}_{\hat A}{}^D \nonumber \\ 
  \widehat{F}^A{}_{\mu\nu} &= 2\partial_{[\mu} \widehat{A}_{\nu]}{}^A - {F^A}_{BC} \widehat{A}_\mu{}^B \widehat{A}_\nu{}^C \nonumber \\ 
  \widehat{G}_{\mu\nu\rho} &= 3 \partial_{[\mu} B_{\nu\rho]} + \widehat{A}_{[\mu}{}^A \partial_\nu \widehat{A}_{\rho]A} - F_{ABC} \widehat{A}_\mu{}^A \widehat{A}_\nu{}^B \widehat{A}_\rho{}^C\,,
\end{align}
which were discussed in section~\ref{sec:traditionalSS}, we finally arrive at the desired result
\begin{align}
  \mathcal{F}_{\hat a\hat b\hat c} &= e_{\hat a}{}^{\mu} e_{\hat b}{}^{\nu} e_{\hat c}{}^{\rho} \widehat{G}_{\mu\nu\rho} &
  \mathcal{F}_{\hat a\hat b}{}^{\hat c} &= 2 e_{[\hat a}{}^\mu \partial_{\mu} e_{\hat b]}{}^\nu e_{\nu}{}^{\hat c} \nonumber \\ 
  \mathcal{F}_{\hat a\hat b\hat C} &= -e_{\hat a}{}^\mu e_{\hat b}{}^\nu \widehat{E}_{\hat CA} \widehat{F}^A{}_{\mu\nu} &
  \mathcal{F}_{\hat a\hat B\hat C} &= e_{\hat a}{}^\mu \widehat{D}_\mu \widehat{E}_{\hat B}{}^A \widehat{E}_{\hat CA} \nonumber \\
  \mathcal{F}_{\hat A\hat B\hat C} &= \widehat{E}_{\hat A}{}^D \widehat{E}_{\hat B}{}^E \widehat{E}_{\hat C}{}^F \, F_{DEF} &
  \mathcal{F}_{\hat a} &= \tilde f_{ab}^b + 2 e_{\hat a}{}^\mu \partial_\mu \phi \label{eqn:fluxcompWZW}\,.
\end{align}
All gauge covariant objects carry indices such as $A, B, C, \ldots$ instead of $I, J, K, \ldots$ in this context, as opposed to the last section. The reason for this lies in the fact that they have to carry O($n,n$) indices, which are the former in DFT$_\mathrm{WZW}$ (depicted in~\eqref{eqn:groupindices}) and the latter in the original formulation. From this point on, all remaining computations proceed in the same manner as explained in the last section. Consequently, we obtain the effective action
\begin{align}\label{eqn:lowdimeffactionwzw}
  S_\mathrm{eff} = \int d^{D-n}x \sqrt{-g} \, e^{-2\phi} \Big( &R + 4 \partial_\mu \phi \, \partial^\mu \phi - \frac{1}{12} \widehat{G}_{\mu\nu\rho} \widehat{G}^{\mu\nu\rho} \nonumber\\
  &-\frac{1}{4} \widehat{\mathcal H}_{AB} \widehat{\mathcal F}^{A \mu\nu} \widehat{\mathcal F}^B{}_{\mu\nu} + \frac{1}{8} \widehat{\mathcal D}_\mu \widehat{\mathcal H}_{AB} \widehat{\mathcal D}^\mu \widehat{\mathcal H}^{AB} - V \Big)\,,
\end{align}
as well, when plugging our results into the action~\eqref{eqn:actionfluxform} of DFT$_\mathrm{WZW}$'s flux formulation. As previously explained, we just need to replace the indices $I, J, K, \ldots$ by $A, B, C, \ldots$. However, this is for conventional purposes only.
On top of that, the scalar potential in DFT$_\mathrm{WZW}$
\begin{equation}
  V = -\frac{1}{4} F_A{}^{CD} F_{BCD} \widehat{\mathcal H}^{AB} + \frac{1}{2} F_{ACE} F_{BDF} \widehat{\mathcal H}^{AB} \widehat{\mathcal H}^{CD} \widehat{\mathcal H}^{EF}
\end{equation}
lacks the strong constraint violating term $1/6 F_{ABC} F^{ABC}$.
It appears as a cosmological constant in gauged supergravities, even if the strong constraint is not imposed on the background field. We already argued in section~\ref{sec:missingFABCFABCterm} why this term does not occur in our formulation. Anyway, it is totally legitimate to add it by hand, as it was done in the original flux formulation, to the action from a bottom up perspective. It does not spoil any of the theory's symmetries.

This new approach solves an ambiguity of generalized Scherk-Schwarz compactifications: In the DFT${}_\mathrm{WZW}$ framework, the twist is equivalent to the background generalized vielbein $E_A{}^I$. Its construction works in the same fashion as for conventional Scherk-Schwarz reductions. The reason for it lies in the appearance of standard $2D$-diffeomorphisms which are absent in the original DFT formulation. As a result, all mathematical tools known for group manifolds are applicable. However, all these features are immediately lost upon returning to traditional DFT by imposing the extended strong constraint. The extended strong constraint, necessary for this transition, breaks the $2D$-diffeomorphism invariance. Thus, we are left with the issues outlined in section~\ref{sec:traditionalSS}~\cite{Bosque:2015jda}.

All derivations performed in  DFT${}_\mathrm{WZW}$ so far are top down. We began with the full bosonic CSFT~\cite{Blumenhagen:2014gva,Blumenhagen:2015zma} and reduced it step by step until we finally reached the low energy effective action~\eqref{eqn:lowdimeffactionwzw}. Hence, it is immediately possible to check the uplift of solutions for the equation of motion to full string theory. We only have to keep in mind that all results obtained so far are only valid at tree level. Requiring further consistency at loop level, e.g. a modular invariant partition function, puts further additional restrictions upon the theory. We can learn even more from the CSFT perspective: The background fluxes $F_{ABC}$ scale with $1/\sqrt{k}$, where $k$ denotes the level of the Ka\v{c}-Moody algebra on the worldsheet. Moreover, we can decompose them according to
\begin{equation}
  F_{ABC} = \frac{1}{\sqrt{k}} f_{ABC}
\end{equation}
and assume that the structure coefficients $f_{ABC}$ are normalized, e.g.
\begin{equation}
  f_{AC}{}^D f_{BD}{}^C = \frac{1}{2} \delta_{AB}\,.
\end{equation}
As a consequence, the gauge covariant derivative becomes
\begin{equation}
  \widehat{D}_\mu V^A = \partial_\mu V^A - \frac{1}{\sqrt{k}}f^{A}{}_{BC} \widehat{A}_\mu{}^B C^C\,.
\end{equation}
Following this equation, we directly read off the Yang-Mills coupling constant 
\begin{equation}
  g_\mathrm{YM} = \frac{1}{\sqrt{k}}\,.
\end{equation}
It should be noted that the geometric interpretation of DFT${}_\mathrm{WZW}$ only holds for the large level limit $k\gg 1$. The corresponding effective theory is then weakly coupled and hence can be treated perturbatively. Nevertheless, freezing out all fluctuations in the internal directions, which happens for generalized Scherk-Schwarz compactifications, extends our results to $k=1$. This case requires to reduce the number of external directions to cancel the total central charges of the bosons and the ghost system~\cite{Bosque:2015jda}.

\section{Twist construction}\label{sec:twistconstr}

One of the major advantages of DFT${}_\mathrm{WZW}$ is the existence of a straight forward procedure to construct the background vielbein $E_A{}^I$, replacing the twist in the original framework, from embedding tensor solutions when considering generalized Scherk-Schwarz compactifications. In this section, we want to present the according procedure in more detail~\cite{Bosque:2015jda}.

Let us start by assuming that $t_A$ represents $2n$ different $N\times N$ matrices which form the following algebra
\begin{equation}\label{eqn:genalgebra}
  [t_A, t_B] = t_A t_B - t_B t_C = F_{AB}{}^C t_C\,.
\end{equation}
Its structure coefficients are given by an arbitrary solution of the embedding tensor~\eqref{eqn:embeddingtensor}. In this case, the condition
\begin{equation}
  F_{AB}{}^D\eta_{DC} + F_{AC}{}^D\eta_{BD} = 0
\end{equation}
has to hold. On top of that, we need to define a non-degenerate, bilinear, symmetric two-form
\begin{equation}
  \mathcal{K}(t_A, t_B) = \eta_{AB}
\end{equation}
on the vector space spanned by the matrices $t_A$. Later during this section, we are going to explain how the matrices and the two-form are realized. For the moment, these three definitions are sufficient. Equipped with these definitions, it is evident that the background fluxes $F_{ABC}$ can be obtained by
\begin{equation}
  F_{ABC} = \mathcal{K}(t_A, [t_B, t_C])\,.
\end{equation}

The second viable ingredient, we require to derive the background generalized vielbein, is the explicit representation of a group element $g\in G$ of the group $G$ depicting the background. Afterwards, we apply the exponential map
\begin{equation}\label{eqn:expmap}
  g = \exp( t_A X^A ) = \sum\limits_{m=0}^\infty \frac{1}{m !} (t_A X^A)^n
\end{equation}
in order to derive the group element from the generators $t_A$. For compact groups this map is surjective onto the identity component $G_0$ of $G$. We assume that all groups we treat here are path-connected, and thus $G_0$ as well as $G$ are analogous.  If we restrict the domain of the coordinates $X^I$ accordingly, the map~\eqref{eqn:expmap} becomes bijective. Subsequently, each group element is denoted by a unique point in the coordinate space. Thus, the left invariant Maurer-Cartan form is defined through
\begin{equation}\label{eqn:leftinvmc}
  E_{A I} = \mathcal{K}(t_A, g^{-1} \partial_I g)\,.
\end{equation}
We use it to evaluate
\begin{equation}
  \Omega_{ABC} = E_A{}^I \partial_I E_B{}^J E_{C J}\,,
\end{equation}
and obtain
\begin{align}
  \Omega_{ABC} &= -E_A{}^I \big[ \mathcal{K}(t_C, \partial_I g^{-1}\, g g^{-1}\, \partial_J g) + \mathcal{K}(t_C, g^{-1} \partial_I \partial_J g) \big] E_B{}^J \nonumber \\
  &= E_B{}^J \mathcal{K}( g^{-1} \partial_J g, \, t_C \, g^{-1}\partial_I g) E_A{}^I - \mathcal{K}(t_C, g^{-1}\partial_I\partial_J g) E_A{}^I E_B{}^J \nonumber \\
  &= \mathcal{K}(t_B, t_C t_A) - \mathcal{K}(t_C, g^{-1} \partial_I \partial_J g) E_A{}^I E_B{}^J\,.
\end{align}
Hence, the coefficients of anholonomy produce the correct background covariant fluxes, specifically
\begin{equation}
  F_{ABC} = 2\Omega_{[AB]C} = \mathcal{K}(t_A, [t_B, t_C])\,.
\end{equation}
Therefore, we indeed recover the correct relation for the background generalized vielbein $E_A{}^I$ with the left invariant Maurer-Cartan form~\eqref{eqn:leftinvmc}.

As previously mentioned, the generators $t_A$ of the Lie algebra are given by $N\times N$ matrices
\begin{equation}
  t_A = \begin{pmatrix}
   (t_A)_{11} & \cdots & (t_A)_{1N} \\
   \vdots & & \vdots \\
   (t_A)_{N1} & \cdots & (t_A)_{NN}
  \end{pmatrix}\,.
\end{equation}
If we want to evaluate $\mathcal{K}(x, y)$ for arbitrary algebra elements $x, y \in \mathfrak{g}$, we are required to expand them in terms of the generators, e.g.
\begin{equation}
  x = \sum\limits_{A=1}^{2 n} c_A t_A\,,
\end{equation}
where $c_A$ labels the $2n$ expansion coefficients. Furthermore, it is useful to rearrange the matrix $x$ into the vector
\begin{equation}
  x = \begin{pmatrix} 
    x_{11} & \cdots & x_{1N} & x_{2N} & \cdots & x_{NN}
  \end{pmatrix}
\end{equation}
and solve the corresponding linear system
\begin{equation}
  c M = x \quad \text{with} \quad
  M = \begin{pmatrix}
    (t_1)_{11} & \cdots & (t_1)_{1N} & (t_1)_{2N} & \cdots & (t_1)_{NN} \\
    \vdots & & \vdots & \vdots & & \vdots \\
    (t_{2n})_{11} & \cdots & (t_{2n})_{1N} & (t_{2n})_{2N} & \cdots & (t_{2n})_{NN}
  \end{pmatrix}
\end{equation}
\begin{equation}  
  \text{and} \quad 
  c = \begin{pmatrix} c_1 & \cdots c_{2n} \end{pmatrix}
\end{equation}
to compute the individual coefficients. We are only interested in unique solutions, which implies that the $2 n \times N^2$ matrix $M$ has to possess full rank
\begin{equation}\label{eqn:fullrank}
  \rank M = 2n\,.
\end{equation}
On top of~\eqref{eqn:genalgebra}, this equation provides an additional constraint on the generators $t_A$. According to Ado's theorem~\cite{Ado1949}, both conditions can be satisfied for a finite $N$. These representations are called faithful. In the following subsections, we show how one obtains them for semisimple and solvable Lie algebras. They are partly based on~\cite{Hofer2012}. Appendix~\ref{app:twists} applies the presented techniques to all the compact embedding tensor solutions given in table~\ref{tab:solembedding}~\cite{Bosque:2015jda}.

\subsection{Semisimple algebras}\label{sec:semisimple}
For semisimple Lie algebras, the generators
\begin{equation}\label{eqn:gensemisimple}
  (t_A)_{BC} = F_{AB}{}^C
\end{equation}
are directly given by the structure coefficients. Subsequently, we find the adjoint representation
\begin{equation}
  \adj_x y = [x,y] \quad \text{with} \quad x,\,y \in \mathfrak{g}
\end{equation}
in the basis spanned by all abstract generators. It has dimension $N=2n$ and is faithful if the center of the Lie algebra
\begin{equation}\label{eqn:center}
  Z(\mathfrak{g}) = \{x\in \mathfrak{g} \,|\, [x,y]=0 \,\, \text{for all} \,\, y\in \mathfrak{g}\} 
\end{equation}
is trivial. This is the case for semisimple Lie algebras. However, there exist non-semisimple ones, such as e.g. ISO($3$) discussed in appendix~\ref{app:ISO(3)}, with vanishing center as well. The matrix representation of their generators is determined by~\eqref{eqn:gensemisimple}, too.

Generally speaking, the adjoint representation is not the lowest dimensional one. For e.g. in the case of SO($4$), which we demonstrate in appendix~\ref{app:SO4}, the adjoint has $N=6$ dimensions, whereas the fundamental representation is only $4$-dimensional. In the end, both representations work for our purpose. However, taking the smallest dimensional one usually simplifies the calculations considerably~\cite{Bosque:2015jda}.

\subsection{Nilpotent Lie algebras}\label{sec:nilpotent}

For nilpotent Lie algebras,~\eqref{eqn:gensemisimple} does not give rise to linear independent generators $t_A$ anymore. As a consequence, they are neither faithful nor do they satisfy~\eqref{eqn:fullrank}. Before discussing the derivation of correct generators in this case, we first want to present a criterion to identify these algebras. Therefore, let us consider the lower central series
\begin{equation}
  L_{m+1} = [ \mathfrak{g}, L_m ] \quad \text{with} \quad L_0 = \mathfrak{g}\,.
\end{equation}
It yields the series
\begin{equation}\label{eqn:lowersubalg}
  \mathfrak{g} = L_0 \supseteq L_1 \supseteq L_2 \supseteq \dots
\end{equation}
of subalgebras. If this series terminates at a finite $k$ with $L_k=\{0\}$, the algebra $\mathfrak{g}$ is nilpotent of order $k$.

In the following we exploit the infinite dimensional universal enveloping algebra $U(\mathfrak{g})$ of the nilpotent Lie algebra $\mathfrak{g}$. According to the Poincar\'e-Birkhoff-Witt theorem, it is spanned by the ordered monomials
\begin{equation}
  t(\alpha) = t_1^{\alpha_1} t_2^{\alpha_2} \dots t_{2n}^{\alpha_{2n}} \quad \text{with} \quad \alpha \in \mathds{Z}^3_+\,.
\end{equation}
Through left multiplication
\begin{equation}\label{eqn:phi_x}
  \phi_x: U(\mathfrak{g}) \rightarrow U(\mathfrak{g}) \,, \quad 
    \phi_x (y) = x y \quad\text{with}\quad x \in \mathfrak{g}\,,
\end{equation}
algebra elements $x$ act faithful on the universal enveloping algebra. Furthermore, Ado's theorem states that even on the finite dimensional subspace
\begin{equation}\label{eqn:nilpotsubspaceV}
  V^k = \{ t(\alpha) \in U(\mathfrak{g}) \,|\, \ord t(\alpha) \le k \}
\end{equation}
of $U(\mathfrak{g})$, $\phi_x$ still acts faithful. Here, we made use of the order function
\begin{equation}
  \ord t(\alpha) = \sum\limits_m^{2n} \alpha_m \ord t_m\,, \quad
  \ord t_m = \max \{s \,|\, t_m \in L_{s-1} \}
    \quad \text{and} \quad
  \ord 1 = 0
\end{equation}
to fix the subspace. Finally, to be able to find a $N=\dim V$-dimensional, faithful matrix representation of the generators $t_A$, we express the linear operator $\phi_{t_A}$ in the basis which spans $V^k$~\cite{Bosque:2015jda}.

\subsection{Solvable Lie algebras}\label{sec:solvable}

In the case of solvable Lie algebras, we need to apply techniques  from semisimple as well as nilpotent Lie algebras. Solvable Lies algebras are characterized by a derived series
\begin{equation}
  L^{m+1} = [L^m, L^m] \quad \text{with} \quad L^0 = \mathfrak{g}
\end{equation}
which terminates at a finite $k$ with $L^k=\{0\}$. Equivalently to~\eqref{eqn:lowersubalg}, it yields the series
\begin{equation}
  \mathfrak{g} = L^0 \supset L^1 \supset \cdots \supset L^{k-1} \supset \{0\}\,.
\end{equation}
of subalgebras. The first of them, $\mathfrak{n}=L^1$ is nilpotent. Thus, we expand the map~\eqref{eqn:phi_x} for all its generators $t\in\mathfrak{n}$ in the basis~\eqref{eqn:nilpotsubspaceV} to obtain their matrix representation. Moreover, the adjoint representations $\adj_x = [x,y]$,
\begin{equation}
  \adj_x 1 = 0 \quad \text{and} \quad \adj_x y_1 \dots y_l = \sum\limits_{m=1}^l y_1 \dots y_{m-1}[x,y_m] y_{m+1} \dots y_l \quad \text{with} \quad
  y_m \in \mathfrak{n}
\end{equation}
of the remaining generators $x \ni \mathfrak{q}=\mathfrak{g}/\mathfrak{n}$ act faithful on $V^k$, too. On top of $\phi_t$, we also express $\adj_u$, $u\in\mathfrak{q}$, in the basis $V^k$ to complete the $N=\dim V$ dimensional matrix representation of the algebra. It should be noted that all nilpotent Lie algebras turn out to be automatically solvable with $\mathfrak{q}=\{\}$~\cite{Bosque:2015jda}.\clearpage{}
  \clearpage{}\chapter{Generalized Parallelizable Spaces from Exceptional Field Theory}
\label{Kap_5}

In this chapter, we want to generalize the concepts known from Double Field Theory on group manifolds~\cite{Blumenhagen:2014gva}~\cite{Blumenhagen:2015zma,Bosque:2015jda,Blumenhagen:2017noc} to geometric Exceptional Field Theory (gEFT)~\cite{Bosque:2016fpi} by making the U-Duality groups manifest. We start by showing how to implement generalized diffeomorphisms on group manifolds in section~\ref{sec:gendiffonG}. Our goal is it to keep this discussion as general as possible and only to use specific duality groups when it is really required. Furthermore, we work out the differences and similarities with DFT$_{\mathrm{WZW}}$. Simultaneously, we are going to give a short review of all relevant DFT as well as EFT results needed and we set the notation. Afterwards, in~\ref{sec:genLie} we obtain the two linear constraints and the quadratic constraint required for the generalized Lie derivative to close under imposition of the section condition (SC). Solving the linear constraints demands the choice of a specific U-duality group for which we select SL$(5)$ in subsection~\ref{sec:linconstsl5}. Subsequently, we give a detailed picture of the necessary SL($5$) breaking to group manifolds with dim $G< 0$, arising from the $\overline{\mathbf{40}}$ in the embedding tensor solutions. The second part of this chapter is covered by~\ref{sec:sectioncond}. In it, we apply the techniques to solve the SC known from DFT$_{\mathrm{WZW}}$~\cite{Hassler:2016srl} and adapt them to the gEFT framework. The resulting solutions of the SC allow for a Generalized Geometry (GG) description of our theory which we discuss in~\ref{sec:gg}. Using these observations, subsection~\ref{sec:genframe} provides a method to construct the generalized frame field $\mathcal{E}_A$ and its required additional constraints. Finally, we present some explicit examples such as the four-torus with $G$-flux, and the backgrounds contained in its duality chain, as well as the four-sphere with $G$-flux during section~\ref{sec:examples}.

The results in this chapter are based on~\cite{Bosque:2017dfc}.

\section{Generalized Diffeomorphisms on Group Manifolds}\label{sec:gendiffonG}

One of the most important roles in general relativity plays the covariance under diffeomorphisms. For the EFT framework, these diffeomorphisms have to be replaced by generalized diffeomorphisms by combining the diffeomorphisms with gauge transformation on the physical subspace. They emerge after solving the SC. In this context, it becomes important to differentiate between generalized and standard diffeomorphisms on the external space. Of course, they are not identical but they can be made compatible with each other on a group manifold $G$. The word compatible refers to the fact that the generalized Lie derivative transforms covariantly on $G$ in the notion as it is known from general relativity. A similar approach has been proposed by Cederwall for DFT~\cite{Cederwall:2014kxa,Cederwall:2014opa}. Cederwall defines a torsion free, covariant derivative with curvature to find a closing algebra of infinitesimal generalized diffeomorphisms. However, we use an alternative approach here. The covariant derivative in our framework possesses torsion as well as curvature. It is motivated by the works of \DFTwzw{}~\cite{Blumenhagen:2014gva,Blumenhagen:2015zma}. We review the according gauge transformations in subsection~\ref{sec:dftwzwmotiv}. (For a complete review see~\cite{Hassler:2015pea,Blumenhagen:2017noc}.) Successively, we extend the structure known from \DFTwzw{} to EFT. As a result, we observe that the closure of the gauge algebra requires two linear constraints and a quadratic constraint in addition to the SC. Afterwards, we are going to present the solution of the linear constraints in subsection~\ref{sec:linconstsl5} for a specific choice of the U-Duality i.e. SL($5$). Subsequently, we discuss the quadratic constraint in subsection~\ref{sec:quadrconstr} as well. The constraints obtained for this particular U-duality group agree with the results in~\cite{Bosque:2016fpi}.

\subsection{Transition from DFT to EFT}\label{sec:dftwzwmotiv}

We start by reviewing the most important features of generalized and standard diffeomorphisms which we want to combine later on~\cite{Bosque:2017dfc}. In the DFT framework the infinitesimal version of the former is governed by the generalized Lie derivative~\cite{Hohm:2010pp} and takes on the form
\begin{equation}\label{eqn2:genLieDFT}
  \mathcal{L}_\xi V^I = \xi^J \partial_J V^I + ( \partial^I \xi_J - \partial_J \xi^I ) V^J\,. 
\end{equation}
Its corresponding gauge algebra closes according to
\begin{equation}\label{eqn2:genLieDFTclosure}
  [ \mathcal{L}_{\xi_1}, \mathcal{L}_{\xi_2} ] = \mathcal{L}_{[\xi_1, \xi_2]_{\mathrm C}} \,,
\end{equation}
once the SC
\begin{equation}\label{eqn2:SCDFT}
  \partial_I \cdot \partial^I \cdot = 0
\end{equation}
is satisfied~\cite{Hull:2009zb}. The section constraint applies to arbitrary combinations of fields $V^I$ as well as gauge parameters $\xi^I$, and they are denoted by the placeholder $\cdot$\,. Furthermore, we introduced the C-bracket
\begin{equation}\label{eqn2:CbracketDFT}
  [\xi_1, \xi_2]_{\mathrm C} = \frac{1}{2} ( \mathcal{L}_{\xi_1} \xi_2 - \mathcal{L}_{\xi_2} \xi_1 )
\end{equation}
in~\eqref{eqn2:genLieDFTclosure}. For the standard solution of the SC, where all field parameters $\cdot$ only depend on momentum coordinates $x^i$, this bracket reduces to the well-known Courant bracket of GG~\cite{Hull:2009zb}. As can be verified quite easily, the generalized Lie derivative
\begin{equation}\label{eqn2:etainvgenLie}
  \mathcal{L}_\xi \, \eta_{IJ} = 0
\end{equation}
is O($d-1,d-1$) invariant and therefore does not change $\eta_{IJ}$. We use this metric to raise and lower indices $I, J, K, \ldots$ running from $1, \ldots, 2D$. This concludes the short review of the relevant DFT features. In contrast, the infinitesimal diffeomorphisms mediated by the standard Lie derivative are given by
\begin{equation}\label{eqn2:Lie}
  L_\xi V^I = \xi^J \partial_J V^I - V^J \partial_J \xi^I
\end{equation}
which closes in the following way
\begin{equation}
  [ L_{\xi_1}, L_{\xi_2} ] = L_{[\xi_1, \xi_2]}\,.
\end{equation}
The Lie bracket
\begin{equation}
  [\xi_1, \xi_2] = L_{\xi_1} \xi_2 = \frac{1}{2} (L_{\xi_1} \xi_2 - L_{\xi_2} \xi_1 )
\end{equation}
is defined analogous to the C-bracket in~\eqref{eqn2:CbracketDFT}. As opposed to generalized diffeomorphisms, neither does the closure of the algebra demand an additional constraint, nor does it leave the $\eta_{IJ}$ metric invariant.

Making the generalized~\eqref{eqn2:genLieDFT} and standard Lie derivative~\eqref{eqn2:Lie} compatible with each other requires the generalized Lie derivative to transform covariantly under standard diffeomorphisms. In this case
\begin{equation}\label{eqn2:genLiecov}
  L_{\xi_1} \mathcal{L}_{\xi_2} = \mathcal{L}_{L_{\xi_1} \xi_2} + \mathcal{L}_{\xi_2} L_{\xi_1}
    \quad \text{or equivalently} \quad
  [L_{\xi_1}, \mathcal{L}_{\xi_2} ] = \mathcal{L}_{[\xi_1, \xi_2]}
\end{equation}
has to hold. Just assume that $V^I$ and $\xi^I$ in the definition of the generalized Lie derivative would transform covariantly, specifically
\begin{equation}
  \delta_\lambda V^I = L_\lambda V^I
    \quad \text{and} \quad
  \delta_\lambda \xi^I = L_\lambda \xi^I\,,
\end{equation}
then the partial derivative in~\eqref{eqn2:genLieDFT} would spoil~\eqref{eqn2:genLiecov}. This issue can be cured by replacing all partial derivatives with covariant derivatives, i.e.
\begin{equation}
  \nabla_I V^J = \partial_I V^J+ \Gamma_{IL}{}^J V^L\,.
\end{equation}
It gives rise to the generalized Lie derivative
\begin{equation}\label{eqn2:genLieDFTnabla}
  \mathcal{L}_\xi V^I = \xi^J \nabla_J V^I + ( \nabla^I \xi_J - \nabla_J \xi^I ) V^J\,.
\end{equation}
Before we study this generalized Lie derivative in more detail, we have to choose the right connection $\Gamma$. Fixing it requires the imposition of additional constraints. First of all, the covariant derivative needs to be compatible with the $\eta_{IJ}$ metric. As a consequence, it has to satisfy the following relation
\begin{equation}\label{eqn2:metriccomptDFTwzw}
  \nabla_I \, \eta_{JK} = 0\,.
\end{equation}
Otherwise~\eqref{eqn2:etainvgenLie} would be violated. The new generalized Lie derivative still has to close under the SC~\eqref{eqn2:SCDFT} recasted through covariant derivatives. These two constraints are completely sufficient to identify $\Gamma$ with the torsion-free Levi-Civita connection, as was shown in~\cite{Cederwall:2014kxa}. From a purely geometric and symmetry oriented point of view, this is the most natural way to merge generalized with standard diffeomorphisms. However, string theory on group manifolds leads to an interesting subtlety in this construction by providing additional structure~\cite{Bosque:2017dfc}. 

Let us consider a group manifold $G$ to explain the additional constituent for the construction and suppose that $\eta_{IJ}$ is a bi-invariant metric with split signature on it. Thus, $G$ is parallelizable and comes along with the torsion-free, flat derivative
\begin{equation}
  D_A = E_A{}^I \partial_I\,,
\end{equation}
where $E_A{}^I$ (generalized background vielbein) represents the left invariant Maurer-Cartan form on $G$. $D_A$ carries a flat index such as $A, B, C, \ldots$, which runs from  $1, \ldots, 2D$, and is subsequently compatible with the flat metric
\begin{equation}
  \eta_{AB} = E_A{}^I \eta_{IJ} E_B{}^J \,.
\end{equation}
Its corresponding torsion
\begin{equation}\label{eqn2:defFABCDFTWZW}
  [D_A, D_B] = F_{AB}{}^C D_C
\end{equation}
is given by the structure constants of the Lie algebra $\mathfrak{g}$ associated to $G$. As a result, it appears to be more natural to use flat derivatives $D_A$ instead of covariant ones $\nabla_I$ with a torsion-free Levi Civita connection on a group manifold. Indeed, the CSFT calculations for bosonic strings on $G$ imply to rather use flat derivative to express the SC~\cite{Blumenhagen:2014gva}
\begin{equation}\label{eqn2:scDFTWZW}
  D_A \cdot D^A \cdot = 0\,.
\end{equation}
The flat derivatives $D_A$ have a very obvious interpretation as a zero mode of the Ka\v{c}-Moody current algebra on the worldsheet in CSFT and then level matching yields the SC as a direct consequence. Nevertheless, it has the effect that the generalized Lie derivative~\eqref{eqn2:genLieDFTnabla} does not close with only flat derivatives $D_A$ anymore. The only way to prevent these issues is the introduction of two covariant derivatives. One flat derivative required for the SC and one covariant $\nabla$ for everything else. This procedure fixes the covariant derivative
\begin{equation}
  \nabla_A V^B = D_A V^B - \frac{1}{3} F_{CA}{}^B V^C
\end{equation}
entirely and exactly reproduces all the results arising from CSFT~\cite{Blumenhagen:2014gva}. In this context, vectors with flat indices are build by contracting vectors with curved indices with the generalized background vielbein i.e. $V^A = V^I E^A{}_I$.
The Christoffel symbols can then be obtained from the vielbein compatibility condition
\begin{equation}
  \nabla_A \, E_B{}^I = 0\,.
\end{equation}
Extending this structure to EFT, we need to
\begin{itemize}
  \item fix the Lie algebra $\mathfrak{g}$ of the group manifold $G$ by specifying the torsion of the flat derivative
  \item fix the connection of $\nabla$ to obtain a closing generalized Lie derivative
\end{itemize}
These ideas outline the steps we need to execute in the following two subsections. Another guiding principle is that the U-duality groups in tab.~\ref{tab:Udualitygroups} include the T-duality groups O($d-1, d-1$) as a subgroup. Hence, we can always verify our results by considering the \DFTwzw{} limit.

\begin{table}
  \centering
  \begin{tabular}{llllllll}\toprule
    d & 2 & 3 & 4 & 5 & 6 & 7 & 8 \\
    \hline
    U-d. group & SL(2)$\times\mathbb{R}^+$ & SL(3)$\times$SL(2) & SL(5) & Spin(5,5) & E${}_{6(6)}$ & E${}_{7(7)}$ & E${}_{8(8)}$ \\
    coord. irrep & $\mathbf{2}_1$ + $\mathbf{1}_{-1}$ & $(\mathbf{3},\mathbf{2})$ & $\mathbf{10}$ & $\mathbf{16}$ & $\mathbf{27}$ & $\mathbf{56}$ & $\mathbf{248}$ \\
    SC irrep & & $(\overline{\mathbf{3}},\mathbf{1})$ & $\overline{\mathbf{5}}$ & $\mathbf{10}$ & $\overline{\mathbf{27}}$ & $\mathbf{133}$ & $\mathbf{3875}+\mathbf{1}$ \\
    emb. tensor & & & $\mathbf{15} + \overline{\mathbf{40}}$ & $\mathbf{144}$ & $\mathbf{351}$ & $\mathbf{912}$ & $\mathbf{3875}$ \\
    \bottomrule
  \end{tabular}
  \caption{\label{tab:Udualitygroups}U-duality groups~\cite{Hull:1994ys} which have the T-duality groups O($d$-1,$d$-1) as subgroup. Moreover, the coordinate and section condition irreps~\cite{Berman:2011jh,Berman:2012vc} of the corresponding EFTs~\cite{Berman:2015rcc,Hohm:2015xna,Musaev:2015ces,Abzalov:2015ega,Hohm:2013vpa,Hohm:2013uia,Hohm:2014fxa} and the embedding tensor irreps~\cite{deWit:2002vt,deWit:2008ta} after the linear constraint are given~\cite{Bosque:2017dfc}.}
\end{table}

\subsection{Section condition}\label{sec:sectioncond}

For DFT all indices lie in the fundamental representation of the Lie algebra $\mathfrak{o}$($d-1,d-1$). In EFT the setting is a bit more involved. Here, we need to use different indices for the different representations of the U-duality group. We begin with the coordinate irrep labeled by capital letters $I, J, \ldots$. Our U-Duality group of choice in this thesis is SL($5$) EFT for which the irrep is the two index antisymmetric $\mathbf{10}$ of $\mathfrak{sl}(5)$. This allows us to express the SC in the following way~\cite{Berman:2012vc}
\begin{equation}\label{eqn2:sectioncond}
  Y^{MN}{}_{LK} \partial_M \partial_N \cdot = 0\,,
\end{equation}
by using the invariant $Y$-tensor. It projects the symmetric part of the tensor product of two coordinate irreps to the SC irrep. The two irreps are given in tab.~\ref{tab:Udualitygroups}. In the case of SL($5$) the SC irrep is given by the fundamental $\mathbf{5}$ and represented by small lettered indices $a, b, \ldots$. As a consequence, the $Y$-tensor takes on the form~\cite{Musaev:2015ces}
\begin{equation}\label{eqn2:SL5Y}
  Y^{MN}{}_{LK} = \frac{1}{4} \epsilon^{MNa}\epsilon_{LKa}
    \quad\text{with the normalization}\quad
  Y^{MN}{}_{MN} = 30
\end{equation}
where $\epsilon$ is the totally antisymmetric Levi-Civita tensor with five fundamental indices. For the SC itself the normalization plays no important role. However, in the context of expressing the generalized Lie derivative~\eqref{eqn:genlieytensor} using the $Y$-tensor, the normalization becomes crucial. As the flat indices have been defined in analogy to the curved ones, the flat derivative becomes
\begin{equation}
  D_A = E_A{}^I \partial_I
\end{equation}
and has the identical form as in \DFTwzw.  The here used generalized background vielbein $E_A{}^I$ lives in GL($n$) with $n = \dim G$ and describes a non-generate frame field on the group manifold as well. This immediately raises the question to what happens if the dimension of $G$ is not the same as the dimension of the irrep. We postpone the answer to subsection~\ref{sec:linconstsl5}. For the time being let us assume that the dimensions match. Now, the $n$-dimensional standard diffeomorphisms act through the Lie derivative in curved indices and the corresponding SC fulfills the relation
\begin{equation}\label{eqn2:flatSC}
  Y^{CD}{}_{AB} \, D_C \cdot D_D \, \cdot = 0\,.
\end{equation}
Hence, the situations is very much alike as what one would expect from \DFTwzw{} discussed in the previous subsection.

As a consequence, the torsion of the flat derivative, given by the structure coefficients $F_{AB}{}^C$~\eqref{eqn2:defFABCDFTWZW}, lies in the tensor product
\begin{equation}
  \overline{\mathbf 45} \times \overline{\mathbf 10} = \mathbf{10} + \mathbf{14} + \mathbf{40} + \mathbf{175} + \mathbf{210}\,.
\end{equation}
Here, $\overline{\mathbf 45}$ represents the antisymmetric part of $\mathbf{10}\times\mathbf{10}$. This is all we know so far.
Analyzing the closure of the gauge algebra mediated by the generalized Lie derivative in the upcoming subsection will clarify this statement.

\subsection{Generalized Lie derivative}
\label{sec:genLie}
In a similar fashion to the SC~\eqref{eqn2:sectioncond}, one can write the generalized Lie derivative for different EFTs, given in tab.~\ref{tab:Udualitygroups}, using the canonical form
\begin{equation}\label{eqn2:genLieYtensor}
  \mathcal{L}_\xi V^M = L_\xi V^M + Y^{MN}{}_{LK}\partial_N \xi^L V^K\,,
\end{equation}
with the $Y$-tensor and the standard Lie derivative on the extended space~\cite{Bosque:2017dfc}. Once the SC is imposed, the infinitesimal generalized diffeomorphisms governing the gauge algebra close~\cite{Berman:2012vc} according to 
\begin{equation}
  [\mathcal{L}_{\xi_1}, \mathcal{L}_{\xi_2}] V^M= \mathcal{L}_{[\xi_1, \xi_2]_{\mathrm{E}}} V^M
    \quad \text{with} \quad
  [\xi_1, \xi_2]_{\mathrm{E}} = \frac{1}{2} \big(\mathcal{L}_{\xi_1} \xi_2 - \mathcal{L}_{\xi_1} \xi_2 \big)\,.
\end{equation}
It should be mentioned that this formulation of the gauge algebra also includes the DFT results for a specific choice of the $Y$-tensor, i.e. $Y^{MN}{}_{LK} = \eta^{MN}\eta_{LK}$, and therefore naturally extends to the EFT framework with the $Y$-tensor in SL($5$) taking on the form in~\eqref{eqn2:SL5Y}. Thus, it is serving as the logical starting point for our discussion. Furthermore, one should keep the rough structure necessary for the closure in mind, as we have to repeat these steps with a covariant derivative instead of a partial one later on.
Computing
\begin{equation}\label{eqn2:closureterm}
  [\mathcal{L}_{\xi_1}, \mathcal{L}_{\xi_2}] V^M - \mathcal{L}_{[\xi_1, \xi_2]_{\mathrm{E}}} V^M\,,
\end{equation}
we are left with sixteen different terms. All containing two partial derivatives. Although, the partial derivatives acts on the same variable only for four terms. On top of that, the $Y$-tensor exhibits the following properties
\begin{align}
  & \delta_F^{(B} Y^{AC)}{}_{DE} - Y^{(AC}{}_{FG} Y^{B)G}{}_{DE} = 0 \quad \text{and} \nonumber \\
  & \delta_{(F}^{B} Y^{AC}{}_{DE)} - Y^{AC}{}_{G(F} Y^{BG}{}_{DE)} = 0 \label{eqn2:Yzero}
\end{align}
which implies that for $d\le 6$ only terms annihilated by the SC remain. For U-duality groups with $d>6$ the closure calculation becomes much more complicated~\cite{Berman:2012vc}. Here, we are merely interested in a proof of concept for them. As a result, let us concentrate on the simplest cases and adjourn the rest to future work. The generalized and standard Lie derivative coincide for arbitrary scalars which yields
\begin{equation}
  \mathcal{L}_\xi s = L_\xi s\,.
\end{equation}
Applying the Leibniz rule we derive the action of generalized diffeomorphisms on one-forms
\begin{equation}
  \mathcal{L}_\xi V_M = L_\xi V_M - Y^{PQ}{}_{NM} \partial_Q \xi^N V_P\,,
\end{equation}
which are the dual objects of the vector representation. Subsequently, we have to remember that $Y^{MN}{}_{PQ}$ needs to remain invariant under the generalized Lie derivative, i.e.
\begin{equation}
  \mathcal{L}_\xi Y^{MN}{}_{PQ} = 0.
\end{equation}
It is totally analogous to the statement in DFT where $\eta_{IJ}$ stays invariant and therefore the O($d,d$) is preserved. Hence, we have completed the list of requirements necessary to make EFT's generalized diffeomorphisms compatible with standard diffeomorphisms.

Let us take a first step towards this direction. We change to flat indices while replacing all partial derivatives appearing in~\eqref{eqn2:genLieYtensor} by covariant ones and obtain
\begin{equation}\label{eqn2:genLieCov}
  \mathcal{L}_\xi V^A = \xi^B \nabla_B V^A - V^B \nabla_B \xi^A + Y^{AB}{}_{CD} \nabla_B \xi^C V^D\,.
\end{equation}
This equation can be recast using flat derivatives
\begin{equation}
\label{eqn2:covderiv}
  \nabla_A V^B = D_A V^B + \Gamma_{AC}{}^B V^C \quad \text{and} \quad
  \nabla_A V_B = D_A V_B - \Gamma_{AB}{}^C V_C
\end{equation}
by introducing the spin connection $\Gamma_{AB}{}^C$. Inserting this into the generalized Lie derivative gives us
\begin{align}
  \mathcal{L}_\xi V^A &= \xi^B D_B V^A - V^B D_B \xi^A + Y^{AB}{}_{CD} D_B \xi^C V^D + X_{BC}{}^A \xi^B V^C \quad \text{and} \nonumber \\
  \mathcal{L}_\xi V_A &= \xi^B D_B V_A + V_B D_A \xi^B - Y^{CD}{}_{BA} D_D \xi^B V_C - X_{BA}{}^C \xi^B V_C \label{eqn2:genLieVA}
\end{align}
with
\begin{equation}\label{eqn2:XfromGamma}
  X_{AB}{}^C = 2 \Gamma_{[AB]}{}^C + Y^{CD}{}_{BE} \, \Gamma_{DA}{}^E
\end{equation}
where we collected all terms depending on the spin connection. Later, it will turn out that $X_{AB}{}^C$ is closely related to the embedding tensor known from gauged supergravities. The $Y$-tensor should still remain invariant under the modified generalized Lie derivative. It directly translates into the first linear constraint
\begin{equation*}\label{eqn2:linconst1}\tag{C1}
  \nabla_C Y^{AB}{}_{DE} := C_1^{AB}{}_{CDE} = 2 Y^{F(A}{}_{DE} \Gamma_{CF}{}^{B)} - 2 Y^{AB}{}_{(D|F} \Gamma_{C|E)}{}^F = 0
\end{equation*}
on the spin connection $\Gamma$ after imposing $D_A Y^{BC}{}_{DE} = 0$. This constraint is a straight forward generalization of the metric compatibility~\eqref{eqn2:metriccomptDFTwzw} in \DFTwzw{}.

In the next step, we demand closure of this adjusted generalized Lie derivative. All terms~\eqref{eqn2:closureterm} spoiling the closure have to vanish analogously. We start with the ones incorporating no flat derivatives. These only disappear if the quadratic constraint
\begin{equation}\label{eqn2:quadconstr}
  X_{BE}{}^A X_{CD}{}^E - X_{BD}{}^E X_{CE}{}^A + X_{[CB]}{}^E X_{ED}{}^A = 0
\end{equation}
holds. Analyzing these additional conditions makes it necessary to decompose $X_{AB}{}^C$ into a symmetric part $Z^C{}_{AB}$ and an antisymmetric one by
\begin{equation}
  X_{AB}{}^C = Z^C{}_{AB} + X_{[AB]}{}^C\,.
\end{equation}
Furthermore, we see that all terms containing only one flat derivative acting on $V^A$ in~\eqref{eqn2:closureterm} vanish, if we identify the torsion of the flat derivative with
\begin{equation}
\label{eqn2:commutator}
  [D_A, D_B] = X_{[AB]}{}^C D_C\,.
\end{equation}
Here, we have used $D_A \,X_{BC}{}^D = 0$ and $Y^{AB}{}_{BC} = Y^{(AB)}{}_{(BC)}$ which is only valid for $d\le 6$ in all computations. A consistent theory moreover requires the existence of a Bianchi identity. It takes on the form
\begin{equation}\label{eqn2:BianchiFlatD}
  [D_A, [D_B, D_C]] + [D_C, [D_A, D_B]] + [D_B, [D_C, D_A]] = 0\,.
\end{equation}
From explicitly evaluating the commutators above, we observe that this constraint is equivalent to the Jacobi identity
\begin{equation}\label{eqn2:jacobiident}
  \Big(X_{[AB]}{}^E X_{[CE]}{}^D + X_{[CA]}{}^E X_{[BE]}{}^D + X_{[BC]}{}^E X_{[AE]}{}^D\Big) D_D = 0\,.
\end{equation}
Antisymmetrizing~\eqref{eqn2:quadconstr} with respect to $B,C,D$ yields
\begin{equation}
  X_{[BC]}{}^E X_{[CE]}{}^A + X_{[DB]}{}^E X_{[CE]}{}^D + X_{[CD]}{}^E X_{[BE]}{}^D = - Z^A{}_{E[B} X_{CD]}{}^E
\end{equation}
and therefore is not zero. This leaves us with
\begin{equation}\label{eqn2:betterzero}
  Z^A{}_{E[B} \, X_{CD]}{}^E D_A = 0
\end{equation}
which in general does not vanish. For \DFTwzw\, this is not the case since $Z^A{}_{BC}$ vanishes and the issue does not occur. Thus, it is special to gEFT. As we show in subsection~\ref{sec:linconstsl5}, the problem can be circumvented by reducing the dimension of the group manifold representing the extended space.

One of the important properties of the generalized Lie derivatives lies in the fact that the Jacobiator of its E-bracket only vanishes up to trivial gauge transformations. Hence, we want to take a closer look at them
\begin{equation}\label{eqn2:trivialgaugetr}
  \xi^A = Y^{AB}{}_{CD} D_B \chi^{CD}
\end{equation}
in the background of our modified generalized Lie derivative.
Ultimately, we will benefit from it by being better able to organize terms appearing in the closure computation with one flat derivative operating either on $\xi_1$ or $\xi_2$. Plugging~\eqref{eqn2:trivialgaugetr} into the generalized Lie derivative~\eqref{eqn2:genLieVA} yields the following relation
\begin{equation}
  \mathcal{L}_{\xi} V^A = C_{2a}{}^{AB}{}_{CDE} D_B \chi^{CD} V^E + \dots = 0
\end{equation}
where $\dots$ refers to terms which vanish under the SC and as a result of the properties of the $Y$-tensor~\eqref{eqn2:Yzero}. The tensor 
\begin{equation*}\label{eqn2:linconst2a}\tag{C2a}
  C_{2a}^{AB}{}_{CDE} = Y^{BF}{}_{CD} X_{FE}{}^A + \frac{1}{2} Y^{AF}{}_{CD} X_{[FE]}{}^B + \frac{1}{2} Y^{AF}{}_{EH} Y^{GH}{}_{CD} X_{[FG]}{}^B
\end{equation*}
needs to vanish when trivial gauge transformations have the form given in~\eqref{eqn2:trivialgaugetr}.

Terms appearing with two derivatives in~\eqref{eqn2:closureterm} become zero under the SC or due to~\eqref{eqn2:Yzero}. At this point, all terms we are left with contain one flat derivative acting on the gauge parameters $\xi_1$ or $\xi_2$. Since~\eqref{eqn2:closureterm} is antisymmetric with respect to the gauge parameters, it is sufficient to check whether all contributions on one of the terms, e.g. $D_A \xi_1^B$, vanish. They can be written in terms of the tensor
\begin{align*}
  C_{2b}^{AB}{}_{CDE} &= Z^A{}_{DC} \delta_E^B - Z^B{}_{DE} \delta_C^A - Y^{BF}{}_{EC} Z^A{}_{DF} + Y^{AB}{}_{CF}{} Z^F{}_{DE} \nonumber \\
  &+ Y^{AB}{}_{EF} X_{[DC]}{}^F + Y^{AB}{}_{CF} X_{[DE]}{}^F - 2 Y^{F(A}{}_{EC} X_{[DF]}{}^{B)} = 0 \tag{C2b}
\end{align*}
by
\begin{equation}
  \Big(-\frac{1}{2} C_{2a}^{AB}{}_{CDE} + C_{2b}^{AB}{}_{CDE} \Big) \, D_B \xi_1^C \xi_2^D V^E = 0\,,
\end{equation}
which obtains a contribution from trivial gauge transformations~\eqref{eqn2:linconst2a} as well. This is perfectly sensible as the E-bracket also only closes up to trivial gauge transformations. However, generally there exists no explanation why the two contributions have to disappear independently. It implies that only the second linear constraint
\begin{equation*}\label{eqn2:linconst2}\tag{C2}
  -\frac{1}{2} C_{2a}^{AB}{}_{CDE} + C_{2b}^{AB}{}_{CDE} = 0
\end{equation*}
has to be satisfied in combination with the first linear constraint~\eqref{eqn2:linconst1}, and the quadratic constraint~\eqref{eqn2:quadconstr} for closure of generalized diffeomorphisms under the SC. Thus, one has to restrict the connection $\Gamma_{AB}{}^C$ in such a way that all these three restraints are fulfilled. This outlines our steps during the next two subsections.

Right now, we can already perform a first consistency check of our results. Therefore, let us consider the O($d-1$,$d-1$) T-duality group with
\begin{equation}
  Y^{AB}{}_{CD} = \eta^{AB} \, \eta_{CD}\,, \quad
    \Gamma_{AB}{}^C = \frac{1}{3} F_{AB}{}^C
    \quad \text{and} \quad
    X_{AB}{}^C = F_{AB}{}^C\,.
\end{equation}
In this case, the two linear constraints and quadratic constraint reduce to
\begin{align}
  C_1^{AB}{}_{CDE}&= \frac23 \eta_DE F_C{}^{(AB)} - \frac23 \eta^{AB} F_{C(DE)} = 0 \\
  C_{2a}^{AB}{}_{CDE}&= \eta_{CD} ( F^B{}E{}^A + F^A{}_E{}^B ) = 0 \\
  C_{2b}^{AB}{}_{CDE}&= \eta^{AB} ( F_{DCE} + F_{DEC} ) - 2 \eta_{EC} F_D{}^{(AB)} = 0
\end{align}
as a consequence of the total antisymmetry of the structure constants $F_{ABC}$. Thus, this short computation is in perfect agreement with the closure of the gauge algebra known from \DFTwzw{}~\cite{Blumenhagen:2014gva,Bosque:2017dfc}.

\subsection{Linear constraints}\label{sec:linconstsl5}

In gEFT solving the linear constraints turns out to be much more involved than for \DFTwzw{}, our toy example in the last subsection, which serves as a consistency check. As a result, we need to introduce more sophisticated tools known from representation theory. Particularly, we have to procure projection operators filtering out certain irreps of the coordinate irrep in tab.~\ref{tab:Udualitygroups}. For our choice of SL($5$) EFT everything works out neatly. The irreps (or more specifically the projectors onto them) of SL($n$) and their tensor products can be nicely organized through Young tableaux and thus make their representation theory very transparent. We review all the necessary techniques in appendix~\ref{app:SLnrepresentations}. In this context, we use the T-Duality subgroup SL($4$) as an explicit example. It's associated Lie algebra $\mathfrak{sl}(4)$ is isomorphic to $\mathfrak{so}(3,3)$. From previous results we already know the solutions to the linear constraints in this case and therefore allow us to check the methods developed during the appendix~\cite{Bosque:2017dfc}.

Let us begin the discussion with the spin connection $\Gamma_{AB}{}^C$. The indices lie in the $\mathbf{10}$ and $\overline{\mathbf{10}}$ of $\mathfrak{sl}(5)$. We express said indices through the fundamental $\mathbf{5}$ indices, and raised indices are lowered with the totally antisymmetric Levi-Civita tensor. It yields
\begin{equation}
  \Gamma_{a_1 a_2, b_1 b_2, c_1 c_2 c_3} = \Gamma_{a_1 a_2, b_1 b_2}{}^{d_1 d_2} \epsilon_{d_1 d_2 c_1 c_2 c_3}\,.
\end{equation}
For this form, the embedding of the 1000 independent connection components into the tensor product
\begin{equation}\label{eqn2:decomp101010b}
  \mathbf{10} \times \mathbf{10} \times \overline{\mathbf 10} = 3 (\mathbf{10}) + \mathbf{15} + 2 (\mathbf{40}) + 2 (\mathbf{175}) + \mathbf{210} + \mathbf{315}
\end{equation}
is manifest. We can translate this expression into corresponding Young diagrams
\begin{equation}
  \ydiagram{1,1}\times\left(\,\ydiagram{1,1}\times\ydiagram{1,1,1}\,\right) = 3\, \ydiagram{2,2,1,1,1} + \ydiagram{3,1,1,1,1} + 2\,\ydiagram{2,2,2,1} + 2 \,\ydiagram{3,2,1,1} + \ydiagram{3,2,2} + \ydiagram{3,3,1}\,.
\end{equation}
This decomposition looks quite similar to~\eqref{eqn2:P666} in appendix~\ref{app:SLnrepresentations}. However, the $\mathbf{10}$ of $\mathfrak{sl}(5)$ is not self dual as the $\mathbf{6}$ of $\mathfrak{sl}(4)$. Therefore, we have to pick up an additional box in the last irrep on the left hand side. All of these diagrams possess a corresponding projector. Since several irreps occur more than once in the decomposition of the tensor product~\eqref{eqn2:decomp101010b}, we denote them by
\begin{equation}
  \mathbf{10} \times ( \mathbf{10}\times \overline{\mathbf 10} ) = \mathbf{10} \times (\mathbf{1} + \mathbf{24} + \mathbf{75}) = \left\{ \begin{array}{ll}
    \mathbf{10} \times \mathbf{1} &= \mathbf{10}a \\
    \mathbf{10} \times \mathbf{24} &= \mathbf{10}b + \mathbf{15} + \mathbf{40}a + \mathbf{175}a \\
    \mathbf{10} \times \mathbf{75} &= \mathbf{10}c + \mathbf{40}b + \mathbf{210} + \mathbf{315}
  \end{array}\right.
\end{equation}
in order to clearly differentiate between all projectors.

Whereas it is straightforward to solve the first linear constraint~\eqref{eqn2:linconst1} for $\mathfrak{sl}(4)$, things become much more involved in the case of $\mathfrak{sl}(5)$. First, it should be noted that the constraint acts trivially on the index $C$. We suppress this index and write~\eqref{eqn2:linconst1}
\begin{equation}\label{eqn2:C1sl5}
  C_1{}_{a_1 a_2 a_3, b_1 b_2 b_3, d_1 d_2, e_1 e_2} = \sigma_1 \Gamma_{a_1 a_2, a_3 b_1 b_2} \epsilon_{b_3 d_1 d_2 e_1 e_2} = 0
\end{equation}
in terms of the permutations
\begin{gather}
	\sigma_1 = (6\,5\,4\,3) + (3\,5\,2\,4\,1) - (6\,5\,4\,3\,2) - (3\,5\,6\,2\,4\,1) +	(6\,5\,4\,3\,2\,1) - (6\,10\,2\,7\,4\,8\,5\,9\,1) - \nonumber \\
	(6\,10\,5\,9\,4\,8\,2\,7\,1) + (6\,10\,2\,3\,7\,4\,8\,5\,9\,1) + (6\,10\,5\,9\,4\,8\,2\,3\,7\,1) + (3\,5\,1) (4\,6\,2) \nonumber \\- (3\,7\,1)(6\,10\,5\,9\,4\,8,2) - (3\,7\,4\,8\,5\,9\,1)(6\,10\,2)
\end{gather}
which are acting on the ten remaining indices. As a result of this form, the constraint can now be solved by linear algebra techniques. Therefore, let us consider the explicit basis 
\begin{align}\label{eqn2:basis1010b}
  (d_1 d_2),\, (e_1 e_2) \in V_{\mathbf{10}} &= \big\{ (d_1 d_2) \,|\, d_1,\,d_2 \in \{1\, \dots\, 5\} \wedge d_1 < d_2 \big\} \nonumber \\
  (a_1 a_2 a_3),\, (b_1 b_2 b_3) \in V_{\overline{\mathbf 10}} &= \big\{ (a_1 a_2 a_3) \,|\, a_1,\,a_2,\,a_3 \in \{1\, \dots\, 5\} \wedge a_1 < a_2 < a_3 \big\}
\end{align}
for the irreps $\mathbf{10}$ and $\overline{\mathbf 10}$ appearing in~\eqref{eqn2:C1sl5}. If we keep the properties of the totally antisymmetric tensor in mind, we can interpret $\sigma_1$ as a linear map from $\Gamma$ to $C_1$
\begin{equation}
  \sigma_1 : V_{\mathbf{10}} \times V_{\overline{\mathbf 10}} \rightarrow V_{\overline{\mathbf 10}} \times V_{\overline{\mathbf 10}} \times  V_{\mathbf{10}} \times V_{\mathbf{10}}\,.
\end{equation}
Solutions of the first linear constraint must be elements in the kernel of this map and are associated to the projection operators onto the irreps we have discussed before. A straightforward computation proves that
\begin{equation}
  \sigma_1 ( P_{\mathbf{1}} + P_{\mathbf{24}} ) = 0 \quad \text{but} \quad
  \sigma_1 P_{\mathbf{75}} \ne 0
\end{equation}
holds. Thus, the most general solution can be expressed through the projector
\begin{equation}
  P_1 = P_{\mathbf{10}a} + P_{\mathbf{10}b} + P_{\mathbf{15}} + P_{\mathbf{40}a} + P_{\mathbf{175}a}\,.
\end{equation}

Subsequently, we need to verify which of these irreps survive the transition from the connection $\Gamma_{AB}{}^C$ to $X_{AB}{}^C$. Equivalently to~\eqref{eqn2:C1sl5}, we cast~\eqref{eqn2:XfromGamma} in terms of permutations
\begin{equation}\label{eqn2:simgaX}
	\sigma_X = () -(3\,1)(4\,2)+(3\,5\,1)(4\,6\,2)-(3\,5\,1)(4\,6\,7\,2) + (3\,5\,7\,2\,4\,6\,1)
\end{equation}
by using
\begin{equation}
  X_{a_1 a_2, b_1 b_2, c_1 c_2 c_3} = \sigma_X P_1 \Gamma_{a_1 a_2, b_1 b_2, c_1 c_2 c_3}\,.
\end{equation}
It is worth mentioning that the first linear constraint is already implemented in this equation through projector $P_1$. Again, we apply the same techniques demonstrated in appendix~\ref{app:SLnrepresentations} to decompose
\begin{equation}
  \sigma_X P_1 P_{\mathbf{10}\times\mathbf{10}\times\overline{\mathbf 10}} = \frac{12}5 P_{\mathbf{10}ab} + P_{\mathbf{10}c} + 4 P_{\mathbf{15}} + 3 P_{\mathbf{40}a}
\end{equation}
into orthogonal projectors on different $\mathfrak{sl}(5)$ irreps where $P_{\mathbf{10}ab}$ is defined as
\begin{equation}
  P_{\mathbf{10}ab} = \frac5{12} (P_{\mathbf{10}a}-P_{\mathbf{10}b}) \sigma_X (P_{\mathbf{10}a} + P_{\mathbf{10}b})\,.
\end{equation}
This equation merely embeds another ten-dimensional irrep $\mathbf{10}ab$ into $\mathbf{10}a$ and $\mathbf{10}b$. In the following, we restrict ourselves to the $\mathbf{15}$ and $\mathbf{40}$. These are exactly the irreps surviving the linear constraint on the embedding tensor known from seven-dimensional maximal gauged supergravities\footnote{  In~\cite{Samtleben:2005bp} a three index tensor $Z^{ab,c}$ represents the $\overline{\mathbf{40}}$. Here, we use its dual version. Both are connected by~\eqref{eqn2:40bto40} and capture the same information.}~\cite{Samtleben:2005bp}. As is demonstrated in appendix A of~\cite{Berman:2012uy}, the remaining two ten-dimensional irreps can be combined into one $\mathbf{10}$ capturing trombone gaugings as well. Nevertheless, we have limited ourselves to a proof of concept, and thus do not discuss trombone gaugings. They are however considered in~\cite{Bosque:2016fpi} which takes the embedding tensor irreps $\mathbf{10}+\mathbf{15}+\overline{\mathbf{40}}$ as starting point. A priori, we do not restrict the allowed groups $G$. But our attempt to implement generalized diffeomorphisms on them exactly reproduces the correct irreps of the embedding tensor. In the original context, these arise from supersymmetry conditions~\cite{deWit:2002vt}. Here, we did not make any direct contact with supersymmetry. Thus, it is very remarkable that we still replicate this result.

Now, let us turn to the last remaining linear constraint~\eqref{eqn2:linconst2} we require. It proceeds in an analogous fashion as for the first linear constraint and we write
\begin{equation}\label{eqn2:C2sigma2}
  C_2{}_{a_1 a_2 a_3, b_1 b_2 b_3, c_1 c_2, d_1 d_2, e_1 e_2} = \sigma_2 X_{a_1 a_2, a_3 b_1, b_2 b_3 c_1} \epsilon_{c_2 d_1 d_2 e_1 e_2}
\end{equation}
through a sum of permutations represented by $\sigma_2$, being of a similar form as~\eqref{eqn2:simgaX} but containing 54 different terms. Hence, we do not present it explicitly. In the basis~\eqref{eqn2:basis1010b}, $\sigma_2$ generates the linear map
\begin{equation}
  \sigma_2 : V_{\mathbf{10}} \times V_{\mathbf{10}} \times V_{\overline{\mathbf 10}} \rightarrow V_{\overline{\mathbf 10}} \times V_{\overline{\mathbf 10}} \times  V_{\mathbf{10}} \times V_{\mathbf{10}} \times V_{\mathbf{10}}
\end{equation}
whose kernel contains the $\mathbf{15}$, but not the $\mathbf{40}a$. However, we know from maximal gauged supergravities in seven dimensions~\cite{Samtleben:2005bp} that gaugings in the dual $\overline{\mathbf{40}}$ are consistent as well. At first glance this might appear puzzling but we can resolve this contradiction quite easily. We start by implementing the components of this irreps in terms of the tensor $Z^{ab,c}$ and relate it to the $\mathbf{40}a$, discussed above, by
\begin{equation}\label{eqn2:40bto40}
  (X_{\mathbf{40a}})_{a_1 a_2, b_1 b_2, c_1 c_2 c_3} = \epsilon_{a_1 a_2 d_1 d_2 [b_1} Z^{d_1 d_2, e_1} \epsilon_{b_2] c_1 c_2 c_3 e_1} 
\end{equation}
with the expected property
\begin{equation}
  P_{\mathbf{40a}} (X_{\mathbf{40a}})_{a_1 a_2, b_1 b_2, c_1 c_2 c_3} =
    (X_{\mathbf{40a}})_{a_1 a_2, b_1 b_2, c_1 c_2 c_3}\,.
\end{equation}
According to the argumentation given in~\cite{Samtleben:2005bp}, we can interpret $Z^{ab,c}$ as a 10$\times$5 matrix and calculate its rank through
\begin{equation}
  s = \mathrm{rank} ( Z^{ab,c} )\,.
\end{equation}
\begin{figure}[h]
\centering\begin{tikzpicture}[every text node part/.style={align=center}, node distance=10em]
	\node[rectangle split, rectangle split parts=2, 
  	draw, minimum width=3cm, text width=3cm] (SL5)
		{ SL(5)
    	\nodepart{two}
     		D: 10 $\sim$  $\mathbf{10}$ \\
     		E: $\mathbf{15}$ };
	\node[rectangle split, rectangle split parts=2, 
  	draw, minimum width=3cm, text width=3cm, below left of=SL5] (SL4)
		{ SL(4)
    	\nodepart{two}
     		D: 6 $\sim$ $\mathbf{6}$ \\
     		E: $\mathbf{10} + \overline{\mathbf{10}}$ };
	\node[rectangle split, rectangle split parts=3, 
  	draw, minimum width=5.5cm, text width=5.5cm, below right of=SL5] (SL2SL3)
		{ SL(3)$\times$SL(2)
    	\nodepart{two}
     		D: 9 $\sim$ $(\mathbf{3}, \mathbf{2}) + (\overline{\mathbf 3}, \mathbf{1})$  \\
     		E: $(\mathbf{1},\mathbf{3}) + (\mathbf{3},\mathbf{2}) + (\mathbf{6},\mathbf{1}) + (\mathbf{1},\mathbf{2})$
    	\nodepart{three}
     		D: 7 $\sim$ $(\mathbf{1}, \mathbf{1}) + (\mathbf{3}, \mathbf{2})$ \\
     		E: $(\mathbf{1},\mathbf{3}) + (\mathbf{3},\mathbf{2}) + (\mathbf{6},\mathbf{1}) + (\mathbf{8},\mathbf{1})$ };
	\node[rectangle split, rectangle split parts=4, 
  	draw, minimum width=8.5cm, text width=14cm, below left of=SL2SL3, xshift=2em, yshift=-3em] (SL2SL2)
		{ SL(2)$\times$SL(2)
    	\nodepart{two}
        D: 8 $\sim$ $(\mathbf{2}, \mathbf{2}) + (\mathbf{2}, \mathbf{1}) + (\mathbf{1}, \mathbf{2})$  \\
     		E: $(\mathbf{1}, \mathbf{3}) + (\mathbf{1}, \mathbf{2}) + (\mathbf{2}, \mathbf{2}) + (\mathbf{1}, \mathbf{1}) + (\mathbf{2}, \mathbf{1}) + (\mathbf{3}, \mathbf{1}) + (\mathbf{1}, \mathbf{2}) + (\mathbf{2}, \mathbf{1})$
    	\nodepart{three}
        D: 7 $\sim$ $(\mathbf{1}, \mathbf{1}) + (\mathbf{2}, \mathbf{2}) + (\mathbf{2}, \mathbf{1})$  \\
     		E: $(\mathbf{1}, \mathbf{3}) + (\mathbf{1}, \mathbf{2}) + (\mathbf{2}, \mathbf{2}) + (\mathbf{1}, \mathbf{1}) + (\mathbf{2}, \mathbf{1}) + (\mathbf{3}, \mathbf{1}) + (\mathbf{1}, \mathbf{2}) + (\mathbf{1}, \mathbf{1}) + (\mathbf{2}, \mathbf{1}) + (\mathbf{1}, \mathbf{3})$ 
     	\nodepart{four}
        D: 5 $\sim$ $(\mathbf{1}, \mathbf{1}) + (\mathbf{2}, \mathbf{2})$  \\
     		E: $(\mathbf{1}, \mathbf{3}) + (\mathbf{2}, \mathbf{2}) + (\mathbf{3}, \mathbf{1}) + (\mathbf{2}, \mathbf{2}) + (\mathbf{1}, \mathbf{3}) + (\mathbf{3}, \mathbf{1}) + (\mathbf{1}, \mathbf{1})$ };
  \draw[->] (SL5.south) -- (SL4.north);
  \draw[->] (SL5.south) -- (SL2SL3.north);
  \draw[->] (SL4.south) -- (SL2SL2.north);
  \draw[->] (SL2SL3.south) -- (SL2SL2.north);
\end{tikzpicture}
\caption{Solutions of the linear constraints~\eqref{eqn2:linconst1} and~\eqref{eqn2:linconst2}. ``D:'' lists the dimension of the group manifold and the corresponding coordinate irreps. All components of the embedding tensor which are in the kernel of the linear constraints are labeled by ``E:''~\cite{Bosque:2017dfc}.}\label{fig:sollinconst}
\end{figure}
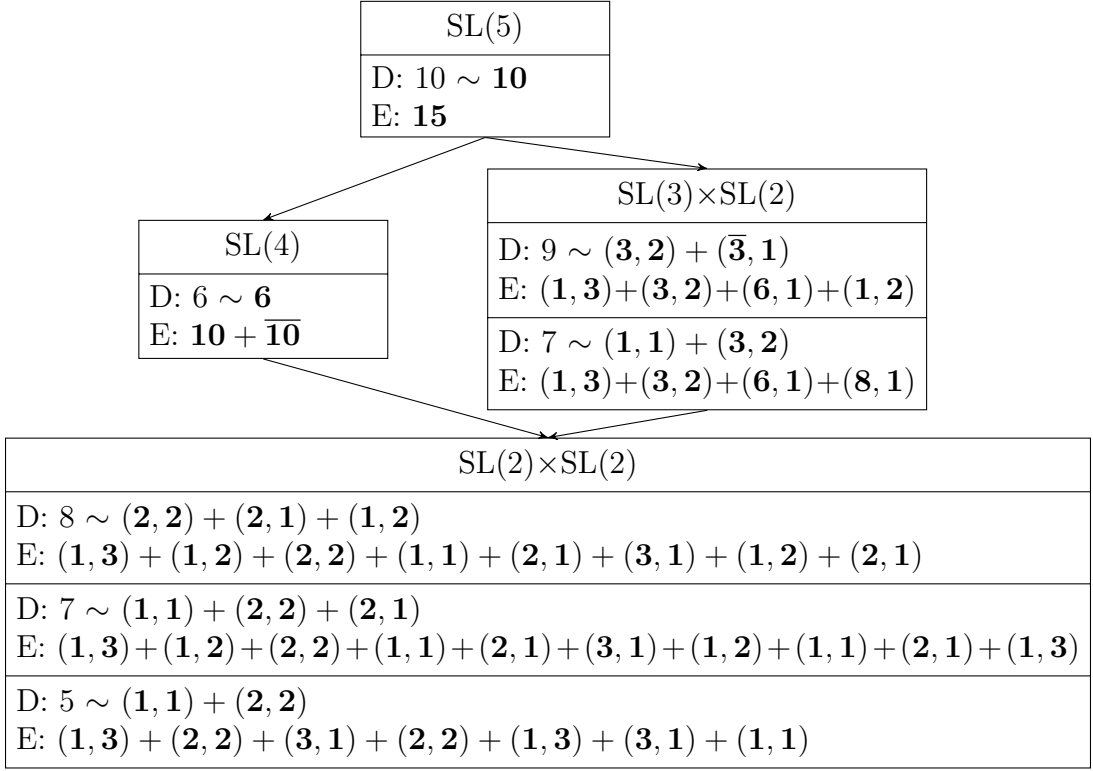
The number of massless vector multiplets in the resulting seven-dimensional gauged supergravity is given by $10-s$. These contain the gauge bosons of the theory and transform in the adjoint representation of the gauge group $G$. As a result, we immediately conclude
\begin{equation}\label{eqn2:dimG40}
  \mathrm{dim} \, G = 10 - s\,.
\end{equation}
For \DFTwzw{} the gauge group of the gauged supergravity, arising after a Scherk Schwarz compactification, is in one-to-one correspondence with the group manifold we are considering~\cite{Bosque:2015jda}. There exists no reason why this should not be the case for gEFT as well. Thus, if we turn on gaugings in the $\overline{\mathbf 40}$, we automatically reduce the dimension of the group manifold representing the extended space. Viable ranks $s$ compatible with the quadratic constraint of the embedding tensor are given by $0 \le s\le 5$. For these cases we have to adapt the coordinates on the group manifold. Therefore, let us consider the possible branching rules of SL(5) to its U-/T-duality subgroups given in tab.~\ref{tab:Udualitygroups}, e.g. SL(4), SL(3)$\times$SL(2), and SL(2)$\times$SL(2)
\begin{align}
  \mathbf{10} &\rightarrow \mathbf{4} + \mathbf{6} \label{eqn2:10branching1}\\
  \mathbf{10} &\rightarrow (\mathbf{1}, \mathbf{1}) + (\mathbf{3}, \mathbf{2}) + (\overline{\mathbf 3}, \mathbf{1}) \label{eqn2:10branching2}\\
  \mathbf{10} &\rightarrow (\mathbf{1}, \mathbf{1}) + (\mathbf{1}, \mathbf{1}) + (\mathbf{2}, \mathbf{1}) + (\mathbf{1}, \mathbf{2}) + (\mathbf{2}, \mathbf{2}) \label{eqn2:10branching3}\,.
\end{align}
For the first case, we obtain a six-dimensional manifold whose coordinates can be identified with the $\mathbf{6}$ of the branching rule~\eqref{eqn2:10branching1} after dropping the $\mathbf{4}$. In the adapted basis
\begin{align}\label{eqn2:V4&V6}
  V_{\mathbf{4}}&= \{15,\,25,\,35,\,45\} & V_{\mathbf{6}} &= \{12,\,13,\,14,\,23,\,24,\,34\} \\
  V_{\overline{\mathbf 4}}&= \{234,\,134,\,124,\,123\} & V_{\overline{\mathbf 6}} &= \{345,\,245,\,235,\,145,\,135,\,125\}\,,
\end{align}
$\sigma_2$ is now restricted to
\begin{equation}\label{eqn2:SL4sigma2}
  \sigma_2 : V_{\mathbf{6}} \times V_{\mathbf{6}} \times V_{\overline{\mathbf 6}} \rightarrow V_{\overline{\mathbf 6}} \times V_{\overline{\mathbf 6}} \times  V_{\mathbf{6}} \times V_{\mathbf{6}} \times V_{\mathbf{6}}\,,
\end{equation}
while the irreps $\mathbf{15}$ and $\mathbf{40}$ split into
\begin{align}
  \mathbf{15} &\rightarrow \xcancel{\mathbf{1}} + \xcancel{\mathbf{4}} + \mathbf{10} \\
  \mathbf{40} &\rightarrow \xcancel{\overline{\mathbf 4}} + \mathbf{6} + \overline{\mathbf 10} + \xcancel{\mathbf{20}}\,.
\end{align}
All crossed out irreps at least partially depend on $V_4$ or its dual basis which is not available as coordinate irrep anymore. Clearly, the $\mathbf{10}$ coming from the $\mathbf{15}$ still fulfills all linear constraints. But now the $\mathbf{6}$ gets excluded by the second linear constraint~\eqref{eqn2:C2sigma2} with~\eqref{eqn2:SL4sigma2}, while the $\overline{\mathbf 10}$ lies in its kernel. This result agrees with the SL(4) case we examined in appendix~\ref{app:SLnrepresentations}. Thus, turning on specific gaugings in the $\mathbf{40}$ indeed breaks the U-duality group into one of its subgroups. An alternative approach~\cite{Bosque:2016fpi} works by keeping the full SL(5) covariance of the embedding tensors and not solving its associated linear constraints. However, this technique obscures the interpretation of the extended space as a group manifold. It is crucial for constructing the generalized frame $\mathcal{E}_A$ in the next section. Furthermore, the breaking of symmetries by non-trivial background expectation values for fluxes is a well-known paradigm. Therefore, only a torus without fluxes has the maximal amount of abelian isometries and should allow for the full U-duality group. All \DFTwzw{} results are naturally embedded as a subset of the EFT formalism, when restricting ourselves to a T-duality subgroup to solve the linear constrains. For the remaining branchings~\eqref{eqn2:10branching2} and~\eqref{eqn2:10branching3}, we perform the same analysis in appendix~\ref{app:linconst2additional} again. All results are summarized in figure~\ref{fig:sollinconst}~\cite{Bosque:2017dfc}.

\subsection{Quadratic constraint}\label{sec:quadrconstr}
Finally, we turn to the quadratic constraint~\eqref{eqn2:quadconstr} which simplifies drastically to the form
\begin{equation}
  X_{[BC]}{}^E X_{[ED]}{}^A + X_{[DB]}{}^E X_{[EC]}{}^{A} + X_{[CD]}{}^E X_{[EB]}{}^A = 0
\end{equation}
after solving the linear constraints which result in $Z^C{}_{AB} = 0$ for the remaining coordinates on the group manifold $G$. Moreover, it is completely identical to the Jacobi identity obtained before~\eqref{eqn2:jacobiident} which is always satisfied for the Lie algebra $\mathfrak{g}$. Thus, the flat derivative has to fulfill the corresponding first Bianchi identity~\eqref{eqn2:BianchiFlatD}. For the covariant derivative~\eqref{eqn2:covderiv}, we are able to calculate the curvature and the torsion by evaluating the commutator
\begin{equation}
\left[ \nabla_A, \nabla_B \right] V_C = R_{ABC}{}^D V_D - T_{AB}{}^D \nabla_D V_C\,.
\end{equation}
Thus, we obtain the curvature given by
\begin{equation}
R_{ABC}{}^D =  2 \Gamma_{[A|C}{}^E \, \Gamma_{|B]E}{}^D + X_{[AB]}{}^E \, \Gamma_{EC}{}^D\,,
\end{equation}
where we used that $\Gamma_{AB}{}^C$ is constant due to~\eqref{eqn2:XfromGamma} and the torsion takes on the form
\begin{equation}
T_{AB}{}^C = - X_{[AB]}{}^C + 2 \Gamma_{[AB]}{}^C = Y^{CD}{}_{[A|E} \, \Gamma_{D|B]}{}^E\,,
\end{equation}
for which we successively applied~\eqref{eqn2:XfromGamma} and~\eqref{eqn2:commutator}. In general both are zero. Using these equations, we ultimately compute the first Bianchi identity
\begin{align}
  R_{[ABC]}{}^D + & \nabla_{[A} T_{BC]}{}^D - T_{[AB}{}^E T_{C]E}{}^D = \nonumber \\
  & 2 X_{[AB]}{}^E \, X_{[CE]}{}^D + 2 X_{[CA]}{}^E X_{[BE]}{}^D + 2 X_{[BC]}{}^E X_{[AE]}{}^D = 0
\end{align}
for $\nabla$. Again, it is fulfilled as a consequence of the Jacobi identity~\eqref{eqn2:jacobiident}. These results are in perfect agreement with the ones of \DFTwzw{}. It is now straightforward to verify that all gaugings, given in tab. 3 of~\cite{Samtleben:2005bp}, can be reproduced in the framework we presented in the first part of this paper. Explicit examples of these gaugings with ten-dimensional groups CSO(1,0,4), SO(5) as well as a nine-dimensional group are discussed in section~\ref{sec:examples}.

\section{Solving the section condition}\label{sec:solSC}

Until now, we have introduced generalized diffeomorphisms on group manifolds $G$ which allow for embeddings into one of the U-duality groups with $d \le 4$ as shown in tab.~\ref{tab:Udualitygroups}. Even though, they only close to form a consistent gauge algebra once the SC~\eqref{eqn2:flatSC} is imposed. Thus, a full comprehension of these SC solutions is imperative and focus of this section. We are going to modify a technique known from \DFTwzw{}~\cite{Hassler:2016srl} to the gEFT framework to obtain the most general solutions of the SC. It is based upon an $H$-principle bundle over the physical subspace $M = G/H$. $H$ is a ($\dim G$-$\dim M$)-dimensional subgroup of $G$ which possesses certain traits. They are explained in subsection~\ref{sec:HPrincipal}. The construction presented in this section is very similar to the steps partaken in \DFTwzw{}. When necessary, we introduce generalizations of specific ideas and notions. Furthermore, in subsection~\ref{sec:gg} we prove that every SC solution yields a GG on $M$ accommodating two basic constituents: a twisted generalized Lie derivative and a generalized frame field. Both of them act on the generalized tangent bundle $T M \oplus \Lambda^2 T^* M$. In certain instances the choice of the subgroup $H$ is not unique for a given $G$. However, different subgroups give rise to dual background in these cases. During subsection~\ref{sec:dualbg} we provide a systematic procedure to analyze these different dual backgrounds. It works in the same fashion as in~\cite{Hassler:2016srl} and we therefore restrict ourselves to a brief discussion. Once the SC solution has been derived, we are able to construct the generalized frame field satisfying the relation $\widehat{\mathcal{L}}_{\mathcal{E}_A} \mathcal{E}_B = X_{AB}{}^C \mathcal{E}_C$ in subsection~\ref{sec:genframe}. Moreover, it is necessary to impose an additional constraint on the structure coefficients $X_{AB}{}^C$ for this construction to function~\citep{Bosque:2017dfc}.

\subsection{Reformulation as \texorpdfstring{$H$}{H}-principal bundle}\label{sec:HPrincipal}

Following the steps outlined in~\cite{Hassler:2016srl}, we start by replacing the quadratic version~\eqref{eqn2:flatSC} of the SC by an equivalent linear constraint~\cite{Berman:2012vc}
\begin{equation}\label{eqn2:sclinear}
  v_a \, \epsilon^{aBC} D_B \, \cdot = 0\,,
\end{equation}
involving a vector field $v_a$ in the fundamental (SC irrep) of SL(5)~\cite{Bosque:2017dfc}. This field can take different values on each point of $G$. Note that relating different points requires the translations on $G$ to be generated by the Lie algebra $\mathfrak{g}$. Specifically, we are interested in the operation of its generators on the representations
\begin{equation}\label{eqn2:genfund&adj}
  \mathbf{5}: \quad (t_A)_b{}^c = X_{A,b}{}^c \quad \text{and} \quad \mathbf{10}: \quad (t_A)_B{}^C = X_{AB}{}^C = 2 X_{A, [b_1}{}^{[c_1} \delta_{b_2]}{}^{c_2]} = 2 (t_A)_{[b_1}{}^{[c_1} \delta_{b_2]}{}^{c_2]}\,.
\end{equation}
Both of them are governed by the embedding tensor. The corresponding group elements can now be obtained by performing an exponential map. At this point, let us assume there exists a set of fields $f_i$ with a coordinate dependence in such a way that they solve the linear constraint~\eqref{eqn2:sclinear} for a particular choice of $v_a$. Subsequently, there can be found another set of fields $f_i'$ depending on different coordinates, i.e.
\begin{equation}
  D_A f_i' = (\mathrm{Ad}_g)_A{}^B D_B f_i
    \quad \text{and} \quad
 (\mathrm{Ad}_g)_A{}^B t_B = g\, t_A g^{-1}
\end{equation}
which solve the linear constraint after transforming $v_a$ according to
\begin{equation}\label{eqn2:shiftvag}
  v'_a = (g)_a{}^b v_b\,.
\end{equation}
In the given context, $(g)_a{}^b$ denotes the left action of a group element $g$ on the vector $v_b$. This property of the linear constraint~\eqref{eqn2:sclinear} is a result of the totally antisymmetric tensor's SL$(5)$ invariance.

The situation resembles the one of \DFTwzw{} a lot. Only the groups and their corresponding representations are different. A slight deviation from~\cite{Hassler:2016srl} lies in the segregation of the $\mathbf{10}$ indices into two sets of subindices. If we want to implement the section condition, we need to define a vector $v_a^0$ yielding
\begin{equation}\label{eqn2:sclinearv0}
  v_a^0 \epsilon^{a\beta C} t_\beta = 0 \quad \text{and} \quad
  v_a^0 \epsilon^{a\tilde\beta C} t_{\tilde\beta} \ne 0\,.
\end{equation}
It splits the generators $t_A$ of $\mathfrak{g}$ in the following fashion
\begin{equation}\label{eqn2:splittingtA}
  t_A = \begin{pmatrix} t_\alpha & t_{\tilde\alpha} \end{pmatrix} 
    \quad \text{and} \quad
  t_{\alpha} \in \mathfrak{m}\,,\quad t_{\tilde \alpha} \in \mathfrak{h}\,,
\end{equation}
where $\mathfrak{h}$ represents the subalgebra and $\mathfrak{m}$ the complement with $\alpha$=1,\,\ldots,\,$\dim G/H$ as well as $\tilde\alpha$=$\tilde 1$,\,\dots,\,$\dim H$. The decomposition of $\mathfrak{g}$ can be made manifest by splitting the $\mathbf{10}$ index $A$ into two non-intersecting subindices $\alpha, \tilde{\alpha}$. Furthermore, the generators $t_{\tilde\alpha}$ generate the stabilizer subgroup $H \subset G$. $v^0_a$ is left invariant by its elements under the transformation~\eqref{eqn2:shiftvag}. It suggests to decompose each group element $g\in G$ according to
\begin{equation}
  g = m h \quad \text{with} \quad h \in H
\end{equation}
while $m$ is a coset representative of the left coset $G/H$. Since the action of $h$ is free and transitive, we can identify $G$ as a $H$-principal bundle
\begin{equation}\label{eqn2:Hprincipal}
  \pi: G \rightarrow G/H = M
\end{equation}
over the physical manifold $M$.

We are now interested in analyzing this bundle in greater detail. Our discussion is closely related to the one in~\cite{Hassler:2016srl}. Thus, we keep it short but concise. An arbitrary group element $g\in G$ is given by the coordinates $X^I$. Introducing the decomposition~\eqref{eqn2:splittingtA} requires us to assign the coordinates $x^i$ to the coset representative $m$ and the coordinates $x^{\tilde i}$ to the elements $h\in H$ (generated by $t_{\tilde \alpha}$). Hence, we obtain
\begin{equation}\label{eqn2:splittingcoord}
  X^I = \begin{pmatrix} x^i & x^{\tilde i} \end{pmatrix}
    \quad \text{with} \quad
  I = 1,\,\dots,\,\mathrm{dim}\,G\,, \quad
  i = 1,\,\dots,\,\mathrm{dim}\,G/H \quad \text{and} \quad
  \tilde i = \tilde 1,\,\dots,\,\mathrm{dim}\, H \,.
\end{equation}
$\pi$ removes the $x^{\tilde i}$ part of the coordinates $X^I$ in this adapted basis, i.e.
\begin{equation}\label{eqn2:piX}
  \pi(X^I) = x^i\,.
\end{equation}
One should also note that the associated differential map is given by
\begin{equation}\label{eqn2:pi*X}
  \pi_*( V^I \partial_I ) = V^i \partial_i\,.
\end{equation}
Each element of the Lie algebra $\mathfrak{g}$ generates a fundamental vector field on $G$ accordingly. Relating the two of them can be achieved by defining the map
\begin{equation}\label{eqn2:sharp}
  t_A^\sharp = E_A{}^I \partial_I
\end{equation}
which assigns a left invariant vector field to each $t_A\in\mathfrak{g}$. It possesses the important feature $\omega_L ( t_A^\sharp ) = t_A$ where
\begin{equation}\label{eqn2:leftinvMCF}
  (\omega_L)_g = g^{-1} \partial_I g \, d X^I = t_A E^A{}_I d X^I
\end{equation}
is the left invariant Maurer-Cartan form given on $G$. Both~\eqref{eqn2:leftinvMCF} and~\eqref{eqn2:sharp} are completely determined by the generalized background vielbein $E_A{}^I$ and its inverse transposed $E^A{}_I$. After taking the decomposition of the generators~\eqref{eqn2:splittingtA} and the coordinates~\eqref{eqn2:splittingcoord} into account, they can be expressed through
\begin{equation}\label{eqn2:compbgvielbein}
  E^A{}_I = \begin{pmatrix} E^\alpha{}_i & 0 \\
    E^{\tilde\alpha}{}_i & E^{\tilde\alpha}{}_{\tilde i}
  \end{pmatrix} \quad \text{and} \quad
  E_A{}^I = \begin{pmatrix} E_\alpha{}^i & E_\alpha{}^{\tilde i} \\
    0 & E_{\tilde\alpha}{}^{\tilde i}
  \end{pmatrix}\,.
\end{equation}

Further, it is possible to equip the principal bundle with the $\mathfrak{h}$-valued connection one-form $\omega$. This splits the tangent bundle $T G$ into a horizontal/vertical bundle $H G$/$V G$. Whereas the horizontal part 
\begin{equation}
  H G = \{ X \in T G \,|\, \omega(X) = 0 \}
\end{equation}
can be directly obtained from the connection one-form, the vertical part is given by the kernel of the differential map $\pi_*$. On top of that, we have to impose two consistency conditions
\begin{equation}\label{eqn2:constrconnection}
  \omega( t_{\tilde\alpha}^\sharp ) = t_{\tilde\alpha} \quad \text{and} \quad
  R_h^* \omega = Ad_{h^{-1}} \omega
\end{equation}
on $\omega$. Here, $R_g$ represents right translations on $G$ by the group element $g\in G$. Equivalently to \DFTwzw{}, the connection one-form is chosen in a way that the bundle $H G$ solves the linear version~\eqref{eqn2:sclinear} of the SC. Sticking to~\cite{Hassler:2016srl}, we define the projector $P_m$ at each point $m$ of the coset space $G/H$ as a map
\begin{equation}
  P_m: \mathfrak{g} \rightarrow \mathfrak{h}, \quad P_m = t_{\tilde\alpha} (P_m)^{\tilde\alpha}{}_B \theta^B
\end{equation}
where the dual one-form of the generator $t_A$ is denoted as $\theta^A$. $P_m$ cannot be chosen completely arbitrary. It has to fulfill the property
\begin{equation}\label{eqn2:Pmprop}
  P_m t_{\tilde\alpha} = t_{\tilde\alpha}\, \quad \forall t_{\tilde\alpha} \in \mathfrak{h}\,.
\end{equation}
So far, we have only defined the projection operator for coset representatives $m$ but not for arbitrary group elements $g$. However, it can be extended to the full group manifold $G$ through
\begin{equation}\label{eqn2:Pg}
  P_g = P_{m h} = \mathrm{Ad}_{h^{-1}} P_m \mathrm{Ad}_h\,.
\end{equation}
Subsequently, we derive the connection-one form
\begin{equation}\label{eqn2:omegag}
  \omega_g = P_g \, (\omega_L)_g
\end{equation}
with $(\omega_L)_g$ being the left invariant Maurer-Cartan from~\eqref{eqn2:leftinvMCF}.  As a consequence of~\eqref{eqn2:Pmprop}, it satisfies all the constraints given in~\eqref{eqn2:constrconnection}.

Finally, the $H$-principal bundle~\eqref{eqn2:Hprincipal} splits into sections $\sigma_i$ which are only defined on patches $U_i \subset M$. They possess the form
\begin{equation}\label{eqn2:sigmai}
  \sigma_i (x^j) = \begin{pmatrix} \delta^j_k x^k & f_i^{\tilde j} \end{pmatrix}
\end{equation}
in the coordinates~\eqref{eqn2:splittingcoord} and are determined by the functions $f^{\tilde j}_i$. As was the case in \DFTwzw, we choose these functions such that the pull back of the connection one-form $A_i = {\sigma_i}^* \omega$ vanishes in every patch $U_i$~\cite{Hassler:2016srl}. The only way to achieve this is if the corresponding field strength
\begin{equation}\label{eqn2:F_i=0}
  F_i (X,Y) = d A_i (X,Y) + [A_i(X), A_i(Y)] = 0
\end{equation}
vanishes. Then $A_i$ is of pure gauge and can be locally ``gauged away''. It is very important to note that the field strength describing the tangent bundle $T_M$ is a different one. For instance, take the four sphere $S^4 \cong$SO(5)/SO(4). The space cannot be parallelized and therefore its associated tangent bundle is not trivial. However, this statement is not in any relation with the field strength introduced in~\eqref{eqn2:F_i=0}.

\subsection{Connection and three-form potential}\label{sec:linkvomega}

In \DFTwzw{} the projection operator $P_m$ is in close connection with the NS/NS two-form field $B_{ij}$. Subsequently, we show that this result can be extended to the three-form $C_{ijk}$ for SL($5$) gEFT as well~\cite{Bosque:2017dfc}. Therefore, we analyze solutions of the SC's linear version~\eqref{eqn2:sclinearv0} more carefully. Using an appropriate SL($5$) rotation, it is always possible to cast $v^0_a$ into the canonical form
\begin{equation}\label{eqn2:v0acanonical}
  v^0_a = \begin{pmatrix} 1 & 0 & 0 & 0 & 0 \end{pmatrix}\,.
\end{equation}
This can be achieved by choosing an explicit basis 
\begin{equation}\label{eqn2:decompSL510}
  \alpha = \{12,\,13,\,14,\,15\} \quad \text{and} \quad
  \tilde\alpha = \{23,\,24,\,25,\,34,\,35,\,45\}
\end{equation}
for the indices used in our construction. Furthermore, we define the following tensor
\begin{equation}\label{eqn2:etatensor}
  \eta^{\alpha\beta,\tilde\gamma} = \frac12 \epsilon^{1 \hat\alpha \hat\beta \tilde\gamma}
\end{equation}
where $\hat\beta$ denotes the second fundamental index of the antisymmetric pair (e.g. $\beta=13$ and $\hat\beta=3$). The version of $\eta$ with lowered indices is defined in the same manner by
\begin{equation}
  \eta_{\alpha\beta,\tilde\gamma} = \epsilon_{1 \hat\alpha \hat\beta \tilde\gamma}
\end{equation}
and the normalization is chosen in a way that the conditions
\begin{equation}\label{eqn2:etasl5props}
  \eta^{\alpha\beta, \tilde\alpha} \eta_{\alpha\beta, \tilde\beta} = \delta^{\tilde\alpha}_{\tilde\beta}
    \quad \text{and} \quad
  \eta^{\alpha\beta, \tilde\alpha} \eta_{\gamma\delta, \tilde\alpha} = \delta^{[\alpha}_{[\gamma} \delta^{\beta]}_{\delta]}
\end{equation}
hold. With the help of this tensor, it is now possible to express the projector
\begin{equation}\label{eqn2:PmC}
  (P_m)^{\tilde\alpha}{}_B = \begin{pmatrix} \eta^{\gamma\delta,\tilde\alpha} C_{\beta\gamma\delta} & \delta^{\tilde\alpha}_{\tilde\beta} \end{pmatrix}
\end{equation}
through the totally antisymmetric field $C_{\alpha\beta\gamma}$ on $M$. As we will see in the next subsection, we can relate it to the three-form flux
\begin{equation}\label{eqn2:CfromC}
  C = \frac16 C_{\alpha\beta\gamma} E^\alpha{}_i E^\beta{}_j E^\gamma{}_k\, d x^i \wedge d x^j \wedge d x^k
\end{equation}
on the background by considering the SC solution's GG. It should be noted that the projector~\eqref{eqn2:PmC} can be chosen such that its kernel contains all the solutions of the linear SC~\eqref{eqn2:sclinear} for a fixed $v_a$. At this point, it is straightforward to identify
\begin{equation}\label{eqn2:v->C}
  C_{\alpha\beta\gamma} = \frac{1}{v_1} \sum_{\delta} \epsilon_{1 \hat\alpha\hat\beta\hat\gamma\hat\delta} v_{\hat\delta}\,.
\end{equation}
However, this equation is only valid for $v_1 \ne 0$. As rescaling leaves~\eqref{eqn2:sclinear} invariant all values of $v_a$ specifying a distinct solution of the section condition are elements of $\mathbb{RP}^4$. This projective space consists of five patches $U_a=\{ v_a \in \mathbb{R}^5 | v_a = 1\}$ in so-called homogeneous coordinates. From~\eqref{eqn2:v->C}, it follows that the projector and thereby the connection only cover the subset $U_1$ for possible SC solutions. As previously explained during the last subsection, a solution of the SC is characterized through a vanishing connection $A_i$. Therefore, it is possible to use~\eqref{eqn2:PmC} and~\eqref{eqn2:CfromC} to compute the three-from flux
\begin{equation}\label{eqn2:CfromE}
  C = - \frac16 \eta_{\alpha\beta,\tilde\gamma} E^\alpha{}_i E^\beta{}_j E^{\tilde\gamma}{}_k \, d x^i \wedge d x^j \wedge d x^k
\end{equation}
which emerges in the GG of this framework.

Again, it is a great consistency check to verify our results by considering the symmetry breaking of SL(5) to SL(4), discussed in subsection~\ref{sec:linconstsl5}. Now, the index of $v_a$ runs only from $a=1,\dots,4$ and the linear constraint takes on the form
\begin{equation}\label{eqn2:SL4SC}
  v_a^0 \epsilon^{a\beta c} = 0
\end{equation}
with $\epsilon$ being the four-dimensional totally antisymmetric tensor and the explicit basis
\begin{equation}
  \alpha = \{12,\,13,\,14\}
    \quad \text{and} \quad
  \tilde\alpha = \{23,\,24,\,34\}\,,
\end{equation}
if we assume $v^0_a = \begin{pmatrix} 1 & 0 & 0 & 0\end{pmatrix}$. At this point, it is essential to restrict $C$ used in our previous discussion to the two-from field
\begin{equation}
  C_{\alpha\beta4} = B_{\alpha\beta}
\end{equation}
as we want to describe SC solutions with $v_5=0$. Applying this reduction to~\eqref{eqn2:PmC} and~\eqref{eqn2:v->C} yields
\begin{equation}
  (P_m)^{\tilde \alpha}{}_B = \begin{pmatrix}\eta^{\gamma,\tilde\alpha} B_{\beta \gamma} & \delta^{\tilde\alpha}_{\tilde\beta} \end{pmatrix}
    \quad \text{and} \quad
  B_{\alpha\beta} = \frac{1}{v_1} \sum\limits_{\gamma} \epsilon_{1\hat\alpha\hat\beta\hat\gamma} v_{\hat\gamma}
\end{equation}
where
\begin{align}\label{eqn2:etasl4}
  \eta^{\alpha,\tilde\beta} = \epsilon^{\alpha \tilde\beta}
    \quad &\text{and} \quad
  \eta_{\alpha,\tilde\beta} = \epsilon_{\alpha \tilde\beta}\,.
\intertext{In this context, we choose the normalization for the $\eta$-tensor in a way that the analogous relations}
  \eta^{\alpha,\tilde\alpha} \eta_{\alpha,\tilde\beta} = \delta^{\tilde\alpha}_{\tilde\beta}
    \quad &\text{and} \quad
  \eta^{\alpha,\tilde\alpha} \eta_{\beta, \tilde\alpha} = \delta^\alpha_\beta
\end{align}
to~\eqref{eqn2:etasl5props} hold. Moreover, the same comments apply as before, but this time we have to consider $\mathbb{RP}^3$ instead of $\mathbb{RP}^4$. Thus, our results are in  perfect agreement with the ones of \DFTwzw{} in~\cite{Hassler:2016srl}. Particularly, the $\eta$-tensor gives rise to the O(3,3) invariant flat metric
\begin{equation}
  \eta^{AB} = \epsilon^{AB} = \begin{pmatrix} 0 & \eta^{\alpha,\tilde\beta} \\
    \eta^{\beta,\tilde\alpha} & 0 \end{pmatrix}
\end{equation}
with indices $A$, $B$ in the coordinate irrep $\mathbf{6}$ of $\mathfrak{sl}(4)$. As opposed to~\cite{Hassler:2016srl}, we use a different basis for the Lie algebra resulting in an off-diagonal form while $\eta^{\alpha,\tilde \beta}$ and $\eta^{\beta,\tilde \alpha}$ are not diagonal.

In general, it can get quite difficult to obtain a vanishing connection $A_i = 0$ for a given SC. Although, if $\mathfrak{m}$ and $\mathfrak{h}$ appearing in the decomposition~\eqref{eqn2:splittingtA} form a symmetric pair with the defining property
\begin{equation}\label{eqn2:symmetricpair}
  [\mathfrak{h},\mathfrak{h}] \subset \mathfrak{h} \,, \quad
  [\mathfrak{h},\mathfrak{m}] \subset \mathfrak{m} \quad \text{and} \quad
  [\mathfrak{m},\mathfrak{m}] \subset \mathfrak{h}\,,
\end{equation}
there exists an explicit construction procedure. It was already worked out for \DFTwzw{} in~\cite{Hassler:2016srl} and we modify it to gEFT in the remainder. Let us begin with the observation that the connection $A$ vanishes if 
\begin{equation}
  C_{ijk} = - \eta_{\alpha\beta,\tilde\gamma} E^\alpha{}_i E^\beta{}_j E^{\tilde\gamma}{}_k
\end{equation}
is totally antisymmetric in the indices $i$, $j$, $k$. This constraint can be rewritten as
\begin{equation}\label{eqn2:Ctotantisym}
  2 C_{ijk} - C_{kij} - C_{jki} = D_{ijk} = 0\,.
\end{equation}
We are now going to study it further. Therefore, it is convenient to introduce the following shortcut notation 
\begin{equation}\label{eqn2:def(...)}
  ( t_A , t_B , t_C ) = 2 \eta_{\alpha\beta,\tilde\gamma} - \eta_{\gamma\alpha,\tilde\beta} 
    - \eta_{\beta\gamma,\tilde\alpha}
\end{equation}
through which we can recast~\eqref{eqn2:Ctotantisym} as
\begin{equation}
  D_{ijk} = ( m^{-1} \partial_i m,\, m^{-1} \partial_j m,\, m^{-1} \partial_k m )
\end{equation}
after taking into account that $E^\alpha{}_i$ and $E^{\tilde\alpha}{}_i$ are particular components of the left invariant Maurer-Cartan form~\eqref{eqn2:leftinvMCF} with a section where $h$ is identified as the identity element of $H$. Following~\cite{Hassler:2016srl}, we work with the coset representative
\begin{equation}
  m = \exp\left(- f( x^i )\right)
\end{equation}
which yields the expansion
\begin{equation}
  m^{-1} \partial_i m = \sum\limits_{n=0}^\infty \frac1{(n+1)!} [f, \partial_i f]_n
    \quad \text{with} \quad
  [f,t]_n = [\underbrace{f [\dots, [f}_{n\text{ times}},t] \dots ]]\,.
\end{equation}
Hence, we are left with showing that
\begin{equation}\label{eqn2:hastobezero}
  D_{ijk} = \sum_{n_1=0}^\infty \, \sum_{n_2=0}^\infty \, \sum_{n_3=0}^\infty \frac{1}{
    (n_1 + 1)! (n_2 + 1)! (n_3 + 1)!} ([f,\partial_i f]_{n_1}, [f,\partial_j f]_{n_2}, 
    [f,\partial_k f]_{n_3} )
\end{equation}
vanishes under the constraint~\eqref{eqn2:symmetricpair}. Let us first simplify the notation using the abbreviation
\begin{equation}
  \langle n_1, n_2, n_3 \rangle_{ijk} := ([f,\partial_i f]_{n_1}, [f,\partial_j f]_{n_2}, [f,\partial_k f]_{n_3} )
\end{equation}
and rearrange the terms in~\eqref{eqn2:hastobezero} accordingly. It results in
\begin{equation}
\label{eqn2:Dijk}
  D_{ijk} = \sum\limits_{m=0}^\infty \, \sum\limits_{n_1+n_2+n_3=m} \,
    \frac{\langle n_1, n_2, n_3 \rangle_{ijk}}{(n_1 + 1)! (n_2 + 1)! (n_3 + 1)!}
    = \sum\limits_{m=0}^\infty S^m_{ijk}
\end{equation}
and allows us to verify this equation. The expression is zero if $S^m_{ijk}$ disappears for all $m$. Thus, we perform the computation order by order. Starting with
\begin{equation}
  S^0_{ijk} = \langle 0, 0, 0\rangle_{ijk} = 0\,.
\end{equation}
It vanishes since $(t_A, t_B, t_C)$ only gives a contribution if two of its arguments are in $\mathfrak{m}$ and one is in $\mathfrak{h}$ which is evident from the definition~\eqref{eqn2:def(...)}. Consequently, all arguments are in $\mathfrak{m}$. For the next order, we find
\begin{equation}
  S^m_{ijk} = \frac1{2!} \left( \langle 1, 0, 0 \rangle_{ijk} + \langle 0, 1, 0 \rangle_{ijk} + 
    \langle 0, 0, 1 \rangle_{ijk} \right) = 0
\end{equation}
which imposes a linear constraint on the structure constants $X_{AB}{}^C$. It is completely analogous to
\begin{equation}\label{eqn2:linconstflatconn}
  ( [t, \mathfrak{m}], \mathfrak{m}, \mathfrak{m} ) + 
  ( \mathfrak{m}, [t, \mathfrak{m}], \mathfrak{m} ) + 
  ( \mathfrak{m}, \mathfrak{m}, [t, \mathfrak{m}] ) = 0
\end{equation}
where $t$ represents a generator in the algebra $\mathfrak{sl}(5)$. Its components lie in the adjoint irrep $\mathbf{24}$. Furthermore, it should be noted that the splitting of the flat coordinate index $A$ into $\alpha$ and $\tilde\alpha$ singles out the specific direction $v_a^0$ in~\eqref{eqn2:v0acanonical}. As a consequence, it breaks SL(5) to SL(4) with the branching rule
\begin{equation}\label{eqn2:sollinconstflatconn}
  \mathbf{24} \rightarrow \xcancel{\mathbf{1}} + \mathbf{4} + \overline{\mathbf{4}} + \mathbf{15}
\end{equation}
of the adjoint irrep.  The only generator violating~\eqref{eqn2:linconstflatconn} corresponds to the canceled out irrep. In quadratic order, we obtain
\begin{equation}
  S^2_{ijk} = \frac14 \left( \langle 1, 1, 0 \rangle_{ijk} + \langle 0, 1,1 \rangle_{ijk} 
    + \langle 1, 0, 1 \rangle_{ijk} \right) + \frac16 \left( \langle 2, 0, 0 \rangle_{ijk} + 
    \langle 0, 2, 0 \rangle_{ijk} + \langle 0, 0, 2 \rangle_{ijk} \right) = 0
\end{equation}
which sets a quadratic constraint on the structure constants. A solution is given by the symmetric pair~\eqref{eqn2:symmetricpair}. It indicates that the first three terms are of the form $( \mathfrak{h}, \mathfrak{h}, \mathfrak{m})$, plus cyclic permutations, while the last three terms are covered by $(\mathfrak{m}, \mathfrak{m}, \mathfrak{m})$. As previously noticed, all of them vanish independently. More generally speaking, we are now left with
\begin{equation}\label{eqn2:[f,df]_nevenodd}
  [f, \partial_i f]_n \subset \begin{cases} \mathfrak{h} & n \text{ odd} \\ \mathfrak{m} & n \text{ even}
  \end{cases}
\end{equation}
which implies 
\begin{equation}
  \langle n_1, n_2, n_3 \rangle_{ijk} = 0 \quad \text{if} \quad
  n_1 \,\mathrm{mod}\, 2 + n_2 \,\mathrm{mod}\, 2 + n_3 \,\mathrm{mod}\, 2 = 1\,.
\end{equation}
Choose a specific contribution $\langle n_1, n_2, n_3 \rangle_{ijk}$ to $S^m{}_{ijk}$ in~\eqref{eqn2:Dijk} which is constrained by $n_1 + n_2 + n_3 = m$. If $m$ is even, then either two of the integers $n_1$, $n_2$, $n_3$ are odd while the third one is even or they are all even. For both cases $\langle n_1, n_2, n_3 \rangle_{ijk}$ must vanish and hence $S_m$ for even $m$ as well. Combining this observation with~\eqref{eqn2:[f,df]_nevenodd},~\eqref{eqn2:linconstflatconn} gives rise to
\begin{equation}\label{eqn2:shuffle<...>1}
  \langle n_1 + 1, n_2, n_3 \rangle_{ijk} + \langle n_1, n_2 + 1, n_3 \rangle_{ijk}  + 
    \langle n_1, n_2, n_3 + 1 \rangle_{ijk} = 0 \quad \text{for } n_1, n_2, n_3 \text{ even}\,.
\end{equation}
It allows us to simplify the cubic contribution
\begin{equation}
  S^3_{ijk} = - \frac1{4!} \left( \langle 3, 0, 0\rangle_{ijk} + \langle 0, 3, 0\rangle_{ijk} +
    \langle 0, 0, 3\rangle_{ijk}\right) = 0
\end{equation}
which is equivalent to~\eqref{eqn2:shuffle<...>1} after substituting $1$ with $3$. Iterating this step over and over again for $S^m_{ijk}$ with odd $m$, we are finally left with the conditions
\begin{equation}\label{eqn2:shuffle<...>2l+1}
  \langle n_1 + 2 l + 1, n_2, n_3 \rangle_{ijk} + 
    \langle n_1 , n_2 + 2 l + 1, n_3 \rangle_{ijk} + \langle n_1, n_2, n_3 + 2 l + 1 \rangle_{ijk} = 0
    \quad \forall \,\, l \in \mathbb{N}
\end{equation}
(again with $n_1$, $n_2$, $n_3$ even) which yield the desired result~\eqref{eqn2:Ctotantisym} and show $A_i = 0$. Their proof requires a generalization of~\eqref{eqn2:linconstflatconn} and uses that the generator $t$ in this equation lies within $\mathfrak{m}$. Subsequently, the commutator algebra for the symmetric pair~\eqref{eqn2:symmetricpair} restrains $t$ to the $\mathbf{4}$ and $\overline{\mathbf{4}}$ appearing during the decomposition~\eqref{eqn2:sollinconstflatconn}. Thus, we observe that~\eqref{eqn2:linconstflatconn} cannot be seen as an independent constraint, but follows directly from possessing a symmetric pair. It is further possible to represent two remaining, dual irreps by $x_i$ and $y^i$, with $i=1,\,\dots,\,4$, which yields
\begin{equation}
  [t, \partial_i t]_{2l + 1} = [t, \partial_i t]_1 \left( \frac{x_i y^i}4 \right)^l\,.
\end{equation}
It reduces~\eqref{eqn2:shuffle<...>2l+1} to~\eqref{eqn2:shuffle<...>1} and completes the prove. At last, note that there exists another possible case
\begin{equation}
  [ \mathfrak{m}, \mathfrak{m} ] \subset \mathfrak{m}
\end{equation}
for which one immediately obtains a flat connection. Then, all terms in~\eqref{eqn2:Dijk} are of the form $(\mathfrak{m},\mathfrak{m},\mathfrak{m})$ and disappear~\cite{Bosque:2017dfc}.

\subsection{Generalized Geometry}\label{sec:gg}

All of the SC solutions discussed in the previous two subsection are closely related with GG~\cite{Bosque:2017dfc}. As we want to show this connection explicitly, we have to introduce a map between $\mathfrak{h}$ and the vector space of two-forms $\Lambda^2 T_p^* M$ at each point $p\in M$. In particular, we utilize the $\eta$-tensor~\eqref{eqn2:etatensor} to define the isomorphism $\eta_p: \mathfrak{h} \rightarrow \Lambda^2 T_p^* M$ by
\begin{equation}\label{eqn2:etamap}
  \eta_p( t_{\tilde\gamma} ) = \frac12 \left. \eta_{\alpha\beta,\tilde\gamma} E^{\alpha}{}_i E^{\beta}{}_j dx^i \wedge dx^j\right|_{\sigma(p)} \,.
\end{equation}
Its inverse, given through
\begin{equation}
  \eta_p^{-1}( \nu ) = \left. \eta^{\alpha\beta, \tilde\gamma} t_{\tilde\gamma} \iota_{E_\alpha} \iota_{E_\beta} \nu \right|_{\sigma(p)}\,,
\end{equation}
follows directly from the properties of the $\eta$-tensors and the vectors $E_a = E_a{}^i \partial_i$. Equipped with this map as well as $\pi_*$~\eqref{eqn2:pi*X}, and $\omega_g(X)$~\eqref{eqn2:omegag} from subsection~\ref{sec:HPrincipal}, we are able to construct the generalized frame~\cite{Hull:2007zu,Coimbra:2011ky,Lee:2014mla}
\begin{equation}
  \hat E_A{}(p) = \pi_{*\,p} (t_A^\sharp) + \eta_p\,\omega_{\sigma(p)} ( t_A^\sharp)
\end{equation}
at each point $p$ of the physical space $M$. This map goes from a Lie algebra element $t_A$ to a vector in the generalized tangent space $T_p M \oplus \Lambda^2 T_p^* M$ of $M$ at $p$. It should be noted that we suppress the index denoting the patch dependence of the section for brevity here. However, the generalized frame $\hat E_A$ is explicitly section dependent. In case of a non-trivial $H$-principal bundle we obtain different frame fields in each patch and have to derive transition functions between them accordingly.

Exploiting the features of these maps
\begin{equation}
  \pi_*(t_{\tilde\alpha}^\sharp) = 0 \,, \quad
  \omega \sigma_* = \sigma^* \omega = A = 0\,, \quad
  \pi_* \sigma_*  = \mathrm{id}_{T M} \quad \text{and} \quad
  \omega(t_{\tilde\alpha}^\sharp) = t_{\tilde\alpha}\,,
\end{equation}
we conclude for the dual frame
\begin{equation}\label{eqn2:vielbeinhatinv}
  \hat E^A{} (p, v, \tilde v) = \theta^A \Big( \eta_p^{-1} (\tilde v) + \iota_{\sigma_{*\,p}(v)} \,(\omega_L)_{\sigma(p)} \Big) \,.
\end{equation}
where elements of the generalized tangent bundle are denoted by $V = v + \tilde v$ with $v \in T M$ and $\tilde v \in \Lambda^2 T^* M$. Finally, it is possible to expand the generalized frame and its dual into components, i.e.
\begin{equation}\label{eqn2:genframe&dualform}
  \hat E_A = \begin{pmatrix}
    E_{\alpha}{}^i \partial_i + C_{\alpha\beta\gamma} E^{\beta}{}_i E^{\gamma}{}_j \, d x^i \wedge d x^j \\
     \eta_{\beta\gamma, \tilde\alpha} E^{\beta}{}_i E^{\gamma}{}_j \, d x^i\wedge d x^j \end{pmatrix}
    \quad \text{and} \quad
    \hat E^A (v,\tilde v) = \begin{pmatrix} E^{\alpha}{}_i v^i \\
      \eta^{\beta\gamma,\tilde\alpha} ( E_{\beta}{}^i E_{\gamma}{}^j \tilde v_{ij}
      - C_{\beta\gamma\delta} E^{\delta}{}_i v^i )
    \end{pmatrix}
\end{equation}
which depend on $p$ and the indices representing the patch are suppressed. For the computation deriving the dual frame one has to bear in mind that
\begin{equation}\label{eqn2:Cterm}
  \theta^{\tilde\alpha} \Big( \omega_L ( \sigma_* v ) \Big) = - C_{\beta\gamma\delta} \eta^{\gamma\delta,\tilde\alpha} E^{\beta}{}_i v^i\,,
\end{equation}
stemming from $\sigma^* \omega = 0$. This result is in total alignment with the expectations, as it recovers the canonical vielbein of an SL($5$) theory~\cite{Hull:2007zu}
\begin{equation}
  \mathcal{V}_{\hat A}{}^{\hat I} = \begin{pmatrix} E_\alpha{}^i & E_\alpha{}^k C_{ijk} \\
    0 & E^\alpha{}_{[i} E^\beta{}_{j]}
  \end{pmatrix}
\end{equation}
and its inverse transposed. The three-form $C_{ijk}$ appearing during this equation is related to ours~\eqref{eqn2:Cterm} through $C_{ijk} = C_{\alpha\beta\gamma} E^\alpha{}_i E^\beta{}_j E^\gamma{}_k$.

As we have already determined the generalized frame and its inverse, we are now at the point to transport the generalized Lie derivative~\eqref{eqn2:genLieCov} to the generalized tangent bundle with its constituents
\begin{equation}\label{eqn2:VhatI}
  V^{\hat I} = \begin{pmatrix} v^i & \tilde v_{ij} \end{pmatrix} = V^A \hat E_A{}^{\hat I}
    \quad \text{and the dual} \quad
  V_{\hat I} = \begin{pmatrix} v_i & \tilde v^{ij} \end{pmatrix} V_A \hat E^A{}_{\hat I} \,.
\end{equation}
Here, we have introduced hatted indices to distinguish the generalized tangent bundle from the tangent bundle of the group manifold. Using said index convention,~\eqref{eqn2:genframe&dualform} takes on the form
\begin{equation}\label{eqn2:genframe&dual}
  \hat E_A{}^{\hat I} = \begin{pmatrix} E_\alpha{}^i & E_\alpha{}^k C_{kij} \\
    0 & \eta_{ij,\tilde\alpha} \end{pmatrix}
    \quad \text{and} \quad
   \hat E^A{}_{\hat I} = \begin{pmatrix} E^\alpha{}_i & 0 \\
    - C_{imn} \eta^{mn,\tilde\alpha} & \eta^{ij,\tilde\alpha} \end{pmatrix}
\end{equation}
with
\begin{equation}
  \eta^{ij,\tilde\alpha} = \eta^{\beta\gamma,\tilde\alpha} E_\beta{}^i E_\gamma{}^j
    \quad \text{and} \quad
  \eta_{ij,\tilde\alpha} = \eta_{\beta\gamma,\tilde\alpha} E^\beta{}_i E^\gamma{}_j\,.
\end{equation}
Applying the dual frame on the flat derivative, we find
\begin{equation}
  \partial_{\hat I} = \hat E^A{}_{\hat I} D_A = \begin{pmatrix} \partial_i & 0 \end{pmatrix}\,.
\end{equation}
For the infinitesimal parameter of a generalized diffeomorphism $\xi^{\hat J}$, we use the analogous convention as for $V^{\hat I}$~\eqref{eqn2:VhatI}. At this point, it is advantageous to decompose the generalized Lie derivative into two parts. First, we derive
\begin{equation}\label{eqn2:genlie0}
  \widehat{\mathcal{L}}_\xi V^{\hat I} = \xi^{\hat J} \partial_{\hat J} V^{\hat I} - V^{\hat J} \partial_{\hat J} \xi^{\hat I} + Y^{\hat I\hat J}{}_{\hat K\hat L} \partial_{\hat J} \xi^{\hat K} V^{\hat L}\,.
\end{equation}
Second, there exists the curved version $\mathcal{F}_{\hat I\hat J}{}^{\hat K} = \mathcal{F}_{AB}{}^C \hat E^A{}_{\hat I} \hat E^B{}_{\hat J} \hat E_C{}^{\hat K}$ of
\begin{equation}\label{eqn2:scFABC}
  \mathcal{F}_{AB}{}^C = X_{AB}{}^C - \widehat{\mathcal{L}}_{\hat E_A} \hat E_B{}^{\hat I} \hat E^C{}_{\hat I}\,.
\end{equation}
Together, they combine to the generalized Lie derivative
\begin{equation}\label{eqn2:gengeometr}
  \mathcal{L}_\xi V^{\hat I} = \widehat{\mathcal{L}}_\xi V^{\hat I} + \mathcal{F}_{\hat J\hat K}{}^{\hat I} \xi^{\hat J} V^{\hat K}\,.
\end{equation}
In the remaining part, we prove that $\widehat{\mathcal{L}}$ reproduces the untwisted generalized Lie derivative of GG and $\mathcal{F}_{\hat I\hat J}{}^{\hat K}$ realizes its twist with the non-vanishing form and vector components
\begin{align}
  \mathcal{F}^{ijkl}{}_{mn} &= X_{\tilde\alpha\tilde\beta}{}^{\tilde\gamma} \eta^{ij,\tilde\alpha}
    \eta^{kl,\tilde\beta} \eta_{mn,\tilde\gamma}
    \nonumber \\
  \mathcal{F}_i{}^{jkl} &= X_{\alpha\tilde\beta}{}^\gamma E^\alpha{}_i \eta^{jk,\tilde\beta}
    E_\gamma{}^l \nonumber \\
    \mathcal{F}^{ij}{}_k{}^l &= X_{\tilde\alpha\beta}{}^\gamma \eta^{ij,\tilde\alpha} E^\beta{}_k
    E_\gamma{}^l \nonumber \\
    \mathcal{F}_i{}^{jk}{}_{lm} &= \mathcal{F}_{\alpha\tilde\beta}{}^{\tilde\gamma} E^\alpha{}_i 
    \eta^{jk,\tilde\beta} \eta_{lm,\tilde\gamma} + \mathcal{F}_i{}^{jkn} C_{lmn} - \mathcal{F}^{nojk}{}_{lm}
    C_{ino} \nonumber \\
  \mathcal{F}^{ij}{}_{klm} &= \mathcal{F}_{\tilde\alpha\beta}{}^{\tilde\gamma} \eta^{ij,\tilde\alpha}
    E^\beta{}_k \eta_{lm,\tilde\gamma} + \mathcal{F}^{ij}{}_k{}^n C_{lmn} - \mathcal{F}^{ijno}{}_{lm}
    C_{kno} \nonumber \\
  \mathcal{F}_{ij}{}^k &= \mathcal{F}_{\alpha\beta}{}^\gamma E^\alpha{}_i E^\beta{}_j E_\gamma{}^k -
    \mathcal{F}_i{}^{lmk} C_{jlm} - \mathcal{F}^{lm}{}_j{}^k C_{ilm} \nonumber \\
  \mathcal{F}_{ijkl} &= \mathcal{F}_{\alpha\beta}{}^{\tilde\gamma} E^\alpha{}_i E^\beta{}_j
    \eta_{kl,\tilde\gamma} + \mathcal{F}_{ij}{}^m C_{klm} - \mathcal{F}_i{}^{mn}{}_{kl} C_{jmn} -
    \mathcal{F}^{mn}{}_{jkl} C_{imn} + \nonumber \\
    & \qquad
    \mathcal{F}_i{}^{mno} C_{jmn} C_{klo} + \mathcal{F}^{mn}{}_j{}^o C_{imn} C_{klo} -
    \mathcal{F}^{mnop}{}_{kl} C_{imn} C_{jop} \label{eqn2:twistcontrib}
\end{align}
with
\begin{align}
  \mathcal{F}_{\alpha\beta}{}^\gamma &= X_{\alpha\beta}{}^\gamma - f_{\alpha\beta}{}^\gamma &
  \mathcal{F}_{\alpha\beta}{}^{\tilde\gamma} &= X_{\alpha\beta}{}^{\tilde\gamma} - G_{ijkl} E_\alpha{}^i 
    E_\beta{}^j \eta^{kl,\tilde\gamma}  \nonumber \\
  \mathcal{F}_{\alpha\tilde\beta}{}^{\tilde\gamma} &= X_{\alpha\tilde\beta}{}^{\tilde\gamma} 
    + 2 f_{\alpha\beta}{}^\gamma \eta_{\delta\gamma,\tilde\beta} \eta^{\delta\beta,\tilde\gamma} &
    \mathcal{F}_{\tilde\alpha\beta}{}^{\tilde\gamma} &= - \mathcal{F}_{\beta\tilde\alpha}{}^{\tilde\gamma} + f_{\alpha\gamma}{}^{\delta} \eta_{\beta\delta,\tilde\alpha} \eta^{\alpha\gamma,\tilde\gamma} \,. \label{eqn2:genframetwist}
\end{align}
Here, we made use of
\begin{equation}\label{eqn2:fandGfluxEaI}
  f_{\alpha\beta}{}^\gamma = 2 E_{[\alpha}{}^i \partial_i E_{\beta]}{}^j E^\gamma{}_j
    \quad\text{and}\quad
  G = d C = \frac1{4!} G_{ijkl}\, d x^i\wedge d x^j \wedge d x^k \wedge d x^l\,.
\end{equation}
These are the geometric and four-form fluxes induced by the generalized frame~\eqref{eqn2:genframe&dual}.

Explicitly verifying that $\widehat{\mathcal{L}}$ is equivalent to the well-known generalized Lie derivative~\cite{Coimbra:2011ky,Berman:2011cg} of exceptional GG, we compute its individual components. The derivation of the first two terms in~\eqref{eqn2:gengeometr} is straightforward. However, the evaluation of the term containing the $Y$-tensor is more elaborate. Thus, we perform it componentwise and begin with
\begin{equation}
  Y^{AB}{}_{CD} \hat E_A{}^i \hat E_B{}^j = Y^{\alpha\beta}{}_{CD} \hat E_{\alpha}{}^i \hat E_{\beta}{}^j = 0\,.
\end{equation}
The last step exploits that the indices $\alpha$ and $\beta$ are by construction solutions of the SC. Hence, the vector components for the first two indices of the $Y$-tensor must vanish. Additionally, we already know that the form part of the partial derivative $\partial_{\hat I}$ disappears ($\partial^{ij} = 0$). Subsequently, the only contributing $Y$-tensor components are $Y_{ij}{}^{k}{}_{\hat{L}\hat{M}}$ which we calculate now. Therefore, let us consider
\begin{equation}
  Y_{ij}{}^k{}_{\hat L\hat M} = -\delta_{[i}^k E_{j]}{}^{a5} \epsilon_{a B C} E^B{}_{\hat L} E^C{}_{\hat M}
\end{equation}
and apply the dual generalized frame~\eqref{eqn2:genframe&dual} to find the non-vanishing component
\begin{equation}
  Y_{ij}{}^k{}_l{}^{mn} = -2 \delta_{[i}{}^k \delta_{j]}{}^{[m} \delta^{n]}{}_l\,.
\end{equation}
As a consequence of the $Y$-tensor's symmetry, we are able to compute the third term in~\eqref{eqn2:genlie0} and obtain
\begin{equation}
  Y_{ij}{}^k{}_{\hat L\hat M} \partial_k \xi^{\hat L} V^{\hat M} = 
    \partial_i \xi^k \tilde v_{kj} + \partial_j \xi^k \tilde v_{ik} - \partial_i \tilde\xi_{jk} v^k - \partial_j \tilde\xi_{ki} v^k \,.
\end{equation}
Combining this result with the other two terms, this finally gives rise to
\begin{equation}\label{eqn2:genLieGG}
  \widehat{\mathcal{L}}_\xi V^{\hat{I}} = \widehat{\mathcal{L}}_\xi \begin{pmatrix}
v^i \\ \tilde v_{ij}
\end{pmatrix} =  \begin{pmatrix}
 L_\xi v^i \\ L_\xi \tilde v_{ij} - 3v^k \partial_{[k} \tilde\xi_{ij]}
\end{pmatrix}\,,
\end{equation}
the generalized Lie derivative of exceptional GG~\cite{Coimbra:2011ky,Berman:2011cg}.

Similar to subsection~\ref{sec:linkvomega}, we probe our results by reducing the U-Duality group to the T-duality subgroup SL(4). It is now necessary to modify the map $\eta_p: \mathfrak{h} \rightarrow T_p^* M$ in the following manner
\begin{equation}
  \eta_p(t_{\tilde\beta}) = \left. \eta_{\alpha, \tilde\beta} E^\alpha{}_i d x^i \right|_{\sigma(p)}\,.
\end{equation}
This relation takes the altered $\eta$-tensor~\eqref{eqn2:etasl4} for our particular duality group into account. If we reprise all the steps given above, we obtain the generalized frame
\begin{equation}\label{eqn2:genFrameDFT}
  \hat E_A = \begin{pmatrix}
    E_{\alpha}{}^i \partial_i + B_{\alpha\beta} E^{\beta}{}_i \, d x^i \\
     \eta_{\beta, \tilde\alpha} E^{\beta}{}_i \, d x^i \end{pmatrix}
    \quad \text{its dual} \quad
    \hat E^A (v,\tilde v) = \begin{pmatrix} E^{\alpha}{}_i v^i \\
      \eta^{\beta,\tilde\alpha} ( E_{\beta}{}^i \tilde v_i
      - B_{\beta\gamma} E^{\gamma}{}_i v^i )
    \end{pmatrix}
\end{equation}
 and its corresponding GG generalized Lie derivative. It takes on the form given in~\eqref{eqn2:gengeometr} with
\begin{equation}
  \widehat{\mathcal{L}}_\xi V^{\hat I} =  \widehat{\mathcal{L}}_\xi \begin{pmatrix}
v^i \\ \tilde v_{i} \end{pmatrix} = \begin{pmatrix}
  L_\xi v^i \\ L_\xi \tilde v_i - 2 v^k \partial_{[k} \tilde \xi_{i]}
  \end{pmatrix}\,,
\end{equation}
whereas the twist in~\eqref{eqn2:scFABC} has now to be evaluated for the generalized frame in~\eqref{eqn2:genFrameDFT}~\cite{Bosque:2017dfc}. Performing a suitable basis change this result matches the one found in~\cite{Hassler:2016srl}.

\subsection{Lie algebra cohomology and dual backgrounds}\label{sec:dualbg}

Clearly, the SC allows for more than just one solution. These solutions emerge as a result of the freedom of choice regarding $v^0_a$~\eqref{eqn2:sclinearv0} and produce a distinguished splitting of the Lie algebra $\mathfrak{g}$ into the coset part $\mathfrak{m}$ and the subalgebra $\mathfrak{h}$.
Nevertheless, it is always possible to restore the canonical form of $v^0_a$~\eqref{eqn2:v0acanonical} by performing an SL($5$) rotation. In this special case, the index decomposition~\eqref{eqn2:decompSL510} remains intact and we only have to verify whether the generators $t_{\tilde \alpha}$ still form a Lie algebra $\mathfrak{h}$. Of course, this situation is very closely related to the \DFTwzw{} case studied in~\cite{Hassler:2016srl}. Furthermore, we also make use of Lie algebra cohomology to explore possible subgroups created by the Lie group $\mathfrak{g}$~\cite{Bosque:2017dfc}.

Let us now review the most important elements of this construction. We start by considering transformations in the coset SO(5)/SO(4)$\subset$ SL(5). All others leave the subalgebra $\mathfrak{h}$ invariant as they scale at most with $v^0_a$. An arbitrary coset element
\begin{equation}\label{eqn2:deformrepr}
  \mathcal{T}_A{}^B = \exp ( \lambda \, t_A{}^B )
\end{equation}
can be generated by performing the exponential map to a $\mathfrak{so}(5)$ generator $t$ acting on the coordinate irrep $\mathbf{10}$. As a consequence, the embedding tensor receives modifications according to
\begin{equation}
  X'_{AB}{}^C = \mathcal{T}_A{}^D \mathcal{T}_B{}^E X_{DE}{}^F \mathcal{T}_F{}^C\,.
\end{equation}
Expanding this expression in $\lambda$ yields
\begin{equation}
  X'_{AB}{}^C = X_{AB}{}^C + \lambda \delta X_{AB}{}^C + \lambda^2 \delta^2 X_{AB}{}^C + \dots
\end{equation}
and we read off the $\mathfrak{g}$-valued two-forms
\begin{equation}
  c_n = t_C ( \delta^n X_{AB}{}^C ) \theta^A \wedge \theta^B\,.
\end{equation}
It only permits for transformations with $\delta^n X_{\tilde\alpha \tilde\beta}{}^{\gamma} = 0$. Otherwise $\mathfrak{h}$ fails to be a subalgebra. Finally, we need to verify whether the restricted forms
\begin{equation}
  c_n = t_{\tilde\gamma} ( \delta^n X_{\tilde\alpha \tilde\beta}{}^{\tilde\gamma} ) \theta^{\tilde \alpha} \wedge \theta^{\tilde \beta} 
\end{equation}
are elements of the Lie algebra cohomology $H^2(\mathfrak{h},\mathfrak{h})$. If so, they generate infinitesimal non-trivial deformations of $\mathfrak{h}$. Obstructions to the integrability of this deformation lie in $H^3(\mathfrak{h}, \mathfrak{h})$~\cite{Bosque:2017dfc}.

\subsection{Generalized frame field}\label{sec:genframe}
An important application of the here presented formalism lies in the construction of the frame fields $\mathcal{E}_A{}^{\hat I}$ of generalized parallelizable manifolds $M$. During this subsection, we are going to prove that
\begin{equation}\label{eqn2:genparaframe}
  \mathcal{E}_A{}^{\hat I} = - M_A{}^B \hat E'_B{}^{\hat I}
\end{equation}
satisfies the defining equation 
\begin{equation}
\label{eqn2:genparaframeintro}
\widehat{\mathcal{L}}_{\mathcal{E}_A} \mathcal{E}_B = X_{AB}{}^C \mathcal{E}_C
\end{equation}
once an additional linear constraint on the structure constants $X_{AB}{}^C$ holds~\cite{Bosque:2017dfc}. The procedure is performed step by step and we begin with the frame $\hat E'_A{}^{\hat I}$. It deviates from~\eqref{eqn2:genframe&dual} as we use a three-from $\mathcal{C}$ instead of $C$. We start by computing
\begin{equation}
  X'_{AB}{}^C = \widehat{\mathcal{L}}_{\hat E'_A} \hat E'_B{}^{\hat I} {E'}^C{}_{\hat I}
\end{equation}
which possesses the following non-trivial components
\begin{align}
  X'_{\alpha\beta}{}^\gamma &= f_{\alpha\beta}{}^\gamma &
  X'_{\alpha\beta}{}^{\tilde\gamma} &= \mathcal{G}_{ijkl} E_\alpha{}^i E_\beta{}^j \eta^{kl,\tilde\gamma}
    \nonumber \\
  X'_{\alpha\tilde\beta}{}^{\tilde\gamma} &= 2 f_{\alpha\beta}{}^\gamma \eta_{\delta\gamma,\tilde\beta}
    \eta^{\delta\beta,\tilde\gamma} &
  X'_{\tilde\alpha\beta}{}^{\tilde\gamma} &= - X'_{\beta\tilde\alpha}{}^{\tilde\gamma} - f_{\alpha\gamma}{}^{\delta} \eta_{\beta\delta,\tilde\alpha} \eta^{\alpha\gamma,\tilde\gamma}\,.
\end{align}
As was the case before, $f_{\alpha\beta}{}^\gamma$ represents the geometric flux~\eqref{eqn2:fandGfluxEaI} and
\begin{equation}
  \mathcal{G} = d \mathcal{C} = \frac{1}{4!} \mathcal{G}_{ijkl}\, d x^i \wedge d x^j \wedge d x^k \wedge d x^l
\end{equation}
is the field strength associated to $\mathcal{C}$. In equation~\eqref{eqn2:genparaframe}, $\hat E'_A{}^{\hat I}$ gets twisted by the SL(5) rotation
\begin{equation}\label{eqn2:SL5rot}
  M_B{}^A t_A = m^{-1} t_B m = (\mathrm{Ad}_{m^{-1}})_B{}^A t_A
\end{equation}
with its inverse transposed
\begin{equation}
  t_A M^A{}_B = m t_B m^{-1}\,.
\end{equation}
Now, we combine the two and obtain
\begin{equation}
  X''_{AB}{}^C = \widehat{\mathcal{L}}_{M_A{}^D \hat E'_D} (M_B{}^E \hat E'_E{}^{\hat I})
    M^C{}_F \hat{E'}{}^F{}_{\hat I}\,.
\end{equation}
It is convenient to simplify the result by writing
\begin{equation}\label{eqn2:X''ABC}
  X''_{AB}{}^C = X'''_{DE}{}^F M_A{}^D M_B{}^E M^C{}_F
    \quad \text{with} \quad
  X'''_{AB}{}^C = X'_{AB}{}^C + 2 T_{[AB]}{}^C + Y^{CD}{}_{EB} T_{DA}{}^E
\end{equation}
and
\begin{equation}
  T_{AB}{}^C = - \hat E'_A{}^{\hat I} \partial_{\hat I} M^D{}_B M_D{}^C\,.
\end{equation}
As a consequence of the particular form of $M_B{}^A$ in~\eqref{eqn2:SL5rot}, this tensor can evaluated to be
\begin{equation}
  T_{AB}{}^C = - \hat E_A{}^i E^D{}_i X_{DB}{}^C = 
    \begin{pmatrix} - X_{\alpha B}{}^C +
    \eta^{\delta\epsilon,\tilde\delta} C_{\alpha\delta\epsilon} X_{\tilde\delta B}{}^C & 0\end{pmatrix}\,.
\end{equation}
Secondly, we note that for a SC solution the connection $A$ must vanish. Therfore, we identify $E_\alpha{}^i E^{\tilde\beta}{}_i = - \eta^{\gamma\delta,\tilde\beta} C_{\alpha\gamma\delta}$. 
Inserting the solution for $T_{AB}{}^C$ into~\eqref{eqn2:X''ABC} gives rise to the non-vanishing components
\begin{align}
  X'''_{\tilde\alpha\tilde\beta}{}^{\tilde\gamma} &= - X_{\tilde\alpha\tilde\beta}{}^{\tilde\gamma} &
  X'''_{\alpha\tilde\beta}{}^{\gamma} &= - X_{\alpha\tilde\beta}{}^{\gamma} &
  X'''_{\tilde\alpha\beta}{}^{\gamma} &= - X_{\tilde\alpha\beta}{}^{\gamma} &
    \nonumber \\
  X'''_{\alpha\tilde\beta}{}^{\tilde\gamma} &= -2 X'''_{\alpha\beta}{}^\gamma
    \eta_{\delta\gamma,\tilde\beta} \eta^{\delta\beta,\tilde\gamma} &
  X'''_{\tilde\alpha\beta}{}^{\tilde\gamma} &= - X'''_{\beta\tilde\alpha}{}^{\tilde\gamma}
    \nonumber \\
  X'''_{\alpha\beta}{}^\gamma &= - 2 X_{\alpha\beta}{}^\gamma + 
    2 X_{\tilde\alpha[\beta}{}^\gamma C_{\alpha]\delta\epsilon} \eta^{\delta\epsilon,\tilde\alpha}
    + f_{\alpha\beta}{}^\gamma
\end{align}
and
\begin{gather}
  X'''_{\alpha\beta}{}^{\tilde\gamma} = - 2 X_{\alpha\beta}{}^{\tilde\gamma} +
      2 X_{\gamma\alpha}{}^{\tilde\alpha} \eta_{\delta\beta,\tilde\alpha} \eta^{\gamma\delta,\tilde\gamma} -
      ( 2 X_{\gamma\beta}{}^{\delta} C_{\delta\epsilon\alpha} -
        4 X_{\gamma\alpha}{}^{\delta} C_{\delta\epsilon\beta} ) \eta^{\gamma\epsilon,\tilde\gamma} - 
          \nonumber \\
      2 X_{\alpha\beta}{}^\gamma C_{\delta\epsilon\gamma} \eta^{\delta\epsilon,\tilde\gamma}
      + \mathcal{G}_{ijkl} E_\alpha{}^i E_\beta{}^j \eta^{kl,\tilde\gamma} \label{eqn2:X'''ABCcomp}
\end{gather}
after imposing the constraints
\begin{equation*}\label{eqn2:linconst3}\tag{C3}
  X_{A\tilde\beta}{}^{\tilde\gamma} = -2 X_{A\beta}{}^\gamma
    \eta_{\delta\gamma,\tilde\beta} \eta^{\delta\beta,\tilde\gamma}
    \quad \text{and} \quad
  X_{\alpha\gamma}{}^\delta \eta_{\beta\delta,\tilde\alpha} \eta^{\alpha\gamma,\tilde\gamma} = 0\,.
\end{equation*}
At this stage, it appears that~\eqref{eqn2:SL5rot} was a good choice. Up to a sign, many components are already as expected. Taking the explicit form
\begin{equation}
  f_{\alpha\beta}{}^\gamma = X_{\alpha\beta}{}^\gamma - 2 X_{\tilde\alpha[\beta}{}^\gamma C_{\alpha]\delta\epsilon} \eta^{\delta\epsilon,\tilde\alpha}
\end{equation}
for the geometric flux into account, the situation improves even further. It yields
\begin{equation}
  X'''_{\alpha\tilde\beta}{}^{\tilde\gamma} = - X_{\alpha\tilde\beta}{}^{\tilde\gamma}\,, \quad
  X'''_{\tilde\alpha\beta}{}^{\tilde\gamma} = - X_{\tilde\alpha\beta}{}^{\tilde\gamma}
    \quad \text{and} \quad
  X'''_{\alpha\beta}{}^{\gamma} = - X_{\alpha\beta}{}^\gamma
\end{equation}
when imposing the constraints~\eqref{eqn2:linconst3}. On top of that, we are left with the last contribution~\eqref{eqn2:X'''ABCcomp} which should evaluate to $-X_{\alpha\beta}{}^{\tilde\gamma}$. However, it requires to find an appropriate choice for the four-form
\begin{equation}
  \mathcal{G}_{ijkl} = f(x^1, x^2, x^3, x^4) \epsilon_{ijkl}\,.
\end{equation}
As the four-form is the top-form on $M$ it can only possess one degree of freedom. It is captured by the function $f$. Applying this particular ansatz, the last term in~\eqref{eqn2:X'''ABCcomp} reduces to
\begin{equation}
  \mathcal{G}_{ijkl} E_\alpha{}^i E_\beta{}^j \eta^{lk,\tilde\gamma} = f \det (E_\rho{}^i) \epsilon_{1\hat\alpha\hat\beta\hat\gamma\hat\delta} \eta^{\gamma\delta,\tilde\gamma}\,.
\end{equation}
If we choose $f = \lambda \det (E^\rho{}_i)$ for an appropriate, constant $\lambda$, something spectacular occurs and we obtain $X'''_{\alpha\beta}{}^{\tilde\gamma} = -X_{\alpha\beta}{}^{\tilde\gamma}$. The main reason for this result is that the structure constants $X_{AB}{}^C$ cannot be chosen arbitrarily, but are highly constrained by the linear conditions~\eqref{eqn2:linconst1},~\eqref{eqn2:linconst2} and~\eqref{eqn2:linconst3}. We solved the first two in subsection~\ref{sec:linconstsl5} and at the end of this subsection we demonstrate the solution to the last one. For the time being, let us continue with
\begin{equation}\label{eqn2:correctX'''}
  X'''_{AB}{}^C = - X_{AB}{}^C \quad \text{under~\eqref{eqn2:linconst1} -~\eqref{eqn2:linconst3}}\,.
\end{equation}
In general, the structure constants of a Lie algebra are preserved under the adjoint action~\eqref{eqn2:SL5rot}. Subsequently, we immediately conclude
\begin{equation}
  X''_{AB}{}^C = X'''_{AB}{}^C = - X_{AB}{}^C\,.
\end{equation}
Up to the minus sign, it is exactly the result we have been seeking. As we want to get rid of the remaining minus sign, we insert an additional minus in the generalized frame field $\mathcal{E}_A{}^{\hat I}$~\eqref{eqn2:genparaframe}. The result is now what we expected, i.e.~\eqref{eqn2:genparaframeintro}. We already stated above, the three-from $\mathcal{C}$ it accommodates needs to be chosen in a way that
\begin{equation}\label{eqn2:G=lambdavol}
  \mathcal{G} = d \mathcal{C} = \lambda \det(E^\rho{}_i) d x^1 \wedge d x^2 \wedge d x^3 \wedge d x^4 = 
    \lambda \mathrm{vol}\,,
\end{equation}
where $\mathrm{vol}$ denotes the volume form on $M$ induced by the frame field $E^\alpha{}_i$.

Finally, we are left with obtaining the solutions of the linear constraint~\eqref{eqn2:linconst3}. Otherwise the previous does not hold. Identifying these solutions requires us to analyze the embedding tensor components of the $\mathbf{15}$~\cite{Samtleben:2005bp}
\begin{equation}\label{eqn2:emb1055b}
  X_{abc}{}^d = \delta_{[a}^d Y_{b]c}
\end{equation}
parameterized by the symmetric matrix $Y_{ab}$, and of the $\overline{\mathbf{40}}$~\cite{Samtleben:2005bp}
\begin{equation}\label{eqn2:embeddingt40}
  X_{abc}{}^d = -2 \epsilon_{abcef} Z^{ef,d}
\end{equation}
given through the tensor $Z^{ab,c}$ with $Z^{ab,c}=Z^{[ab],c}$ and $Z^{[ab,c]}=0$. The structure constants of the associated Lie algebra $\mathfrak{g}$ are given from the further embedding into $\mathbf{10}\times\mathbf{10}\times\overline{\mathbf{10}}$~\cite{Samtleben:2005bp}
\begin{equation}\label{eqn2:emb101010b}
  X_{AB}{}^C = X_{a_1 a_2, b_1 b_2}{}^{c_1 c_2} = 2 X_{a_1 a_2[b_1}{}^{[c_1} \delta_{b_2]}^{c_2]}\,.
\end{equation}
If all the contributions are only originating in the $\mathbf{15}$, this expression equals the structure coefficients since the corresponding group manifold is ten-dimensional. We are going to start by studying this case. Performing a segregation of the indices $A$, $B$, $C$, \dots{} into a coset component $\alpha$ and a subalgebra part $\tilde\alpha$, according to~\eqref{eqn2:decompSL510}, identifies one special direction in the fundamental irrep of SL($5$). This direction is determined by $v^0_a$ in~\eqref{eqn2:v0acanonical} and gives rise to the branching rule
\begin{equation}\label{eqn2:sollinconst3_15}
  \mathbf{15} \rightarrow \mathbf{1} + \xcancel{\mathbf{4}} + \mathbf{10} 
\end{equation}
from SL(5) to SL(4). Here, the crossed out irreps would violate the linear constraint~\eqref{eqn2:linconst3}. Taking only the remaining irreps into account, all terms accommodating $C_{\alpha\beta\gamma}$ in $X'''_{\alpha\beta}{}^{\tilde\gamma}$ vanish. As we want~\eqref{eqn2:correctX'''} to hold, it is essential that the relation
\begin{equation}
  X_{\alpha\beta}{}^{\tilde\gamma} - 2 X_{\gamma\alpha}{}^{\tilde\alpha} \eta_{\delta\beta,\tilde\alpha} \eta^{\gamma\delta,\tilde\gamma} = \lambda \epsilon_{1\hat\alpha\hat\beta\hat\gamma\hat\delta} \eta^{\gamma\delta,\tilde\gamma}
\end{equation}
is fulfilled. This is indeed the case, if we conclude
\begin{equation}\label{eqn2:lambdafromY11}
  \lambda = -\frac34 \, Y_{11}\,.
\end{equation}
In principle, it should be possible to construct generalized parallelizable spaces $M$ for all the remaining gaugings in~\eqref{eqn2:sollinconst3_15}. However, the construction procedure relies on obtaining a flat connection $A$ to solve the SC. Yet, deriving such a vanishing connection can be quite cumbersome. Although, as we explained at the end of~\ref{sec:linkvomega}, once we have a symmetric space $M$ it is straightforward to solve this task immediately. Fortunately, all remaining irreps in~\eqref{eqn2:sollinconst3_15} allow us to find symmetric pairs $\mathfrak{m}$ as well as $\mathfrak{h}$ and we can solve the SC right away. On top of that, the solutions of the quadratic constraint~\eqref{eqn2:quadconstr} are also known. Therefore, the resulting group manifolds highly depend on the eigenvalues of the symmetric, real matrix $Y_{ab}$.
If we assume $p$ of them to be positive, $q$ to be negative, and $r$ to be zero, we obtain
\begin{equation}
  G = \text{CSO}(p,q,r) = SO(p,q) \ltimes \mathbb{R}^{(p+q) r}
    \quad \text{with} \quad
  p + q + r = 5\,.
\end{equation}
Our construction algorithm applies to all corresponding generalized frames $\mathcal{E}_A$. These have also been constructed in~\cite{Hohm:2014qga} by exploiting a clever ansatz in a particular coordinate system. Previous to this work,~\cite{Lee:2014mla} already derived the generalized frame for SO(5) ($p$=5, $q$=$r$=0), the four-sphere with $G$-flux.

Only the gaugings in the $\mathbf{40}$ for group manifolds with $\dim G<10$ are relevant. As we observed in subsection~\ref{sec:linconstsl5}, the irreps of the embedding tensor branch into the individual U-duality subgroups. In this case, $v^0_a$ in~\eqref{eqn2:v0acanonical} again singles out a specific direction and gives rise to an additional branching. We consider the SL(3)$\times$SL(2) solutions in figure~\ref{fig:sollinconst} to see how this works. For $\dim G=9$ the relevant components of the embedding tensor
\begin{equation}\label{eqn2:sollinconst3_40_9}
  (\mathbf{1},\mathbf{3}) + (\mathbf{3},\mathbf{2}) + (\mathbf{6},\mathbf{1}) + (\mathbf{1},\mathbf{2}) \rightarrow (\mathbf{1},\mathbf{3}) + \xcancel{(\mathbf{1},\mathbf{2})} + (\mathbf{2},\mathbf{2}) + 
  (\mathbf{1},\mathbf{1}) + \xcancel{(\mathbf{2},\mathbf{1})} + (\mathbf{3},\mathbf{1}) + (\mathbf{1},\mathbf{2})
\end{equation}
branch from SL(3)$\times$SL(2) to SL(2)$\times$SL(2). The crossed out irreps originate in the $\mathbf{4}$ of~\eqref{eqn2:sollinconst3_15}. Only the last irrep $(\mathbf{1},\mathbf{2})$ stems from the $\mathbf{40}$. However, it does not allow for a symmetric pair. Nevertheless, it is possible to construct a generalized frame field for the four-torus with geometric flux which we do in subsection~\ref{sec:T4Gflux}. We realize it through a gauging in this irrep. The relation~\eqref{eqn2:lambdafromY11} is still valid for the scaling factor $\lambda$ in~\eqref{eqn2:G=lambdavol}. Furthermore, one can continue this discussion for group manifolds with $\dim G < 9$.
It is not necessary to present it in this context, as all the examples we are going to provide in the next section are already covered by the cases above~\cite{Bosque:2017dfc}.

\section{Examples}\label{sec:examples}

In this section, we want to demonstrate some illustrative examples for the construction prescribed in the previous sections. We start with the four-torus with $G$-flux and its dual backgrounds. Afterwards, we turn to the four-sphere with $G$-flux. The former is already well-known from the conventional EFT description, but in our framework we are also going to observe how naturally the dual backgrounds arise. On top of that, we can now study group manifolds $G$ with $\dim G<10$ originating in the gaugings of the 
$\mathbf{40}$. For this case SL($5$) breaks down to SL($3$)$\times$SL($2$). A much more elaborate configuration presents the four-sphere with $G$-flux. It is associated with the group manifold SO($5$). This example has already been analyzed in~\cite{Lee:2014mla,Hohm:2014qga} and therefore provides an additional comparison of our resulting generalized frame field $\mathcal{E}_A$ with the literature.

\subsection{Duality-chain of the four-torus with \texorpdfstring{$G$}{G}-flux}\label{sec:T4Gflux}

In string theory there exists the famous duality chain~\eqref{int:dualitychain}~\cite{Shelton:2005cf}
\begin{equation}
 H_{ijk} \leftrightarrow f_{ij}{}^k \leftrightarrow Q_i{}^{jk} \leftrightarrow R^{ijk} \,,
\end{equation}
where the adjoining backgrounds are related by a single T-duality mapping IIA $\leftrightarrow$ IIB string theory. During the remainder of this section, we are interested in showing how parts of this duality chain result from different SC solutions on a ten- and a nine-dimensional group manifold~\cite{Bosque:2017dfc}. If we want to uplift these examples to M-theory, it is only necessary to consider IIA backgrounds and two T-duality transformations connecting IIA $\leftrightarrow$ IIA string theory. Subsequently, the previously mentioned duality chain decomposes into two distinct duality chains
\begin{equation}
  H_{ijk} \leftrightarrow Q_i{}^{jk}
\end{equation}
and
\begin{equation}
 f_{ij}{}^k \leftrightarrow R^{ijk} \,.
\end{equation}
Contemplating this situation in M-theory works quite similar. We apply three U-duality transformations to guarantee we map M-theory onto itself. It can be thought of as considering a $T^3$ in the limit of vanishing volume. Indeed, an $S^1$ with vanishing volume would have given us weakly-coupled IIA string theory, whereas for a $T^2$ with vanishing volume we would have obtained IIB string theory (One can think of taking repeated small radii limits regarding the two circles of $T^2$). However, in our case we get weakly coupled IIA compactified on a small circle. Performing a T-Duality transformation on this circle yields IIB in the decompactification limit. Hence, we open a new dual direction for every two-cycle of vanishing volume. As a consequence, a $T^3$ with vanishing volume implies that we lose three directions but open up three new ones (One for each of the three two-cycles in $T^3$). Ergo, we arrive at an eleven-dimensional background once again. Another approach works by identifying two directions of the U-Duality transformation with the two directions of the T-Duality transformation and the third one with the M-theory circle. This ansatz also takes care of the proper dilaton transformation. The argumentation makes it evident that for M-theory the $T^4$ duality chain also decomposes while we find \begin{equation}\label{eqn2:chain1}
 G_{ijkl} \leftrightarrow Q_{i}{}^{jkl}
\end{equation}
and
\begin{equation}\label{eqn2:chain2}
 f_{ij}{}^k \leftrightarrow R^{i,jklm} \,.
\end{equation}
Although, it is only possible to realize the former duality within our framework. We cannot perceive the second duality as the $R$-flux background does not admit a maximally isotropic subalgebra $\mathfrak{h}$. This observation is in perfect agreement with the \DFTwzw{} case found in~\cite{Hassler:2016srl}.

The decomposition~\eqref{eqn2:chain1} and~\eqref{eqn2:chain2} of the duality chain is manifest in the embedding tensor as well~\cite{Blair:2014zba}. For SL($5$) it possesses two irreducible representations (We do not count the trombone as we neglect it in our framework). Each individual chain corresponds to one of these irreps. The duality transformations are then implemented by SL($5$) rotations which do not mix different irreps~\cite{Bosque:2017dfc}. 

\subsubsection{Gaugings in the \texorpdfstring{$\mathbf{15}$}{15}}\label{sec:T4_15}

Now, we begin with the first duality chain. It is fully covered by the the irrep $\mathbf{15}$~\cite{Blair:2014zba} which can be expressed through the symmetric tensor $Y_{ab}$. As a consequence, we obtain the embedding tensor~\eqref{eqn2:emb1055b} and the corresponding structure coefficients emerge from~\eqref{eqn2:emb101010b}. Furthermore, by applying a SO($5$) rotation we can always diagonalize the symmetric matrix $Y_{ab}$. The gaugings in the $\mathbf{15}$ automatically satisfy the quadratic constraint. Then, the four-torus with $\mathbf{g}$ units of $G$-flux can be cast in the form
\begin{equation}\label{eqn2:gaugingT4Gflux}
  Y_{ab} = - 4 \mathbf{g} \, \diag( 1,\,0,\,0,\,0,\,0 )\,.
\end{equation}
This specific solution is consistent with the vector $v_0^a$ in~\eqref{eqn2:v0acanonical} and the decomposition~\eqref{eqn2:decompSL510} of the $\mathbf{10}$ index $A = (\alpha, \tilde\alpha)$. It yields the group manifold $G =\text{CSO}(1,0,4)$ with an abelian subgroup $H$ being generated by all infinitesimal translations in $\mathbb{R}^6$. We work with the $21$-dimensional, faithful representation of $\mathfrak{g}$ derived in appendix~\ref{app:faithfulrepr} to acquire the matrix representation
\begin{align}
  m &= \exp( t_1 x^1 ) \exp( t_2 x^2 ) \exp( t_3 x^3 ) \exp{ t_4 x^4} \quad \text{and} \label{eqn2:matrixexpm}\\
  h &= \exp( t_{\tilde 1} x^{\tilde 1}) \exp( t_{\tilde 2} x^{\tilde 2}) \exp( t_{\tilde 3} x^{\tilde 3})
  \exp(t_{\tilde 4} x^{\tilde 4}) \exp( t_{\tilde 5} x^{\tilde 5}) \exp(t_{ \tilde 6} x^{\tilde 6}) \label{eqn2:matrixexph}
\end{align}
of the Lie group $G$. Unfortunately, the resulting group is not yet compact and therefore does not depict the background we are interested in (Clearly a torus is compact). Hence, we need to mod out the discrete subgroup CSO(1,0,4,$\mathds{Z}$) from the left. 
It is equivalent to imposing certain coordinate identifications~\eqref{eqn2:coordident1} and~\eqref{eqn2:coordident2} which we obtained in appendix~\ref{app:faithfulrepr}.

For this particular setup, the connection $A = A^{\tilde \alpha} t_{\tilde \alpha}$ takes on the form
\begin{align}
  A^{\tilde 1} &= \Big[ ( \mathbf{g} x^2 + C_{134} ) d x^1 + C_{234} \, d x^2 \Big] \,, &
  A^{\tilde 2} &= \Big[ ( \mathbf{g} x^3 - C_{124} ) d x^1 + C_{234} \, d x^3 \Big] \,,
    \nonumber \\
  A^{\tilde 3} &= \Big[ ( \mathbf{g} x^3 - C_{124} ) d x^2 - C_{134} \, d x^3 \Big] \,, &
  A^{\tilde 4} &= \Big[ ( \mathbf{g} x^4 + C_{123} ) d x^1 + C_{124} \, d x^4 \Big] \,,
    \nonumber \\
  A^{\tilde 5} &= \Big[ ( \mathbf{g} x^4 + C_{123} ) d x^2 - C_{134} \, d x^4 \Big] \,, &
  A^{\tilde 6} &= \Big[ ( \mathbf{g} x^4 + C_{123} ) d x^3 + C_{124} \, d x^4 \Big]
\end{align}
in the patch we are studying. The field strength $F = d A$ vanishes for the three-form field
\begin{equation}\label{eqn2:CT4Gflux}
  C = \frac{\mathbf{g}}2 ( x^1\,d x^2 \wedge d x^3 \wedge d x^4 - x^2\,d x^1 \wedge d x^3 d \wedge x^4 + x^3\,d x^1 \wedge d x^2 \wedge d x^4 - x^4\,d x^1\wedge x^2 \wedge x^3 )\,,
\end{equation}
with flux contribution
\begin{equation}\label{eqn2:GfieldhatE}
  G_{\hat E} = d C = 2 \mathbf{g}\, d x^1\wedge d x^2 \wedge d x^3 \wedge d x^4
\end{equation}
to the generalized frame field $\hat E_A$. As we are interested in setting $A=0$ within the current patch, we perform the transformation $g\rightarrow g \exp( t_{\tilde\alpha} \lambda^{\tilde\alpha} )$ on all group elements with
\begin{align}
  \lambda^{\tilde 1} &= - \frac{\mathbf{g}}2 x^1 x^2\,, &
  \lambda^{\tilde 2} &= - \frac{\mathbf{g}}2 x^1 x^3\,, &
  \lambda^{\tilde 3} &= - \frac{\mathbf{g}}2 x^2 x^3\,, \nonumber \\
  \lambda^{\tilde 4} &= - \frac{\mathbf{g}}2 x^1 x^4\,, &
  \lambda^{\tilde 5} &= - \frac{\mathbf{g}}2 x^2 x^4\,, &
  \lambda^{\tilde 6} &= - \frac{\mathbf{g}}2 x^3 x^4\,. \label{eqn2:gaugetrafoGflux}
\end{align}
Executing this transformation yields the desired $A=0$ and gives rise to the background generalized vielbein
\begin{equation}\label{eqn2:EAiGflux}
  E^{\alpha}{}_i = \begin{pmatrix}
    1 & 0 & 0 & 0 \\ 0 & 1 & 0 & 0 \\ 0 & 0 & 1 & 0 \\ 0 & 0 & 0 & 1
  \end{pmatrix}\,, \quad
  E^{\tilde\alpha}{}_i = \frac{\mathbf{g}}2 \begin{pmatrix}
    x^2 & - x^1 & 0 & 0 \\
    x^3 & 0 & -x^1 & 0 \\
    0 & x^3 & -x^2 & 0 \\
    x^4 & 0 & 0 & -x^1 \\
    0 & x^4 & 0 & -x^2 \\
    0 & 0 & x^4 & -x^3
  \end{pmatrix}
    \quad \text{and} \quad
  E^{\tilde\alpha}{}_{\bar i} = \begin{pmatrix}
    1 & 0 & 0 & 0 & 0 & 0 \\
    0 & 1 & 0 & 0 & 0 & 0 \\
    0 & 0 & 1 & 0 & 0 & 0 \\
    0 & 0 & 0 & 1 & 0 & 0 \\
    0 & 0 & 0 & 0 & 1 & 0 \\
    0 & 0 & 0 & 0 & 0 & 1
  \end{pmatrix}\,.
\end{equation}
Moreover, we observe that this gauging represents a symmetric space. Therefore, it implies that we also could have worked with the coset representative
\begin{equation}
  m = \exp ( t_1 x^1 + t_2 x^2 + t_3 x^3 + t_4 x^4 )
\end{equation}
instead of~\eqref{eqn2:matrixexpm} to derive the same result. However, we want to present the full technique at least once. With~\eqref{eqn2:EAiGflux} at hand, we compute the generalized frame field $\hat E_A{}^{\hat I}$, its dual and finally the twist $\mathcal{F}_{\hat I\hat J\hat K}$ of the generalized Lie derivative~\eqref{eqn2:gengeometr}. It receives only contributions~\eqref{eqn2:twistcontrib} from the four-form
\begin{equation}
  G_{\mathcal{F}} = \frac{1}{4!} \mathcal{F}_{ijkl} \, d x^i\wedge d x^j\wedge d x^k\wedge d x^l =
    - \mathbf{g}\, d x^1 \wedge d x^2 \wedge d x^3 \wedge d x^4\,.
\end{equation}
Thus, we find in total the expected $\mathbf{g}$ units of $G$-flux
\begin{align}
  G = G_{\hat E} + G_{\mathcal{F}} &= \frac{1}{4!} X_{\alpha\beta}{}^{\tilde \gamma} E^\alpha{}_i E^\beta{}_j \eta_{\gamma\delta,\tilde \gamma} E^\gamma{}_k E^\delta{}_l \, d x^i \wedge d x^j \wedge d x^k \wedge d x^l \nonumber \\
  &=  \mathbf{g}\, d x^1 \wedge d x^2 \wedge d x^3 \wedge d x^4 \,
\end{align}
on the background after combining this contribution with $G_{\hat E}$ from the generalized frame.

The obtained result looks very similar to the one found for the torus with $H$-flux in~\cite{Hassler:2016srl}. Here, the flux decomposes between the twist and the frame field in a certain particular fashion as well. Nevertheless, it should be noted that this splitting arises as a natural consequence from the principle bundle construction. In order to analyze how this works, we compute the flux contribution coming from the frame field
\begin{equation}\label{eqn2:GhatEfromC}
  G_{\hat E} = d C = - \frac16 E^\alpha{}_i E^\beta{}_j d ( \eta_{\alpha\beta,\tilde\gamma} E^{\tilde \gamma}{}_j d x^j) \wedge d x^i \wedge d x^j
\end{equation}
by using relation~\eqref{eqn2:CfromE}. Furthermore, we identify
\begin{equation}
  \mathcal{A}_{\alpha\beta} = \eta_{\alpha\beta,\tilde \gamma} E^{\tilde\gamma}{}_i \, d x^i
\end{equation}
with the connection of a $T^6$ bundle over the tours. Hence, every independent $\mathcal{A}_{\alpha\beta}$ component, such as e.g. $\mathcal{A}_{12}$, describes the connection of a circle bundle. The first Chern class of these bundles is defined as
\begin{equation}
  c_{\alpha\beta} = d \mathcal{A}_{\alpha\beta}\,.
\end{equation}
By inserting the result~\eqref{eqn2:EAiGflux} for $E^{\tilde\alpha}{}_i$ into this equation, we acquire the independent classes
\begin{align}
  c_{21} &= \mathbf{g}\, d x^3 \wedge d x^4 \,, &
  c_{13} &= \mathbf{g}\, d x^2 \wedge d x^4 \,, &
  c_{41} &= \mathbf{g}\, d x^2 \wedge d x^3 \,, \nonumber \\
  c_{32} &= \mathbf{g}\, d x^1 \wedge d x^4 \,, &
  c_{24} &= \mathbf{g}\, d x^1 \wedge d x^3 \,, &
  c_{43} &= \mathbf{g}\, d x^1 \wedge d x^2\,,
\end{align}
explicitly. Everyone of them portrays a class in the integer valued cohomology $H^2(\mathcal{S}_{\alpha\beta}, M)=\mathbb{Z}$ of the circle bundle $\mathcal{S}_{\alpha\beta}$ over $M=T^4$. Moreover, they are not trivial. It proves that the principal bundle we constructed is non-trivial as well. If we identify the cohomology class of a closed form $\omega$ by $[ \omega ]$, we can recast~\eqref{eqn2:GhatEfromC} as
\begin{equation}\label{eqn2:[GhatE]fromChern}
  [ G_{\hat E} ] = \frac13 ( [c_{21}] + [c_{13}] + [c_{41}] + [c_{32}] + [c_{24}] + [c_{43}] )\,.
\end{equation}
Since $G_{\hat E}$ describes a top form on the $T^4$ it is an element of the integer valued de Rham cohomology $H^4_{\mathrm{dR}}(M)$ and it is isomorph to $H^2(\mathcal{S}_{\alpha\beta}, M)$. Subsequently, there exists no obstruction in comparing the Chern numbers with $[G_{\hat E}]$ and~\eqref{eqn2:[GhatE]fromChern} is absolutely sensible. All different $S^1$ factors in the $H$-principal bundle give the same contribution. Thus, it is quite natural that they all share the same Chern number, i.e. one. Therefore,~\eqref{eqn2:[GhatE]fromChern} forces
\begin{equation}
  [G_{\hat E}] = 2 \mathbf{g}
\end{equation}
which is in perfect alignment with our result~\eqref{eqn2:GfieldhatE}.

It is worth mentioning that although the field strength $F = d A$ for the $H$-principal bundle vanishes everywhere on $M$, it is still not possible to completely gauge away the connection $A$.
The reason for this lies in the fact that the gauge transformation $\lambda^{\tilde a}$ in~\eqref{eqn2:gaugetrafoGflux} is not globally well-defined on $M$.
This is a remainder of modding out the discrete subgroup from the left to make $G$ compact. However, the effect is not connected to the topological non-trivial $G$-flux in this background as one might think. This proves the four-sphere with $G$-flux considered in the next subsection. There, it is possible to everywhere get rid of the connection. Nevertheless, locally we are always able to solve the SC and construct the according generalized frame field
\begin{equation}\label{eqn2:genframeT4Gflux}
  \mathcal{E}_\alpha = - E_\alpha{}^i \partial_i + \iota_{E_\alpha}  \mathcal{C}' \,, \quad
  \mathcal{E}_{\tilde\alpha} = - \frac{1}{2} \eta_{ij,\tilde\alpha} \, d x^i \wedge d x^j
\end{equation}
where $E_\alpha{}^i$ denotes the inverse transpose of the frame in~\eqref{eqn2:EAiGflux} and
\begin{equation}
  \mathcal{C}' = \mathbf{g} \left(
    2 x^4\, d x^1 \wedge d x^2 \wedge d x^3 
    + x^3\, d x^1 \wedge d x^2 \wedge d x^4
    - x^2\, d x^1 \wedge d x^3 \wedge d x^4
    + x^1\, d x^2 \wedge d x^3 \wedge d x^4 \right)\,.
\end{equation}
with
\begin{equation}
  G = d \mathcal{C}' = \mathbf{g} \, d x^1 \wedge d x^2 \wedge d x^3 \wedge d x^4\,.
\end{equation}
It should be noted that the gauging~\eqref{eqn2:gaugingT4Gflux} portrays the irrep $\mathbf{1}$ in the solution~\eqref{eqn2:sollinconst3_15} of the third linear constraint. As a consequence, this frame is a result of the construction presented in subsection~\ref{sec:genframe} with $\lambda = 3 \mathbf{g}$ and 
\begin{equation}
  \mathcal{C} = -3 \mathbf{g} \, x^4 \, d x^1 \wedge d x^2 \wedge d x^3 \,,
\end{equation}
giving rise to the required
\begin{equation}
  \mathcal{G} = d \mathcal{C} = 3 \mathbf{g} \, d x^1 \wedge d x^2 \wedge d x^3 \wedge d x^4\,.
\end{equation}

At last, we consider a deformation of this solution by applying $\mathcal{T}_A{}^B$ generating the SO(5) rotation
\begin{equation} \label{eqn2:rotGtoQflux}
  \mathcal{T}_a{}^b = \begin{pmatrix}
    0  & 1 & 0 & 0 & 0 \\
    -1 & 0 & 0 & 0 & 0 \\
    0  & 0 & 1 & 0 & 0 \\
    0  & 0 & 0 & 1 & 0 \\
    0  & 0 & 0 & 0 & 1
  \end{pmatrix}
    \quad \text{as} \quad
  \mathcal{T}_{a_1 a_2}{}^{b_1 b_2} =  2 \delta_{[a_1}{}^{[b_1} \mathcal{T}_{a_2]}{}^{b_2]}\,.
\end{equation}
Following this rotation the subalgebra $\mathfrak{h}$ becomes non-abelian and is governed by the non-vanishing commutator algebra
\begin{equation}
  [t_{\tilde 1}, t_{\tilde 2} ] = \mathbf{g}\, t_{\tilde 3}\,, \quad
  [t_{\tilde 1}, t_{\tilde 4} ] = \mathbf{g}\, t_{\tilde 5} \quad \text{and} \quad
  [t_{\tilde 2}, t_{\tilde 4} ] = \mathbf{g}\, t_{\tilde 6}\,.
\end{equation}
Solving the SC is much simpler for this situation in comparison to the one stated above. The reason for this stems from the fact that the field strength $A$ vanishes automatically for $C=0$. This causes a trivial vielbein, i.e. it being the identity, with $E^{\tilde a}{}_i$ vanishing whereas the remaining components $E^\alpha{}_i$ and $E^{\tilde \alpha}{}_{\tilde i}$ are equivalent to the previous results in~\eqref{eqn2:EAiGflux}. Moreover, the generalized frame field $\hat E_A$ does not provide any additional contributions to the fluxes of the background. Therefore, the only non-vanishing contribution originates in the twist~\eqref{eqn2:twistcontrib}
\begin{equation}\label{eqn2:QformX}
  Q_i{}^{jkl} = \mathcal{F}_i{}^{jkl} - \mathcal{F}^{jk}{}_i{}^l = 2 X_{\alpha\tilde\beta}{}^{\gamma} E^\alpha{}_i \eta^{jk,\tilde\beta} E_\gamma{}^l
\end{equation}
which is totally antisymmetric in the indices $i$, $j$, $l$. It is quite convenient to rewrite this quantity as
\begin{equation}
  Q_{ij} = \frac1{3!} Q_i{}^{klm} \epsilon_{klmj} =  -\mathbf{g} \diag( 1,\, 0,\,  0,\, 0 )
\end{equation}
where $\epsilon_{klmj}$ denotes the totally antisymmetric tensor in four dimensions. Thus, we conclude that this background possesses $\mathbf{g}$ units of $Q$-flux. Furthermore, as it emerges by a SO(5) transformation from the previous background with $\mathbf{g}$ units of $G$-flux, we found a direct realization of the duality chain~\eqref{eqn2:chain1}.

The gauging lies in the $\mathbf{10}$ of~\eqref{eqn2:sollinconst3_15}. Hence, we can construct the generalized frame
\begin{equation}
  \mathcal{E}_\alpha = - E_\alpha{}^i \partial_i \,, \quad
  \mathcal{E}_{\tilde\alpha} = \eta_{ij,\tilde\alpha} \beta^{ijk} \partial_k - \frac12 \eta_{ij,\tilde\alpha}\, d x^i \wedge d x^j
\end{equation}
with $\mathcal{C}=0$ and the totally antisymmetric $\beta^{ijk}$ whose only non-vanishing components take on the form
\begin{equation}
  \beta^{234} = - \frac{\mathbf{g}}2 x^1\,.
\end{equation}
Finally, it gives rise to the $Q$-flux
\begin{equation}
  Q_i{}^{jkl} = - 2 \partial_i \beta^{jkl}
\end{equation}
in~\eqref{eqn2:QformX}. Another approach to find a generalized frame with the same properties works by rotating the generalized frame field of the previous duality frame~\eqref{eqn2:genframeT4Gflux} with $\mathcal{T}_A{}^B$ in~\eqref{eqn2:rotGtoQflux}~\cite{Bosque:2017dfc}.

\subsubsection{Gaugings in the \texorpdfstring{$\mathbf{40}$}{40}}\label{sec:T4_40}

Realizing the twisted four-torus from which the second duality chain~\eqref{eqn2:chain2} emerges requires us to consider the embedding tensor solution~\eqref{eqn2:embeddingt40}. It demands the following non-vanishing components~\cite{Blair:2014zba}
\begin{equation}\label{eqn2:ZfFlux}
  Z^{23,3} = - Z^{32,3} = \frac{\mathbf{f}}2
\end{equation}
in order to obtain $\mathbf{f}$ units of geometric flux. The structure coefficients of the Lie algebra $\mathfrak{g}$ originate from~\eqref{eqn2:emb101010b} as above. However, it should be noted that this algebra is not ten-dimensional anymore. As we previously discussed in subsection~\ref{sec:linconstsl5}, the gaugings in the $\mathbf{40}$ reduce the dimension of the group manifold corresponding to~\eqref{eqn2:dimG40}. Subsequently, the here discussed group manifold $G$ possesses only nine dimensions and allows for a SL(3)$\times$SL(2) structure as presented in figure~\ref{fig:sollinconst}. The coordinates then split into the two irreps
\begin{equation}
  (\mathbf{3},\mathbf{2}): \{1,\,2,\,\tilde 1,\,\tilde 2,\,\tilde 3,\,\tilde 4\}
    \quad \text{as well as} \quad
  (\overline{\mathbf{3}}, \mathbf{1}): \{3,\,4,\,\tilde 5\}
\end{equation}
with the adapted basis version~\eqref{eqn2:decompSL510}
\begin{equation}
  \alpha = \{12,\,13,\,14,\,15\}
    \quad \text{and} \quad
  \tilde\alpha = \{24,\,25,\,34,\,35,\,45\}
\end{equation}
for the components of the $\mathbf{10}$ indices $\alpha$ and $\tilde\alpha$.  For this basis the non-vanishing commutator algebra of the Lie algebra $\mathfrak{g}$ takes on the form
\begin{equation}\label{eqn2:defgexample40}
  [t_{\tilde 5}, t_3] = \mathbf{f}\, t_{\tilde 2}\,,\quad
  [t_{\tilde 5}, t_4] = \mathbf{f}\, t_{\tilde 4} \quad \text{and} \quad
  [t_3, t_4] = \mathbf{f}\, t_2\,.
\end{equation}
In combination, the six generators arising in these three relations form the algebra $\mathfrak{cso}(1,0,3)$ with the center $\{t_2,\,t_{\tilde 2},\,t_{\tilde 4}\}$. The remaining generators $t_1$, $t_{\tilde 1}$ and $t_{\tilde 3}$ source a three-dimensional abelian factor. Furthermore, there exists a 16-dimensional faithful representation for $\mathfrak{g}$ we presented in appendix~\ref{app:faithfulrepr}. Subsequently, we can derive the coset elements $m$ according to~\eqref{eqn2:matrixexpm}, whereas the elements of the subgroup $H$ are given by
\begin{equation}
  h = \exp( t_{\tilde 1} x^{\tilde 1}) \exp( t_{\tilde 2} x^{\tilde 2}) \exp( t_{\tilde 3} x^{\tilde 3})
  \exp(t_{\tilde 4} x^{\tilde 4}) \exp( t_{\tilde 5} x^{\tilde 5})\,.
\end{equation}
Equivalently to the duality chain discussed in the last subsubsection, the identifications~\eqref{eqn2:coordident3} and~\eqref{eqn2:coordident4} on the coordinates of the group manifold are required to hold here as well. Otherwise, we would not be able to obtain a compact background. It describes a fibration
\begin{equation}
  T^2=F \hookrightarrow M \rightarrow B=T^2
\end{equation}
where a point on the fiber $F$ is denoted by the coordinates $x^1$, $x^2$, while the base $B$ is parameterized by the remaining coordinates $x^3$ and $x^4$. This fiber is contained in the coordinate irrep $(\mathbf{2},\mathbf{3})$ and the base is part of $(\mathbf{1},\overline{\mathbf{3}})$. Again, the gauge potential $A$ vanishes for $C=0$ by construction. Hence, there exists a solution of the SC with the generalized background vielbein
\begin{equation}\label{eqn2:EAifflux}
  E^{\alpha}{}_i = \begin{pmatrix}
    1 & 0 & 0 & 0 \\ 0 & 1 & f x^4 & 0 \\ 0 & 0 & 1 & 0 \\ 0 & 0 & 0 & 1
  \end{pmatrix}\,, \quad
  E^{\tilde\alpha}{}_i = - \mathbf{f} \begin{pmatrix}
    0 & 0 & 0 & 0 \\
    0 & 0 & x^{\tilde 5} & 0 \\
    0 & 0 & 0 & 0 \\
    0 & 0 & 0 & x^{\tilde 5} \\
    0 & 0 & 0 & 0
  \end{pmatrix}
    \quad \text{and} \quad
  E^{\tilde\alpha}{}_{\bar i} = \begin{pmatrix}
    1 & 0 & 0 & 0 & 0 \\
    0 & 1 & 0 & 0 & 0 \\
    0 & 0 & 1 & 0 & 0 \\
    0 & 0 & 0 & 1 & 0 \\
    0 & 0 & 0 & 0 & 1 \\
  \end{pmatrix}\,.
\end{equation}
It gives rise to the non-vanishing geometric flux
\begin{equation}
  f^2 = \partial_{[i} E^2{}_{j]} d x^i \wedge d x^j = - \mathbf{f}\, d x^3 \wedge d x^4
\end{equation}
as it did in the \DFTwzw{} example, the three-torus with $f$-flux~\cite{Hassler:2016srl}. As was observed for \DFTwzw{}, the twist term in the generalized Lie derivative~\eqref{eqn2:gengeometr} vanishes for this background.

It is quite informative to analyze the GG of this setup even further. Since the group manifold does not possess the full ten-dimensional structure anymore the situation becomes more subtle. Let us remind ourselves that in general the SC of SL(3)$\times$SL(2) EFT allows for two distinct solutions. First, there exist solutions reproducing eleven-dimensional supergravity with three internal directions and secondly, there exist the ones resulting in ten-dimensional type IIB (Only two internal directions)\cite{Hohm:2015xna}. It is manifest from the SL(5) point of view we take. Every individual solution of~\eqref{eqn2:SL4SC} is assigned a distinct $v^0_a$ in the $\mathbf{5}$ of SL(5) which branches in the following way
\begin{equation}
  \mathbf{5}\rightarrow (\mathbf{1},\mathbf{2}) + (\mathbf{3}, \mathbf{1})
\end{equation}
to SL(3)$\times$SL(2). The first irrep appearing in this equation corresponds to SC solutions with an eleven-dimensional SUGRA description, whereas the second irrep captures type IIB. The latter is implemented on the two-dimensional fiber $F$. Moreover, the decomposition of $M$ into a base $B$ and a fiber $F$ admits three distinctive two-forms on $\Lambda^2 T^* M$. The two-forms with all the legs on the base or on the fiber as well as the ones with one leg on the base and the other leg on the fiber. Above each point $p$ of $M$, $\Lambda^2 T_p^* M$ lies a six-dimensional vector space. Although, $\mathfrak{h}$ is only five-dimensional.
Thus, the map $\eta_p$ in~\eqref{eqn2:etamap} is not describing an isomorphism anymore. It poses an issue as this property is an essential ingredient to our construction presented in subsection~\ref{sec:gg}. However, this property can be restored by removing all two-forms whose legs are completely on the base of the codomain $\eta_p$. These are not part of the resulting GG. Despite this fact,~\eqref{eqn2:genLieGG} still holds. Specifically, we are now in the position to construct the generalized frame field $\mathcal{E}_A$ as the gauging for this case is the surviving $(\mathbf{1},\mathbf{2})$ of~\eqref{eqn2:sollinconst3_40_9}. For the commutator algebra provided in~\eqref{eqn2:defgexample40}, we observe that the emerging physical manifold $M$ is not describing a symmetric space anymore because $[\mathfrak{h},\mathfrak{m}]\subset\mathfrak{m}$ and $[\mathfrak{m},\mathfrak{m}]\subset\mathfrak{h}$ are violated. Nevertheless, we are still are able to obtain a SC solution, as we did, since $\mathfrak{m}$ is a subalgebra of $\mathfrak{g}$ with $[\mathfrak{m},\mathfrak{m}]\subset \mathfrak{m}$. The associated generalized frame field takes on the form
\begin{equation}
  \mathcal{E}_\alpha = E'_\alpha{}^i \partial_i\,, \quad
  \mathcal{E}_{\tilde\alpha} = \frac12 \eta_{\beta\gamma,\tilde\alpha} 
    {E'}^\beta{}_i {E'}^\gamma{}_i \, d x^i \wedge d x^j
\end{equation}
with the frame
\begin{equation}
  E'_\alpha{}^i = \begin{pmatrix} -1 & 0 & 0 & 0 \\
    0 & -1 & 0 & 0 \\
    0 & 0 & -1 & 0 \\
    0 & x^3 \mathbf{f} & 0 & - 1
  \end{pmatrix}
    \quad \text{and the dual} \quad
  {E'}^\alpha{}^i = \begin{pmatrix} 
    -1 &  0 &  0 &  0 \\
     0 & -1 &  0 & - x^3 \mathbf{f} \\
     0 &  0 & -1 &  0 \\
     0 &  0 &  0 & -1
  \end{pmatrix}\,.
\end{equation}
However, it should be noted that this step is redundant as the twist $\mathcal{F}_{\hat I\hat J}{}^{\hat K}$ already vanished for $\hat E_A$.

Now, let us finally turn to the dual background with $R$-flux in~\eqref{eqn2:chain2}. For the specific choice of $v^0_a$ in~\eqref{eqn2:v0acanonical} we have made, it is completely fixed by the four independent components $Z^{a1,1}$ ($a$=1, \dots, 4) of the $\mathbf{40}$ in the embedding tensor~\eqref{eqn2:embeddingt40}~\cite{Blair:2014zba}. Naturally, the SO(5) transformation
\begin{equation}\label{eqn2:trafof->R}
  \mathcal{T}_a{}^b = \begin{pmatrix}
    0  & 0 & 1 & 0 & 0 \\
    0  & 1 & 0 & 0 & 0 \\
    -1 & 0 & 0 & 0 & 0 \\
    0  & 0 & 0 & 1 & 0 \\
    0  & 0 & 0 & 0 & 1
  \end{pmatrix}
\end{equation}
casts~\eqref{eqn2:ZfFlux} into this form. However, there exist two issues with the resulting setup. First, the generators $t_{\tilde\alpha}$ do not source a subalgebra $\mathfrak{h}$ after the rotation $\mathcal{T}$ anymore. In \DFTwzw{}, we are confronted with the same problem. It perfectly agrees with the completely non-geometric nature of the $R$-flux. If we would have obtained a SC solution for the $R$-flux with our procedure, there would have existed a geometric interpretation in terms of a manifold $M$ equipped with a GG. This is definitely not the case. But we are faced with another subtlety which cannot be observed in \DFTwzw{}. We remind ourselves that SL(5) is being broken down to SL(3)$\times$SL(2) for the torus with geometric flux as the associated structure constants emerge from the $\mathbf{40}$. Yet, the rotation~\eqref{eqn2:trafof->R} is not an element of this reduced symmetry group. Subsequently, the second background appearing in the duality chain~\eqref{eqn2:chain2} does not allow the most general SC solutions we consider in this thesis. Although, there still exist solutions with constant fluctuations~\cite{Bosque:2017dfc}.

\subsection{Four-sphere with \texorpdfstring{$G$}{G}-flux}

Finding the solution of the SC for the four-sphere with radius $R$ as the physical manifold requires us to consider the group manifold SO(5). It results from the $\mathbf{15}$ in the embedding tensor solution and we identify it with the symmetric matrix
\begin{equation}
  Y_{ab} = -\frac4R \, \diag(1,\,1,\,1,\,1,\,1)\,.
\end{equation}
As opposed to the former examples presented in subsection~\ref{sec:T4Gflux} it is much easier to derive a faithful representation for the corresponding Lie algebra $\mathfrak{g}=\mathfrak{so}$(5). The most convenient choice are the antisymmetric matrices
\begin{equation}
  (t_A)_b{}^c = -\frac12 X_{Ab}{}^c
\end{equation}
which are a direct consequence of the embedding tensor~\eqref{eqn2:emb1055b} and operate on the fundamental irrep of $\mathfrak{g}$. In contrast to our previous choice~\eqref{eqn2:matrixexpm}, we now parameterize coset representatives by
\begin{align}\label{eqn2:mS4}
  m = \exp\Big[ R \, \phi^1  &\left(\cos(\phi^2) t_1 + \sin(\phi^2)\cos(\phi^3) t_2 + \right.\nonumber \\ 
  &\left. \sin(\phi^2) \sin(\phi^3) \cos(\phi^4) t_3 + \sin(\phi^2) \sin(\phi^3) \sin(\phi^4) t_4\right) \Big] \,,
\end{align}
where the angels are associated with spherical coordinates
\begin{equation}
  \phi^1\,,\, \phi^2\,,\,  \phi^3 \in [0, \pi] \quad \text{and} \quad
  \phi^4 \in [0, 2\pi)\,.
\end{equation}
However, the elements of the subgroup are still constructed by~\eqref{eqn2:matrixexpm}. Combining $\mathfrak{m}$ and $\mathfrak{h}$, they form a symmetric pair. As we demonstrated at the end of subsection~\ref{sec:linkvomega}, this particular choice~\eqref{eqn2:mS4} for $m$ has the advantage that the gauge potential $A$ vanishes by construction for
\begin{equation}
  C = R^3 \tan\left(\frac{\phi^1}2\right) \sin^3(\phi^1) \sin^2(\phi^2) \sin(\phi^3) \, d \phi^2 \wedge d \phi^3 \wedge d \phi^4 \,.
\end{equation}
The corresponding field strength given by
\begin{equation}
  G_{\hat E} = d C = 4 R^3 \cos\left(\frac{\phi^1}2\right)\sin^3\left(\frac{\phi^1}2\right)\left((1+3\cos(\phi^1)\right) \sin^2 (\phi^2) \sin (\phi^3 ) \, d \phi^1 \wedge d \phi^2 \wedge d \phi^3 \wedge d \phi^4
\end{equation}
lies in the trivial cohomology class of $H^4_{\mathrm{dR}(S^4)}$ as the integral
\begin{equation}
  \int_{S^4} G_{\hat E} = 0
\end{equation}
is zero. Nevertheless, the three-form $C$ and the connection $A^{\tilde\alpha}$ are globally well-defined. It is now possible to globally gauge away the connection even though the background possesses $G$-flux in a non-trivial cohomology class as well. Another intriguing quantity to calculate is the first Pontryagin class of the connection
\begin{equation}
  \mathcal{A}^{\tilde\alpha} = E^{\tilde\alpha}{}_i \, d x^i\,.
\end{equation}
It provides a totally analogous classification of the Chern classes, we derived for the $T^6$-bundle in the $T^4$ with $G$-flux background and therefore vanishes entirely.

Constructing the generalized frame field~\eqref{eqn2:genframe&dual} makes it necessary to obtain the background vielbein
\begin{equation}
  E^\alpha{}_i = R \begin{pmatrix}
    c_2 & - s_1 s_2 & 0 & 0 \\
    c_3 s_2 & c_2 c_3 s_1 & - s_1 s_2 s_3 & 0 \\
    c_4 s_2 s_3 & c_2 c_4 s_1 s_3 & c_3 c_4 s_1 s_2 & - s_1 s_2 s_3 s_4 \\
    s_2 s_3 s_4 & c_2 s_1 s_3 s_4 & c_3 s_1 s_2 s_4 & c_4 s_1 s_2 s_3
  \end{pmatrix}
\end{equation}
with $c_i = \cos(\phi^i)$ and $s_i = \sin(\phi^i)$. This vielbein is part of the left invariant Maurer-Cartan form $E^A{}_I$ given in~\eqref{eqn2:compbgvielbein}. Subsequently, it yields the metric
\begin{equation}
  d s^2 = E^\alpha{}_i \delta_{\alpha\beta} E^\beta{}_j \, d \phi^i d \phi^j = R^2 \left( (d \phi^1)^2 +
    s_1^2  (d \phi^2)^2 + s_1^2 s_2^2 (d \phi^3)^2 + s_1^2 s_2^2 s_3^2 (d \phi^3)^2 \right)\,,
\end{equation}
a round sphere with radius $R$. Once we have found the SC solution for $G =$ SO($5$), we can execute the construction procedure demonstrated in subsection~\ref{sec:genframe} and derive the generalized frame field $\mathcal{E}_A$ with $\mathcal{C}$ such that
\begin{equation}\label{eqn2:mcG_S^4}
  \mathcal{G} = d \mathcal{C} = 3 R^3 \sin^3(\phi^1) \sin^2(\phi^2) \sin(\phi^3) \, 
  d \phi^1 \wedge d \phi^2 \wedge d \phi^3 \wedge d \phi^4 = \frac3R \mathrm{vol}\,.
\end{equation}
As the full result is too bulky, we leave it out. Instead, we present an alternative parameterization of the group elements $m$ in terms of Cartesian coordinates
\begin{align}
  y^1 &= R \cos(\phi^1)  &
  y^2 &= R \sin(\phi^1) \cos(\phi^2) \nonumber \\
  y^3 &= R \sin(\phi^1) \sin(\phi^2) \cos(\phi^3) &
  y^4 &= R \sin(\phi^1) \sin(\phi^2) \sin(\phi^3) \cos(\phi^4) \nonumber \\
  y^5 &= R \sin(\phi^1) \sin(\phi^2) \sin(\phi^3) \sin(\phi^4)\,.
\end{align}
These have the benefit that they give rise to a straightforward coset representative
\begin{equation}
  m = \frac{1}{R} \begin{pmatrix}
    y^1 & -y^2 & -y^3 & -y^4 & -y^5 \\
    y^2 & y^{22} & y^{23} & y^{24} & y^{25} \\
    y^3 & y^{23} & y^{33} & y^{34} & y^{35} \\
    y^4 & y^{24} & y^{34} & y^{44} & y^{45} \\
    y^5 & y^{25} & y^{35} & y^{45} & y^{55}
  \end{pmatrix}
    \quad \text{with} \quad
  y^{ij} = R \delta^{ij} - \frac{y^i y^j}{R + y^1}
\end{equation}
and enable us to compare our results with the ones found in~\cite{Lee:2014mla}. However, it requires us to implement the additional constraint
\begin{equation}
  \sum_{i=1}^5 (y^i)^2 = R
\end{equation}
in all remaining equations. As above, we compute the following part of the left invariant Maurer-Cartan form
\begin{equation}
  E^\alpha{}_i = \frac1R \begin{pmatrix}
    -y^2 & y^{22} & y^{23} & y^{24} & y^{25} \\
    -y^3 & y^{23} & y^{33} & y^{34} & y^{35} \\
    -y^4 & y^{24} & y^{34} & y^{44} & y^{45} \\
    -y^5 & y^{25} & y^{35} & y^{45} & y^{55}
  \end{pmatrix} = E_\alpha{}^i \,.
\end{equation}
Obtaining the vectors $E_\alpha{}^i$ is a bit more involved in this context than it was before, since $E^\alpha{}_i$ is not a quadratic matrix and thus not invertible. However, the condition $E_\alpha{}^i E^\beta{}_i = \delta_\alpha^\beta$ fixes it completely and we moreover require all vectors $E_\alpha{}^i$ to be perpendicular to the radial direction $\vec r$=$(y^1\,y^2\,y^3\,y^4\,y^5)^T$. Subsequently, we are now in the position to derive the vector part $\mathcal{E}_A{}^i$ of the generalized frame, which we label as $V_A{}^i$, in order to allow for a direct comparison of our results with the ones given in~\cite{Lee:2014mla}. Its components take on the form
\begin{equation}
  V_A{}^i = \frac1R \left( \delta_{a1}^i y^{a2}  - \delta_{a2}^i y^{a1} \right)
\end{equation}
where the $\mathbf{10}$ index $A$ has been decomposed into the two fundamental indices $a_1$ and $a_2$. One can easily verify that they produce the algebra $\mathfrak{so}(5)$, governed by the Lie derivative $L$. Specifically, they give rise to
\begin{equation}\label{eqn2:so5vectors}
  L_{V_A} V_B = X_{AB}{}^C V_C\,.
\end{equation}
Furthermore, it is interesting to take a closer look at the two-form
\begin{equation}
  \sigma_A = \frac12 \mathcal{E}_A{}_{ij} \, d y^i \wedge d y^j
\end{equation}
which yields
\begin{equation}
  \sigma_A = - \frac1R \epsilon_{a_1 a_2 ij} d y^i \wedge d y^j\,.
\end{equation}
In the same fashion as~\eqref{eqn2:so5vectors}, they generate the Lie algebra $\mathfrak{g}$ under the Lie derivative
\begin{equation}
  L_{V_A} \sigma_B = X_{AB}{}^C \sigma_C\,.
\end{equation}
Finally, we have to evaluate the volume form
\begin{align}
  \mathrm{vol} &= \frac{1}{4!} \epsilon_{1\hat\alpha\hat\beta\hat\gamma\hat\delta} 
    E^\alpha{}_i E^\beta{}_j E^\gamma{}_k E^\delta{}_l \, d y^i \wedge d y^j \wedge d y^k \wedge d y^l
     \nonumber \\
    & = \frac1{4! R} \epsilon_{ijklm} y^i d y^j \wedge d y^k \wedge d y^l \wedge d y^m
\end{align}
which satisfies the condition\footnote{ As opposed to~\cite{Lee:2014mla}, we work with structure coefficients $X_{AB}{}^C$ which have the opposite sign. For instance, $X_{\tilde 1\tilde 2}{}^{\tilde 3}=X_{23,24}{}^{34} = R^{-1}$ whereas from (2.5) in~\cite{Lee:2014mla} follows that $X_{23,24}{}^{34}=-R^{-1}$. Thus, the vectors $V_A$ and the forms $\sigma_A$ we derived also possess the opposite sign compared to their results. However,~\eqref{eqn2:connectionVAsigmaA} takes on the same form.}~\cite{Lee:2014mla}
\begin{equation}\label{eqn2:connectionVAsigmaA}
  \iota_{V_A} \mathrm{vol} = \frac{R}3 d \sigma_A\,.
\end{equation}
Therefore, we have produced all ingredients which have been discussed in~\cite{Lee:2014mla} to prove that the $S^4$ with four-form flux is parallelizable. Following their paper, we insert the generalized frame field
\begin{equation}
  \mathcal{E}_A = V_A + \sigma_A + \iota_{V_A} \mathcal{C} 
\end{equation}
into the generalized Lie derivative
\begin{equation}
  \widehat{\mathcal{L}}_{\mathcal{E}_A} \mathcal{E}_B = L_{V_A} V_B + L_{V_A} \sigma_B + \iota_{[V_A,V_B]} \mathcal{C} - \iota_{V_B} ( d \sigma_A  - \iota_{V_A} d \mathcal{C} )
\end{equation}
where the last term disappears for a $\mathcal{C}$ governed by~\eqref{eqn2:mcG_S^4}. Although, it is still possible to rescale 
$\sigma_A$ and $\mathcal{C}$ by the same constant factor to find a different generalized frame field which would satisfy~\eqref{eqn2:genparaframeintro}. This factor has been fixed in~\cite{Lee:2014mla} by imposing the equations of motion. If one takes a closer look at these equations, it turns out that they emerge from eleven-dimensional SUGRA with the action
\begin{equation}\label{eqn2:S11dSUGRA}
  S = \frac{1}{2 \kappa_{11}^2} \int d^{11} x \sqrt{-G} \left( \mathcal{R} - \frac{1}2 | d \mathcal{C} |^2 \right)
\end{equation}
for the bosonic subsector. Here, $G$ denotes the metric in eleven dimensions, $\mathcal{R}$ labels the associated curvature scalar and $\mathcal{C}$ represents a three-form gauge field. Performing the Freund-Rubin ansatz~\cite{Freund:1980xh} to find a solution to the field equations of this action on an AdS$_7 \times$S$^4$ spacetime, we obtain
\begin{equation}
  \mathcal{R}_{S^4} = \frac{12}{R^2} = \frac{4}{3} | d \mathcal{C} |^2
    \quad \text{or} \quad
  |\mathcal{G}|^2 = \frac9{R^2}\,.
\end{equation}
We conclude that this result perfectly agrees with~\eqref{eqn2:mcG_S^4}, once we impose the following relations
\begin{equation}
  \mathcal{G} \wedge \star \,\mathcal{G} = | F |^2 \, \mathrm{vol} \quad \text{and} \quad \star \mathrm{vol} = 1\,.
\end{equation}
$\star$ is the Hodge star operator on the $S^4$. Obviously, this result highly depends on the relative factor between the two terms in the action. These are however fixed by supersymmetry. In SL(5) EFT, this relation between the gravity sector and the form-field is a direct consequence of the generalized frame field being an SL(5) element. Generally, $\mathcal{E}_A{}^{\hat I}$ possesses 100 independent components. They are part of the branching rule
\begin{equation}
  \mathbf{10} \times \overline{\mathbf{10}} = \xcancel{\mathbf{1}} + \mathbf{24} + \xcancel{\mathbf{75}}\,.
\end{equation}
Yet, only the components contained in the adjoint irrep are non-vanishing. This feature is by construction implemented in our framework and can be seen from~\eqref{eqn2:genparaframe}. Hence, the frame field $E_\beta{}^i$ and the three-form $\mathcal{C}$ are constituting the generalized frame field $\hat E'_B{}^{\hat I}$. They occupy the irreps $\mathbf{1}+\mathbf{15}$ and $\mathbf{4}$ of SL(4) which are a result of the branching
\begin{equation}
  \mathbf{24} \rightarrow \mathbf{1} + \mathbf{4} + \overline{\mathbf{4}} + \mathbf{15}
\end{equation}
in SL(5). Subsequently, $\hat E'_B{}^{\hat I}$ must be an element of SL(5). $M_A{}^B$ has to fulfill this property by construction as well. As a consequence, $\mathcal{E}_A{}^{\hat I}$ is also an SL(5) element, as it results from multiplying the two. Finally, we obtain the correct scaling factor of the four-form flux~\cite{Bosque:2017dfc}.\clearpage{}

  \clearpage{}\chapter{Conclusion and Outlook}
\label{conclusion}

In the beginning a review of the most important ideas and notions of original DFT in both its generalized metric formulation as well as its flux formulation was made. In this context, we introduced crucial concepts and definitions for later convenience. Moreover, we observed that a consistent formulation of this theory requires the implementation of a SC which can be seen as a level matching condition for scattering processes on the worldsheet. This constraint emerged as an inevitable restraint when trying to introduce a gauge algebra and requiring it to close. It is governed by the C-bracket. Furthermore, the SC is essential for the action to be invariant under generalized diffeomorphisms.

During the next three chapters, we used these principles as motivation to develop DFT$_{\mathrm{WZW}}$. The theory is governed by a WZW model based on a group manifold instead of a torus while the doubling of the coordinates refers to the left-
and right-moving currents. Performing CSFT computations at tree level up to cubic order in fields and leading order in $\alpha'$ made it possible to obtain an action and its associated gauge transformations. Afterwards, we extrapolated these results to all orders in fields by introducing a generalized metric formulation of DFT$_{\mathrm{WZW}}$. Equipped with this powerful tool, we were able to show that the corresponding gauge algebra closes as well as the theory's invariance under generalized diffeomorphisms by imposing a modified SC for the fluctuations, whereas the background fields only require a weaker Jacobi identity. It is in stark contrast to toroidal DFT where all fields need to fulfill the SC. At this point, we can admire how DFT$_{\mathrm{WZW}}$ generalizes the structures of original DFT in a fascinating way. It expressed itself e.g. by the appearance of structure coefficients in the entire theory. These have been absent in the traditional formulation. On top of that, we have observed the emergence of an additional $2D$-diffeomorphism invariance of the theory which cannot be found in the original framework. This fact can be explained with the extended strong constraint. It reduces DFT$_{\mathrm{WZW}}$ down to original DFT and thereby breaks this particular invariance. As a result, it connects background and fluctuation fields with one another. However, the extended strong constraint is not required for a consistent theory. Therefore, we have found a true generalization of traditional DFT and not just a mere rewriting. With these results at hand, we put the theory in a flux formulation by introducing covariant fluxes. Here, we found a double Lorentz symmetry. 

All of these discoveries allowed us to address two major issues connected to generalized Scherk-Schwarz compactifications in the next chapter. We can now construct the twist, characterizing the compactifications, in the same manner as it is constructed in ordinary Scherk-Schwarz reductions. The reason for this lies in the fact that the twist is no longer restricted to be an element of O($D,D$) and that all tools from Riemannian geometry are available. It enables us to construct a corresponding twist for each embedding tensor solution. Subsequently, an appropriate generalized Scherk-Schwarz compactification ansatz made it possible to recover half-maximal, electrically gauged SUGRA from DFT$_{\mathrm{WZW}}$.

At last, we extended our concepts to the gEFT framework. By making the Lie group $G$ on the extended space manifest, we were able to obtain a procedure to construct generalized frame fields for generalized parallelizable coset spaces $M=G/H$ in four dimensions. As a result of the linear and quadratic constraints, there exist several restrictions on this Lie group $G$. These originate from the embedding tensor of the U-duality group SL($5$). It should be noted that in this context the extended space served only as a tool during the entire discussion.	Eventually, all of the unphysical directions in this space need to be removed. It is achieved by solving the SC. Naturally, the SC of EFT gets modified in gEFT as well. Finally, we solved this condition by selecting a particular embedding of the physical subspace $M$ in $G$. Every solution of the SC is accompanied by a canonical frame field $\hat{E}_A$ and a GG with a twisted generalized Lie derivative. Although, the frame field is defined through an untwisted generalized Lie derivative for a generalized parallelizable space. Therefore, it is necessary to modify $\hat{E}_A$. In the end, it has to absorb the twist part and consequently turns into $\mathcal{E}_A$ for which~\eqref{eqn2:genparaframeintro} holds. Nevertheless, there are three linear constraints which have to be fulfilled. Once these are solved, we, e.g., observe that our construction is viable for all gaugings contained in the $\mathbf{15}$. Yet, the associated generalized frame fields were already known~\cite{Lee:2014mla,Hohm:2014qga}. However, we can treat gaugings from the $\mathbf{40}$ as well but their dimension is generally smaller than ten as the U-duality group SL($5$) gets broken. Of course, this does not obstruct our technique.

As we have seen during the course of this thesis, DFT$_\mathrm{WZW}$ /gEFT generalize DFT/EFT even further and the obtained results open up new perspectives on said theories. One of the major disadvantages of DFT/EFT is the combination of background and fluctuation fields into a single object governed by a double/extended geometry. Firstly, the double/extended geometry suffers several major problems, such as undetermined components of the connection and Riemann tensor. Secondly, through the indistinction between background and fluctuations it becomes impossible to produce a construction procedure for the twist originating from the embedding tensor. These twists have to be guessed which slightly frustrates the physicist. With our work, we offer a way out of these issues. Our solution is based on two main elements: A $2D$/$11$-dimensional geometric background space governed by a group manifold $N$ and a corresponding $D$-dimensional subspace $M$, induced by the SC, on which the physical fields live. The group manifold $N$ is accompanied by a non-constant $\eta$ metric. Moreover, there exists no unique solution to the SC and as a consequence there are several different subspaces $M_i$ embedded into $N$. This fact gives rise to totally different spaces in target space. However, they are all unified by T/U-duality in the doubled/extended space $N$ for a given background. Thus, it captures the initial concept of unification. Whereas DFT/EFT is only able to seize this idea for the torus/U-duality groups, we have extended this notion to arbitrary group manifolds, which naturally includes the previous cases.

Nevertheless, there is still lots to do. So, let us finally collect some ideas for future projects to further expand the DFT$_\mathrm{WZW}$/gEFT framework to address open and unanswered question.

\begin{itemize}

\item One could for instance compute $\alpha'$ corrections of DFT$_\mathrm{WZW}$. The results could be very fascinating and more profound than the ones found on the torus for original DFT~\cite{Hohm:2014xsa,Hohm:2013jaa,Hohm:2014eba,Bedoya:2014pma,Hohm:2015mka,Marques:2015vua,Hohm:2016yvc,Baron:2017dvb}. At non-vanishing $\alpha'$ there is a back reaction between the closed string and the curvature of the target space it probes~\cite{Hassler:2015pea}. Furthermore, it could even lead to the discovery of fuzzy spheres, the generalization of spheres to a non-commutative space. E.g. the fuzzy two-sphere comes along with non-commutativity, whereas higher dimensional fuzzy spheres i.e. the fuzzy $S^3$ include non-associativity as well~\cite{Ramgoolam:2001zx}. Therefore, $\alpha'$ corrections of DFT$_\mathrm{WZW}$ should in principle automatically implement non-associativity and non-commutativity target space geometries within our framework.

\item As was seen in~\cite{Hassler:2016srl}, DFT$_\mathrm{WZW}$ implements T-duality on group manifolds. For the toroidal case, they are implemented by the Buscher rules~\ref{sec:buscherrules} and when gauging an U($1$) isometry they give rise to abelian T-duality. Although, the situation becomes more involved for non-abelian group manifolds. Of course, these are also covered by our theory. Here, the barely understood non-abelian T-duality needs to be carried out~\cite{Giveon:1993ai,Alvarez:1994np}. Yet, it should emerge naturally through different SC solutions within our framework.

\item On the worldsheet, non-abelian T-duality can for instance be introduced with a gauged WZW model~\cite{Sfetsos:1994vz,Polychronakos:2010fg,Polychronakos:2010hd} while their corresponding current algebras emerge from a GKO construction~\cite{Goddard:1984vk}. Thus, it would be intriguing to extend our construction from group manifolds to coset spaces. The sector of the coset CFTs required to obtain a tree-level, cubic, and low energy effective action can still be managed analytically and therefore it should be possible to reach this goal.

\item Another intriguing possibility would be to extend the presented formalism for gEFT to other U-duality groups as well. Fortunately, there does not seem to be an obstruction to do so for other dimensions, see tab.~\ref{tab:Udualitygroups}. From analyzing the necessary linear constraints one should find a large class of generalized parallelizable spaces for $\dim M \neq 4$. Due to the very close relationship between these spaces, maximally gauged supergravities, and the embedding tensor formalism it might even be possible to obtain a full classification of them.

\end{itemize}

\clearpage{}

  \begin{appendices}
  \clearpage{}\chapter{Embedding Tensor Solutions}\label{app:twists}
\label{AppendixC}
During this part of the appendix, we apply the procedure outlined in section~\ref{sec:twistconstr} to obtain the background vielbeins $E_A{}^I$ for all $n=3$ compact solutions of the embedding tensor~\cite{Bosque:2015jda}, see table~\ref{tab:solembedding}. Therefore, we start by computing the structure coefficients $F_{ABC}$ from equation~\eqref{eqn:SL(4)toSO(3,3)}. For most cases, to further simplify the results, we are going to apply additional O($3,3$) rotations $R_A{}^B$
\begin{equation}
  F'_{ABC} = R_A{}^D R_B{}^E R_C{}^F F_{DEF} \quad \text{and} \quad
  \eta'_{AB} = R_A{}^D R_B{}^E \eta_{DE}\,.
\end{equation}
Furthermore, the six generators are being assigned symbols e.g.
\begin{equation}
  t_A = \{a, b, c, d, e, f\}\,,
\end{equation}
and subsequently we read off their corresponding Lie algebra~\eqref{eqn:genalgebra}. The algebra allows us to derive an $N$-dimensional matrix representation for the generators by pursuing the techniques presented in~\ref{sec:semisimple}-\ref{sec:solvable}. This enables us to obtain the group elements $g$ through performing an exponential map~\eqref{eqn:expmap}. Finally, we can use the group elements to compute the left invariant Maurer-Cartan form~\eqref{eqn:leftinvmc}.

\section*{SO(4)/SO(3)}\label{app:SO4}
Applying the rotation
\begin{equation}
  R_A{}^B = \begin{pmatrix}
    \phantom{-}0 &  \phantom{-}0  & \phantom{-}0 &  \phantom{-}0 &  \phantom{-}0 &  \phantom{-}1 \\
    \phantom{-}0 &  \phantom{-}0  & \phantom{-}0 &  \phantom{-}0 &  \phantom{-}1 &  \phantom{-}0 \\
    \phantom{-}0 &  \phantom{-}0  & \phantom{-}0 &  \phantom{-}1 &  \phantom{-}0 &  \phantom{-}0 \\
    \phantom{-}0 &  \phantom{-}0  & -1 &  \phantom{-}0  & \phantom{-}0  & \phantom{-}0 \\
    \phantom{-}0  & -1  & \phantom{-}0  & \phantom{-}0  & \phantom{-}0 & \phantom{-}0 \\
    -1  & \phantom{-}0  & \phantom{-}0  & \phantom{-}0  & \phantom{-}0 &  \phantom{-}0
\end{pmatrix}\,,
\end{equation}
results in
\begin{equation}
  \eta'_{AB} = \begin{pmatrix} 
    -1 & \phantom{-}0 & \phantom{-}0 & \phantom{-}0 & \phantom{-}0 & \phantom{-}0\\
     \phantom{-}0 & -1 & \phantom{-}0 & \phantom{-}0 & \phantom{-}0 & \phantom{-}0\\
     \phantom{-}0 & \phantom{-}0 & -1 & \phantom{-}0 & \phantom{-}0 & \phantom{-}0\\
     \phantom{-}0 & \phantom{-}0 &  \phantom{-}0 & \phantom{-}1 & \phantom{-}0 & \phantom{-}0 \\
     \phantom{-}0 & \phantom{-}0 &  \phantom{-}0 & \phantom{-}0 & \phantom{-}1 & \phantom{-}0 \\
     \phantom{-}0 & \phantom{-}0 &  \phantom{-}0 & \phantom{-}0 & \phantom{-}0 & \phantom{-}1
  \end{pmatrix}
\end{equation}
and the semisimple Lie algebra
\begin{equation}
\label{eqn:so4generators}
  [ s_a ,s_b ] = \alpha_-\, \varepsilon_{ab}{}^c \, s_c 
    \quad \text{and} \quad
  [ \bar s_a ,\bar s_b ] = \alpha_+\,\varepsilon_{ab}{}^c \, \bar s_c
\end{equation}
with
\begin{equation}
  \alpha_+ = - \sqrt{2} \big( \cos(\alpha) + \sin(\alpha) \big)
    \quad \text{and} \quad
  \alpha_- =  \sqrt{2} \big( \cos(\alpha) - \sin(\alpha) \big)\,,
\end{equation}
after assigning the symbols
\begin{equation}
  t_A = \{ s_1, s_2, s_3, \bar s_1, \bar s_2, \bar s_3 \}
\end{equation}
for the generators. In this context, $\varepsilon_{ab}{}^c$ is the totally antisymmetric Levi-Civita tensor in three dimensions with the signature $\varepsilon_{12}{}^3=1$. For $\alpha = 0$ the Lie algebra reduces to $\mathfrak{so}(4)$ and in the case $\alpha = \pi / 4$ it degenerates to $\mathfrak{so}(3)$. The decomposition
\begin{equation}
  \mathfrak{so}(4)_\alpha = \mathfrak{so}(3)_{\alpha_+} \oplus \mathfrak{so}(3)_{\alpha_-}
\end{equation}
is manifest in the basis we have chosen.

The Lie algebra's adjoint representation allows us to construct the group elements.

However, as opposed to the exponential map~\eqref{eqn:expmap}, we work with
\begin{equation}\label{eqn:prodexpmap}
  g =  \exp(t_6 X^6) \exp(t_5 X^5) \cdots \exp(t_1 X^1)\,,
\end{equation}
which directly enables us to read off the inverse group element
\begin{equation}
  g^{-1} = \exp(-t_1 X^1) \exp(-t_2 X^2) \cdots \exp(-t_6 X^6)\,.
\end{equation}
The coordinates
\begin{equation}
  X^I = \{ \phi_1, \phi_2, \phi_3, \bar\phi_1, \bar\phi_2, \bar\phi_3 \}
\end{equation}
split into twice the set of three angles, describing a rotation $\mathds{R}^3$ each. Finally, we derive the left invariant Maurer-Cartan form
\begin{equation}
  E^A{}_I = \begin{pmatrix}
    A_{\alpha-}(\phi_1, \phi_2, \phi_3) & 0 \\ 0 & A_{\alpha+}(\bar\phi_1, \bar\phi_2, \bar\phi_3)
\end{pmatrix}
\end{equation}
and its inverse transposed
\begin{equation}
  E_A{}^I = \begin{pmatrix}
    A_{\alpha-}^{-T}(\phi_1, \phi_2, \phi_3) & 0 \\ 0 & A_{\alpha+}^{-T}(\bar\phi_1, \bar\phi_2, \bar\phi_3)
  \end{pmatrix}
\end{equation}
where $A_\alpha$ corresponds to the matrix
\begin{equation}\label{eqn:Aalpha}
A_\alpha(\phi_1,\phi_2,\phi_3) = \begin{pmatrix}
    \phantom{-}1 & 0 & -\sin(\phi_2\alpha) \\
    0 & \phantom{-}\cos(\phi_1\alpha) & \phantom{-}\cos(\phi_2\alpha) \sin(\phi_1\alpha) \\
    0 & -\sin(\phi_1\alpha) & \phantom{-}\cos(\phi_1\alpha) \cos(\phi_2\alpha)
  \end{pmatrix}
\end{equation}
while its inverse transpose reads
\begin{equation}
A_\alpha^{-T}(\phi_1,\phi_2,\phi_3)= \begin{pmatrix}
    \phantom{-}1 & 0 & \phantom{-} 0 \\
    \phantom{-} \sin(\phi_1\alpha) \tan(\phi_2\alpha) & \phantom{-}\cos(\phi_1\alpha) & \phantom{-} \sec(\phi_2\alpha) \sin(\phi_1\alpha) \\
    \phantom{-} \cos(\phi_1\alpha) \tan(\phi_2 \alpha) & - \sin(\phi_1\alpha) & \phantom{-}\cos(\phi_1\alpha) \sec(\phi_2\alpha)
  \end{pmatrix}\,.
\end{equation}
Setting $\alpha = 0$, wee see that this background vielbein describes an $S^3$ with $H$-flux~\cite{Bosque:2015jda}.

\section*{ISO(3)}\label{app:ISO(3)}
Applying the rotation
\begin{equation}
  R_A{}^B = \frac{1}{\sqrt{2}} \begin{pmatrix}
    \phantom{-}0  & \phantom{-}0  & \phantom{-}1  & \phantom{-}0  & \phantom{-}0  & \phantom{-}1 \\
    \phantom{-}0  & \phantom{-}1 & \phantom{-}0 & \phantom{-}0 & \phantom{-}1  & \phantom{-}0 \\
    -1 & \phantom{-}0 & \phantom{-}0 & -1 & \phantom{-}0 & \phantom{-}0 \\ \phantom{-}0 & \phantom{-}0 & -1 & \phantom{-}0 & \phantom{-}0 & \phantom{-}1 \\ \phantom{-}0 & -1 & \phantom{-}0 & \phantom{-}0 & \phantom{-}1 & \phantom{-}0 \\
    -1 & \phantom{-}0 & \phantom{-}0 & \phantom{-}1 & \phantom{-}0 & \phantom{-}0
  \end{pmatrix}\,,
\end{equation} 
results in
\begin{equation}  
  \eta'_{AB} = \begin{pmatrix} 0 && 0 && 0 && 1 && 0 && 0 \\
    0 && 0 && 0 && 0 && 1 && 0 \\
    0 && 0 && 0 && 0 && 0 && 1 \\
    1 && 0 && 0 && 0 && 0 && 0 \\
    0 && 1 && 0 && 0 && 0 && 0 \\
    0 && 0 && 1 && 0 && 0 && 0 \\
  \end{pmatrix}
\end{equation}
and the non-semisimple Lie algebra
\begin{equation}\label{eqn:iso3generators}
  [ s_a ,s_b ] =   \cos(\alpha) \varepsilon_{ab}{}^c \, s_c + \sin(\alpha) \varepsilon_{ab}{}^c  \, t_c\,, \quad 
  [ s_a ,t_b ] =   \cos(\alpha) \varepsilon_{ab}{}^{c} \, t_{c} 
    \quad \text{and} \quad
  [ t_a ,t_b ] = 0\,, 
\end{equation}
after assigning the symbols
\begin{equation}
  t_A = \{s_1, s_2, s_3, t_1, t_2, t_3\}
\end{equation}
for the generators. This algebra is equivalent to $\mathfrak{iso}(3)$ for $\alpha = 0$. It arises from a Lie algebra contraction of $\mathfrak{so}(4)$~\cite{Subag2012:contract}.

The algebra possesses a trivial center~\eqref{eqn:center}. Hence, we have a faithful adjoint representation for the six generators. For us to derive the group elements $g$, we use the same exponential map~\eqref{eqn:prodexpmap} as for the case of SO($4$), but we identify the coordinates as
\begin{equation}
  X^I = \{\phi_1, \phi_2, \phi_3, x_1, x_2, x_3\}\,.
\end{equation}

Finally, we are able to derive the left invariant Maurer-Cartan form for the case $\alpha=0$. We chose this case as otherwise the results would become too voluminous. Naturally, the technique works for all values of $\alpha$ in the same fashion. The evaluation of~\eqref{eqn:leftinvmc} yields~\cite{Bosque:2015jda}
\begin{equation}\label{eqn:iso3vielbein}
  E^A{}_I = \begin{pmatrix}
    A_1(\phi_1,\phi_2,\phi_3) & 0 \\
    0 & B(\phi_1,\phi_2,\phi_3)
      \end{pmatrix}
\end{equation}
and its inverse transposed
\begin{equation}
  E_A{}^I = \begin{pmatrix}
    A_1^{-T}(\phi_1,\phi_2,\phi_3) & 0 \\
    0 & B(\phi_1,\phi_2,\phi_3)
  \end{pmatrix}
\end{equation}
where $A_1(\phi_1,\phi_2,\phi_3)$ is given by~\eqref{eqn:Aalpha} and $B(\phi_1,\phi_2,\phi_3)$ is defined through
\begin{equation}
  B(\phi_1,\phi_2,\phi_3) = \begin{pmatrix}
   \phantom{-}\text{c}_2 \, \text{c}_3 & \phantom{-} \text{c}_2 \, \text{s}_3 & -\text{s}_2 \\
   \phantom{-}\text{c}_3 \, \text{s}_1 \,\text{s}_2 - \text{c}_1 \, \text{s}_3 & \phantom{-}\text{c}_1 \, \text{c}_3 + \text{s}_1 \, \text{s}_2 \, \text{s}_3  & \phantom{-} \text{c}_2 \, \text{s}_1 \\
   \phantom{-} \text{c}_1 \, \text{c}_3 \, \text{s}_2 + \text{s}_1 \, \text{s}_3 & - \text{c}_3 \, \text{s}_1 + \text{c}_1 \, \text{s}_2 \, \text{s}_3  & \phantom{-}\text{c}_1 \, \text{c}_2
  \end{pmatrix} 
\end{equation}
with
\begin{equation}
    \text{s}_i = \sin \phi_i \quad \text{and} \quad
    \text{c}_i = \cos \phi_i\,.
\end{equation}

\section*{CSO(2,0,2)/$\mathfrak{f}_1$}\label{app:cso202}
Applying the rotation
\begin{equation}
  R_A{}^B = \frac{1}{\sqrt{2}} \begin{pmatrix}
    \phantom{-}1 & \phantom{-}0 & \phantom{-}0 & \phantom{-}1 & \phantom{-}0 & \phantom{-}0 \\
    \phantom{-}0 & \phantom{-}0 & \phantom{-}0 & \phantom{-}0 & \phantom{-}0 & \phantom{-}\sqrt{2} \\
    \phantom{-}0 & \phantom{-}0 & \phantom{-}0 & \phantom{-}0 & \phantom{-}\sqrt{2} & \phantom{-}0 \\
    -1 & \phantom{-}0 & \phantom{-}0 & \phantom{-}1 & \phantom{-}0 & \phantom{-}0 \\
    \phantom{-}0 & \phantom{-}0 & \phantom{-}\sqrt{2} & \phantom{-}0 & \phantom{-}0 & \phantom{-}0 \\
    \phantom{-}0 & \phantom{-}\sqrt{2} & \phantom{-}0 & \phantom{-}0 & \phantom{-}0 & \phantom{-}0
  \end{pmatrix}\,,
\end{equation}
results in
\begin{equation}
  \eta'_{AB} =  \begin{pmatrix}
    \phantom{-}0 & \phantom{-}0 & \phantom{-}0 & \phantom{-}1 & \phantom{-}0 & \phantom{-}0 \\
    \phantom{-}0 & -1 & \phantom{-}0 & \phantom{-}0 & \phantom{-}0 & \phantom{-}0 \\
    \phantom{-}0 & \phantom{-}0 & -1 & \phantom{-}0 & \phantom{-}0 & \phantom{-}0 \\
    \phantom{-}1 & \phantom{-}0 & \phantom{-}0 & \phantom{-}0 & \phantom{-}0 & \phantom{-}0 \\
    \phantom{-}0 & \phantom{-}0 & \phantom{-}0 & \phantom{-}0 & \phantom{-}1 & \phantom{-}0 \\
    \phantom{-}0 & \phantom{-}0 & \phantom{-}0 & \phantom{-}0 & \phantom{-}0 & \phantom{-}1
  \end{pmatrix}
\end{equation}
and the solvable Lie algebra
\begin{align}\label{eqn:cso202generators}
  [ t_0 ,t_a ] &= \alpha_+ \, \varepsilon_a{}^b \, t_b\,, & & &
  [ t_0 ,s_a ] &= \phantom{-} \alpha_- \, \varepsilon_a{}^b \, s_b\,, \nonumber \\
  [ t_a ,t_b ] &= \alpha_+ \, \varepsilon_{ab} \, z  & &\text{and}&
  [ s_a ,s_b ] &= -\alpha_- \, \varepsilon_{ab} \, z
\intertext{with}
  \alpha_+ &= -\cos(\alpha) - \sin(\alpha) 
    & &\text{and}&
  \alpha_- &=  \cos(\alpha) - \sin(\alpha)
\end{align}
after assigning the symbols
\begin{equation}
  t_A = \{t_0, s_1, s_2, z, t_1, t_2\}
\end{equation}
for the generators. Here, $\varepsilon_a{}^b$ denotes again the totally antisymmetric Levi-Civita tensor but in two dimensions with the signature $\varepsilon_1{}^2 = 1$, while the indices $a, b, c, \dots \in \{1,2\}$. This algebra is equivalent to $\mathfrak{cso}(2,0,2)$ for $\alpha = 0$. The derived series is given by
\begin{equation}
  L^0 = \{t_0, t_1, t_2, s_1, s_2, z\} \supset \{t_1, t_2, s_1, s_2, z\} \supset \{z\} \supset \{0\}
\end{equation}
for every $\alpha\ne \pi/4$ and $z$ is the non-trivial center. As a result, we have an unfaithful adjoint representation. We read off the nilpotent subalgebra by using the method outlined in~\ref{sec:solvable}
\begin{equation}
  \mathfrak{n} = L^1 = \{s_1, s_2, z, t_1, t_2\} \quad \text{and the remaining generators} \quad \mathfrak{q} = \{t_0\}\,.
\end{equation}
The subalgebra yields the lower central series
\begin{equation}
  L_0 = \mathfrak{n} = \{s_1, s_2, z, t_1, t_2\} \supset
  \{z\} \supset \{0\}\,,
\end{equation}
and shows that $\mathfrak{n}$ is indeed nilpotent of order $k=2$.

With the data at hand, we are able to construct the $N=16$-dimensional subspace of the universal enveloping algebra
\begin{align}
  V^2 = \{&s_1^2,\, s_1 s_2,\, s_2^2,\, t_1^2,\, t_1 t_2,\, s_1 t_1,\, s_2 t_1,\, t_2^2,\, s_1 t_2,\, s_2 t_2, z, &&\ord \cdot = 2 \nonumber \\
    &t_1,\, t_2,\, s_1,\, s_2, &&\ord \cdot = 1 \nonumber \\
    &1 \} &&\ord \cdot = 0\,.
\end{align}
This allows us to derive the generators by using the technique from section~\ref{sec:solvable}.

Again, we obtain the group elements by using the exponential map~\eqref{eqn:prodexpmap} with the coordinates
\begin{equation}\label{eqn:cso202coord}
  X^I = \{ \phi, x_1, x_2, z, y_1, y_2 \}\,.
\end{equation}
Equation~\ref{eqn:leftinvmc} allows us to compute the background vielbein and we find
\begin{equation}
  E^A{}_I = \begin{pmatrix}
    \phantom{-}1 & \phantom{-}0 & \phantom{-}0 & \phantom{-}0 & \phantom{-}0 & \phantom{-}0 \\
    \phantom{-}0 & \phantom{-}\cos(\alpha_- \phi) & \phantom{-}\sin(\alpha_-  \phi) & \phantom{-}0 & \phantom{-}0 & \phantom{-}0 \\
    \phantom{-}0 & -\sin(\alpha_- \phi) & \phantom{-}\cos(\alpha_- \phi) & \phantom{-}0 & \phantom{-}0 & \phantom{-}0 \\
    \phantom{-}0 & \phantom{-}0 & \alpha_- \, x_1 & \phantom{-}1 & \phantom{-}0 & -\alpha_+ \, y_1 \\ 
    \phantom{-}0 & \phantom{-}0 & \phantom{-}0 & \phantom{-}0 & \phantom{-}\cos(\alpha_+ \, \phi) & \phantom{-}\sin(\alpha_+ \, \phi) \\
    \phantom{-}0 & \phantom{-}0 & \phantom{-}0 & \phantom{-}0 & -\sin(\alpha_+ \, \phi) & \phantom{-}\cos( \alpha_+ \, \phi)
  \end{pmatrix}
\end{equation}
with the inverse transposed
\begin{equation}
E_A{}^I = \begin{pmatrix}
    1 & \phantom{-}0 & 0 & \phantom{-}0 & \phantom{-}0 & 0 \\
    0 & \phantom{-}\cos(\alpha_- \phi) & \sin(\alpha_-  \phi) & -\alpha_- \, x_1 \, \sin(\alpha_- \phi) & \phantom{-}0 & 0 \\
    0 & -\sin(\alpha_- \phi) & \cos(\alpha_- \phi) & -\alpha_- \, x_1 \, \cos(\alpha_- \phi) & \phantom{-}0 & 0 \\
    0 & \phantom{-}0 & 0 & \phantom{-}1 & \phantom{-}0 & 0 \\ 
    0 & \phantom{-}0 & 0 & \phantom{-}\alpha_+ \, y_1 \, \sin(\alpha_+ \phi) & \phantom{-}\cos(\alpha_+ \, \phi) & \sin(\alpha_+ \, \phi) \\
    0 & \phantom{-}0 & 0 & \phantom{-}\alpha_+ \, y_1 \, \cos(\alpha_+ \phi) & -\sin(\alpha_+ \, \phi) & \cos( \alpha_+ \, \phi)
  \end{pmatrix}\,.
\end{equation}
The given background describes a twisted torus. The base is given by a circle with coordinate $\phi$. A two dimensional torus is fibered over this circle. Then, the monodromy arising after one complete succession around the base is expressed through the complex structure / Kähler parameter of the fibered torus. We distinguish between two important cases: For $\alpha=0$ we find a geometric solve manifold which is also called single elliptic case. For $\alpha\ne 0$ it corresponds to the double elliptic cases~\cite{Hassler:2014sba}, a background which is not T-Dual to any geometric setup. For $\alpha =\pm \pi/4$ the group reduces to $\mathfrak{f}_1$ as either $\alpha_+$ or $\alpha_-$ vanishes~\cite{Bosque:2015jda,Dibitetto:2012rk}.

\section*{$\mathfrak{h}_1$}
Applying the rotation
\begin{equation}
  R_A{}^B = \frac{1}{\sqrt{2}} \begin{pmatrix}
    -1 & \phantom{-}0 & \phantom{-}0 & \phantom{-}1 & \phantom{-}0 & \phantom{-}0 \\
    \phantom{-}1 & \phantom{-}0 & \phantom{-}0 & \phantom{-}1 & \phantom{-}0 & \phantom{-}0 \\
    \phantom{-}0 & -1 & \phantom{-}0 & \phantom{-}0 & \phantom{-}0 & \phantom{-}1 \\
    \phantom{-}0 & \phantom{-}0 & -1 & \phantom{-}0 & -1 & \phantom{-}0 \\
    \phantom{-}0 & \phantom{-}1 & \phantom{-}0 & \phantom{-}0 & \phantom{-}0 & \phantom{-}1 \\
    \phantom{-}0 & \phantom{-}0 & -1 & \phantom{-}0 & \phantom{-}1 & \phantom{-}0
  \end{pmatrix}\,,
\end{equation}
results in
\begin{equation}  
  \eta'_{AB} = \begin{pmatrix}
    \phantom{-}0 & \phantom{-}0 & \phantom{-}0 & -1 & \phantom{-}0 & \phantom{-}0 \\
    \phantom{-}0 & \phantom{-}0 & \phantom{-}0 & \phantom{-}0 & \phantom{-}0 & \phantom{-}1 \\
    \phantom{-}0 & \phantom{-}0 & \phantom{-}0 & \phantom{-}0 & -1 & \phantom{-}0 \\ -1 & \phantom{-}0 & \phantom{-}0 & \phantom{-}0 & \phantom{-}0 & \phantom{-}0 \\
    \phantom{-}0 & \phantom{-}0 & -1 & \phantom{-}0 & \phantom{-}0 & \phantom{-}0 \\
    \phantom{-}0 & \phantom{-}1 & \phantom{-}0 & \phantom{-}0 & \phantom{-}0 & \phantom{-}0
  \end{pmatrix}
\end{equation}
and the solvable Lie algebra
\begin{align}
  [t_0, t_a] &= \phantom{-}\cos(\alpha) \, \varepsilon_a{}^b \, t_b\,, & &&
  [t_0, s_a] &= \phantom{-}\cos(\alpha) \, \varepsilon_a{}^b \, s_b - \sin(\alpha) \, t_a\,, \\
  [s_a, s_b] &= -\sin(\alpha)\, \varepsilon_{ab} \, z   & &\text{and}&
  [t_a, s_b] &=  -\cos(\alpha) \, \delta_{ab} \, z\,,
\end{align}
after assigning the symbols
\begin{equation}
  t_A = \{t_0, s_1, s_2, z, t_1, t_2\}\,.
\end{equation}
The derived series as well as the lower central series of its nilpotent subalgebra $\mathfrak{n}$ equal the $\mathfrak{cso}(2,0,2)$ discussed before. Hence, deriving its $N=16$-dimensional matrix representation works exactly as in appendix~\ref{app:cso202}. 

Group elements arise again using the exponential map~\eqref{eqn:prodexpmap} with the coordinates previously given in~\eqref{eqn:cso202coord}. In this context, we only present the background vielbein for $\alpha=0$. We exactly recover the $\mathfrak{h}_1$ algebra given in~\cite{Dibitetto:2012rk}. Nevertheless, this restriction is not imperative. It however significantly simplifies the results of the left invariant Maurer-Cartan form~\cite{Bosque:2015jda}
\begin{equation}
  E^A{}_I = \begin{pmatrix}
  \phantom{-}1 & \phantom{-}0 & \phantom{-}0 & \phantom{-}0 & \phantom{-}0 & \phantom{-}0 \\ \phantom{-}0 & \phantom{-}\cos(\phi) & \phantom{-}\sin(\phi) & \phantom{-}0 & \phantom{-}0 & \phantom{-}0 \\ \phantom{-}0 & -\sin(\phi) & \phantom{-}\cos(\phi) & \phantom{-}0 & \phantom{-}0 & \phantom{-}0 \\ \phantom{-}0 & \phantom{-}0 & \phantom{-}0 & \phantom{-}1 & -x_1 & -x_2 \\ \phantom{-}0 & \phantom{-}0 & \phantom{-}0 & \phantom{-}0 & \phantom{-}\cos(\phi) & \phantom{-}\sin(\phi) \\ \phantom{-}0 & \phantom{-}0 & \phantom{-}0 & \phantom{-}0 & -\sin(\phi) & \phantom{-}\cos(\phi) \\ 
  \end{pmatrix}
\end{equation}
and its inverse transposed
\begin{equation}
  E_A{}^I = \begin{pmatrix}
  \phantom{-}1 & \phantom{-}0 & \phantom{-}0 & \phantom{-}0 & \phantom{-}0 & \phantom{-}0 \\ \phantom{-}0 & \phantom{-}\cos(\phi) & \phantom{-}\sin(\phi) & \phantom{-}0 & \phantom{-}0 & \phantom{-}0 \\ \phantom{-}0 & -\sin(\phi) & \phantom{-}\cos(\phi) & \phantom{-}0 & \phantom{-}0 & \phantom{-}0 \\ \phantom{-}0 & \phantom{-}0 & \phantom{-}0 & \phantom{-}1 & \phantom{-}0 & \phantom{-}0 \\ \phantom{-}0 & \phantom{-}0 & \phantom{-}0 & x_1 \, \cos(\phi) + x_2 \, \sin(\phi) & \phantom{-}\cos(\phi) & \phantom{-}\sin(\phi) \\ \phantom{-}0 & \phantom{-}0 & \phantom{-}0 & x_2 \, \cos(\phi) - x_1 \, \sin(\phi) & -\sin(\phi) & \phantom{-}\cos(\phi) \\
  \end{pmatrix} \,.
\end{equation}

\section*{CSO(1,0,3)/$\mathfrak{l}$}
Applying the rotation
\begin{equation}
  R_A{}^B = \frac{1}{\sqrt{2}} \begin{pmatrix}
    \phantom{-}0 & -1 & \phantom{-}0 & 0 & \phantom{-}1 & \phantom{-}0 \\
    -1 & \phantom{-}0 & \phantom{-}0 & 1 & \phantom{-}0 & \phantom{-}0 \\
    \phantom{-}0 & \phantom{-}0 & \cos(\alpha) + \sin(\alpha) & 0 & \phantom{-}0 & -\cos(\alpha) + \sin(\alpha)
\\
    \phantom{-}0 & \phantom{-}1 & \phantom{-}0 & 0 & \phantom{-}1 & \phantom{-}0 \\ \phantom{-}1 & \phantom{-}0 & \phantom{-}0 & 1 & \phantom{-}0 & \phantom{-}0 \\
    \phantom{-}0 & \phantom{-}0 & -\cos(\alpha) + \sin(\alpha) & 0 & \phantom{-}0 & -\cos(\alpha) - \sin(\alpha)
  \end{pmatrix}\,,
\end{equation}
results in
\begin{equation}\label{eqn:etacso(103)}
  \eta'_{AB} = \begin{pmatrix}
    \phantom{-}0 & \phantom{-}0 & \phantom{-}0 & -1 & \phantom{-}0 & \phantom{-}0 \\
    \phantom{-}0 & \phantom{-}0 & \phantom{-}0 & \phantom{-}0 & -1 & \phantom{-}0 \\
    \phantom{-}0 & \phantom{-}0 & \phantom{-}\sin(2\alpha) & \phantom{-}0 & \phantom{-}0 & -\cos(2\alpha) \\
    -1 & \phantom{-}0 & \phantom{-}0 & \phantom{-}0 & \phantom{-}0 & \phantom{-}0 \\
    \phantom{-}0 & -1 & \phantom{-}0 & \phantom{-}0 & \phantom{-}0 & \phantom{-}0 \\
    \phantom{-}0 & \phantom{-}0 & -\cos(2\alpha) & \phantom{-}0 & \phantom{-}0 & -\sin(2\alpha)
  \end{pmatrix}
\end{equation}
and the nilpotent Lie algebra
\begin{equation}
\label{eqn:cso103commutator}
  [t_1, t_2 ] = \cos(2\alpha) \, z_3 - \sin(2\alpha) \, t_3\,, \quad [t_2, t_3] = z_1 \quad \text{and} \quad [t_3, t_1] =  z_2
\end{equation}
after assigning the symbols
\begin{equation}
  t_A = \{t_1, t_2, t_3, z_1, z_2, z_3\}\,.
\end{equation}
In the case $\alpha=0$, we find the Lie algebra
\begin{equation}
  [t_a, t_b] = \varepsilon_{ab}{}^c \, z_c
\end{equation}
called $\mathfrak{cso}(1,0,3)$~\cite{Dibitetto:2012rk}. The algebra is nilpotent of order 2 and its lower central series is given through
\begin{equation}
  L_0 = \{ t_1, t_2, t_3, z_1, z_2, z_3 \} \supset \{ z_1, z_2, z_3 \} \supset \{0\}\,.
\end{equation}
We find for the center of the algebra $\{ z_1, z_2, z_3 \}$.  Again, we perform the methods presented in section~\ref{sec:nilpotent} and construct the $N=13$-dimensional subspace
\begin{align}
  V^2 = \{&t_1^2,\, t_1 t_2,\, t_1 t_3,\, t_2^2,\, t_2 t_3,\, t_3^2,\, z^1,\, z^2,\, z^3, &&\ord \cdot = 2 \nonumber \\
    &t_1,\, t_2,\, t_3, &&\ord \cdot = 1 \nonumber \\
    &1 \} &&\ord \cdot = 0
\end{align}
of the universal enveloping algebra. Finally, we are able to derive the matrix representation for the generators $t_A$ by expanding the linear maps $\phi_{t_A}$ in the basis spanned by $V^2$. Group elements are again obtained using the exponential map~\eqref{eqn:prodexpmap} using the coordinates
\begin{equation}
  X^I = \{ x_1, x_2, x_3, z_1, z_2, z_3 \}\,.
\end{equation}
We derive the left invariant Maurer-Cartan form and find
\begin{equation}
  E^A{}_I = \begin{pmatrix}
    \phantom{-}1 & \phantom{-}0 & \phantom{-}0 & \phantom{-}0 & \phantom{-}0 & \phantom{-}0 \\
    \phantom{-}0 & \phantom{-}1 & \phantom{-}0 & \phantom{-}0 & \phantom{-}0 & \phantom{-}0 \\
    \phantom{-}0 & \phantom{-}0 & \phantom{-}1 & \phantom{-}0 & \phantom{-}0 & \phantom{-}0 \\
    \phantom{-}0 & \phantom{-}0 & -x_2 & \phantom{-}1 & \phantom{-}0 & \phantom{-}0 \\ \phantom{-}0 & \phantom{-}0 & \phantom{-}x_1 & \phantom{-}0 & \phantom{-}1 & \phantom{-}0 \\
    \phantom{-}0 & -x_1 & \phantom{-}0 & \phantom{-}0 & \phantom{-}0 & \phantom{-}1 \\
  \end{pmatrix}
\end{equation}
with the inverse transposed
\begin{equation}
  E_A{}^I = \begin{pmatrix}
    \phantom{-}1 & \phantom{-}0 & \phantom{-}0 & \phantom{-}0 & \phantom{-}0 & \phantom{-}0 \\
    \phantom{-}0 & \phantom{-}1 & \phantom{-}0 & \phantom{-}0 & \phantom{-}0 & \phantom{-}x_1 \\
    \phantom{-}0 & \phantom{-}0 & \phantom{-}1 & \phantom{-}x_2 & -x_1 & \phantom{-}0 \\
    \phantom{-}0 & \phantom{-}0 & \phantom{-}0 & \phantom{-}1 & \phantom{-}0 & \phantom{-}0 \\ \phantom{-}0 & \phantom{-}0 & \phantom{-}0 & \phantom{-}0 & \phantom{-}1 & \phantom{-}0 \\
    \phantom{-}0 & \phantom{-}0 & \phantom{-}0 & \phantom{-}0 & \phantom{-}0 & \phantom{-}1 \\
  \end{pmatrix}\,.
\end{equation}
The background describes a $3$-torus with $H$-flux.

For all $\alpha\ne0$, the lower central series alters to
\begin{equation}
  L_0 = \{ t_1, t_2, t_3', z_1, z_2, z_3' \} \supset \{ z_1, z_2, z_3' \} \supset \{z_1, z_2\} \supset \{0\}\,,
\end{equation}
where we used the abbreviations
\begin{equation}\label{eqn:t3z3bar}
  t_3' = \cos(2\alpha) t_3 - \sin(2\alpha) z_3 \quad \text{and} \quad
  z_3' = \sin(2\alpha) t_3 + \cos(2\alpha) z_3\,.
\end{equation}
We identify it as a nilpotent Lie algebra of order $3$. If we want to treat it properly, we have to extend $V^2$ by
\begin{align}
  V^3 = \{& t_1^3,\, t_1^2 t_2,\, t_1^2 t_3',\,t_1 t_2^2,\, t_1 t_2 t_3,\,
    t_1 {t_3}'^2,\, t_2 {t_3}'^2,\, {t_3}'^3,\, t_1 z_3',\, t_2 z_3',\,   t_3' z_3',\, z_1,\, z_2, &&\ord \cdot = 3 \nonumber \\  
    & t_1^2,\, t_1 t_2,\, t_1 t_3',\, t_2^2,\, t_2 t_3',\, {t_3'}^2,\, z_3', &&\ord \cdot = 2 \nonumber \\
    &t_1,\, t_2,\, t_3', &&\ord \cdot = 1 \nonumber \\
    &1 \}\,. &&\ord \cdot = 0
\end{align} 
This gives rise to the adjusted Lie algebra
\begin{align}
  [t_1, t_2]  &= z_3'\,, & [t_1, z_3'] &= \sin(2 \alpha) z_2\,, & [z_3', t_2] &= \sin(2 \alpha) z_1\,, \\
  [t_2, t_3'] &= \cos(2 \alpha) z_1 & & \text{and} & [t_3', t_1] &= \cos(2 \alpha) z_2\,,
\end{align}
used to evaluate the map $\phi_{t_A}$ in the basis $V^3$. It allows us to derive the $N=24$-dimensional matrix representation for the generators of the Lie algebra. We exponentiate them using~\eqref{eqn:prodexpmap}. Finally, we obtain the background vielbein as
\begin{equation}
  E^A{}_I = \begin{pmatrix} \phantom{-}1 & \phantom{-}0 & \phantom{-}0 & \phantom{-}0 & \phantom{-}0 & \phantom{-}0 \\
  \phantom{-}0 & \phantom{-}1 & \phantom{-}0 & \phantom{-}0 & \phantom{-}0 & \phantom{-}0  \\ \phantom{-}0 & \phantom{-}0 & \phantom{-}1 & \phantom{-}0 & \phantom{-}0 & \phantom{-}0  \\ \phantom{-}0 & \phantom{-}0 & -x_2 \, \cos(2\alpha) & \phantom{-}1 & \phantom{-}0 & \phantom{-}x_2 \, \sin(2\alpha)  \\ \phantom{-}0 & \phantom{-}x_1^2 \, \cos(\alpha) \, \sin(\alpha)  & \phantom{-}x_1 \, \cos(2\alpha) & \phantom{-}0 & \phantom{-}1 & -x_1 \, \sin(2\alpha)  \\ \phantom{-}0 & -x_1 & \phantom{-}0 & \phantom{-}0 & \phantom{-}0 & \phantom{-}1  \end{pmatrix}
\end{equation}
by assigning the coordinates
\begin{equation}
  X^I = \{ x_1, x_2, x_3', z_1, z_2, z_3' \}\,.
\end{equation}
The inverse transposed reads
\begin{equation}
E_A{}^I =\begin{pmatrix} \phantom{-}1 & \phantom{-}0 & \phantom{-}0 & \phantom{-}0 & \phantom{-}0 & \phantom{-}0 \\
  \phantom{-}0 & \phantom{-}1 & \phantom{-}0 & -x_1 \, x_2 \, \sin(2\alpha) & \phantom{-}x_1^2 \, \cos(\alpha) \, \sin(\alpha) & \phantom{-}x_1  \\ \phantom{-}0 & \phantom{-}0 & \phantom{-}1 & \phantom{-}x_2 \, \cos(2\alpha) & -x_1 \, \cos(2\alpha) & \phantom{-}0  \\ \phantom{-}0 & \phantom{-}0 & \phantom{-}0 & \phantom{-}1 & \phantom{-}0 & \phantom{-}0  \\ \phantom{-}0 & \phantom{-}0  & \phantom{-}0 & \phantom{-}0 & \phantom{-}1 & \phantom{-}0  \\ \phantom{-}0 & \phantom{-}0 & \phantom{-}0 & -x_2 \, \sin(2\alpha) & \phantom{-}x_1 \, \sin(2\alpha) & \phantom{-}1  \end{pmatrix}\,.
\end{equation}
The flat indices $A,B,C,\ldots$ are lowered with the $\eta_{AB}$ metric 
\begin{equation}
  \eta''_{AB} = \begin{pmatrix}
    \phantom{-}0 & \phantom{-}0 & \phantom{-}0 & -1 & \phantom{-}0 & \phantom{-}0 \\
    \phantom{-}0 & \phantom{-}0 & \phantom{-}0 & \phantom{-}0 & -1 & \phantom{-}0 \\
    \phantom{-}0 & \phantom{-}0 & -\sin(2\alpha) & \phantom{-}0 & \phantom{-}0 & -\cos(2\alpha) \\
    -1 & \phantom{-}0 & \phantom{-}0 & \phantom{-}0 & \phantom{-}0 & \phantom{-}0 \\
    \phantom{-}0 & -1 & \phantom{-}0 & \phantom{-}0 & \phantom{-}0 & \phantom{-}0 \\
    \phantom{-}0 & \phantom{-}0 & -\cos(2\alpha) & \phantom{-}0 & \phantom{-}0 & \phantom{-}\sin(2\alpha)
  \end{pmatrix}\,.
\end{equation}
In the case $\alpha=\pi/4$, we obtain the algebra $\mathfrak{l}$ given in~\cite{Dibitetto:2012rk}, after performing a further rotation of the $\mathfrak{cso}(1,0,3)$ structure coefficients~\eqref{eqn:cso103commutator} with
\begin{equation}
  R''_A{}^B = \begin{pmatrix}
    -1 & \phantom{-}0 & \phantom{-}0 & \phantom{-}0 & \phantom{-}0 & \phantom{-}0 \\
    \phantom{-}0 & \phantom{-}1 & \phantom{-}0 & \phantom{-}0 & \phantom{-}0 & \phantom{-}0 \\
    \phantom{-}0 & \phantom{-}0 & \phantom{-}0 & \phantom{-}1 & \phantom{-}0 & \phantom{-}0 \\
    \phantom{-}0 & \phantom{-}0 & \phantom{-}0 & \phantom{-}0 & \phantom{-}0 & \phantom{-}1 \\
    \phantom{-}0 & \phantom{-}0 & \phantom{-}0 & \phantom{-}0 & \phantom{-}1 & \phantom{-}0 \\
    \phantom{-}0 & \phantom{-}0 & \phantom{-}1 & \phantom{-}0 & \phantom{-}0 & \phantom{-}0
  \end{pmatrix}
\end{equation}
resulting in
\begin{equation}  
  \eta'''_{AB} = \begin{pmatrix}
    \phantom{-}0 & \phantom{-}0 & \phantom{-}0 & \phantom{-}0 & \phantom{-}0 & \phantom{-}1  \\
    \phantom{-}0 & \phantom{-}0 & \phantom{-}0 & \phantom{-}0 & -1 & \phantom{-}0  \\
    \phantom{-}0 & \phantom{-}0 & -1 & \phantom{-}0 & \phantom{-}0 & \phantom{-}0  \\
    \phantom{-}0 & \phantom{-}0 & \phantom{-}0 & \phantom{-}1 & \phantom{-}0 & \phantom{-}0 \\
    \phantom{-}0 & -1 & \phantom{-}0 & \phantom{-}0 & \phantom{-}0 & \phantom{-}0  \\
    \phantom{-}1 & \phantom{-}0 & \phantom{-}0 & \phantom{-}0 & \phantom{-}0 & \phantom{-}0 
  \end{pmatrix}
\end{equation}
and the commutation relations
\begin{equation}
  \big[t_1, t_2 \big] = t_4 \quad \big[t_1, t_4 \big] =  t_5 \quad \big[t_2, t_4 \big] =  t_6\,,
\end{equation}
where we assigned
\begin{equation}
  t_A = \{ t_1, t_2, t_3, t_4, t_5, t_6 \}
\end{equation}
for the generators~\cite{Bosque:2015jda}.\clearpage{}
  \clearpage{}\chapter{SL(\texorpdfstring{$n$}{n}) Representation Theory}
\label{AppendixD}
\label{app:SLnrepresentations}

We start this part of the appendix by reviewing the construction of projectors on $\mathfrak{sl}(N)$ irreps from Young symmetrizers. Moreover, we demonstrate how it is possible to decompose tensor products with the help of these projectors into direct sums. As a first application of these concepts, we solve the linear constraints found in subsection~\ref{sec:genLie} for the T-duality group SL($4$)~\cite{Bosque:2017dfc}.

\section*{Theory: Young Tableaux and Projectors on Irreps}

At first, it is essential to set some conventions: A Young diagram is a set of $n$ boxes arranged in rows and columns, beginning from the left. The amount of boxes contained in each row may not increase when going from the top to the bottom of the diagram~\cite{Bosque:2017dfc}. An illustrative example for $n=6$ is
\begin{equation}\label{eqn2:ex321}
  \ydiagram{3,2,1}\,.
\end{equation}
This diagram is analogous to the partition $(3,2,1)$ of $6$. Such a diagram turns into a Young tableau, once we start writing numbers from one to $n$ into the boxes. Generally, there exist $n!$ distinct ways of doing so. If the numbers of the tableaux are increasing in every row and column simultaneously, it is called a standard tableau. Counting the number of standard tableaux for a given diagram is achieved by using the hook length formula: For each box in a diagram $\lambda$ we need to count the number of boxes in the same row $i$ from the left to the right and the number of boxes in the same column $j$ below. For the box itself, obtaining the hook length $h_\lambda(i,j)$ requires adding an extra one to the result. With this information, we are now able to compute the number of standard tableaux
\begin{equation}
  d_{\mathrm{std}} = \frac{n!}{\prod h_\lambda(i,j) }\,.
\end{equation}
Choose for instance example~\ref{eqn2:ex321}, it yields
\begin{equation*}
  \ytableaushort{531,31,1} \quad\text{for each box and }
  d_{\mathrm{std}} = \frac{6!}{5\cdot3^2} = 16\,.
\end{equation*}
Coming from a Young tableaux $t$ we combine all permutations from the symmetric group $S_n$. These only shuffle elements within each row into a new row group $R_t$. Equivalently, all permutations which only shuffle elements in columns can be fused into the column group $C_t$. Together $R_t$ and $C_t$ generate the Young symmetrizer
\begin{equation}
  e_t = \sum_{\pi\in R_t,\,\sigma\in C_t} \mathrm{sign}(\sigma) \sigma\circ\pi\,.
\end{equation}
A demonstrative example for it is given by
\begin{equation}\label{eqn2:ytabexample}
  t = \ytableaushort{12,3} \quad\text{and}\quad
  e_t = \big( () - (1\,3) \big)\big( () + (1\,2) \big) = () + (1\,2) - (1\,3) - (3\,2\,1)\,,
\end{equation}
where we use cyclic notation for elements in $S_3$. We are interested in applying the Young symmetrizer $e_t$ to tensors such as $X_{a_1 \dots a_n}$ as well, where the permutations act on the individual indices. Take for example the tableau $t$ from~\eqref{eqn2:ytabexample}, we obtain
\begin{equation}\label{eqn2:youngsym21}
  e_t X_{a_1 a_2 a_3} = X_{a_1 a_2 a_3} + X_{a_2 a_1 a_3} - X_{a_3 a_2 a_1} - X_{a_2 a_3 a_1}\,.
\end{equation}
It is straightforward to verify that the resulting tensor is antisymmetric in the first two indices $a_1$, $a_2$ and that furthermore the total antisymmetrization $X_{[a_1 a_2 a_3]}$ vanishes. If the indices $a_i=1,\ldots,N$ live in the fundamental of $\mathfrak{sl}(N)$, the induced tensor $e_t X_{a_1 a_2 a_3}$ is an irrep of the Lie algebra. Hence, the Young symmetrizer $e_t$ is proportional to the projector of a tensor product onto this irrep. This observation also succeeds for all other Young tableaux. Computing the dimension of the irrep we project from the tableaux $t$ makes it necessary to assign the number $N$ to the top left corner of the diagram $\lambda$ associated to $t$. In every column to the right we increase this number and in every row below we reduce it. Using again the diagram~\ref{eqn2:ex321} as an instructive example, we find
\begin{equation*}
  \ytableausetup{mathmode, boxsize=2em}
  \begin{ytableau}
    \scriptstyle N & \scriptstyle N+1 & \scriptstyle N+2 \cr
    \scriptstyle N-1 & \scriptstyle N  \cr
    \scriptstyle N-2 \cr
  \end{ytableau}\,.
  \ytableausetup{boxsize=1em,aligntableaux=center}
\end{equation*}
These numbers are in correspondence to the hook length with $f_\lambda(i,j)$. Finally, the dimension of the irrep associated to $t$ can be obtained through
\begin{equation}
  d_{\mathrm{irrep}} = \frac{\prod f_\lambda(i,j)}{\prod h_\lambda(i,j)}\,,
\end{equation}
which reproduces the dimension $N(N^2-1)/3$ for the Young symmetrizer~\eqref{eqn2:youngsym21}. For $N=5$ it yields $\overline{\mathbf{40}}$, exactly one of the two irreps contained in the embedding tensor.

As previously mentioned $e_t$ is only proportional to a projector and satisfies the relation
\begin{equation}
  e_t e_t = k_t e_t\,,
\end{equation}
where $k_t$ is a constant which depends on the tableaux $t$. We use this observation to introduce the projector onto $t$ through
\begin{equation}
  P_t = \frac{1}{k_t} e_t \quad \text{with} \quad P_t^2 = P_t\,.
\end{equation}
Those projectors are accompanied by the following properties:
\begin{itemize}
  \item Projectors of tableaux corresponding to different diagrams are orthogonal.
  \item Projectors of standard tableaux are linear independent. They can be combined into a system of orthogonal projectors $P_{\lambda, i}$. Here, $\lambda$ denotes the diagram they decent from.
  \item The total sum of all these projectors for all diagrams with $n$ boxes is the identity element of $S_n$.
\end{itemize}
Now, assume that we possess a projector $P$ onto a reducible representation and want to decompose it into a sum of orthogonal projectors $P_{\lambda, i}$ onto irreps by
\begin{equation}\label{eqn2:decompP}
  P = \sum_\lambda \sum_i P_{\lambda, i}\,.
\end{equation}
As we already mentioned, these orthogonal projectors emerge from a sum
\begin{equation}
  P_{\lambda, i} = \sum\limits_t (c_{\lambda, i})_t e_t \circ P
\end{equation}
over different projectors arising from standard tableaux for a specific diagram $\lambda$. However, the coefficients $(c_{\lambda, i})_t$ appearing in this expansion still have to be fixed. It can be achieved by demanding that the commutator
\begin{equation}
  [P, P_{\lambda, i}] = \sum_t (c_{\lambda, i})_t [e_t \circ P, P] = 0
\end{equation}
of $P$ with each $P_{\lambda, i}$ vanishes. For the resulting null space we choose an orthonormal basis, i.e.~\cite{Bosque:2017dfc}
\begin{equation}
  P_{\lambda, i} \circ P_{\lambda, j} = \begin{cases} P_{\lambda, i} & i = j\\ 0 & i \ne j\,. \end{cases}
\end{equation}

\section*{Application: Linear Constraints for SL(4)}

In the following, we want to solve the linear constraints from subsection~\ref{sec:genLie}. Subsequently, we need to decompose the connection $\Gamma_{AB}{}^C$ on which the constraints are working into several irreps. During this part of the appendix, we work with the Lie algebra $\mathfrak{sl}(4)$~\cite{Bosque:2017dfc}. Thus, we denote indices in the irrep $\mathbf{6}$ with capital letters and small letters represent the fundamental representation $\mathbf{4}$. As we explained in the first part, Young symmetrizer only operate on the latter. Hence, we identify
\begin{equation}
  \Gamma_{AB}{}^C \rightarrow \Gamma_{[a_1 a_2], [b_1 b_2]}{}^{[c_1 c_2]}\,.
\end{equation}
At this state, it becomes important to differentiate between raised and lowered indices. The former live in the $\mathbf{6}$, whereas the latter are elements of the dual $\overline{\mathbf 6}$\footnote{ Note that for $\mathfrak{sl}(4)$ the six-dimensional representation is real, e.g. $\mathbf{6} = \overline{\mathbf 6}$. Therefore, it is generally not crucial to distinguish between them. However, it will help during the disucssion of $\mathfrak{sl}(5)$.}. Switching between these irreps can be performed by contracting with the totally antisymmetric tensor
\begin{equation}\label{eqn2:lowerdouleind}
  \Gamma_{a_1 a_2, b_1 b_2, c_1 c_2} = \Gamma_{a_1 a_2, b_1 b_2}{}^{d_1 d_2} \epsilon_{d_1 d_2 c_1 c_2}\,.
\end{equation}
The connection possesses in total 216 independent components. These are given by the following irreps
\begin{equation}\label{eqn2:666}
  \mathbf{6}\times\mathbf{6}\times\overline{\mathbf 6} = 3 (\mathbf{6}) + \mathbf{10} + \overline{\mathbf 10} + \mathbf{50} + 2 (\mathbf{64})\,.
\end{equation}
Each of them owns a corresponding Young diagram
\begin{equation}
  \ydiagram{1,1}\times\left(\,\ydiagram{1,1}\times\ydiagram{1,1}\,\right) = 3\, \ydiagram{2,2,1,1} + \ydiagram{3,1,1,1} + \ydiagram{2,2,2} + \ydiagram{3,3} + 2\, \ydiagram{3,2,1}\,.
\end{equation}
On the right hand side of this equation we find the projector
\begin{equation}\label{eqn2:P666}
  P_{\mathbf{6}\times\mathbf{6}\times\overline{\mathbf 6}} = \frac18 \big( () - (1\,2) \big) \big( () - (3\,4) \big) \big( () - (5\,6) \big)
\end{equation}
on a reducible representation. Obtaining the complete decomposition of this projector into a sum~\eqref{eqn2:decompP}, we additionally have to take the diagrams $(1,1,1,1,1,1)$ and $(2,1,1,1,1)$ into account. Even though they clearly vanish in the case of $\mathfrak{sl}(4)$. Nevertheless, they still pay contributions to the full decomposition into irreps of the symmetric group $S_6$. Whereas the first one only generates one projector, the second one induces two additional orthogonal projectors. As we did for~\eqref{eqn2:P666}, we neglect their contributions in the remainder of this appendix. If a diagram surfaces more than once in a decomposition, there exist different schemes to organize the associated projectors. Here, we work with the following approach
\begin{equation}\label{eqn2:P666names}
  \mathbf{6} \times ( \mathbf{6}\times \overline{\mathbf 6} ) = \mathbf{6} \times (\mathbf{1} + \mathbf{15} + \mathbf{20}') = \left\{ \begin{array}{ll}
    \mathbf{6} \times \mathbf{1} &= \mathbf{6}a \\
    \mathbf{6} \times \mathbf{15} &= \mathbf{6}b + \mathbf{10} + \overline{\mathbf 10} + \mathbf{64}a \\
    \mathbf{6} \times \mathbf{20}' &= \mathbf{6}c + \mathbf{50} + \mathbf{64}b
  \end{array}\right. \,.
\end{equation}
Thus, we are finally able to write down the resulting decomposition
\begin{equation}\label{eqn2:P6662}
  P_{\mathbf{6}\times\mathbf{6}\times\overline{\mathbf 6}} = P_{\mathbf{6}a} + P_{\mathbf{6}b} + P_{\mathbf{6}c} + P_{\mathbf{10}} + P_{\overline{\mathbf 10}} + P_{\mathbf{50}} + P_{\mathbf{64}a} + P_{\mathbf{64}b}\,.
\end{equation}
Now, we are at the point to analyze the first linear constraint~\eqref{eqn2:linconst1}. In fundamental indices, it takes on the form
\begin{align}
  C_{a_1 a_2, b_1 b_2, c_1 c_2, d_1 d_2, e_1 e_2} =&\, \epsilon_{a_1 a_2 b_1 b_2} ( - \Gamma_{c_1 c_2, d_1 d_2, e_1 e_2} - \Gamma_{c_1 c_2, e_1 e_2, d_1 d_2 } ) +  \nonumber \\
  &\, \epsilon_{d_1 d_2 e_1 e_2} ( \Gamma_{c_1 c_2, b_1 b_2, a_1 a_2} - \Gamma_{c_1 c_2, a_1 a_2, b_1 b_2 } )
\end{align}
after inserting the $Y$-tensor
\begin{equation}
  Y^{a_1 a_2, b_1 b_2}{}_{c_1 c_2, d_1 d_2} = \frac{1}{4}\epsilon^{a_1 a_2 b_1 b_2} \epsilon_{c_1 c_2 d_1 d_2}
\end{equation}
and lowering all indices with the antisymmetric tensor as described in~\eqref{eqn2:lowerdouleind}. Clearly, for this express to vanish all the terms appearing in the brackets need to cancel each other. However, they are not linearly independent. Subsequently, we need to solve the constraint
\begin{equation}
  \Gamma_{a_1 a_2, b_1 b_2, c_1 c_2} + \Gamma_{a_1 a_2, c_1 c_2, b_1 b_2} = 0\,,
\end{equation}
which can be recast through a projector
\begin{equation}
  2 P_1 \Gamma_{a_1 a_2, b_1 b_2, c_1 c_2} = 0 \quad \text{with} \quad
  P_1 = \frac{1}{2} \big( () + (3\,5) (4\,6) \big)\,.
\end{equation}
All irreps of the decomposition~\eqref{eqn2:P6662} which are not in the kernel of this projector and therefore violate~\eqref{eqn2:linconst1} have to vanish. As a consequence, we replace~\eqref{eqn2:P6662} by
\begin{equation}
  (1 - P_1) P_{\mathbf{6}\times\mathbf{6}\times\overline{\mathbf 6}} = 
    P_{\mathbf{6}b} + P_{\mathbf{10}} + P_{\overline{\mathbf 10}} +    P_{\mathbf{64}a}\,.
\end{equation}
Astonishingly, this equations exactly reproduces the embedding tensor components of half-maximal, electrically gauged supergravities in seven dimensions. Although, not all of these irreps survive the linear constraint~\cite{Samtleben:2005bp}. Let us verify whether this is the case for our setup as well. Thus,  we compute $X_{AB}{}^C$ according to~\eqref{eqn2:XfromGamma}. In components, this equation yields
\begin{equation}
  X_{a_1 a_2, b_1 b_2, c_1 c_2} = \Gamma_{a_1 a_2, b_1 b_2, c_1 c_2} - \Gamma_{b_1 b_2, a_1 a_2, c_1 c_2} + \Gamma_{c_1 c_2,a_1 a_2, b_1 b_2}
\end{equation}
or when written in terms of permutations
\begin{equation}
  \sigma_X = () - (1\,3)(2\,4) + (1\,3\,5)(2\,4\,6)
		\quad \text{as} \quad
	X_{a_1 a_2, b_1 b_2, c_1 c_2} = 
	\sigma_X (1 - P_1) \Gamma_{a_1 a_2, b_1 b_2, c_1 c_2}\,.
\end{equation}
Again, we rework $\sigma_X$ in terms of orthogonal irrep projectors and finally obtain
\begin{equation}\label{eqn2:sollinSL4}
  \sigma_X (1 - P_1) P_{\mathbf{6}\times\mathbf{6}\times\overline{\mathbf 6}} = 3 P_{\mathbf{10}} + 3 P_{\overline{\mathbf 10}}\,.
\end{equation}
These two irreps give rise to the 20 independent components of the totally antisymmetric tensor $F_{ABC}$ (structure coefficients).

From our previous consideration, we already know that this case solves all remaining linear constraint as well. Moreover, it reproduces the correct factor, i.e. 3, between the connection  $\Gamma_{AB}{}^C$ and the embedding tensor $X_{AB}{}^C$~\cite{Bosque:2017dfc}.

\clearpage{}
  \clearpage{}\chapter{Additional Solutions of the Linear Constraint}
\label{AppendixE}
\label{app:linconst2additional}

In this part of the appendix, we want to present the remaining solutions of the group manifolds given in table~\ref{fig:sollinconst}~\cite{Bosque:2017dfc}. We start to continue with the case of SL(3)$\times$SL(2). The coordinates are determined through the branching rule~\eqref{eqn2:10branching2}
\begin{equation}
  \mathbf{10} \rightarrow (\mathbf{1}, \mathbf{1}) + (\mathbf{3}, \mathbf{2}) + \xcancel{(\overline{\mathbf 3}, \mathbf{1})}
\end{equation}
after dropping the last term. Counting the dimensions of the surviving irreps, we conclude that the manifold must have seven independent directions. Again, it is possible to choose an appropriate basis for the vector space
\begin{align}
  V_{(\mathbf{1},\mathbf{1})} &= \{12\} &
  V_{(\mathbf{3},\mathbf{2})} &= \{13,\,14,\,15,\,23,\,24,\,25\} &
  V_{(\overline{\mathbf 3}, \mathbf{1})} &= \{34,\,35,\,45\} \\
  V_{\overline{(\mathbf{1},\mathbf{1})}} &= \{345\} &
  V_{\overline{(\mathbf{3},\mathbf{2})}} &= \{245,\,235,\,234,\,145,\,135,\,134\} &
  V_{\overline{(\overline{\mathbf 3}, \mathbf{1})}} &= \{125,\,124,\,123\}
\end{align}
and examine the ramifications on the representations of the embedding tensor
\begin{align}\label{eqn2:branching15}
  \mathbf{15} &\rightarrow (\mathbf{1}, \mathbf{3}) + (\mathbf{3}, \mathbf{2}) + (\mathbf{6},\mathbf{1}) \\
  \mathbf{40} &\rightarrow \xcancel{(\mathbf{1}, \mathbf{2})} + \xcancel{(\overline{\mathbf 3}, \mathbf{1})} + \xcancel{(\mathbf{3},\mathbf{2})} + \xcancel{(\overline{\mathbf 3}, \mathbf{3})} + \xcancel{(\overline{\mathbf 6}, 2)} + (\mathbf{8}, \mathbf{1})\,.\label{eqn2:81survive}
\end{align}
We do not observe any restriction on the irreps originating in the branching of the $\mathbf{15}$ here. However, the second linear constraint~\eqref{eqn2:linconst2} only permits  the $(\mathbf{8},\mathbf{1})$ contribution from the $\mathbf{40}$.
These are exactly the expected gaugings from the gauged supergravity perspective~\cite{Samtleben:2005bp}.
An alternative decomposition of the coordinates takes on the form
\begin{equation}
  \mathbf{10} \rightarrow \xcancel{(\mathbf{1}, \mathbf{1})} + (\mathbf{3}, \mathbf{2}) + (\overline{\mathbf 3}, \mathbf{1})\,.
\end{equation}
It generates a nine-dimensional group manifold. Again, it does not exclude any of the irreps in~\eqref{eqn2:branching15} and the
\begin{equation}\label{eqn2:sl3sl2case2}
  \mathbf{40} \rightarrow (\mathbf{1}, \mathbf{2}) + \xcancel{(\overline{\mathbf 3}, \mathbf{1})} + \xcancel{(\mathbf{3},\mathbf{2})} + \xcancel{(\overline{\mathbf 3}, \mathbf{3})} + \xcancel{(\overline{\mathbf 6}, 2)} + \xcancel{(\mathbf{8}, \mathbf{1})}
\end{equation}
is confined to the $(\mathbf{1},\mathbf{2})$ components. Clearly, one could also consider the branching~\eqref{eqn2:10branching2} by discarding both $(\mathbf{1},\mathbf{1})$ and $(\overline{\mathbf 3}, \mathbf{1})$. It would produce a six-dimensional group manifold. Yet, none of the irreps survive when executing the explicit computation.

Now, we extend this procedure to the T-duality subgroup SL(2)$\times$SL(2). Therefore, we consider the following basis for the vector space
\begin{align}
  V_{(\mathbf{1},\mathbf{1})} &= \{12\} &
  V_{(\mathbf{1},\mathbf{2})} &= \{15,\,25\} &
  V_{(\mathbf{2},\mathbf{2})} &= \{13,\,14,\,23,\,24\} \nonumber \\
  V_{(\mathbf{1}, \mathbf{1})} &= \{34\} &
  V_{(\mathbf{2},\mathbf{1})} &= \{35,\,45\}\\
  V_{\overline{(\mathbf{1},\mathbf{1})}} &= \{345\} &
  V_{\overline{(\mathbf{1},\mathbf{2})}} &= \{234,\,134\} &
  V_{\overline{(\mathbf{2},\mathbf{2})}} &= \{245,\,235,\,145,\,135\} \nonumber \\
  V_{\overline{(\mathbf{1}, \mathbf{1})}} &= \{125\} &
  V_{\overline{(\mathbf{2},\mathbf{1})}} &= \{124,\,123\}
\end{align}
adjusted to the coordinate branching
\begin{equation}
  \mathbf{10} \rightarrow (\mathbf{1}, \mathbf{1}) + (\mathbf{1}, \mathbf{2}) + (\mathbf{2}, \mathbf{2}) + (\mathbf{1}, \mathbf{1}) + (\mathbf{2}, \mathbf{1})\,.
\end{equation}
At this point, we remove the corresponding irreps
\begin{equation}\label{eqn2:SL410}
  \mathbf{10} \rightarrow (\mathbf{1}, \mathbf{1}) + \xcancel{(\mathbf{1}, \mathbf{2})} + (\mathbf{2}, \mathbf{2}) + \xcancel{(\mathbf{1}, \mathbf{1})} + (\mathbf{2}, \mathbf{1})
\end{equation}
from this decomposition and we observe a seven-dimensional group manifold with the following allowed gaugings
\begin{align}
  \mathbf{15} \rightarrow &\;\;\;\;\,(\mathbf{1}, \mathbf{3}) + (\mathbf{1}, \mathbf{2}) + (\mathbf{2}, \mathbf{2}) + (\mathbf{1}, \mathbf{1}) + (\mathbf{2}, \mathbf{1}) + (\mathbf{3}, \mathbf{1})\,, \\
  \mathbf{40} \rightarrow &\;\;\;\;\,(\mathbf{1}, \mathbf{2}) + \xcancel{(\mathbf{1}, \mathbf{2})} + \xcancel{(\mathbf{2}, \mathbf{2})} +  (\mathbf{1}, \mathbf{1}) + (\mathbf{2}, \mathbf{1}) + (\mathbf{1}, \mathbf{3}) + \xcancel{(\mathbf{2}, \mathbf{3})} \nonumber \\ &+ \xcancel{(\mathbf{1}, \mathbf{2})} + \xcancel{(\mathbf{2}, \mathbf{2})} + \xcancel{(\mathbf{3}, \mathbf{2})} + \xcancel{(\mathbf{1}, \mathbf{1})} + \xcancel{(\mathbf{2}, \mathbf{1})} + \xcancel{(\mathbf{2}, \mathbf{1})} + \xcancel{(\mathbf{3}, \mathbf{1})} \,.\label{eqn2:SL2SL2case1}
\end{align}
Right now, it is essential to identify the irreps which have to be canceled. In this context, one should note that the linear constraint for the $\mathbf{40}$ possesses an eight-dimensional solution space. As opposed to the previous case, we are not in the position to identify these eliminated irreps by their dimension alone. However, it is possible to compare the projectors of the linear constraint solutions with the ones for the SL(4) case and observe that they share three independent directions. From the branching
\begin{equation}
\label{eqn2:SL410B}
  \overline{\mathbf{10}} \rightarrow (\mathbf{2},\mathbf{2}) + (\mathbf{3},\mathbf{1}) + ( \mathbf{1},\mathbf{3})
\end{equation}
of SL(4) to SL(2)$\times$SL(2), this induces that it must be one of the two irreps $(\mathbf{3},\mathbf{1})$ or $(\mathbf{1},\mathbf{3})$. Moreover, the solution does not overlap with the ($\mathbf{8},\mathbf{1}$) from~\eqref{eqn2:81survive} which branches according to
\begin{equation}
\label{eqn2:SL3SL281}
  (\mathbf{8}, \mathbf{1}) \rightarrow (\mathbf{1}, \mathbf{1}) + 2 (\mathbf{2}, \mathbf{1}) + (\mathbf{3}, \mathbf{1})\,.
\end{equation}
Hence, $(\mathbf{1},\mathbf{3})$ is the only viable choice. An analogous argumentation holds after taking the $(\mathbf{1},\mathbf{2})$ of the SL(3)$\times$SL(2) case in~\eqref{eqn2:sl3sl2case2} into account. It shares two common directions with the solution of the linear constraint. The branching of SL(3)$\times$SL(2) to SL(2)$\times$SL(2) for this irrep is trivial and yields
\begin{equation}
\label{eqn2:SL3SL221}
(\mathbf{1},\mathbf{2}) \rightarrow (\mathbf{1},\mathbf{2})\,.
\end{equation}
Now, we are left with three remaining, unidentified directions. These can be fixed by their dimension. As a result, we find the branching~\eqref{eqn2:SL2SL2case1}. This gauging has been predicted by gauged supergravity as well~\cite{Samtleben:2005bp}.

Furthermore, we obtain two more very intriguing solutions. None of them lies in one of the previous cases, i.e. SL(3)$\times$SL(2) and SL(4). The first one generates an eight-dimensional group manifolds with the coordinate irreps
\begin{equation}
  \mathbf{10} \rightarrow \xcancel{(\mathbf{1}, \mathbf{1})} + (\mathbf{1}, \mathbf{2}) + (\mathbf{2}, \mathbf{2}) + \xcancel{(\mathbf{1}, \mathbf{1})} + (\mathbf{2}, \mathbf{1})\,.
\end{equation}
In this case, the solution space for the part of the linear constraints possesses four independent directions. These are partially contained in the $(\mathbf{1}, \mathbf{2})$ and $(\mathbf{8},\mathbf{1})$ of SL(3)$\times$SL(2). Both solutions share two directions each. Corresponding to~\eqref{eqn2:SL3SL281} and~\eqref{eqn2:SL3SL221}, we recognize them as the irreps $(\mathbf{1}, \mathbf{2})$ and $(\mathbf{2}, \mathbf{1})$. These are the only allowed irreps. There are no restrictions on the $\mathbf{15}$ part from the linear constraints. Hence, we find
\begin{align}
  \mathbf{15} \rightarrow &\;\;\;\;\,(\mathbf{1}, \mathbf{3}) + (\mathbf{1}, \mathbf{2}) + (\mathbf{2}, \mathbf{2}) + (\mathbf{1}, \mathbf{1}) + (\mathbf{2}, \mathbf{1}) + (\mathbf{3}, \mathbf{1}) \,, \\
  \mathbf{40} \rightarrow &\;\;\;\;\,(\mathbf{1}, \mathbf{2}) + \xcancel{(\mathbf{1}, \mathbf{2})} + \xcancel{(\mathbf{2}, \mathbf{2})} +  \xcancel{(\mathbf{1}, \mathbf{1})} + \xcancel{(\mathbf{2}, \mathbf{1})} + \xcancel{(\mathbf{1}, \mathbf{3})} + \xcancel{(\mathbf{2}, \mathbf{3})} \nonumber \\ &+ \xcancel{(\mathbf{1}, \mathbf{2})} + \xcancel{(\mathbf{2}, \mathbf{2})} + \xcancel{(\mathbf{3}, \mathbf{2})} + \xcancel{(\mathbf{1}, \mathbf{1})} + \xcancel{(\mathbf{2}, \mathbf{1})} + (\mathbf{2}, \mathbf{1}) + \xcancel{(\mathbf{3}, \mathbf{1})} \,.
\end{align}

Finally, there also exist five-dimensional group manifolds with the coordinate irreps
\begin{equation}
  \mathbf{10} \rightarrow (\mathbf{1}, \mathbf{1}) + \xcancel{(\mathbf{1}, \mathbf{2})} + (\mathbf{2}, \mathbf{2}) + \xcancel{(\mathbf{1}, \mathbf{1})} + \xcancel{(\mathbf{2}, \mathbf{1})}\,.
\end{equation}
Here, the solution space of the $\mathbf{40}$ part allows for $11$ independent directions. They are partially contained\footnote{ There are four directions in the $(\mathbf{8},\mathbf{1})$ of SL(3)$\times$SL(2), but only one of them is not contained in the $\overline{\mathbf{10}}$ of SL(4).} in the $(\mathbf{8},\mathbf{1})$ of SL(3)$\times$SL(2) and lie entirely in the $\overline{\mathbf{10}}$ of SL(4). As a consequence, we only observe a new $(\mathbf{1},\mathbf{1})$ from~\eqref{eqn2:SL3SL281} and the right hand side of~\eqref{eqn2:SL410B}. Contrary to the previous cases, only ten directions of the linear constraint's $\mathbf{15}$ part can be turned on. The solution of the $\mathbf{15}$ lives completely in the $\mathbf{10}$ of SL(4). Taking the branching rule~\eqref{eqn2:SL410} into account, we obtain
\begin{align}
  \mathbf{15} \rightarrow &\;\;\;\;\,(\mathbf{1}, \mathbf{3}) + \xcancel{(\mathbf{1}, \mathbf{2})} + (\mathbf{2}, \mathbf{2}) + \xcancel{(\mathbf{1}, \mathbf{1})} + \xcancel{(\mathbf{2}, \mathbf{1})} + (\mathbf{3}, \mathbf{1}) \,, \\
  \mathbf{40} \rightarrow &\;\;\;\;\,\xcancel{(\mathbf{1}, \mathbf{2})} + \xcancel{(\mathbf{1}, \mathbf{2})} + (\mathbf{2}, \mathbf{2}) +  \xcancel{(\mathbf{1}, \mathbf{1})} + \xcancel{(\mathbf{2}, \mathbf{1})} + (\mathbf{1}, \mathbf{3}) + \xcancel{(\mathbf{2}, \mathbf{3})} \nonumber \\ &+ \xcancel{(\mathbf{1}, \mathbf{2})} + \xcancel{(\mathbf{2}, \mathbf{2})} + \xcancel{(\mathbf{3}, \mathbf{2})} + (\mathbf{1}, \mathbf{1}) + \xcancel{(\mathbf{2}, \mathbf{1})} + \xcancel{(\mathbf{2}, \mathbf{1})} + (\mathbf{3}, \mathbf{1}) \,.
\end{align}

All other solutions of the linear constraints lie completely in one of the previously discussed cases, i.e. SL(4) or SL(3)$\times$SL(2)~\cite{Bosque:2017dfc}.\clearpage{}
  \clearpage{}\chapter{Faithful Representations and Identifications}\label{app:faithfulrepr}
\label{AppendixF}

Let us first consider the Lie algebra of CSO(1,0,4) which can be expressed through the non-vanishing commutator algebra~\cite{Bosque:2017dfc}
\begin{equation}
\label{eqn2:appendixcliealgebra}
[t_{\hat\alpha}, t_{\hat\beta}] = \mathbf{g} \, t_{\hat\alpha\hat\beta}\,,
\end{equation}
where we identified the generators
\begin{equation}
  t_A = \Big( t_1,\, t_2,\, t_3,\, t_4,\,
      t_{\tilde 1},\, t_{\tilde 2},\, t_{\tilde 3},\, t_{\tilde 4},\, t_{\tilde 5},\, t_{\tilde 6} \Big)\,.
\end{equation}
The given algebra is essential for the first duality chain~\eqref{eqn2:chain1} in subsection~\ref{sec:T4Gflux} and possesses the lower central series
\begin{equation}
  L_0= \{ t_1,\, t_2,\, t_3,\, t_4,\, t_{\tilde 1},\, t_{\tilde 2},\, t_{\tilde 3},\, t_{\tilde 4},\, t_{\tilde 5},\, t_{\tilde 6}\} \supset \{ t_{\tilde 1},\, t_{\tilde 2},\, t_{\tilde 3},\, t_{\tilde 4},\, t_{\tilde 5},\, t_{\tilde 6} \} \supset \{0\}\,.
\end{equation}
Performing the technique sketched in chapter \ref{Kap_4}~\cite{Bosque:2015jda}, we construct the $N=21$-dimensional subspace
\begin{align}
  V^2 = \{&t_1^2,\, t_1 t_2,\, t_1 t_3,\, t_1 t_4,\, t_2^2,\, t_2 t_3,\, t_2 t_4,\, t_3^2,\, t_3 t_4,\, t_4^2,\, t_{\tilde 1},\, t_{\tilde 2},\, t_{\tilde 3},\, t_{\tilde 4},\, t_{\tilde 5},\, t_{\tilde 6},  &&\ord \cdot = 2 \nonumber \\
    &t_1,\, t_2,\, t_3,\, t_4, &&\ord \cdot = 1 \nonumber \\
    &1 \} &&\ord \cdot = 0
\end{align}
of the universal enveloping algebra. The center of this algebra takes on the form \\ $\{$ $t_{\tilde 1}$, $t_{\tilde 2}$, $t_{\tilde 3}$, $t_{\tilde 4}$, $t_{\tilde 5}$, $t_{\tilde 6}\}$. These six generators produce an abelian subalgebra $\mathfrak{h}$. Using this information, we are in a position to derive the matrix representation for the generators $t_A$ by expanding the linear maps $\phi_{t_A}$ in the basis $V^2$. Finally, the exponential maps~\eqref{eqn2:matrixexpm} and~\eqref{eqn2:matrixexph} yield group elements
\begin{equation}
g = m h = \left(
\begin{array}{*{21}c} 
 1 & 0 & 0 & 0 & 0 & 0 & 0 & 0 & 0 & 0 & 0 & 0 & 0 & 0 & 0 & 0 & 0 & 0 & 0 & 0 & x^1 \\
 0 & 1 & 0 & 0 & 0 & 0 & 0 & 0 & 0 & 0 & 0 & 0 & 0 & 0 & 0 & 0 & 0 & 0 & 0 & 0 & x^2 \\
 0 & 0 & 1 & 0 & 0 & 0 & 0 & 0 & 0 & 0 & 0 & 0 & 0 & 0 & 0 & 0 & 0 & 0 & 0 & 0 & x^3 \\
 0 & 0 & 0 & 1 & 0 & 0 & 0 & 0 & 0 & 0 & 0 & 0 & 0 & 0 & 0 & 0 & 0 & 0 & 0 & 0 & x^4 \\
 x^1 & 0 & 0 & 0 & 1 & 0 & 0 & 0 & 0 & 0 & 0 & 0 & 0 & 0 & 0 & 0 & 0 & 0 & 0 & 0 & (x^1)^2/2 \\
 x^2 & x^1 & 0 & 0 & 0 & 1 & 0 & 0 & 0 & 0 & 0 & 0 & 0 & 0 & 0 & 0 & 0 & 0 & 0 & 0 & x^1 x^2 \\
 x^3 & 0 & x^1 & 0 & 0 & 0 & 1 & 0 & 0 & 0 & 0 & 0 & 0 & 0 & 0 & 0 & 0 & 0 & 0 & 0 & x^1 x^3 \\
 x^4 & 0 & 0 & x^1 & 0 & 0 & 0 & 1 & 0 & 0 & 0 & 0 & 0 & 0 & 0 & 0 & 0 & 0 & 0 & 0 & x^1 x^4 \\
 0 & x^2 & 0 & 0 & 0 & 0 & 0 & 0 & 1 & 0 & 0 & 0 & 0 & 0 & 0 & 0 & 0 & 0 & 0 & 0 & (x^2)^2/2 \\
 0 & x^3 & x^2 & 0 & 0 & 0 & 0 & 0 & 0 & 1 & 0 & 0 & 0 & 0 & 0 & 0 & 0 & 0 & 0 & 0 & x^2 x^3 \\
 0 & x^4 & 0 & x^2 & 0 & 0 & 0 & 0 & 0 & 0 & 1 & 0 & 0 & 0 & 0 & 0 & 0 & 0 & 0 & 0 & x^2 x^4 \\
 0 & 0 & x^3 & 0 & 0 & 0 & 0 & 0 & 0 & 0 & 0 & 1 & 0 & 0 & 0 & 0 & 0 & 0 & 0 & 0 & (x^3)^2 \\
0 & 0 & x^4 & x^3 & 0 & 0 & 0 & 0 & 0 & 0 & 0 & 0 & 1 & 0 & 0 & 0 & 0 & 0 & 0 & 0 & x^3 x^4 \\
0 & 0 & 0 & x^4 & 0 & 0 & 0 & 0 & 0 & 0 & 0 & 0 & 0 & 1 & 0 & 0 & 0 & 0 & 0 & 0 & (x^4)^2/2 \\
- \mathbf{g} \, x^2 & 0 & 0 & 0 & 0 & 0 & 0 & 0 & 0 & 0 & 0 & 0 & 0 & 0 & 1 & 0 & 0 & 0 & 0 & 0 & x^{\tilde 1} \\
- \mathbf{g} \, x^3 & 0 & 0 & 0 & 0 & 0 & 0 & 0 & 0 & 0 & 0 & 0 & 0 & 0 & 0 & 1 & 0 & 0 & 0 & 0 & x^{\tilde 2} \\
- \mathbf{g} \, x^4 & 0 & 0 & 0 & 0 & 0 & 0 & 0 & 0 & 0 & 0 & 0 & 0 & 0 & 0 & 0 & 1 & 0 & 0 & 0 & x^{\tilde 3} \\
0 & - \mathbf{g} \, x^3 & 0 & 0 & 0 & 0 & 0 & 0 & 0 & 0 & 0 & 0 & 0 & 0 & 0 & 0 & 0 & 1 & 0 & 0 & x^{\tilde 4} \\
0 & - \mathbf{g} \, x^4 & 0 & 0 & 0 & 0 & 0 & 0 & 0 & 0 & 0 & 0 & 0 & 0 & 0 & 0 & 0 & 0 & 1 & 0 & x^{\tilde 5} \\
0 & 0 & - \mathbf{g} \, x^4 & 0 & 0 & 0 & 0 & 0 & 0 & 0 & 0 & 0 & 0 & 0 & 0 & 0 & 0 & 0 & 0 & 1 & x^{\tilde 6} \\
0 & 0 & 0 & 0 & 0 & 0 & 0 & 0 & 0 & 0 & 0 & 0 & 0 & 0 & 0 & 0 & 0 & 0 & 0 & 0 & 1 \\
\end{array}
\right)
\end{equation}
with $\mathbf{g}$ denoting the number of $G$-flux carried by the background. It is very inefficient to work with such large matrices. Therefore, we represent $g$ instead by the ten tuple
$($ $x^1$, $x^2$, $x^3$, $x^4$, $x^{\tilde 1}$, $x^{\tilde 2}$, $x^{\tilde 3}$, $x^{\tilde 4}$, $x^{\tilde 5}$, $x^{\tilde 6}$ $)$. Subsequently, the group multiplication is given by
\begin{align}\label{eqn2:groupmult}
  (x^1\,,x^2\,,x^3\,, x^4\,,x^{\tilde 1}\,,x^{\tilde 2}\,, x^{\tilde 3}\,, x^{\tilde 4}\,, &x^{\tilde 5}\,, 
    x^{\tilde 6}) (y^1\,,y^2\,,y^3\,, y^4\,,y^{\tilde 1}\,,y^{\tilde 2}\,, y^{\tilde 3}\,, y^{\tilde 4}\,,
    y^{\tilde 5}\,, y^{\tilde 6}) = \\ 
  (x^1+y^1\,,x^2+y^2\,,x^3+y^3\,,x^4+y^4\,,&- \mathbf{g} \,  x^2 y^1 + x^{\tilde 1} + y^{\tilde 1}\,, 
    - \mathbf{g} \, x^3 y^1 + x^{\tilde 2} + y^{\tilde 2} \nonumber \\
  - \mathbf{g} \, x^3 y^2 + x^{\tilde 3} + y^{\tilde 3}\,, - \mathbf{g} \,  &x^4 y^1 + x^{\tilde 4} + y^{\tilde 4}\,,
    - \mathbf{g} \, x^4 y^2 + x^{\tilde 5} + y^{\tilde 5}\,,- \mathbf{g} \,  x^4 y^3 + x^{\tilde 6} + y^{\tilde 6} )\,. \nonumber 
\end{align}
We now want to verify that this indeed generates a group. The identity element can be identified by $e$=$(0$, $0$, $0$, $0$, $0$, $0$, $0$, $0$, $0$, $0)$ and satisfies
\begin{equation}
  g e = e g = g\,.
\end{equation}
Moreover, there exists the inverse element
\clearpage
\begin{align}
  g^{-1}=(-x^1\,,-x^2\,,-x^3\,,-x^4\,, &- \mathbf{g} \, x^1 x^2 -  x^{\tilde 1}\,, - \mathbf{g} \, x^1 x^3 - x^{\tilde 2}\,, \\ 
 &- \mathbf{g} \, x^2 x^3 - x^{\tilde 3}\,, - \mathbf{g} \, x^1 x^4 - x^{\tilde 4}\,, - \mathbf{g} \,  x^2 x^4 -  x^{\tilde 5}\,, - \mathbf{g} \, x^3 x^4 - x^{\tilde 6}) \nonumber
\end{align}
fulfilling
\begin{equation}
  g^{-1} g = g g^{-1} = e\,.
\end{equation}
Since $\mathbf{g}$ is an integer, the group multiplication~\eqref{eqn2:groupmult} does not only close over the real numbers, but for $x^i$ and $x^{\tilde i}$ being integers as well. Hence, CSO(1,0,4,$\mathbb{Z}$) is a subgroup of CSO(1,0,4) and we need to mod it out by considering the right coset CSO(1,0,4,$\mathbb{Z}$)\textbackslash CSO(1,0,4). It results in the equivalence relation
\begin{equation}
  g_1 \sim g_2 \quad \text{if and only if} \quad g_1 = k g_2 \quad \text{with} \quad
  g_1\,, g_2 \in \mathrm{CSO}(1,0,4) \quad \text{and} \quad
  k \in \mathrm{CSO}(1,0,4,\mathbb{Z})\,.
\end{equation}
After substituting $k=(n^1\,,n^2\,,n^3\,,n^4\,,n^{\tilde 1}\,,n^{\tilde 2}\,,n^{\tilde 3}\,,n^{\tilde 4}\,,n^{\tilde 5}\,,n^{\tilde 6})$ with $n^i$, $n^{\tilde i} \in\mathbb{Z}$, we find the following identifications
\begin{align}
  (x^1\,,x^2\,,x^3\,,x^4\,,& x^{\tilde 1}\,,x^{\tilde 2}\,, x^{\tilde 3}\,, x^{\tilde 4}\,, 
    x^{\tilde 5}\,, x^{\tilde 6}) \sim \\ 
  (x^1+n^1\,,x^2+n^2\,,x^3+n^3\,,x^4+n^4\,,&- \mathbf{g} \, x^1 n^2 + x^{\tilde 1} + n^{\tilde 1}\,, 
    - \mathbf{g} \, x^1 n^3 + x^{\tilde 2} + n^{\tilde 2} \nonumber \\
  - \mathbf{g} \, x^2 n^3 + x^{\tilde 3} + n^{\tilde 3}\,,  - \mathbf{g} \, & x^1 n^4 + x^{\tilde 4} + 
    n^{\tilde 4}\,, - \mathbf{g} \, x^2 n^4 + x^{\tilde 5} + n^{\tilde 5}\,,- \mathbf{g} \,
    x^3 n^4 + x^{\tilde 6} + n^{\tilde 6} ) \nonumber
\end{align}
from~\eqref{eqn2:groupmult}. Particularly, we observe
\begin{align}
\label{eqn2:coordident1}
  (x^1\,,x^2\,,x^3\,, x^4\,,& x^{\tilde 1}\,,x^{\tilde 2}\,, x^{\tilde 3}\,, x^{\tilde 4}\,, x^{\tilde 5}\,, x^{\tilde 6})  \sim (x^1 + 1\,,x^2\,,x^3\,, x^4\,,x^{\tilde 1}\,,x^{\tilde 2}\,, x^{\tilde 3}\,, x^{\tilde 4}\,, x^{\tilde 5}\,, x^{\tilde 6}) \nonumber \\
  & \sim (x^1\,,x^2 + 1\,,x^3\,, x^4\,,x^{\tilde 1} - \mathbf{g} \, x^1 \,,x^{\tilde 2}\,, x^{\tilde 3}\,, x^{\tilde 4}\,, x^{\tilde 5}\,, x^{\tilde 6})  \\
  & \sim (x^1\,,x^2\,,x^3 + 1\,, x^4\,,x^{\tilde 1}\,,x^{\tilde 2} - \mathbf{g} \, x^1\,, x^{\tilde 3} - \mathbf{g} \, x^2\,, x^{\tilde 4}\,, x^{\tilde 5}\,, x^{\tilde 6})\nonumber  \\
  & \sim (x^1\,,x^2\,,x^3\,, x^4 + 1\,,x^{\tilde 1}\,,x^{\tilde 2}\,, x^{\tilde 3}\,, x^{\tilde 4} - \mathbf{g} \, x^1\,, x^{\tilde 5} - \mathbf{g} \, x^2 \,, x^{\tilde 6} - \mathbf{g} \, x^3) \nonumber  
\end{align}
for the physical coordinates and
\begin{align}
\label{eqn2:coordident2}
  (x^1\,,x^2\,,x^3\,, x^4\,,x^{\tilde 1}\,,x^{\tilde 2}\,, x^{\tilde 3}\,, x^{\tilde 4}\,, x^{\tilde 5}\,, x^{\tilde 6})  & \sim (x^1\,,x^2\,,x^3\,, x^4\,,x^{\tilde 1} + 1\,,x^{\tilde 2}\,, x^{\tilde 3}\,, x^{\tilde 4}\,, x^{\tilde 5}\,, x^{\tilde 6}) \nonumber   \\
  & \sim (x^1\,,x^2\,,x^3\,, x^4\,,x^{\tilde 1}\,,x^{\tilde 2} + 1\,, x^{\tilde 3}\,, x^{\tilde 4}\,, x^{\tilde 5}\,, x^{\tilde 6})  \nonumber \\
  & \sim (x^1\,,x^2\,,x^3\,, x^4\,,x^{\tilde 1}\,,x^{\tilde 2}\,, x^{\tilde 3} + 1\,, x^{\tilde 4}\,, x^{\tilde 5}\,, x^{\tilde 6})  \nonumber \\
  & \sim (x^1\,,x^2\,,x^3\,, x^4\,,x^{\tilde 1}\,,x^{\tilde 2}\,, x^{\tilde 3}\,, x^{\tilde 4} + 1\,, x^{\tilde 5}\,, x^{\tilde 6})  \nonumber \\
  & \sim (x^1\,,x^2\,,x^3\,, x^4\,,x^{\tilde 1}\,,x^{\tilde 2}\,, x^{\tilde 3}\,, x^{\tilde 4}\,, x^{\tilde 5} + 1\,, x^{\tilde 6})  \nonumber \\
  & \sim (x^1\,,x^2\,,x^3\,, x^4\,,x^{\tilde 1}\,,x^{\tilde 2}\,, x^{\tilde 3}\,, x^{\tilde 4}\,, x^{\tilde 5}\,, x^{\tilde 6} + 1)  
\end{align}
for the remaining ones. Finally, taking these identifications  into account, the left invariant Maurer-Cartan form
\begin{equation}
\label{eqn2:appendixvielbein}
E^A{}_I = \begin{pmatrix}
1 & 0 & 0 & 0 & 0 & 0 & 0 & 0 & 0 & 0 \\
0 & 1 & 0 & 0 & 0 & 0 & 0 & 0 & 0 & 0 \\
0 & 0 & 1 & 0 & 0 & 0 & 0 & 0 & 0 & 0 \\
0 & 0 & 0 & 1 & 0 & 0 & 0 & 0 & 0 & 0 \\
\mathbf{g} \, x_2 & 0 & 0 & 0 & 1 & 0 & 0 & 0 & 0 & 0 \\
\mathbf{g} \, x_3 & 0 & 0 & 0 & 0 & 1 & 0 & 0 & 0 & 0 \\
0 & \mathbf{g} \, x_3 & 0 & 0 & 0 & 0 & 1 & 0 & 0 & 0 \\
\mathbf{g} \, x_4 & 0 & 0 & 0 & 0 & 0 & 0 & 1 & 0 & 0 \\
0 & \mathbf{g} \, x_4 & 0 & 0 & 0 & 0 & 0 & 0 & 1 & 0 \\
0 & 0 & \mathbf{g} \, x_4 & 0 & 0 & 0 & 0 & 0 & 0 & 1
\end{pmatrix}\,,
\end{equation}
is globally well defined~\cite{Bosque:2017dfc} and specifically
\begin{align}
 E_1 &= d x^1  \\
 E_2 &= d x^2 \nonumber \\
 E_3 &= d x^3 \nonumber \\
 E_4 &= d x^4 \nonumber \\
  E^1 &= d x^{\tilde 1} + \mathbf{g} \, x^2 d x^1 = d (x^{\tilde 1}- \mathbf{g} \, x^1) + (x^2 + 1) \, \mathbf{g} \,  d x^1 \nonumber \\
  E^2 &= d x^{\tilde 2} + \mathbf{g} \, x^3 d x^1 = d (x^{\tilde 2} - \mathbf{g} \, x^1) + (x^3 + 1) \, \mathbf{g} \, d x^1 \nonumber \\
  E^3 &= d x^{\tilde 3} + \mathbf{g} \, x^3 d x^2 = d (x^{\tilde 3} - \mathbf{g} \, x^2) + (x^3 + 1) \, \mathbf{g} \, d x^2 \nonumber \\
  E^4 &= d x^{\tilde 4} + \mathbf{g} \, x^4 d x^1 = d (x^{\tilde 4} - \mathbf{g} \, x^1) + (x^4 + 1) \, \mathbf{g} \, d x^1 \nonumber \\
  E^5 &= d x^{\tilde 5} + \mathbf{g} \, x^4 d x^2 = d (x^{\tilde 5} - \mathbf{g} \, x^2) + (x^4 + 1) \, \mathbf{g} \, d x^2 \nonumber \\
  E^6 &= d x^{\tilde 6} + \mathbf{g} \, x^4 d x^3 = d (x^{\tilde 6} - \mathbf{g} \, x^3) + (x^4 + 1) \, \mathbf{g} \, d x^3\,. \nonumber 
\end{align}
\newpage
For the second duality chain~\eqref{eqn2:chain2}, we need to consider the nine-dimensional Lie algebra $\mathfrak{g}$ in~\eqref{eqn2:defgexample40}~\cite{Bosque:2017dfc}. Therefore, we execute the exponential maps~\eqref{eqn2:matrixexpm} as well as~\eqref{eqn2:matrixexph}, to derive the group element
\begin{equation}
g = m h = \left(
\begin{array}{cccccccccccccccc}
 1 & 0 & 0 & 0 & 0 & 0 & 0 & 0 & 0 & 0 & 0 & 0 & 0 & 0 & 0 & x^{\tilde 5} \\
 0 & 1 & 0 & 0 & 0 & 0 & 0 & 0 & 0 & 0 & 0 & 0 & 0 & 0 & 0 & x^3 \\
 0 & 0 & 1 & 0 & 0 & 0 & 0 & 0 & 0 & 0 & 0 & 0 & 0 & 0 & 0 & x^4 \\
 0 & 0 & 0 & 1 & 0 & 0 & 0 & 0 & 0 & 0 & 0 & 0 & 0 & 0 & 0 & x^1 \\
 0 & 0 & 0 & 0 & 1 & 0 & 0 & 0 & 0 & 0 & 0 & 0 & 0 & 0 & 0 & x^{\tilde 1} \\
 0 & 0 & 0 & 0 & 0 & 1 & 0 & 0 & 0 & 0 & 0 & 0 & 0 & 0 & 0 & x^{\tilde 3} \\
 x^{\tilde 5} & 0 & 0 & 0 & 0 & 0 & 1 & 0 & 0 & 0 & 0 & 0 & 0 & 0 & 0 & ({x^{\tilde 5}})^2 /2 \\
 x^3 & {x^{\tilde 5}} & 0 & 0 & 0 & 0 & 0 & 1 & 0 & 0 & 0 & 0 & 0 & 0 & 0 & x^3 {x^{\tilde 5}} \\
 x^4 & 0 & {x^{\tilde 5}} & 0 & 0 & 0 & 0 & 0 & 1 & 0 & 0 & 0 & 0 & 0 & 0 & x^4 {x^{\tilde 5}} \\
 0 & x^3 & 0 & 0 & 0 & 0 & 0 & 0 & 0 & 1 & 0 & 0 & 0 & 0 & 0 & (x^3)^2 / 2 \\
 0 & x^4 & x^3 & 0 & 0 & 0 & 0 & 0 & 0 & 0 & 1 & 0 & 0 & 0 & 0 & x^3 x^4 \\
 0 & 0 & x^4 & 0 & 0 & 0 & 0 & 0 & 0 & 0 & 0 & 1 & 0 & 0 & 0 & (x^4)^2 / 2 \\
 0 & -\mathbf{f} x^4 & 0 & 0 & 0 & 0 & 0 & 0 & 0 & 0 & 0 & 0 & 1 & 0 & 0 & x^2 \\
 -\mathbf{f} x^4 & 0 & 0 & 0 & 0 & 0 & 0 & 0 & 0 & 0 & 0 & 0 & 0 & 1 & 0 &
   x^{\tilde 4} -\mathbf{f} x^4 {x^{\tilde 5}} \\
 -\mathbf{f} x^3 & 0 & 0 & 0 & 0 & 0 & 0 & 0 & 0 & 0 & 0 & 0 & 0 & 0 & 1 &
   x^{\tilde 2}-\mathbf{f} x^3 {x^{\tilde 5}} \\
 0 & 0 & 0 & 0 & 0 & 0 & 0 & 0 & 0 & 0 & 0 & 0 & 0 & 0 & 0 & 1 \\
\end{array}
\right)
\end{equation}
with $\mathbf{f}$ representing the number of $F$-flux carried by the background. Again, we represent $g$ through the nine tuple $($ $x^1$, $x^2$, $x^3$, $x^4$, $x^{\tilde 1}$, $x^{\tilde 2}$, $x^{\tilde 3}$, $x^{\tilde 4}$, $x^{\tilde 5}$ $)$ instead of handling these huge matrices. Then, the group multiplication is performed by
\begin{align}
\label{eqn2:groupmult2}
  (x^1\,,x^2\,,x^3\,, x^4\,,x^{\tilde 1}\,,x^{\tilde 2}\,, x^{\tilde 3}\,, x^{\tilde 4}\,, x^{\tilde 5}) (&y^1\,,y^2\,,y^3\,, y^4\,,y^{\tilde 1}\,,y^{\tilde 2}\,, y^{\tilde 3}\,, y^{\tilde 4}\,, y^{\tilde 5}) = \\ 
  (x^1+y^1\,,- \mathbf{f} \, x^4 y^3 + x^2+y^2 \,,x^3+y^3\,,x^4+y^4\,,& \, x^{\tilde 1} + y^{\tilde 1}\,, \mathbf{f} \, x^{\tilde 5} y^3 + x^{\tilde 2} + y^{\tilde 2} \nonumber \\ x^{\tilde 3} + y^{\tilde 3}\,, & \, \mathbf{f} \,  x^{\tilde 5} y^4 + x^{\tilde 4} + y^{\tilde 4}\,, x^{\tilde 5} + y^{\tilde 5} )\,. \nonumber 
\end{align}
Next, we verify that $g$ forms a group. First, consider the identity element \\ $e$=$(0$, $0$, $0$, $0$, $0$, $0$, $0$, $0$, $0)$ which fulfills
\begin{equation}
  g e = e g = g\,.
\end{equation}
Furthermore, the inverse element can be obtained through
\begin{align}
  g^{-1}=(-x^1\,,- \mathbf{f} \, x^3 x^4 -x^2\,,-x^3\,,-x^4\,, &-  x^{\tilde 1}\,, \mathbf{f} \, x^3 x^{\tilde 5} - x^{\tilde 2}\,, - x^{\tilde 3}\,, \mathbf{f} \, x^4 x^{\tilde 5} - x^{\tilde 4}\,, -  x^{\tilde 5}) 
\end{align}
satisfying
\begin{equation}
  g^{-1} g = g g^{-1} = e\,.
\end{equation}
In this case, the group multiplication~\eqref{eqn2:groupmult2} does not only close over the real numbers either, but for $x^i$ and $x^{\tilde i}$ being integers as well, as $\mathbf{f}$ is an integer. Consequently, we need to mod out the discrete subgroup $G_{\mathbb{Z}}$ formed by restricting all coordinates to integers from the left. As a result, we obtain the equivalence relation
\begin{equation}
  g_1 \sim g_2 \quad \text{if and only if} \quad g_1 = k g_2 \quad \text{with} \quad
  g_1\,, g_2 \in G \quad \text{and} \quad
  k \in G_{\mathbb{Z}}\,.
\end{equation}
Finally, by substituting $k=(n^1\,,n^2\,,n^3\,,n^4\,,n^{\tilde 1}\,,n^{\tilde 2}\,,n^{\tilde 3}\,,n^{\tilde 4}\,,n^{\tilde 5})$ with $n^i$, $n^{\tilde i} \in\mathbb{Z}$ gives rise to the identifications
\begin{align}
  (x^1\,,x^2\,,x^3\,, x^4\,,x^{\tilde 1}\,,x^{\tilde 2}\,, x^{\tilde 3}\,, x^{\tilde 4}\,, &x^{\tilde 5}) \sim \\ 
  (x^1+n^1\,,-\mathbf{f} \, x^3 n^4 + x^2+n^2\,,x^3+n^3\,,x^4+n^4\,,&x^{\tilde 1} + n^{\tilde 1}\,, \mathbf{f} \, x^3 n^{\tilde 5} + x^{\tilde 2} + n^{\tilde 2} \nonumber \\ x^{\tilde 3} + n^{\tilde 3}\,,  &\mathbf{f} \, x^4 n^{\tilde 5} + x^{\tilde 4} + n^{\tilde 4}\,, x^{\tilde 5} + n^{\tilde 5}) \nonumber
\end{align}
from~\eqref{eqn2:groupmult2}. Specifically, for the physical coordinates
\begin{align}
\label{eqn2:coordident3}
  (x^1\,,x^2\,,x^3\,, x^4\,, x^{\tilde 1}\,,x^{\tilde 2}\,, x^{\tilde 3}\,, x^{\tilde 4}\,, x^{\tilde 5})  &\sim (x^1 + 1\,,x^2\,,x^3\,, x^4\,,x^{\tilde 1}\,,x^{\tilde 2}\,, x^{\tilde 3}\,, x^{\tilde 4}\,, x^{\tilde 5})  \\
  & \sim (x^1\,,x^2 + 1\,,x^3\,, x^4\,,x^{\tilde 1}\,,x^{\tilde 2}\,, x^{\tilde 3}\,, x^{\tilde 4}\,, x^{\tilde 5}) \nonumber  \\
  & \sim (x^1\,,x^2\,,x^3 + 1\,, x^4\,,x^{\tilde 1}\,,x^{\tilde 2}\,, x^{\tilde 3}\,, x^{\tilde 4}\,, x^{\tilde 5})\nonumber  \\
  & \sim (x^1\,,x^2 - \mathbf{f} x^3 \,,x^3\,, x^4 + 1\,,x^{\tilde 1}\,,x^{\tilde 2}\,, x^{\tilde 3}\,, x^{\tilde 4}\,, x^{\tilde 5})  \nonumber
\end{align}
and for the remaining coordinates
\begin{align}\label{eqn2:coordident4}
  (x^1\,,x^2\,,x^3\,, x^4\,,x^{\tilde 1}\,,x^{\tilde 2}\,, x^{\tilde 3}\,, x^{\tilde 4}\,, x^{\tilde 5})  & \sim (x^1\,,x^2\,,x^3\,, x^4\,,x^{\tilde 1} + 1\,,x^{\tilde 2}\,, x^{\tilde 3}\,, x^{\tilde 4}\,, x^{\tilde 5}) \\
  & \sim (x^1\,,x^2\,,x^3\,, x^4\,,x^{\tilde 1}\,,x^{\tilde 2} + 1\,, x^{\tilde 3}\,, x^{\tilde 4}\,, x^{\tilde 5})  \nonumber \\
  & \sim (x^1\,,x^2\,,x^3\,, x^4\,,x^{\tilde 1}\,,x^{\tilde 2}\,, x^{\tilde 3} + 1\,, x^{\tilde 4}\,, x^{\tilde 5})  \nonumber \\
  & \sim (x^1\,,x^2\,,x^3\,, x^4\,,x^{\tilde 1}\,,x^{\tilde 2}\,, x^{\tilde 3}\,, x^{\tilde 4} + 1\,, x^{\tilde 5})  \nonumber \\
  & \sim (x^1\,,x^2\,,x^3\,, x^4\,,x^{\tilde 1}\,,x^{\tilde 2} + \mathbf{f} \, x^3 \,, x^{\tilde 3}\,, x^{\tilde 4} + \mathbf{f} \, x^4\,, x^{\tilde 5} + 1) \nonumber \,. 
\end{align}
After taking these identifications into account, we derive the left invariant Maurer-Cartan form
\begin{equation}
\label{eqn2:appendixvielbein2}
E^A{}_I = \begin{pmatrix}
1 & 0 & 0 & 0 & 0 & 0 & 0 & 0 & 0 \\
0 & 1 & \mathbf{f} \, x^4 & 0 & 0 & 0 & 0 & 0 & 0 \\
0 & 0 & 1 & 0 & 0 & 0 & 0 & 0 & 0 \\
0 & 0 & 0 & 1 & 0 & 0 & 0 & 0 & 0  \\
0 & 0 & 0 & 0 & 1 & 0 & 0 & 0 & 0 \\
0 & 0 & - \mathbf{f} \, x^{\tilde 5} & 0 & 0 & 1 & 0 & 0 & 0 \\
0 & 0 & 0 & 0 & 0 & 0 & 1 & 0 & 0 \\
0 & 0 & 0 & - \mathbf{f} \, x^{\tilde 5} & 0 & 0 & 0 & 1 & 0 \\
0 & 0 & 0 & 0 & 0 & 0 & 0 & 0 & 1
\end{pmatrix}\,.
\end{equation}
Under consideration of the identifications~\eqref{eqn2:coordident3} and~\eqref{eqn2:coordident4}, it is straightforward to verify that $E^A{}_I$ is globally well defined~\cite{Bosque:2017dfc}, i.e.
\begin{align}
 E_1 &= d x^1 \\
 E_2 &= d x^2 + \mathbf{f} \, x^4 dx^3 = d( x^2 - \mathbf{f} \, x^3 ) + ( x^4 + 1 ) \, \mathbf{f} \, dx^3 \nonumber \\
 E_3 &= d x^3 \nonumber \\
 E_4 &= d x^4 \nonumber \\
  E^1 &= d x^{\tilde 1} \nonumber \\
  E^2 &= d x^{\tilde 2} - \mathbf{f} \, x^{\tilde 5} dx^3 = d( x^{\tilde 2} + \mathbf{f} \, x^3 ) - ( x^{\tilde 5} + 1 ) \, \mathbf{f} \, dx^3 \nonumber \\
  E^3 &= d x^{\tilde 3} \nonumber \\
  E^4 &= d x^{\tilde 4} - \mathbf{f} \, x^{\tilde 5} dx^4 = d( x^{\tilde 4} + \mathbf{f} \, x^4 ) - ( x^{\tilde 5} + 1 ) \, \mathbf{f} \, dx^4 \nonumber \\
  E^5 &= d x^{\tilde 5}\,. \nonumber 
\end{align}

\clearpage{}  
  \end{appendices}

  \bibliographystyle{JHEP}
  \bibliography{literatur1}

  \cleardoubleplainpage
  
\end{document}